\documentclass[11pt]{article}
\usepackage{graphicx}
\usepackage{enumitem,booktabs,needspace,array,longtable}
\usepackage{tikz}
\usetikzlibrary{arrows.meta}
\usetikzlibrary{calc}
\usepackage[T1]{fontenc}
\usepackage{tcolorbox}
\tcbuselibrary{breakable,skins}
\newtcolorbox[auto counter,number within=section]{constructionbox}[2]{%
    enhanced,breakable,sharp corners,lines before break=40,
    colback=navy!2!white,colframe=navy!65!white,
    colbacktitle=navy!10!white,coltitle=navy,
    boxrule=0.5pt,left=9pt,right=9pt,top=7pt,bottom=7pt,
    fonttitle=\bfseries,fontupper=\small,
    title={Construction \thetcbcounter\enspace #1},label={#2},
    title after break={Construction \thetcbcounter\enspace (continued)},
    before skip=11pt,after skip=11pt,
    before upper={\setlength{\parindent}{0pt}\setlength{\parskip}{4pt}}}
\newcommand{\cbingredients}{\par\noindent\textbf{Ingredients from earlier parts.}\ }
\newcommand{\cbsetup}{\par\smallskip\noindent\textbf{Public parameters and fixed setup.}\ }
\newcommand{\cbinput}{\par\smallskip\noindent\textbf{Input.}\ }
\newcommand{\cbsteps}{\par\smallskip\noindent\textbf{Evaluation steps.}}
\newcommand{\cboutput}{\par\smallskip\noindent\textbf{Output.}\ }
\newcommand{\cbproofonly}{\par\smallskip\noindent\textbf{Not evaluated.}\ }
\newlist{constructionsteps}{enumerate}{1}
\setlist[constructionsteps]{label=\textbf{\arabic*.},leftmargin=1.7em,
  itemsep=4pt,topsep=4pt,parsep=0pt}

\usepackage{lmodern}
\usepackage[letterpaper,margin=1in,headheight=15pt]{geometry}
\usepackage{amsmath,amssymb,amsthm,mathtools,microtype}
\usepackage{enumitem,booktabs,needspace,array,longtable}
\usepackage{xcolor}
\definecolor{navy}{RGB}{25,50,77}
\usepackage[colorlinks=true,linkcolor=navy,citecolor=navy,urlcolor=navy,bookmarksnumbered=true,bookmarksdepth=3]{hyperref}
\usepackage{bookmark}

\setlist{itemsep=3pt,topsep=5pt}
\newcommand{\zo}{\bits}

\def\calX{{\mathcal X}}

\newcommand{\anm}{\mathsf{anmExt}}

\theoremstyle{definition}

\newcommand{\e}{\varepsilon}

\newcommand{\nmExt}{\mathsf{nmExt}}

\newcommand{\BI}{\begin{itemize}}
\newcommand{\EI}{\end{itemize}}
\newcommand{\BE}{\begin{enumerate}}
\newcommand{\EE}{\end{enumerate}}

\newtheorem{thm}{Theorem}      
\newcommand{\BT}{\begin{theorem}}   \newcommand{\ET}{\end{theorem}}
\newcommand{\BD}{\begin{definition}}   \newcommand{\ED}{\end{definition}}
\newcommand{\BCR}{\begin{corollary}} \newcommand{\ECR}{\end{corollary}}
\newtheorem{constr}[thm]{Construction}
\newcommand{\BCT}{\begin{constr}} \newcommand{\ECT}{\end{constr}}
\newcommand{\BL}{\begin{lemma}}   \newcommand{\EL}{\end{lemma}}

\newcommand{\BP}{\begin{proposition}}   \newcommand{\EP}{\end{proposition}}
\newcommand{\BCM}{\begin{claim}}   \newcommand{\ECM}{\end{claim}}
\newcommand{\BF}{\begin{fact}}   \newcommand{\EF}{\end{fact}}
\newcommand{\BA}{\begin{assumption}}   \newcommand{\EA}{\end{assumption}}

\def\le{\leqslant} \def\ge{\geqslant}

\makeatletter
\def\ExtendSymbol#1#2#3#4#5{\ext@arrow 0099{\arrowfill@#1#2#3}{#4}{#5}}
\def\RightExtendSymbol#1#2#3#4#5{\ext@arrow 0359{\arrowfill@#1#2#3}{#4}{#5}}
\def\LeftExtendSymbol#1#2#3#4#5{\ext@arrow 6095{\arrowfill@#1#2#3}{#4}{#5}}
\makeatother

\numberwithin{equation}{section}
\newtheorem{theorem}{Theorem}[section]
\newtheorem{lemma}[theorem]{Lemma}
\newtheorem{proposition}[theorem]{Proposition}
\newtheorem{corollary}[theorem]{Corollary}

\theoremstyle{definition}\newtheorem{definition}[theorem]{Definition}
\theoremstyle{remark}\newtheorem{remark}[theorem]{Remark}
\newcommand{\F}{\mathbb F}\newcommand{\E}{\mathbb E}
\newcommand{\bits}{\{0,1\}}\newcommand{\Tr}{\operatorname{Tr}}
\newcommand{\rk}{\operatorname{rank}}\newcommand{\rank}{\operatorname{rank}}
\newcommand{\spn}{\operatorname{span}}\newcommand{\Span}{\operatorname{span}}
\newcommand{\supp}{\operatorname{supp}}\newcommand{\cp}{\operatorname{cp}}
\newcommand{\TV}{\Delta}\newcommand{\Hin}{H_\infty}
\newcommand{\Ext}{\operatorname{Ext}}
\newcommand{\U}{\mathsf U}
\newcommand{\aff}{\operatorname{aff}}

\newcommand{\LExt}{\operatorname{LExt}}
\newcommand{\codim}{\operatorname{codim}}
\newcommand{\twExt}{\operatorname{TExt}}
\newcommand{\1}{\mathbf 1}
\newcommand{\Hav}{\widetilde H_\infty}
\newcommand{\eps}{\varepsilon}\newcommand{\polylog}{\operatorname{polylog}}
\newcommand{\seedcost}[2]{\left[\log^2(#1+2)+(#2+1)\log\!\left(2+\frac{#1}{#2+1}\right)\right]}

\newcommand{\Prb}{\mathbb P}\newcommand{\Cond}{\operatorname{Cond}}

\newcommand{\ip}[2]{\langle #1,#2\rangle}
\newcommand{\SumExt}{\operatorname{SumExt}}
\newcommand{\Kcal}{\mathcal K}

\title{Asymptotically Optimal Two Source Extractors for Arbitrary Constant Entropy Rate}
\author{Xin Li\thanks{
Department of Computer Science, Johns Hopkins University, \texttt{lixints@cs.jhu.edu}. 
}}
\date{}
\begin{document}
\maketitle\thispagestyle{plain}
\begin{abstract}
We construct explicit two-source extractors for any constant entropy rate with asymptotically optimal parameters. Specifically, for sources on $n$ bits with min-entropy $k =\delta n$ for any constant $\delta \in (0, 1]$, and error $\eps \geq 2^{-b_{\delta}n}$ where $b_{\delta}$ is a constant depending on $\delta$, our extractor outputs $2k-O(\log(1/\eps)+1)$ bits that are $\eps$ close to uniform. In particular, this gives two-source extractors with $\Omega(n)$ output bits and $2^{-\Omega(n)}$ error for any constant entropy rate. Previously, explicit two source extractors with such error are only known to work for entropy rate close to $1/3$ (Lewko 2026). We also extend our constructions to give (strong) non-malleable two source extractors and sumset extractors for any constant entropy rate with asymptotically optimal parameters. 

Our extractors can be extended to handle slightly sub-constant entropy rate, with sub-optimal error. Using a different construction and analysis, we give explicit (non-malleable) two source extractors for entropy $k \geq C n/\log n$ for some constant $C>1$ with $\Omega(k)$ output bits and error close to $2^{-\Omega(\sqrt{n})}$. 

An independent work of Doron and Goodman constructed explicit two-source extractors for entropy $k \geq n^{1-\alpha}$ with error $2^{-k^{\alpha}}$ for some constant $\alpha>0$. Their extractors work for smaller entropy than ours, while our extractors have better error and output length for any constant entropy rate. Our extractors for entropy $k \geq C n/\log n$ are obtained independently before we were aware of their results, while our extractors for constant entropy rate with asymptotically optimal parameters are obtained after we learned their unpublished results at the Simons Institute, using different ideas and techniques.
\end{abstract}
\thispagestyle{empty}
\newpage\tableofcontents \thispagestyle{empty}
\newpage
\pagenumbering{arabic}

\section{Introduction}
The theory of \emph{randomness extraction} concerns the process of extracting almost uniform random bits from \emph{weak random sources}, which are probability distributions with various defects and bias. Through decades of study, many different kinds of \emph{randomness extractors}, which are explicitly constructed objects that perform randomness extraction, have been designed. These objects have since become fundamental objects in the area of pseudorandomness, with wide applications in other areas such as cryptography, complexity theory, graph theory, and so on. In this paper, we use the standard \emph{min-entropy} to measure the amount of randomness in a weak random source. Specifically, the min-entropy of a random variable~$X$ is defined as $H_\infty(X)=\min_{x \in \supp(X)}\log_2(1/\Pr[X=x])$. A source $X$ with support in $\bits^n$ is said to be an $(n,H_\infty(X))$-source, with entropy rate $H_\infty(X)/n$. 

 It is well known that a single arbitrary weak source with only a min-entropy guarantee admits no deterministic extraction in general, thus motivating models with structures. 

\BD
Let $\calX$ be a family of distributions over $\zo^n$. A function $\Ext : \zo^n \to \zo^m$ is a deterministic extractor for $\calX$ with error $\e$ if for every distribution $X \in \calX$, we have
\[ \TV(\Ext(X),U_{m})\le \e,\]
where $U_m$ stands for the uniform distribution over $\zo^m$, and $\TV( )$ denotes the statistical distance. Sometimes we also write $\Ext(X) \approx_{\e} U_m$. We say $\Ext$ is explicit if it is computable by a polynomial-time algorithm.
\ED
 Again, many such models have been studied. Some of the most well studied ones include for example two source extractors, where the input to the extractor consists of two independent $(n, k)$ sources $X,Y$;  affine extractors, where the input source is the uniform distribution over some unknown affine subspace; sumset extractors, where the input is $X+Y$ for two independent $(n, k)$ sources $X,Y$ (here $+$ is bitwise addition over $\F_2$); and non-malleable two source extractors, where the extractor has an additional security guarantee against tamperings on the input sources. In this paper we study two source extractors, sumset extractors, and non-malleable two source extractors. We briefly survey the history of these extractors below.


\paragraph{Two-source extractors.}
Chor and Goldreich~\cite{ChorG88} initiated the formal study of weak random sources and explicit two-source extractors. Their construction gives an explicit two source extractor for entropy rate above $1/2$ with exponentially small error, while the probabilistic method shows the existence of two source extractors with error $\eps$, output length $m=2k-2\log(1/\eps)-O(1)$ when the entropy $k$ of each source is slightly larger than $\log n+2\log(1/\eps)+O(1)$. In particular, this implies the existence of two source extractors with output length $\Omega(k)$ and error $2^{-\Omega(k)}$ when $k \geq c \log n$ for some constant $c>1$. Dodis, Elbaz, Oliveira, and Raz~\cite{DEOR} improved the output length of the extractor in \cite{ChorG88}, but the entropy rate remains above $1/2$. Raz~\cite{Raz05} constructed an explicit two source extractor where one source is required to have entropy rate above $1/2$, while the other one can have logarithmic entropy. Breaking the entropy rate $1/2$ barrier has since been a major open problem in the study of two source extractors.

A 2005 work by Bourgain~\cite{Bourgain05} first breaks the entropy rate $1/2$ barrier, by using sum-product phenomena and estimates for exponential sums. However Bourgain's extractor only works for entropy rate slightly below one half. Subsequently, in a series of works ~\cite{lewko_2019, Lewko2020Rectangles, Lewko2026SzemerediTrotter}, Lewko revisited this approach and improved the min-entropy rate to be arbitrarily close to $1/3$. A recent work by Chattopadhyay and
Gurumukhani \cite{ChattopadhyayGurumukhani2027Vinogradov} also constructed a two source extractor based on this approach, that works for every entropy rate above $3/7$. All of these extractors have error $2^{-\Omega(n)}$.  

On the other hand, a long line of works \cite{BarakIW04, BarakKSSW05, Raz05, Bourgain05, Rao06, BarakRSW06, Li11b, Li12b, Li13a, Li13b, Li15, Cohen15, DBLP:conf/stoc/Cohen16, CZ15, Li16, Coh16b, CL16, Coh16, BDT16, Cohen16, DBLP:conf/stoc/Cohen17, Li17, Li19, Li23} studied the more general model of several independent sources, and gradually reduced the number of sources. These works finally led to the breakthrough of Chattopadhyay and Zuckerman~\cite{CZ15}, which obtained an explicit one-bit two-source extractor for entropy $k\ge\log^C n$ for some constant $C>1$ with polynomially small error. Later works refined the constructions and culminated in Li's 2023 work~\cite{Li23}, which constructed explicit two source extractors for asymptotically optimal entropy of $O(\log n)$, with one bit of output and constant error. These extractors also give currently the best known explicit constructions of \emph{Ramsey graphs}. 

However, the two source extractors constructed recently following the framework of \cite{CZ15} only have polynomially small error at the best, which is insufficient for cryptographic tasks, where one naturally requires negligible error. Given that one of the main applications of randomness extractors is in cryptography, the quest for small error two source extractors remains an important open problem, where the previously best known construction due to Lewko~\cite{Lewko2026SzemerediTrotter} still requires entropy rate above $1/3$.

\paragraph{Sumset Extractors.}
Chattopadhyay and Li~\cite{ChattopadhyayL16} introduced and initiated the study of extractors for sumset sources, which unifies several previously studied models such as independent sources, affine sources, and small space sources. Here, unlike the case of two source extractors, a simple application of the probabilistic method fails to show the existence of such extractors for small entropy (e.g., $O(\log n)$). Nevertheless, \cite{ChattopadhyayL16} constructed explicit extractors for the sum of a constant number of independent sources with poly-logarithmic entropy, and achieved a polynomially small error. Chattopadhyay and Liao~\cite{ChattopadhyayL22} later improved this to work for the \emph{sum of two} independent sources, with entropy $k=\polylog(n)$ and polynomially small error, or entropy $k=\log n(\log\log n)^{1+o(1)}$ with constant error. Li~\cite{Li23} subsequently obtained sumset extractors for the sum of two independent sources with asymptotically optimal entropy of $O(\log n)$ and constant error. 

However, if one requires negligible error, then since sumset extractors are stronger than two source extractors, the entropy requirement of the former is at least that of the latter. One known construction is the Paley graph extractor in \cite{ChorG88}, although the sum there is over a field with odd characteristic. The extractor requires one
source to have entropy rate above $1/2$ while the other can have entropy $k$ as small as $O(\log n)$, and the extractor has error $2^{-\Omega(k)}$. Thus for $k=\omega(\log n)$ the error is negligible. The analysis is based on character sum estimates
of Karatsuba \cite{Kar71, Kar91}. 


\paragraph{Non malleable extractors.} Many of the standard models of randomness extractors can also be extended to a stronger version, known as \emph{non-malleable extractors}. This notion arises from cryptographic applications, such as privacy amplification with an active adversary \cite{DW09} and non-malleable codes \cite{DPW10}. Due to their cryptographic applications and recent surprising connections to constructing standard extractors \cite{Li12b, CZ15}, such extractors have also been the focus of much study in the past 15 years or so. In this setting, the input to the extractor is tampered with by a certain kind of adversary, and the non-malleable extractor guarantees that the output of the extractor on the original inputs is still close to uniform even conditioned on the tampered output. We begin the discussion of non-malleable extractors by defining tampering functions.

\begin{definition}[Tampering Function]
For any function $f:S \rightarrow S$, we say $f$ has no fixed points if $f(s) \neq s$ for all $s \in S$. For any $n>0$, let $\mathcal{F}_n$ denote the set of all functions $f: \{ 0,1\}^n \rightarrow \{0,1\}^n$. Any subset of $\mathcal{F}_n$ is a family of  tampering functions. 
\end{definition}

As in recent literature, for non-malleable extractors we only consider tampering functions with no fixed points, since if the tampering is the identity function then non-malleability is impossible. A more general definition can be found for example in \cite{CG14b}. We can now extend two source extractors to their corresponding non-malleable version. 

\begin{definition}
A function $\nmExt : (\{ 0,1\}^{n})^2 \rightarrow \{ 0,1\}^m$ is a $(k, \e)$ $C$-non-malleable two source extractor, if it satisfies the following property: Let $X, Y$ be two independent $(n, k)$ sources, and $\{(f_i, g_i)\}_{i \in [C]}$ be $C$ arbitrary pairs of tampering functions such that for each $i \in [C]$, at least one of $f_i, g_i$ has no fixed points, then  $$ (\nmExt(X, Y) , (\nmExt(f_i(X),g_i(Y)))_{i=1}^C) \approx_{\e} (U_m , (\nmExt(f_i(X),g_i(Y)))_{i=1}^C),$$ 
where $U_m$ is independent of $(\nmExt(f_i(X),g_i(Y)))_{i=1}^C$. We say the non-malleable two source extractor is (source) strong if the above property holds with the additional conditioning of the source $X$ or $Y$. 
\end{definition}

\begin{remark}
    In the special case of $C=0$, this gives a standard two source extractor, and the strong version gives a \emph{strong two source extractor}.
\end{remark}

The probabilistic method can be used to prove the existence of such non-malleable extractors with excellent parameters. For example, \cite{CG14b} showed that for $C=1$, such extractors exist for $(n, k)$ sources when $k \geq m+\frac{3}{2}\log (1/\eps)+O(1)$ and $k \geq \log n+3\log (1/\eps)+O(1)$. If the error is small, then for non-malleable two source extractors against $C$ tampering functions, a necessary information theoretic bound is that $k \geq (C+1)m/2$. For the strong version, we need $k \geq (C+1)m$. 

Not surprisingly, constructing explicit non-malleable extractors is significantly harder than constructing standard extractors, due to their much stronger properties. Historically, Chattopadhyay, Goyal, and Li~\cite{CGL15} gave the first explicit non-malleable two-source extractors for sources of entropy $n-n^{\Omega(1)}$, with polynomial-size output and exponentially small error. They also developed many-tampering versions in the same very-high-entropy regime. Subsequent works ~\cite{CL16, Li17, Li19, ChungOA21, Li23} improved the constructions in various ways, leading to an explicit construction of Li~\cite{Li23}, which gave non-malleable two-source extractors requiring entropy \((2/3+\gamma)n\) in one source and \(k\ge C\log n\) in the other for some constant $C>1$, with output length \(\Omega(k)\) and error \(2^{-\Omega(k)}\). In the regime of small entropy, Li and Zhong~\cite{LiZhong2024TwoSource} also obtained non-malleable two source extractors that match their standard counterparts: extractors for polylogarithmic entropy with polynomially small error, and for $O(\log n)$ entropy with constant error.

In summary, the question of constructing small error (e.g., negligible error) (non-malleable) two source related extractors remains an important open question. For two source extractors and sumset extractors, the bottleneck is slightly below entropy rate $1/2$ at the best; for non-malleable two source extractors, the bottleneck is at entropy rate $2/3$. 


\subsection{Our results}
In this paper we significantly improve previous results in terms of small error (non-malleable) two source and sumset extractors. We first state our main results of two source extractors. 

\begin{theorem}\label{main:twointro}
There are absolute constants $C_0,C_1>0$ such that, for every constant
$0<\delta\le1$, there are constants $b_\delta>0$ and $n_\delta$ with the following
property. For all $n\ge n_\delta$, integer $\delta n\le k\le n$, and $2^{-b_\delta n}\le\eps\le\tfrac14$, there is an explicit two source extractor 
$\twExt:\bits^n\times\bits^n\longrightarrow\bits^m$
such that, for any two independent $(n, k)$ sources $X,Y$,
\[
 \TV(\twExt(X,Y),U_m)\le\eps,
 \qquad m\ge2k-C_0\log(1/\eps)-C_1.
\]
\end{theorem}

Thus our extractor achieves asymptotically optimal parameters for any constant entropy rate. In particular, this implies a two source extractor for any constant entropy rate with output length $\Omega(n)$ and error $2^{-\Omega(n)}$.

We can extend our results to sumset extractors and non-malleable two source extractors. Specifically, we have the following theorems.
\begin{theorem}\label{main:ssintro}
For every constant $0<\delta\le1$ there are constants $b_\delta>0$ and $n_\delta$ such that the following holds. For all $n\ge n_\delta$, integer $\delta n\le k\le n$, and
$2^{-b_\delta n}\le\eps\le\tfrac14$, there is an explicit sumset extractor 
$\SumExt:\F_2^n\longrightarrow\F_2^m$ such that, for any two independent $(n, k)$ sources $X,Y$,
\[\TV(\SumExt(X+Y),U_m)\le\eps, 
  \qquad m\ge k-2\log(1/\eps)-10.
\]
\end{theorem}

\begin{theorem}\label{main:nmtwointro}
For every constant $0<\delta\le1$ and integer $C\ge0$
there is a constant $b_{\delta,C}>0$ such that, for sufficiently large $n$,
integer $\delta n\le k\le n$, and
$2^{-b_{\delta,C}n}\le\eps\le1/4,$
there is an explicit non malleable two source extractor $\nmExt:\bits^n\times
\bits^n\to\bits^m$ with
\[
 m\ge \frac{k}{C+1}-O(\log(1/\eps)+1)
\]
such that, for any two independent $(n, k)$ sources $X, Y$, and any $C$ pairs of tampering functions $\{(f_i, g_i)\}_{i \in [C]}$ such that for each $i \in [C]$, at least one of $f_i, g_i$ has no fixed points,
\begin{align}
 (\nmExt(X,Y),(\nmExt(f_i(X),g_i(Y)))_{i=1}^C,Y)
 &\approx_\eps (U_m,(\nmExt(f_i(X),g_i(Y)))_{i=1}^C,Y),\\
 (\nmExt(X,Y),(\nmExt(f_i(X),g_i(Y)))_{i=1}^C,X)
 &\approx_\eps (U_m,(\nmExt(f_i(X),g_i(Y)))_{i=1}^C,X).
\end{align}
\end{theorem}

\begin{remark}
Note that in the above theorem, the special case of $C=0$ also gives a standard strong two source extractor with output length $m\ge k-O(\log(1/\eps)+1)$.    
\end{remark}

We can extend our results to slightly sub constant entropy rate. Specifically, we have the following theorem.
\begin{theorem}[Entropy rate $C/\log n$]\label{thm:mainlowintro}
Fix an integer $t\ge0$ and $0<\vartheta<1/2$. There are constants
$C_{t,\vartheta}>0$ and $n_0$ such that, for every $n\ge n_0$,
integer $C_{t,\vartheta}n/\log n\le k\le n$, and
$2^{-n^{1/2-\vartheta}}\le\eps\le1/4$, 
there is an explicit strong $t$-non-malleable two-source
extractor $\nmExt:\bits^n\times\bits^n\to\bits^m$ with
\[
 m\ge\frac{k}{t+1}-O\bigl(\log(1/\eps)+1\bigr).
\]
\end{theorem}

\subsection{Overview of the techniques}\label{un:overview}
Here we give an informal overview of the techniques used in our constructions. Throughout, we will focus more on the constructions of standard two source extractors to illustrate our main ideas. We start by describing the ideas behind our two source extractor for any constant entropy rate with $\Omega(n)$ output length and $2^{-\Omega(n)}$ error. Without loss of generality, we consider flat sources, which are the uniform distributions over some unknown subsets.

Our starting point is the work of Ben-Sasson and Ron-Zewi \cite{BenSZ11}, which shows how to construct two source extractors from a sufficiently good affine extractor.\ We will not be following this specific approach here, but we will use one of their important observations: let $E: \F^n_2 \times \F^n_2 \to \F_2$ be a function and view it as a $2^n \times 2^n$ matrix $M_E$ over $\F_2$ in the natural way, where the $(x, y)$ entry  corresponds to $E(x, y)$. Let $r$ be the rank of $M_E$ over $\F_2$, then there also exist functions $h_X, h_Y: \F^n_2 \to \F^r_2$ such that $E(x, y)=\langle h_X(x), h_Y(y) \rangle$ where $\langle ,  \rangle$ denotes the inner product over $\F_2$. Furthermore, if $E$ is a two source disperser for entropy rate $\rho$ with rank $O(n)$, then for any constant $\gamma>0$, $E$ is also a two source extractor for entropy rate $\rho+\gamma$ with error $2^{-\Omega(n)}$. Here, a two source disperser is a weaker object than a two source extractor, in the sense that it only requires the output to be non-constant (in the case of one bit output). The proof uses \emph{approximate duality}: if instead the function is not a two source extractor with the above parameters, then one can find a relatively large subrectangle in the matrix $M_E$ where the entries are all the same, which contradicts the disperser property. Therefore, our goal is reduced to constructing a two source disperser for any constant entropy rate, with rank $O(n)$. Once we have it, we can use approximate duality to lift it to an extractor with error $2^{-\Omega(n)}$. To get output length $\Omega(n)$, we show that we can actually produce $\Omega(n)$ output bits such that the parity of any non-trivial subset of these bits is still $2^{-\Omega(n)}$ close to uniform. Hence by a standard XOR lemma, these $\Omega(n)$ bits are themselves $2^{-\Omega(n)}$ close to uniform. 


We now explain the ingredients we use to get such a two source extractor. Towards that goal, we first use standard techniques to get a \emph{somewhere high entropy source} from each input weak source. In particular, we use the seeded condenser in \cite{Raz05} which has constant seed length, and can condense any weak source with entropy rate $\delta/2$ to $1-\beta$ with error $\eps$ and output length $\Omega(n)$. Here $\delta, \beta, \eps$ are arbitrary fixed constants, and the seed length $d$ is a constant depending on these. By enumerating all possible seeds, we get a somewhere high entropy source with a constant $r=2^d$ number of rows, such that when averaged over the constant seed space, each row is close to some distribution of entropy rate at least \(1-\beta\). This is in contrast to some of the condensers in e.g., \cite{BarakRSW06, Zuckerman}, which only guarantee that at least one row is close to having high entropy, but with exponentially small error. The error $\eps$ here is larger, but for our purpose of getting a disperser first, this is already enough. For input sources $X, Y$, let the two somewhere high entropy sources be $\widetilde{X}, \widetilde{Y}$. In our analysis, we then show that there exist two \emph{subsources} $X^*, Y^*$ such that when restricted to $X^*, Y^*$, each somewhere high entropy source $\widetilde{X^*}, \widetilde{Y^*}$ has at least $\frac{3r}{4}+1$ rows which have entropy rate at least $1-\tau/2$ for some small constant $\tau>0$. Hence, let $D=\frac{r}{2}$, there are at least $D+1$ rows that simultaneously have entropy rate at least $1-\tau/2$ in both $\widetilde{X^*}$ and $\widetilde{Y^*}$. We now delete an arbitrary row from $\widetilde{X^*}$ and $\widetilde{Y^*}$, and we end up with $r-1$ rows such that at least $D$ rows simultaneously have entropy rate at least $1-\tau/2$ in both $\widetilde{X^*}$ and $\widetilde{Y^*}$. These are the good rows we will use in the analysis, and from now on we'll treat the input sources as $X=X^*, Y=Y^*$.

At this point, our construction goes by applying a specific two source extractor $\twExt$ for high entropy sources to each row $a$ in $\widetilde{X}, \widetilde{Y}$, i.e., compute $\twExt(\widetilde{X}_a, \widetilde{Y}_a)$. Then we use a merger to merge these $r-1$ outputs together. However, in order to carry out our previous plan, we need to construct the two source extractor $\twExt$ and the merger carefully. At first, it seems tempting to just use inner product as $\twExt$. This indeed has rank $O(n)$, but then it is unclear how to merge the outputs together later. To address this, we use a slightly modified two source extractor. Specifically, let $M$ be the length of each row in $\widetilde{X}$ and $\widetilde{Y}$. Let $K=\F_{2^M}$ and identify each row as an element in $K$. For the $a$'th row, denote $U_a(X)=\widetilde{X}_a$ and $V_a(Y)=\widetilde{Y}_a$. Next we choose an appropriate $\ell=\Theta(n) < M$ such that $M=L \ell$, and let $F=\F_{2^{\ell}}$. Finally, choose an integer $J \ge 1$ and compute the following function: 
\[H_{J}(U_a(X),V_a(Y))=\sum_{j=1}^{J} U_a(X)^{2j-1}V_a(Y)^{2j-1}
   =\langle\Psi_J(U_a(X)),\Psi_J(V_a(Y))\rangle_K,
 \quad\Psi_J(z)=(z,z^3,\ldots,z^{2J-1}),\]

 and $W_a(X,Y)=\twExt(U_a(X), V_a(Y))=\Tr_{K/F}H_{J}(U_a(X),V_a(Y))$, where the $\Tr_{K/F}$ linear map to the smaller field $F$ is necessary for the function to be a two source extractor. 

 It is not hard to show that $\twExt(U_a(X), V_a(Y))$ is indeed a two source extractor if each input has high enough entropy, and any parity of the output bits has rank $O(n)$ over $\F_2$, since the function $\Tr_{K/F}H_{J}(U_a(x),V_a(y))$ only has constant rank over $F$. However, it remains to see how to merge these outputs together. Towards that, we prove a \emph{lower bound} on the rank of any non-trivial parity of the output bits when \emph{restricted to} a sufficiently large rectangle $A \times B$, where $A$ and $B$ are the supports of $U_a(X)$ and $V_a(Y)$ respectively. For a good row $a$, we have that $|A|, |B| \geq 2^{(1-\tau)M}$. This step crucially uses Weil bound to argue that any non-trivial parity character of the encoding $\Psi_J(z)$ has small bias over $K$, and the odd exponents in the encoding are used to guarantee the Weil bound in fields with characteristic $2$.  We use this to show that when $|A|, |B| \geq 2^{(1-\tau)M}$, the span of $\Psi_J(A)$ and $\Psi_J(B)$ both have dimension at least $JM-\tau M -1$. Indeed, in the extreme case where $\tau=0$, both spans have full dimension $JM$, for otherwise there will be a non-trivial parity character that is fixed, contradicting the Weil bound. Now any non-trivial parity of the output bits can be written as $h_{J,\lambda}=\Tr_{K/\F_2}\!\bigl(\lambda H_J(U_a(X),V_a(Y))\bigr)$ for some $\lambda \in K, \lambda \neq 0$. Since the binary bilinear form
\[
 B_\lambda(p,q)=\Tr_{K/\F_2}\!\left(\lambda\sum_{j=1}^Jp_jq_j\right)
 \qquad(p,q\in K^J)
\]
is nondegenerate, restricting its arguments to
the spans of $\Psi_J(A)$ and $\Psi_J(B)$ costs at most the sum of
their binary codimensions. Consequently, for all
$|A|,|B|\geq 2^{(1-\tau)M}$, we have
\begin{equation}\label{eq:restricted}
 \boxed{\ \rank_{\F_2}[h_{J,\lambda}]_{A\times B}
 \geq (J-2\tau)M-2,\qquad
 \rank_{\F_2}[h_{J,\lambda}]_{K\times K}\leq JM.\ }
\end{equation} 

This immediately suggests an approach to separate the good rows: we can choose a different number $J_a$ for a different row $a$. Here we use an idea similar to the construction of affine extractors for any constant entropy rate in \cite{Li11a}, by choosing the numbers to be a geometric sequence: $J_a=2^{a-1}$, and thus the summation of all $J_b, b < a$ is $\kappa_a=\sum_{b<a}J_b=J_a-1$.

We use this property to show that on the $D$ good rows, for simplicity of exposition assume row $1$ to $D$, the joint distribution $(W_1(X, Y), \cdots, W_D(X, Y))$ has full support of $F^D$. Towards that, we actually show that the distribution of $(W_1(X, Y), \cdots, W_D(X, Y))$, when restricted to $X^*, Y^*$, is exponentially close to uniform, with error $< 2^{-D\ell}$ where $\ell$ is the output length of each $W_i(X, Y)$. This follows from a hybrid argument: for any $a \in [D]$, let the transcript $\Pi_a=(W_1(X, Y), \cdots, W_{a-1}(X, Y))$, we show that the joint distribution $(W_a(X, Y), \Pi_a)$ is $2^{-\Omega(M)}$ close to $(U_F, \Pi_a)$. To show this, consider row $a$, and let $U=U_a(X), V=V_a(Y)$. By fixing extra randomness, the previous $a-1$ rows can be viewed as deterministic functions of $(U, V)$, and hence the transcript $\Pi_a$ is also a deterministic function $\Pi_a(U, V)$. We now use a standard XOR lemma, and show that for any $\mu \in F, \mu \neq 0$, and any $\beta \in \F^{\ell(a-1)}_2$, the parity character 
\[
\begin{aligned}
\operatorname{Tr}_{F/\mathbb F_2}(\mu W_a)
+\beta\cdot\Pi_a
&=
\operatorname{Tr}_{K/\mathbb F_2}
   \bigl(\mu H_{J_a}(U,V)\bigr)
+\beta\cdot\Pi_a
\end{aligned}
\]
has bias $2^{-\Omega(M)}$. This is one of the key steps in the analysis.

The proof follows from our rank bounds above and another application of the approximate duality lift. Specifically, the previous rank lower and upper bounds together imply that, when restricted to $A \times B$, every such mixed parity character is \emph{nonconstant} because the rank of the parity is at least $(J_a-2\tau)M-2-((J_a-1)M)=(1-2\tau)M-2>1$ for sufficiently large $M$. Recall that in our construction we ensure that in a good row, each $U_a(X)$ and $V_a(Y)$ actually has entropy $(1-\tau/2)M$, we can now apply approximate-duality to lift the bias of the character to be $2^{-\Omega(M)}$. By a careful choice of parameters, the total error is $< 2^{-D\ell}$. This finishes the first part of our analysis.

\paragraph{The merger.} We now describe our final merger, which outputs $\Omega(n)$ bits. It needs to satisfy two properties: first it needs to merge the $2D-1$ rows together, and ensure that any non-trivial parity of the output bits is non-constant. Second, it needs to keep any such parity to have rank $O(n)$, so that we can apply the approximate duality lift again to achieve exponentially small error. 

The second property is actually easier to achieve. Since we are working over a large field $F$, and each output $W_a(x, y) \in F$ can be written as $W_a(x, y)=\sum_{j=1}^{R}A_j(x)B_j(y)$ where $R$ is a constant, and each $A_j, B_j$ is a function from $\F^n_2$ to $F$, applying any constant degree polynomial over $F$ to the $2D-1$ outputs and taking parity will keep the rank to be $O(n)$, by a trace dual basis. To achieve the first property, we design the merger as follows: first apply the following symmetric polynomial on the $s=2D-1$ outputs: $\sigma_D(z_1,\ldots,z_s)=
 \sum_{\substack{S\subseteq[s]\\|S|=D}}\prod_{a\in S}z_a$, then take a fixed surjective binary-linear projection to \(m=\Omega(n)\) bits, where $m$ is sufficiently small so that we can apply the XOR lemma to ensure that the output bits are $2^{-\Omega(n)}$ close to uniform.

To see why this works, recall that previously we showed that there are $D$ good rows which have the full support $F^D$. In particular, this means for any element $\alpha \in F$, there is a positive probability that all the $D$ good rows take the value $\alpha$. Because $D=\frac{r}{2}=2^{d-1}$ is a power of two, in characteristic two we have $(1+\alpha T)^D=1+\alpha^DT^D$. Notice that the coefficient of $T^D$ in the generating function $\prod_{a=1}^{2D-1}(1+W_aT)$ is exactly $\sigma_D(W_1,\ldots,W_{2D-1})$. Thus whenever any $D$ of the $W_a$'s equal $\alpha$, we have $\sigma_D(W_1,\ldots,W_{2D-1})=\alpha^D$, regardless of the remaining $D-1$ inputs. Equivalently, this can be viewed as taking the \emph{majority} of the outputs. Therefore, the output of $\sigma_D$ also has full support in $F$, and thus any non-trivial parity of the linear projection is non-constant. Finally, we can use approximate duality lift again to achieve exponentially small error, and use a XOR lemma again to conclude that the $\Omega(n)$ output bits are $2^{-\Omega(n)}$-close to uniform. In particular, in the analysis we use approximate duality lift twice to reduce the error.

Below is a figure illustrating the construction:
\begin{center}
{\Large\bfseries\color{blue!50!black} Matched-row odd-power two-source extractor}\par
\vspace{.5cm}
\resizebox{\textwidth}{!}{
\begin{tikzpicture}[
  x=1cm,y=1cm, >=Latex,
  node distance=1.0cm,
  every node/.style={font=\sffamily\small,align=center},
  source/.style={draw=black!65,rounded corners=2pt,fill=black!5,
      minimum width=1.15cm,minimum height=0.72cm},
  condenser/.style={draw=blue!65!black,rounded corners=4pt,fill=blue!6,
      text width=2.72cm,minimum height=1.55cm,inner sep=6pt},
  pair/.style={draw=blue!65!black,rounded corners=4pt,fill=blue!3,
      text width=2.75cm,minimum height=1.64cm,inner sep=6pt},
  core/.style={draw=teal!70!black,rounded corners=4pt,fill=teal!6,
      text width=4.55cm,minimum height=2.60cm,inner sep=6pt},
  merger/.style={draw=violet!75!black,rounded corners=4pt,fill=violet!6,
      text width=2.52cm,minimum height=1.50cm,inner sep=6pt},
  projection/.style={draw=orange!70!black,rounded corners=4pt,fill=orange!7,
      text width=1.90cm,minimum height=1.50cm,inner sep=6pt},
  result/.style={draw=black!70,rounded corners=4pt,fill=black!6,
      text width=2.30cm,minimum height=1.40cm,inner sep=6pt},
  wire/.style={-{Latex[length=2.6mm]},line width=1.05pt,draw=black!75},
  hint/.style={font=\sffamily\scriptsize,text=black!65,align=center}
]
 \node[source] (X) at (0,3.15) {$x\in\F_2^n$};
 \node[source] (Y) at (0,-0.10) {$y\in\F_2^n$};
 \node[condenser] (CL) at (3.2,3.15) {\textbf{Strong}\\[2pt]\textbf{condenser}\\[4pt] enumerate $r$ seeds\\[4pt] $U_i(x)\in K$, $i\in[r]$};
 \node[condenser] (CR) at (3.2,-0.10) {\textbf{Strong}\\[2pt]\textbf{condenser}\\[4pt] enumerate $r$ seeds\\[4pt] $V_j(y)\in K$, $j\in[r]$};
 \node[pair] (PA) at (6.45,1.52) {\textbf{Match indices}\\[2pt] $s=r-1=2D-1$\\[2pt] $(U_a,V_a)$, $a\in[s]$};
 \node[core] (CO) at (10.58,1.52) {\textbf{Odd-power}\\\textbf{inner-product blocks}\\[4pt]
    $\displaystyle H_{J_a}(u,v)=\sum_{j=1}^{J_a}u^{2j-1}v^{2j-1}$\\[6pt]
    $\displaystyle W_a=\operatorname{Tr}_{K/F} H_{J_a}(U_a,V_a)$\\[4pt]
    $J_a=2^{a-1}$};
 \node[merger] (ME) at (15.05,1.52) {\textbf{Field merger}\\[5pt]
    $\displaystyle Z=\sigma_D(W_1,\dots,W_s)$};
 \node[projection] (PR) at (18.18,1.52) {\textbf{Linear projection}\\[4pt]
    $\pi:F\to\F_2^m$};
 \node[result] (OUT) at (21.10,1.52) {\textbf{Output}\\ $\operatorname{Ext}(x,y)\in\F_2^m$\\ $m=\Omega_\delta(n)$};
 \draw[wire] (X)--(CL);
 \draw[wire] (Y)--(CR);
 \draw[wire] (CL.east) -- ++(.38,0) |- ($(PA.west)+(0,.52)$);
 \draw[wire] (CR.east) -- ++(.38,0) |- ($(PA.west)+(0,-.52)$);
 \draw[wire] (PA)--(CO);
 \draw[wire] (CO)--(ME);
 \draw[wire] (ME)--(PR);
 \draw[wire] (PR)--(OUT);
 \node[hint] at (10.60,-.57) {$K=\F_{2^M}$, $F=\F_{2^\ell}\subset K$, $M=L\ell$, $\ell=\Theta_\delta(n)$};
 \node[hint] at (15.08,3.45) {$D=r/2$ a power of two};
 \node[hint] at (18.25,3.38) {fixed $m\le\ell$};
\end{tikzpicture}

}
\par\vspace{.6cm}
\begin{minipage}{.95\textwidth}
\small\textbf{Evaluated operations only:} constant-seed strong condenser on each source; evaluation of only the $r-1$ matched odd-power inner products over $K$; trace to $F$; elementary symmetric degree-$D$ merger; fixed linear projection. Here $J_a=2^{a-1}$, $F=\F_{2^\ell}\subset K=\F_{2^M}$, $\ell=\Theta_\delta(n)$, and $D$ is a power of two.
\end{minipage}
\end{center}

\paragraph{Sumset extractor.} Our construction of the sumset extractor roughly has the same structure as the two source extractor, with a couple of modifications. First of all, if we just apply an arbitrary condenser and try every possible seed, this destroys the sumset structure, and the subsequent analysis won't work. Instead, we need a condenser where for each fixed seed, the function is a linear function. We build such a condenser to replace Raz's condenser, by first using the somewhere high entropy linear condenser of Li and Zhong \cite{LZ}, and then merge the rows together by using a curved merger as in \cite{DW, DvirKSS09}. 

Second, because now the only input we get is a sum $X+Y$, we won't be able to apply the function $H_{J}(U_a(X),V_a(Y))=\sum_{j=1}^{J} U_a(X)^{2j-1}V_a(Y)^{2j-1}$ directly. Instead, we use $(U_a(X)+V_a(Y))^{d}$, where $d=2^h-1, h\ge3$. Again let $u=U_a(X), v=V_a(Y)$, in characteristic two, Lucas's theorem implies that all binomial
coefficients of $d$ are odd, so
\begin{equation}\label{eq:lucas}
 \boxed{(u+v)^d=\sum_{j=0}^{d}u^jv^{d-j}.}
\end{equation}
Consequently, for every nonzero $\lambda\in K$,
\begin{equation}\label{eq:globalmoment}
 \rk_{2}\left[
 \Tr_{K/\F_2}(\lambda(u+v)^d)
 \right]_{u,v\in K}
 \le(d+1)M=2^hM.
\end{equation}
As long as $h$ is a constant, the global rank is still $O_h(M)=O(n)$. We then use a moment–trace-code argument and show in Proposition~\ref{ss:prop:rank}, that by choosing $h$ appropriately, we can ensure that the \emph{restricted} rank on a high entropy rectangle $A \times B$ can grow by an arbitrarily large fixed multiple of $M$. Thus we can again use the previous strategy: choose a different $h$ for different rows to separate the good rows, and the remaining argument can follow.

\paragraph{Non-malleable two source extractor.}
On the high level, our $C$ tampering non-malleable two source extractor follows the general framework as our standard two source extractor, by applying a constant-seed strong condenser on each source, computing a suitable \emph{non-malleable} two source extractor for high entropy rate on each aligned pair of rows, and merging the row
outputs with $\sigma_D$. In particular, we will use rank properties of our construction and approximate duality lift. The additional difficulty is that we need to ensure the non-malleable property. Here we briefly explain several new ingredients needed in the new construction. 

First, to prove non-malleability, our strategy is to use a non-malleable version of XOR lemma on the output bits, such as that in \cite{DLWZ11}. This requires to show that any mixed binary character, which corresponds to the linear combination of any non-empty subset of the output bits from the original inputs, and any subset of the output bits from the tampered inputs, has exponentially small bias. In the analysis of the standard two source extractor, we prove this by showing that the joint distribution of the $D$ ``good outputs" has full support of $F^D$, and then use the property that $\sigma_D$ acts like the majority function. This implies the output of $\sigma_D$ has full support of $F$. Thus any non-trivial character is non-constant, and we then reduce the bias to exponentially small by an approximate duality lift. Here, similarly, our strategy to show that the joint distribution of the $D$ good outputs, together with all their $C$ tampered versions, has full support of $F^{D(C+1)}$, hence the joint output after applying $\sigma_D$ has full support of $F^{C+1}$. 

However, in general this is false, since some of the tamperings may not have entropy at all, and some of the tamperings may be the same. To address this, we analyze one mixed binary character at a time. If an active tampering pair has a heavy input value on either side, we restrict to that value on the corresponding original source, which doesn't cause much entropy loss. The contribution of the tampering pair to the character then depends only on the other original source, so we further fix that one-bit contribution at an additional entropy cost of at most one bit, and remove this pair of tamperings from the character. Repeating this process leaves only tamperings whose two input marginals have the required entropy.  For identical tampering copies, we can combine the parities on these outputs together into one parity on just one copy of such outputs. Hence, we can assume that we have at most $C$ tamperings that are all different, and each of them has enough entropy.

\paragraph{Distinguishing the executions.}
As in previous constructions of non-malleable extractors, we need a strategy to distinguish the original inputs and their tampered versions. Here we use a simpler strategy than the advice generators in previous works. Specifically, we
view each full input $x$ as the coefficient vector of a polynomial $P_x$
over a field with a small linear number of bits per element. Thus the polynomial has some constant degree $b$. For each index $a$ of the rows in the condenser output, we choose a different fixed
evaluation point $t_a$, and compute the tag (or advice) $(P_x(t_a),P_y(t_a))$. This approach has the advantage that each $(P_x(t_a)$ or $P_y(t_a))$ is a deterministic function of $X$ and $Y$ respectively, and thus it's convenient to condition on the fixing of such tags while preserving the independence of $X$ and $Y$. 

To apply the $\sigma_D$ as majority argument above, we would like to argue that there are a majority of $D$ good rows that have the \emph{same} indices in every tampered execution of the extractor, as well as the execution on the original inputs. Since we will eventually apply a non-malleable two source extractor for high entropy rate to each row, we want to ensure that in these $D$ rows, \emph{all} inputs are distinguished by the tags. Recall that we assumed before that every tampering is different. Intuitively, we want to argue that every two different set of tags $\{(P_x(t_a),P_y(t_a))\}$ computed from $(x, y)$ and $\{(P_{x'}(t_a),P_{y'}(t_a))\}$ computed from $(x', y')$ can agree on at most $b$ positions (recall $b$ is the degree of the polynomials $P_x, P_y$). So, altogether there are at most $\binom{C+1}{2}b$ positions that can have a collision in tags. We can make sure these positions, together with the bad seeds of the condenser, consist of $<1/2$ fraction of all possible rows, by cycling through the condenser seeds to create enough rows if necessary. 

However, there are a couple of issues with the above argument. First, we only know that any pair of tamperings are different from the original pair of inputs on every possible input strings, but for two different pairs of tamperings, they may still be the same on certain inputs. Second, even if two pairs of tamperings (or one pair of tampering and the original pair of inputs) are different on every possible input strings, the positions where the tags are different can depend on the specific input strings. To address these issues, we \emph{fix} the equality/inequality patterns and move to a subsource. Specifically, let the original inputs be $x_0=x, y_0=y$, and the $C$ tampered inputs be $x_i=f_i(x), y_i=g_i(y)$ for $i \in [C]$. For every $0 \leq i < j \leq C$, we record two bits of whether $x_i=x_j$ and $y_i=y_j$. Altogether these give $2\binom{C+1}{2}$ bits, and we can fix one particular pattern of these bits such that both sources only lose at most $\binom{C+1}{2}$ bits of entropy, and they are still independent. Crucially, conditioned on this fixing, for every \(i,j\), the equality \(x_i=x_j\) is either always true or always false on the retained left support, and the same holds for \(y_i=y_j\) on the retained right support. Furthermore, the two equalities cannot hold simultaneously unless $i, j>0$. If this happens for some $i, j$, then again we can merge these two tamperings in our analysis of the mixed character. Hence, after this step we can assume that the tampering functions are also different on at least one side for every possible input strings. This takes care of the first issue.

Next, we do the same and fix \emph{tag equality indicators}. Suppose we have $R$ rows in total from the previous possible cycling of the condenser seeds. For every different $i, j$, at row $a \in [R]$, their tags are $(P_{x_i}(t_a),P_{y_i}(t_a))$ and $(P_{x_j}(t_a),P_{y_j}(t_a))$. We now record two bits of whether $P_{x_i}(t_a)=P_{x_j}(t_a)$ and $P_{y_i}(t_a)=P_{y_j}(t_a)$. Again there are at most $2R\binom{C+1}{2}$ bits, and we can fix one particular pattern of these bits such that both sources only lose at most $R\binom{C+1}{2}$ bits of entropy, and they are still independent. Now conditioned on this fixing, for any $i, j$, the set of indices $a$ with tag collisions are the \emph{same} for all input strings throughout the retained product support. This takes care of the second issue.

\paragraph{A tagged high entropy non-malleable extractor against one tampering.}
Next, we use a tagged high entropy non-malleable extractor against one tampering as the starter extractor to apply on each pair of rows. We will later modify it to one against $C$ tamperings. In \cite{Li23}, Li constructed explicit non-malleable two source extractors for entropy rate $1-\delta$; however, the rank properties we need here are not supplied by the non-malleable extractor guarantee alone. Therefore, we come up with a new construction tailored for that. 

The new tagged high entropy non-malleable extractor is an asymmetric 
extractor, specifically tailored to satisfy the rank properties. Let $z,w\in\F_2^M$ denote the complete left and right rows, where
$M=6n_0$. Choose binary fields $K\subset F\subseteq L$ with
$L=\F_{2^{n_0}}$ and $[F:K]=2$. Let $\bar{x}$ be a prefix of $z$ with $n_0$ bits and view $\bar{x}$ as an element in $L$.
Parse the complete
right row as $w=(r,t,Y)\in L^2\times K^J$, where
$J\log_2|K|=4n_0$. For a coefficient vector
$d=(d_0,d_1,\ldots)$, write $P_d(T)=\sum_jd_jT^j$. Define
\begin{align}
 \Gamma_c(\bar{x})&=(\bar{x},\bar{x}^{1+2},\ldots,\bar{x}^{1+2^c}),\label{eq:gold}\\
 S_\alpha(\bar{x})&=V_0(\bar{x})+\alpha V_1(\bar{x})=(u,v)\in K^2.\label{eq:taggedseed}
\end{align}
Here $\alpha\in F$ is a fixed numerical tag as we described above, and we identify $F$ with
$K^2$ by using a fixed \(K\)-basis. The functions $V_0,V_1$ come
from Li's near-full-entropy affine non-malleable extractor~\cite{Li23}.
For distinct fixed tags $\alpha\ne\alpha'$ and any affine map
$h:L\to L$ (including the identity), the honest seed
$S_\alpha(U_L)$ is exponentially close to uniform even given
$S_{\alpha'}(h(U_L))$. The tag enters the evaluated starter only
through the coordinates $(u,v)$ of $S_\alpha(\bar{x})$.

Before the final output-shortening trace, the extractor computes the following $K$-valued
function:
\begin{equation}\label{eq:core}
 \mathcal E_\alpha(z,w)=E_\alpha(\bar{x};r,t,Y)
 =\underbrace{\Tr_{L/K}\!\left(tP_{\Gamma_c(\bar{x})}(r)\right)}_{\text{Gold recognition}}
 +\underbrace{vP_Y(u)}_{\text{tagged extraction}},
 \qquad (u,v)=S_\alpha(\bar{x}).
\end{equation}

Thus the extractor ignores the unused suffix of the honest left row,
but its arbitrary left tampering may depend on that suffix. The Gold
coordinates of $\bar{x}$ have binary degree at most two, whereas the
second term computes the right coefficient polynomial at the
seed coordinate $u$ and multiplies by $v$.

The two summands have distinct purposes: the \emph{Gold term}
recognizes when an arbitrary left tampering behaves like an affine tampering, while the
\emph{tagged evaluation term} rules out a constant mixed character once
that affine structure has been found. On the other hand,
the right tampering can be any arbitrary function.

More specifically, consider two independent sources that are uniform on sets $A$ and $B$, and two tampering functions $f$ and $g$. We will argue that for distinct tags $\alpha\ne\alpha'$, 
the mixed binary character
\[
 \chi(z,w)=\Tr_{K/\F_2}\!\left(
   \lambda\mathcal E_\alpha(z,w)
  +\mu\mathcal E_{\alpha'}(f(z),g(w))\right),
 \qquad \lambda\ne0,
\] is non-constant on $A \times B$, and then use approximate duality lift to reduce the bias to exponentially small on slightly larger product sets. The case where $\mu=0$ is easy to analyze, so we focus on the case where $\mu \neq 0$. Towards that, fix the unused left suffix in the analysis, the left tampering function $f$ is then a deterministic function of $\bar{x}$. With slight abuse of notation, let the corresponding prefix of the tampered left row be $f(\bar{x})$, and suppose that $\bar{x}$ is still supported on $A$. If $f$ is an affine function, then $\Gamma_c(f(\bar{x}))$ is also a quadratic function in the bits of $\bar{x}$. We thus measure how far $f$ is from an affine function by the \emph{Gold defect} $d_c(f;A)$, defined to be the dimension of the span of the binary coordinate functions of \(\Gamma_c\circ f\) modulo the space of all Boolean functions of degree at most $2$, restricted to $A$. 

If $d_c(f;A)$ is large, then this implies that the rank of the sum of the Gold terms in $\chi(z,w)$ is already large, because $\Gamma_c(\bar{x})$ are quadratic in the bits of $\bar{x}$, while $d_c(f;A)$ measures the dimension of $\Gamma_c(f(\bar{x}))$ outside of the space of quadratic Boolean functions. On the other hand, the rank of the sum of the tagged extraction part is small, since each of them expands into $J$ separated products over $K$ and has binary matrix rank at most $J\log_2|K|$. Together, this implies that $\chi(z,w)$ has binary matrix rank $>1$, and hence is non-constant.

Conversely, if $\chi(z,w)$ is a constant, then this forces $d_c(f;A)$ to be small.\ Our amortized
recognition theorem (Theorem~\ref{crn:thm:noninjective_recognition})
produces a large subset $A_1\subseteq A$ on which $f$ agrees with an affine map $h$. The crucial feature of our theorem is that the Gold defect \(\rho=d_c(f;A)\) contributes only \(O(\rho/c)\) to the entropy deficiency of the affine-agreement subset, rather than \(O(\rho)\). More precisely, the theorem gives an affine-agreement subset \(A_1\subseteq A\) of ambient entropy deficiency \(O(a+\rho/c+c)\), where \(a=n_0-\log_2|A|\). Thus in our application, we can choose $a$ to be a small enough constant fraction of $n_0$, and \(c\) to be a sufficiently large constant, so that the deficiency of \(A_1\) is an arbitrarily small fraction of \(n_0\). Although \(A_1\) need not be an affine subspace, the affine non-malleable extractor property on the ambient uniform affine subspace implies, by counting, that there exists a value \(z'\) such that
$\{S_\alpha(\bar{x}
):\bar{x}\in A_1,\ S_{\alpha'}(h(\bar{x}))=z'\}$
contains many distinct values. We then use this and a dimension argument to arrive at another contradiction. Roughly, we fix $S_{\alpha'}(h(\bar{x}))=z'$, as well as a pair of $(r, t)$ and its tampered version $(r', t')$, such that the right row still has much entropy. Now in the mixed character $\chi(z,w)$, every term in the summation becomes a local function of either $\bar{x}$ or $w$, except the tagged extraction term $vP_Y(u)$ on the original inputs. Using the bilinear form of $vP_Y(u)$ and the assumption that $\chi(z,w)$ is a constant, we can argue that two subspaces related to $w$ and an encoding of $\bar{x}$ are orthogonal, contradicting the bounds on their dimension.  

\paragraph{Extending to $C$ tamperings.}
The above tagged extractor is left-source-strong against one tampering. We next extend it to handle $C$ tamperings, using a recursive procedure similar to alternating extraction, which doubles the number of tamperings at each intermediate
level. However, we must maintain \emph{two different invariants}: the 
left-source strong extractor property, needed to run the recursion, and bounded
\emph{field separation}, needed to keep the global binary character
rank linear in the input length. In particular, a bound on the binary matrix rank alone
would not suffice for the second invariant, because the intermediate steps perform non-linear operations and may increase the rank.

At level $i+1$, let $Z,W\in\F_2^N$ be the two input rows, and let $\Ext^i$ be the left-source-strong extractor we constructed at the previous level. First we apply $\Ext^i$ to two short prefixes of $Z$ and $W$, say $Z'$ and $W'$, using the same tag throughout the construction. Suppose $\Ext^i$ produces $\mathsf{R}=\mathsf{R}(Z',W')$ of length $w_0$.
Put $q=C+1$ and $v=w_0/(256q)$, let $P$ be a prefix of $\mathsf{R}$ with $64v$ bits. Compute
\[
  \mathsf{L}=H_{8v}(Z;P),\qquad
 \Ext^{i+1}(Z,W)=H_v(\mathsf{R};\mathsf{L}).
\]
Here $H_t$ is a standard strong seeded extractor of a particular form:
for a seed $(u,t_0)$ over $\F_{2^{4t}}$, we treat the source as a coefficient word
$d=(d_j)$, and
\[
 H_t(d;(u,t_0))
 =\Tr_{\F_{2^{4t}}/\F_{2^t}}
       \!\left(t_0\sum_j d_j u^j\right).
\]
In both $H_{8v}$ and $H_v$, the input sources have length a fixed multiple of the relevant
field degree, so the coefficient polynomials have degree $O_C(1)$.

\emph{The statistical induction.}
Suppose $\Ext^i$ handles $C_i$ comparisons, we show that $\Ext^{i+1}$ can handle $C_{i+1}=2C_i$ tamperings. Towards that, divide the
$2C_i$ tamperings into two groups $T,S$ of size $C_i$. Let $\mathsf{R}_0$ and $\mathsf{L}_0$ be the outputs of $\mathsf{R}$ and $\mathsf{L}$ on the original inputs. We will gradually fix a sequence of random variables while preserving the independence between $Z$ and $W$.

First fix the prefixes $Z'$ in all executions, thus all the $\mathsf{R}$ variables produced are deterministic functions of $W$, and $Z$ still has enough entropy if we choose $Z'$ to be short enough. By the fact that $\Ext^i$ handles $C_i$ tamperings, $P_0$ is close to uniform and independent of $\mathbf P_S=(P_j)_{j \in S}$. Thus, the strong extractor property of $H_{8v}$ guarantees that $\mathsf{L}_0$ is close to uniform and independent of $\mathbf L_S=(\mathsf{L}_j)_{j \in S}$. Furthermore, we can fix all the $P$ variables while still preserving this property, and now the $\mathsf{L}$ variables are deterministic functions of $Z$. Notice that at the beginning, again by the fact that $\Ext^i$ handles $C_i$ tamperings and is left source strong, $\mathsf{R}_0$ is close to uniform and independent of all the $\mathsf{R}$-variables in the tampering set $T$. We can therefore fix all these $\mathsf{R}$-variables, in addition to the previous fixing of all the $P$ variables. By our choice of the length parameters, $\mathsf{R}_0$ still has enough entropy. Finally, the strong extractor property of $H_v$ guarantees that the new output of $\Ext^{i+1}$ on the original inputs is close to uniform and independent of all the tampering outputs in $S$, and all the $\mathsf{L}$ variables. Since all the $\mathsf{R}$-variables in $T$ have already been fixed, this means the output of $\Ext^{i+1}$ on the original inputs is actually close to uniform and independent of all the tampered outputs in $S\cup T$, as well as strong in the left source.  

\emph{Achieving the rank properties.} Recall that as in the analysis of the standard two source extractor, we need the output of each row of pairs to satisfy the following two rank properties: (1) every mixed character of the output and the tampered outputs, where the honest-output coefficient is nonzero, can have rank at least $K_* N$ for every prescribed constant $K_*>0$ when restricted to a high entropy rectangle $A \times B$, which we need to merge the final outputs together; and (2) the global rank of any character of the output is bounded by $O(N)$, which we need for the final step of approximate duality lift. This is the place where we use the special form of the strong seeded extractors $H$ above. The idea is that we will work over a suitable common
base field $E=\F_{2^{\Theta_C(N)}}$, so that every field
word appearing in this fixed-depth recursion has only $O_C(1)$
coordinates over $E$. We choose the coordinate bases compatibly over \(E\), and all prefixes and coefficient selections use whole \(E\)-coordinate blocks. Thus these selections and the relative traces are \(E\)-linear, while multiplication has a bounded coordinate expansion. For an $E$-valued function of the two
\emph{unconditioned} inputs, define its separation width by
\[
 \operatorname{sep}_E(U)
 :=\min\left\{r:\ U(z,w)=\sum_{i=1}^r a_i(z)b_i(w),\quad
              a_i,b_i\text{ are $E$-valued}\right\}.
\]
The point is that if the $E$-valued function has constant separation width, then any parity of the output bits has $O(N)$ binary matrix rank. In particular, the $H$ functions above have constant separation width over $E$. The separation width has the following elementary closure rules:
\[
 \begin{aligned}
 \operatorname{sep}_E(U+V)
    &\le \operatorname{sep}_E(U)+\operatorname{sep}_E(V),\\
 \operatorname{sep}_E(UV)
    &\le \operatorname{sep}_E(U)\operatorname{sep}_E(V),\\
 \operatorname{sep}_E(U^{2^j})
    &\le \operatorname{sep}_E(U).
 \end{aligned}
\]
Applying these rules inductively, and noting that we are only doing the recursion for a constant number of steps, we can bound the separation width of all intermediate outputs to be $O_C(1)$.

However, we also need to get a lower bound on the binary rank restricted to a high entropy rectangle $A \times B$. Therefore we change \emph{the last level of the recursion}, and replace the function $H_v(\mathsf{R};\mathsf{L})$ by
\[
 \Tr_{L/K}\!\left(bt(1+ur)^{J-1}\right),
 \qquad \mathsf{L}=(u,b)\in K^2,\quad \mathsf{R}=(r,t)\in L^2,
\]
where $J$ is a sufficiently large fixed power of two. In characteristic $2$, the identity
$(1+ur)^{J-1}=\sum_{i=0}^{J-1}(ur)^i$ expresses this as a
fixed-width polynomial in already separated field words; hence its
\emph{global} binary character rank is still $O_{C,J}(N)$.
But, on the product rectangles obtained from our argument, by a similar argument as in the standard two source extractors, the large supports of the
honest $\mathsf{R}_0$ and $\mathsf{L}_0$ give the independent \emph{restricted} rank
lower bound
\[
 \operatorname{rank}_{\mathrm{restricted}}
 \ge 4Jv-\frac{w_0}{3}-5.
\]
Choosing $J$ after the entropy parameters can thus make this larger than
any prescribed constant times $N$. 

\paragraph{Finishing the extractor.} The rest of the construction and the analysis are similar to the standard two source extractor we described before. The final step is to merge the constant number of outputs together, and in the analysis we make two other uses of approximate duality lift to reduce the error to exponentially small. 

\paragraph{Achieving optimal output length.} We use our extractors with $\Omega(n)$ output length and $2^{-\Omega(n)}$ error to achieve modified extractors with optimal output length for a given error $\e$. Here we illustrate the main ideas by using the standard two source extractor that achieves output length $2k-O(\log(1/\eps)+1)$ as an example. Our previous two source extractor $G(X, Y)$ with error $2^{-\Omega(n)}$ is actually a strong two source extractor, so this suggests the following natural approach: first use that two source extractor to produce a short seed $S=G(X,Y)$ with $d$ bits, then take two strong seeded extractors with optimal parameters $E_L, E_R$, and output $$\Ext(X,Y)=\bigl(S,E_L(X,S),E_R(Y,S)\bigr).$$ The strong property guarantees that either $(S,E_L(X,S))$ or $(S,E_R(Y,S))$ is close to uniform. However, proving that $\bigl(S,E_L(X,S),E_R(Y,S)\bigr)$ is close to uniform is not straightforward.

Let $s=\lceil\log_2(1/\varepsilon)\rceil$, we choose the two output lengths asymmetrically: $E_L(X,S)$ outputs $p_L=k-d-2s-O(1)$ bits, and $E_R(Y,S)$ outputs $p_R=k-O(s+1)$ bits. In the analysis, the main issue is the dependence of $S$ on both sources.
We resolve this by bounding the conditional collision probability
of the first output and seed, weighted by the actual marginal
distribution of $Y$. By using Cauchy Schwarz, one can convert the resulting conditional collision probability bound into statistical distance. This naturally suggests that we need to consider the output of $G(X, Y)$ on different copies of $X$ conditioned on $Y$, which can be captured by the stronger property of \emph{one-sided non-malleability}. That is, we only need $G(X, Y)$ to be non-malleable against one sided tampering. Therefore, we can for example simply take $G(X, Y)$ to be our non-malleable two source extractor described above. With the parameter choices in Section~4.1, this gives
\[
 \bigl(Y,S,E_L(X,S)\bigr)
 \approx_{\varepsilon/4}
 \bigl(Y,U_d,U_{p_L}\bigr),
\]
where the uniform variables on the right are independent of each other
and of $Y$. Once we have this property, we can safely
append the second output of $E_R(Y, S)$ by using the strong seeded extractor $E_R$, and the output length and error follow. 

For strong non-malleable extraction against $C$ tamperings, we
use the symmetric construction
\[
 \nmExt(X,Y)=\bigl(S,E(X,S)\oplus E(Y,S)\bigr),
\]
where $E$ is a strong seeded extractor. Here the seed must withstand $2C+1$ tamperings, because the collision argument
introduces a second input for each original and tampered input. Fortunately, our multi-tampering non-malleable two source extractor with $\Omega(n)$ output length and $2^{-\Omega(n)}$ can still handle this.

For sumset extractors, we use a seed $S=G(Z)$ computable from
$Z=X+Y$, by using our previous sumset extractor, together with a linear seeded extractor $\LExt$. As before, we will need the sumset extractor to have one-sided non-malleability, thus we use a slightly modified construction which achieves this property in Theorem~\ref{ns:thm:seed}.
The construction is
\[
 \operatorname{SumExt}(Z)=\bigl(S,\LExt(Z, S)\bigr).
\]

\paragraph{Achieving $Cn/\log n$ entropy.} Our constructions above can be directly pushed to handle slightly sub-constant entropy rate. However, to achieve $Cn/\log n$ entropy, we use a somewhere high entropy condenser such as those in \cite{BarakRSW06, Zuckerman, LZ}, which outputs a  polylogarithmic number of rows each of length \(n/\operatorname{polylog}n\), such that at least one row is exponentially close to having high entropy. Suppose we get $r=2^d$ rows from each source, now we try every possible pair of rows from the two sources, resulting in $r^2$ pairs such that at least one pair has high entropy from both sources. 

From here, we simply treat each row as a vector in $F^t$ for some appropriately chosen field $F=\F_{2^{\ell}}$, and apply the standard inner product two source extractor over $F$. We show that by choosing the parameters appropriately, the joint distribution of the $r^2$ outputs is exponentially close to a convex combination of \emph{affine somewhere random sources}, which is an affine source on $r^2$ rows such that some row is actually uniform. 

This is motivated by the work of Aggarwal, Dodis, and Lovett \cite{ADL14} on split-state non-malleable codes, where they showed that if $L$ and $R$
are two independent uniformly random strings over $\F^n_p$ for some prime number $p$, and $f, g:  \F^n_p \to \F^n_p$ are two arbitrary functions on $L$ and $R$, then the joint distribution $(\langle L, R \rangle, \langle f(L), g(R) \rangle)$ is “close” to the convex combination of “affine
distributions” $\{(U, aU + b) | a, b \in \F_p\}$, where $U$ is the uniform distribution over $\F_p$. In our situation, instead, the $L$ and $R$ are two independent high entropy sources we obtain from the outputs of the somewhere high entropy condenser on $X$ and $Y$, and by fixing extra randomness, the other outputs we get from the condenser can be viewed as deterministic tampering functions on $L$ and $R$. Furthermore, in our application we have $r^2-1$ pairs of tampering functions instead of just one pair. There is also the technical issue that now we are working over the field $\F_{2^{\ell}}$ instead of a prime field $\F_p$. 

Nevertheless, we can indeed prove such a binary-affine, many-output analogue  of the result in \cite{ADL14} for high-min-entropy sources. The proof again uses the approximate duality formulation in \cite{BenSassonLR14} and roughly goes as follows. First of all, instead of using an arbitrary somewhere high entropy condenser which has non-linear maps, we use the linear condenser in \cite{LZ} as we did in our construction of sumset extractors. The advantage is that in the analysis, every binary parity character of an inner product output is actually a bilinear function in the \emph{original input bits}. This is still true if we consider a joint binary parity character of the $r^2$ outputs. Thus the combined feature dimension is only $2n$, independent of the number of pairs $r^2$, which helps us get better parameters when applying approximate duality. On the other hand, using a non-linear somewhere high entropy condenser could result in a much larger feature dimension that depends on $r^2$. Now, in the analysis we look at the joint distribution of the $r^2$ outputs and consider the affine span of the support. If the distribution is already close to being uniform on the affine subspace, then we are essentially done because the inner product guarantees that one output is close to uniform. Otherwise, some non-trivial character on this affine subspace has large bias under the distribution, and we can invoke approximate duality to find a large rectangle on which that character is constant. We then use this rectangle to divide the original input space into disjoint rectangles (thus maintaining the property of independent sources) and continue this process. Using a potential function, we show that this process terminates in a finite number of steps, and except for a small probability mass, the distributions on the outputs of the rest rectangles are close to being uniform on an affine subspace. Furthermore, we show that we don't lose much entropy in this process, and therefore on any sufficiently large rectangle, by the property of the condenser and the inner product two source extractor, some good row projection of the output affine subspace has full support. The uniform distribution on this affine subspace therefore has an exactly uniform honest row. We can now conclude that the original joint distribution is close to a convex combination of affine somewhere-random sources with $r^2$ rows.

As long as the number of rows $r^2$ here is relatively smaller than the length of each row, there are very good known extractors, e.g., those by Rao \cite{Rao:lowweight} using repeated conditioning and later works using affine correlation breakers \cite{ChattopadhyayGL21, Li23} that can extract random bits with exponentially small error from such affine somewhere random sources. For non-malleable extractors, some extra work needs to be done here, and we eventually apply an advice based correlation breaker. 

However, the above approach only gives us a two source extractor with error $2^{-k^{\Omega(1)}}$ instead of $2^{-\Omega(k)}$, even for constant entropy rate. Indeed, the error comes from two steps in the above construction: the step of converting to an affine somewhere random source, and the later step of extracting from such a source. To get $2^{-\Omega(k)}$ error from the second step, even with a constant number of rows, one would need the row length to be $\Omega(k)$, which will force us to take a field of size $2^{\Omega(k)}$ in the first step and therefore only a constant number of blocks in the inner product. In this case, the analysis of approximate duality breaks. The limitation can be seen more precisely from the
entropy-loss bound in our regularization theorem (Theorems~\ref{lr:thm:reg}
and~\ref{nm:thm:regularize}). For a constant number of rows and constant entropy rate,
if each field inner-product output has $\ell$ bits, then the two principal contributions to the regularization
entropy loss, with rank threshold $R$, are roughly 
$O\left(\frac{k\ell}{R}\right)$ from high-rank approximate-duality, and  $O(\ell R)$ from low-rank exact-message steps.
Balancing these two terms gives $R=\Theta(\sqrt{k})$ and a combined
entropy loss of $O(\ell\sqrt{k})$. This forces us to choose $\ell=O(\sqrt{k})$ with a sufficiently small
constant factor. Hence we only get a square-root scale of the output
length and error exponent by this approach. One can combine with a strong seeded extractor to increase the output length to $\Omega(k)$, but the error will still be $2^{-\Omega(\sqrt{k})}$. At entropy rate \(\Theta(1/\log n)\), the number of rows we get from the condenser further increases to poly-logarithmic, and output-extension losses lead to \(2^{-n^{1/2-\vartheta}}\) error range for every fixed \(\vartheta>0\). Not surprisingly, our extractors for constant entropy rate with asymptotically optimal parameters use genuinely different constructions and techniques. 


\paragraph{Independent work and Acknowledgments.} The work of Doron and Goodman \cite{DG26} constructed explicit two-source extractors for entropy $k \geq n^{1-\alpha}$ with error $2^{-k^{\alpha}}$ for some constant $\alpha>0$. Their error is still $2^{-k^{\alpha}}$ for constant entropy rate. Hence their extractors work for smaller entropy than ours, while our extractors have better error and output length for any constant entropy rate. Their techniques share some similarity with our  extractors for entropy $Cn/\log n$, in that both constructions use approximate duality to first convert the two sources into a convex combination of affine somewhere random sources, and then apply an extractor for such sources. Doron and Goodman went further, and used additional new ideas and techniques to reduce the entropy to $k \geq n^{1-\alpha}$. On the other hand, our asymptotically optimal extractors for constant entropy rate also use new ideas and techniques that are unrelated to  their results.

Doron and Goodman obtained their preliminary results at the end of December 2025. While they were filling in the details and writing up their results, the author independently discovered the extractors for entropy $Cn/\log n$ in this paper around August 2026, without being aware of their results. The author has since been pursuing two directions: reducing the entropy requirement and improving the error. However, while attending the Pseudorandomness and High-Dimensional Expansion program at Simons Institute in September 2026, the author learned their unpublished results and their high level ideas. The author then focused on reducing the error of the extractors, and finally obtained the new constructions of extractors for constant entropy rate with asymptotically optimal parameters. The author thanks
the Simons Institute for its hospitality and support.

Another recent independent work by Lewko \cite{Lewko2026VinogradovSystem} also gives an explicit two source extractor for any constant entropy rate with exponentially small error. His construction is based on a finite-field moment curve, and also uses the theorem of Gowers, Green, Manners and Tao on Marton’s conjecture in abelian groups with bounded torsion \cite{GowersGreenMannersTao2026Marton}.

\paragraph{AI Disclosure.} The author used ChatGPT Astra substantially for exploring proof ideas, literature survey, and drafting the manuscript. Most of the high level ideas belong to the author, while ChatGPT also contributed important technical ideas towards the complete proofs. ChatGPT Astra is also used to help with the tedious work of choosing appropriate parameters. During numerous exchanges with ChatGPT, many intermediate approaches have failed before the final successful constructions emerged. The constructions presented here also include simplifications of previous, more complicated constructions.

\paragraph{Organization of the paper.}
Section~\ref{un:tools} collects the common preliminaries, including source-strong
extractors, rank-sensitive approximate duality, condensers, seeded
primitives, and the shared-feature regularization theorem.
Section~\ref{un:linear} develops the constant-rate constructions with linear output
length and exponentially small error: standard two-source extractors,
sumset extractors, and non-malleable two-source extractors against any
fixed number of tamperings.
Section~\ref{un:optimal} uses the preceding constructions as short seeds and develops
entropy-recycling arguments to obtain optimal output lengths for the above extractors. Finally, Section~\ref{un:lowrate} gives constructions for entropy $k=\Omega(n/\log n)$.

\section{Preliminaries and common tools}\label{un:tools}
A cited theorem is used only under its stated hypotheses. Internal lemmas and their proofs are arranged in the order needed by the constructions. The same symbol may denote a local field length in different subsections; the original input length remains $n$.

\paragraph{Notation and conventions.}
For a nonnegative integer $r$, write $[r]=\{1,\ldots,r\}$, with
$[0]=\varnothing$, and $(x)_+=\max\{x,0\}$. The notation $\1_E$
denotes the indicator of an event or set $E$. For a nonempty finite
set $S$, $U_S$ is its uniform distribution; $U_m$ abbreviates
$U_{\F_2^m}$. A product $U_S^{\otimes r}$ consists of $r$ independent
uniform copies. In an ideal comparison, a newly introduced uniform
variable is independent of the retained variables unless stated otherwise.
We write $\mathcal L(X)$ for the distribution of $X$, $\supp X$ for its
positive-probability support, and $X\approx_\eps Y$ when
$\TV(\mathcal L(X),\mathcal L(Y))\le\eps$, using the statistical
distance defined below. The symbol $\perp$ between random tuples means
independence. For a distribution $P$ on a finite alphabet, its collision
probability is $\cp(P)=\sum_z P(z)^2$; we also write
$\cp(X)=\cp(\mathcal L(X))$. For a subdistribution the same sum
is used, without normalizing its mass.

Unless another field is specified, vector-space spans, dimensions,
codimensions, orthogonality, and matrix ranks are over $\F_2$.
The notation $\rk_2 h$ for a Boolean kernel means the rank of its
matrix of $0/1$ values, not the real rank of its sign matrix.
For a finite set $A$, $\F_2^A$ denotes the space of all functions
$A\to\F_2$. For a nonempty subset $S$ of a binary vector space and
any $s_0\in S$, its affine hull is
\[
 \aff(S)=s_0+\Span_{\F_2}\{s-s_0:s\in S\}.
\]
Its dimension is the dimension of the displayed direction space.
An affine source is uniform on an affine subspace; a tuple is
\emph{jointly affine} when its concatenation has this property.
For a kernel $Q$, the notation $Q(A,B)$ means
$\{Q(x,y):x\in A,\ y\in B\}$, not a probability distribution. All affine
maps between binary-coordinate representations of fields are
$\F_2$-affine unless another field is specified.

For a bit string $z$, $\operatorname{prefix}_d(z)$ denotes its first
$d$ coordinates, with $0\le d\le |z|$. The operation $\oplus$ on bit
strings is coordinatewise addition modulo two, while
$z\mathbin{\|}z'$ denotes concatenation. On vector spaces or codes,
$\oplus$ denotes a direct sum. For a field $K$, both $K^*$ and
$K^\times$ denote $K\setminus\{0\}$.

\begin{definition}[Ordinary two-source and sumset extraction]
An ordinary two-source extractor at component entropy $k$ and error
$\eps$ is a map $E:\bits^n\times\bits^n\to\bits^m$ such
that $E(X,Y)\approx_\eps U_m$ for every independent pair
with min-entropies at least $k$. A sumset extractor is a map
$S:\F_2^n\to\bits^m$ such that $S(X+Y)\approx_\eps U_m$
for every such pair. A guarantee for a sum-only evaluator does not
by itself assert source strength after an original summand is revealed.
\end{definition}
\begin{definition}[Non-malleable two-source extraction]
A map $E:\{0,1\}^n\times\{0,1\}^n\to\{0,1\}^m$ is a
$t$-non-malleable two-source extractor at entropy $k$ and error
$\eps$ if, for independent sources $X,Y$ of entropy at least
$k$ and arbitrary separate maps $f_j,g_j$ satisfying
$\Pr[(f_j(X),g_j(Y))=(X,Y)]=0$, one has
\[
 \TV\bigl((E(X,Y),(E(f_j(X),g_j(Y)))_{j=1}^t),
          (U_m,(E(f_j(X),g_j(Y)))_{j=1}^t)\bigr)\le\eps .
\]
The fresh uniform variable is independent of the actual joint
tampered-output distribution. No entropy assumption is made on tampered inputs.
For $t=0$ the comparison tuple is empty and the definition is ordinary
two-source extraction. All fixed-$C$ statements use this convention.
\end{definition}

\subsection{Source-strong extraction and entropy-slack transfer}\label{sec:sourcestrong}
\begin{definition}[Separate source strength]\label{pre:def:sourcestrong}
Let $Z=F(X,Y)$, and, for an admissible tuple of $t$ separate
tamperings, put $T_j=F(f_j(X),g_j(Y))$ and
$\mathbf T=(T_1,\ldots,T_t)$. A $t$-non-malleable function $F$
is \emph{source strong with error $\xi$} at component entropy $k$
if, for all independent $k$-sources and all such tamperings,
\[
 \TV((Z,\mathbf T,X),(U_m,\mathbf T,X))\le\xi,
 \qquad
 \TV((Z,\mathbf T,Y),(U_m,\mathbf T,Y))\le\xi.
\]
The two guarantees are separate; neither discloses $(X,Y)$ jointly.
For ordinary source-strong extraction, set $t=0$ and omit
$\mathbf T$.
\end{definition}

\begin{lemma}[A source-strong output cannot exceed one source's entropy]
\label{pre:lem:strong-output-cap}
Let $k$ be an integer with $0\le k\le n$. If a deterministic
two-source function is source strong with error
$\eps<1$ at component entropy $k$ and output length $m$, then
\[
 \eps\ge1-2^{k-m},
 \qquad\text{hence}\qquad
 m\le k+\log\frac1{1-\eps}.
\]
In particular, for $\eps<1/2$, a source-strong output cannot
contain more than $k$ bits.
\end{lemma}
\begin{proof}
Take $Y$ uniform on exactly $2^k$ points and choose any
independent $k$-source $X$. For each fixed value $X=x$, the
conditional honest output is a function of only the $2^k$
points in the support of $Y$, hence has support size at most
$2^k$. Its distance from uniform on $2^m$ outputs is at least
$1-2^{k-m}$. Average over $x$ to obtain the strong-distance
lower bound. Rearranging proves the claim.
\end{proof}

\paragraph{Finite initial lengths.}\label{pre:finite-lengths}
When a linear-output theorem is stated for every positive input
length but its field construction is given only for $n\ge n_0$, the
following deterministic completion suffices. For $1\le n<n_0$,
return one constant zero bit. Even with an arbitrary comparison
tuple and either original source retained, replacing that bit by an
independent uniform bit has statistical distance exactly $1/2$.
Every output character has binary truth-table rank zero. If the
large-length output and error constants are $c,a>0$, replace them by
$\min\{c,1/n_0\}$ and $\min\{a,1/n_0\}$. Then the one-bit completion
satisfies both $m(n)\ge cn$ and error at most $2^{-an}$ throughout
the finite initial range. This convention is invoked explicitly
below; it does not extend an additive-loss theorem beyond its
stated nontrivial error range.

\begin{lemma}[Atomwise entropy-slack strongification]
\label{pre:lem:strongify-general}
Let $F$ be a $t$-non-malleable two-source extractor at component
entropy threshold $h$, output length $m_0$, and error $\gamma$,
where $t=0$ denotes ordinary extraction. Fix $k\ge h$, let
$0\le r\le m_0$, and let $F_r$ be the first $r$ output bits.
For every independent $k$-source pair and every admissible tuple of
$t$ tamperings, both source-strong comparisons for $F_r$ have
error at most
\begin{equation}\label{pre:eq:strongify-general}
 2^{(t+1)r}\bigl(\gamma+2^{h-k}\bigr).
\end{equation}
The statement remains true if a fixed binary-linear surjection of
rank $r$ replaces coordinate truncation, with all actual tampered
outputs projected through the same map.
\end{lemma}
\begin{proof}
Truncation and any fixed surjective linear projection preserve the
initial $t$-non-malleability error by data processing, so work with
the $r$-bit outputs. Let $\mathbf T$ denote the actual comparison
tuple. Fix an atom $(z,\mathbf v)$ of the $(t+1)r$ output bits and,
for each numerical $x$ in the original left support, define
\[
 D_{z,\mathbf v}(x)
 =\Pr[Z=z,\mathbf T=\mathbf v\mid X=x]
 -2^{-r}\Pr[\mathbf T=\mathbf v\mid X=x].
\]
Write $b=k-h$. If the set
$E=\{x:D_{z,\mathbf v}(x)>\gamma\}$ had probability
at least $2^{-b}$, conditioning $X$ on $E$ would leave
min-entropy at least $h$ (each conditional atom is at most
$2^{-k}/\Pr[E]$). The independent source $Y$ would remain
unchanged, and the original tampering maps would remain admissible
on the smaller product support. On that source, the positive
excess of this single joint atom would exceed $\gamma$, contrary
to non-malleability. Thus $\Pr[E]<2^{-b}$.
Because $|D_{z,\mathbf v}(x)|\le1$,
\[
 \E_X[D_{z,\mathbf v}(X)]_+
 \le\gamma+2^{-b}.
\]
The conditional signed atom differences sum to zero. Hence the
conditional total variation equals the sum of their positive parts.
Summing over the $2^{(t+1)r}$ atoms proves the first comparison of
\eqref{pre:eq:strongify-general}. Repeating the argument after
exchanging $X$ and $Y$ proves the second. No independence or entropy
is assumed for the tampered images.
\end{proof}

\begin{remark}[Recycling parameters and conditional entropy]\label{pre:rem:recycling-conventions}
For source-strong output extension, $C\ge0$ counts the final tampered
outputs, while the short seed is required to handle $R=2C+1$ auxiliary
tamperings. This distinction is maintained throughout Section~\ref{rec:sec:main}.
For finite variables, write
\[
 \widetilde H_\infty(A\mid T)
 =-\log\sum_t\Pr[T=t]\max_a\Pr[A=a\mid T=t].
\]
Revealing a variable with at most $2^b$ values costs at most $b$ in
this average conditional min-entropy. Markov's inequality then gives
pointwise entropy at least the average bound minus $\log(1/\xi)$,
outside mass $\xi$. For subdistributions we use half the $\ell_1$
distance as well; discarded mass is charged explicitly in the centered recycling proof. Conditional source independence is used only on
rectangular communication leaves or in an explicitly independent
reference experiment.
\end{remark}

\subsection{Elementary probability, Fourier analysis, and rank}\label{sec:prelim}
\begin{definition}[Sources and extraction]
For a finite random variable $X$, its min-entropy is
$\Hin(X)=-\log\max_x\Pr[X=x]$.
A \emph{flat $k$-source}, for integral $k$, is uniform on a set of
exactly $2^k$ points. We use the statistical distance
$\TV(P,Q)=\frac12\sum_z|P(z)-Q(z)|$ between finite distributions $P,Q$.

A strong seeded $(k,\eps)$ extractor $E$ satisfies
\[
 \TV((U_d,E(X,U_d)),U_{d+m})\le\eps
\]
for every source of min-entropy at least $k$, with $U_d$ independent of $X$.
An ordinary two-source extractor uses independent $X,Y$; separate strength
means the two guarantees retaining $X$ or retaining $Y$, separately.
A Boolean two-source \emph{disperser} at side threshold $k$ is nonconstant
on every product of supports of sizes at least $2^k$.
\end{definition}

\begin{lemma}[Flat decomposition and elementary conditioning]\label{pre:flat}
Every distribution with point probabilities at most $1/K$, for an integer
$K\ge1$, is a convex combination of uniform distributions on $K$-element
sets. Independent such decompositions produce a convex combination of
product distributions. If an event $A$ depends only on $X$, then for every
deterministic view $T$,
\[
 \Hin(T(X)\mid A)\ge\Hin(T(X))+\log\Pr[A].
\]
Separate events in independent sources preserve independence.
\end{lemma}
\begin{proof}
The polytope $\{p:\sum_xp_x=1,\ 0\le p_x\le1/K\}$ has only the
uniform distributions on $K$-element sets as extreme points. Indeed, if two
coordinates are strictly between their bounds, shifting a sufficiently small
amount between them expresses the point as a nontrivial midpoint. If only
one coordinate were fractional, the equation $\sum p_x=1$ and integrality
of $K$ would make it a bound coordinate too. A finite-dimensional polytope is
the convex hull of its extreme points. Tensor the two convex decompositions
for the independence assertion. The conditioning bound follows from
$\Pr[T(X)=z\mid A]\le\Pr[T(X)=z]/\Pr[A]$. The joint conditional distribution under
separate events factors directly.
\end{proof}
Integer floors in flat decompositions cost at most one entropy bit; every
later use reserves a fixed positive linear entropy margin.

\begin{lemma}[Source-local disclosure]\label{pre:disclosure}
If $\Hin(X)\ge k$ and $L(X)$ takes at most $2^v$ values, then, except on
labels of total probability at most $2^{-t}$,
\[
 \Hin(X\mid L(X)=l)\ge k-v-t.
\]
\end{lemma}
\begin{proof}
Discard labels of probability below $2^{-(v+t)}$. There are at most $2^v$
of them. On each remaining label, apply Lemma~\ref{pre:flat}.
\end{proof}

\ifdefined\transcriptlemma\else
  \begingroup
    \theoremstyle{plain}
    \newtheorem{transcriptlemma}{Lemma}
  \endgroup
  \renewcommand{\thetranscriptlemma}{\thesection.9a}
\fi

\begin{transcriptlemma}[Interactive-transcript disclosure]\label{pre:interactive-transcript}
Let $X,Y$ be independent finite random variables. Let $T$ be the
transcript of a finite deterministic binary communication protocol:
at each internal node the speaker is determined by the preceding
transcript, and the next bit is a function of that speaker's input
and the preceding transcript. Suppose the left and right parties send
at most $b_L$ and $b_R$ bits, respectively, on every execution.
Every supported transcript leaf preserves independence of $X,Y$.
For every $u>0$, outside transcript probability at most $2^{1-u}$,
simultaneously for every deterministic view $V(X)$ and $W(Y)$,
\begin{align*}
 \Hin(V(X)\mid T=\tau)&\ge \Hin(V(X))-b_L-u,\\
 \Hin(W(Y)\mid T=\tau)&\ge \Hin(W(Y))-b_R-u.
\end{align*}
The same good leaves work for all views: conditioning on such a leaf
increases the probability of any event depending only on $X$ by at most
$2^{b_L+u}$, and of any event depending only on $Y$ by at most
$2^{b_R+u}$. Moreover, with the average conditional min-entropy
convention of Remark~\ref{pre:rem:recycling-conventions},
\[
 \Hav(V(X)\mid T)\ge\Hin(V(X))-b_L,\qquad
 \Hav(W(Y)\mid T)\ge\Hin(W(Y))-b_R.
\]
\end{transcriptlemma}
\begin{proof}
Each supported transcript has a rectangular preimage
$A_\tau\times B_\tau$. Put
$p_\tau=\Pr[X\in A_\tau]$ and $q_\tau=\Pr[Y\in B_\tau]$.
Independence gives $\Pr[T=\tau]=p_\tau q_\tau$, and the conditional
distribution on this leaf is the product of $X\mid A_\tau$ and $Y\mid B_\tau$.

Fix a numerical right input $y$. Pruning the protocol tree to that
input leaves only left-message branching: each right message is
determined by $y$ and the preceding transcript. The pruned binary
tree has at most $b_L$ branching bits along a path, hence at most
$2^{b_L}$ leaves. Averaging over $y$ gives
\[
 \sum_\tau q_\tau
 =\E_Y\bigl|\{\tau:Y\in B_\tau\}\bigr|\le2^{b_L}.
\]
The symmetric argument gives $\sum_\tau p_\tau\le2^{b_R}$.
Consequently,
\[
 \Pr_T[p_T<2^{-b_L-u}]
 =\sum_{\tau:p_\tau<2^{-b_L-u}}p_\tau q_\tau
 \le2^{-b_L-u}\sum_\tau q_\tau\le2^{-u}.
\]
Similarly, $\Pr_T[q_T<2^{-b_R-u}]\le2^{-u}$. On every leaf
outside the union of these two exceptional sets, every left-local
event $E$ satisfies
\[
 \Pr[E\mid T=\tau]
 =\Pr[E\mid X\in A_\tau]
 \le\frac{\Pr[E]}{p_\tau}\le2^{b_L+u}\Pr[E].
\]
This gives the left event and min-entropy bounds, simultaneously for
all left-local views. The same argument proves the right bounds.
No union bound over the number of views is used.

Finally, write $a_V=\max_v\Pr[V(X)=v]$. The same conditional-atom
bound, now averaged over all supported leaves, gives
\[
 \sum_\tau\Pr[T=\tau]\max_v\Pr[V(X)=v\mid T=\tau]
 \le\sum_\tau p_\tau q_\tau\frac{a_V}{p_\tau}
 \le2^{b_L}a_V.
\]
Taking negative binary logarithms proves the average conditional
min-entropy assertion on the left; the right is identical.
The proof does not require either source to be flat or any view to
be injective.
\end{proof}

\begin{lemma}[Conditional representative tables]\label{pre:tables}
For any finite pair $(U,X)$, its distribution is a convex combination of distributions
$(U,f(U))$, where the random choice of the deterministic table $f$ is
independent of the subsequent $U$ and every component has the original
$U$ marginal. This operation can be performed independently on two sides
of a product distribution.
\end{lemma}
\begin{proof}
Independently for every $u$ of positive probability, draw $f(u)$ according
to $X\mid U=u$, and then draw $U$ from its original distribution. Averaging over
all tables reproduces the pair. Fill zero-probability entries arbitrarily.
If $U$ is a function of $X$, every sampled representative is in its correct
fiber. Independence of the two table selections proves the last assertion.
\end{proof}

\begin{lemma}[Fourier distance]\label{pre:fourier}
Let $P,Q$ be distributions on $\F_2^m$, and write
$\widehat P(\lambda)=\E_{Z\sim P}(-1)^{\lambda\cdot Z}$. Then
\[
 \TV(P,Q)\le\frac12
 \left(\sum_{\lambda\in\F_2^m}
       |\widehat P(\lambda)-\widehat Q(\lambda)|^2\right)^{1/2}.
\]
In particular, if every nonzero character of $P$ has bias at most $\beta$,
then $\TV(P,U_m)\le2^{m/2}\beta/2$. If $(A,B)$ is compared with
$(U_a,B)$, only characters with a nonzero $A$ coefficient contribute.
\end{lemma}
\begin{proof}
Parseval gives $2^m\sum_z(P(z)-Q(z))^2=
\sum_\lambda|\widehat P(\lambda)-\widehat Q(\lambda)|^2$.
Apply Cauchy--Schwarz to $\frac12\sum_z|P(z)-Q(z)|$.
The assertions about uniform marginals follow by orthogonality.
\end{proof}
Statistical distance contracts under deterministic maps and stochastic
kernels, and is convex under mixtures with common weights. These facts
follow directly from the triangle inequality for the defining $\ell_1$ sum.

\begin{definition}[Separation and common features]
A field-valued kernel has separation length at most $R$ over $F$ if
$H(x,y)=\sum_{i=1}^R a_i(x)b_i(y)$. For a Boolean kernel this is its
truth-table rank over $\F_2$. A vector output has \emph{common binary
features of dimension $R$} if there are fixed $u(x),v(y)\in\F_2^R$ such
that every output character is a bilinear form in this same pair of maps.
\end{definition}
Ranks are subadditive under addition and cannot increase under arbitrary
separate precomposition. Separation lengths multiply under products.
For $F=\F_{2^\ell}$, a separated $F$-product has absolute-trace rank at
most $\ell$. More generally, trace-dual bases for a degree-$a$ extension
$K/F$ give
\[
 \Tr_{K/F}(xy)=\sum_{j=1}^a x_jy_j.
\]
Thus relative trace costs at most $a$ separated products. Concatenating all
left and right factors of finitely many field-valued rows produces one
common binary feature pair for the entire table, not merely independent
rank bounds for its separate characters.

\paragraph{Explicit fields.}
We use fixed binary bases and deterministic finite-field construction in
fixed characteristic, as in Shoup~\cite{Shoup}. A relative extension degree
that is required only to exceed a fixed constant can be chosen to be a
power of two; a fixed-height quadratic tower then supplies $K/F$ and its
relative trace. All such arithmetic is polynomial-time in the displayed
binary field degree. Field constants are fixed independently of the sources.

\subsection{Odd-power features and dense spans}

\begin{lemma}[Rank under restriction]\label{cr2:lem:rankrestriction}
Let $H:V\times W\to\F_2$ be bilinear, and let $V_0\le V$, $W_0\le W$.
Then
\[
 \rk(H|_{V_0\times W_0})\ge
 \rk(H)-\codim V_0-\codim W_0.
\]
If $A\subseteq V$ and $B\subseteq W$ span $V_0$ and $W_0$, respectively,
the matrix $[H(a,b)]_{a\in A,b\in B}$ has exactly the rank of
$H|_{V_0\times W_0}$.
\end{lemma}
\begin{proof}
Restriction of one argument deletes at most its codimension in independent
rows or columns. This proves the inequality. For the second assertion,
select spanning bases from $A$ and $B$. The corresponding submatrix
represents the restricted bilinear form, and every other row or column is
a linear combination of these.
\end{proof}

\begin{lemma}[Odd-degree additive-character bound]
\label{pre:oddweil}\label{ss:ext:weil}
\textnormal{\emph{Reference.} \cite[Lemma~1]{MW} (the characteristic-two odd-degree case of the additive Weil bound).}
Let $K=\F_{2^N}$ and let $P\in K[z]$ have positive odd degree $d$.
Then
\[
 \left|\E_{z\sim U_K}(-1)^{\Tr_{K/\F_2}P(z)}\right|
 \le(d-1)2^{-N/2}.
\]
\end{lemma}
This is the additive Weil bound in characteristic two; positive odd
leading degree excludes the Artin--Schreier exceptional form. We use
it at fixed degree and sufficiently large $N$; see
\cite[Lemma 1]{MW}. Linear nonconstant phases have bias zero.

\begin{lemma}[Small character bias forces a dense feature span]
\label{pre:smallbias-span}
Let $\Phi:\Omega\to\F_2^r$ and suppose every nonzero character of
$\Phi(U_\Omega)$ has absolute bias at most $\beta$.
If $A\subseteq\Omega$ has density at least $\alpha>0$ and
$\beta\le\alpha/2$, then
\[
 \codim_{\F_2}\Span\Phi(A)\le\log(1/\alpha)+1.
\]
\end{lemma}
\begin{proof}
Write $V=\F_2^r$, and regard $V^*=\operatorname{Hom}_{\F_2}(V,\F_2)$
as the space of all binary-linear characters, using the standard dot
product to identify $V^*$ with $V$.  Let
\[
 S=\Span_{\F_2}\{\Phi(\omega):\omega\in A\}\le V,
 \qquad c=\codim_{\F_2}S.
\]
The annihilator of $S$ is
\[
 S^\perp=\{a\in V:a\cdot s=0\text{ for all }s\in S\}.
\]
Elementary linear algebra gives $\dim S^\perp=c$, so
$|S^\perp|=2^c$.  We first justify the character-orthogonality
identity, for every $v\in V$,
\begin{equation}\label{pre:eq:span-orthogonality}
 \mathbf1[v\in S]=2^{-c}\sum_{a\in S^\perp}(-1)^{a\cdot v}.
\end{equation}
If $v\in S$, every summand is $1$.  If $v\notin S$, the double
annihilator identity $(S^\perp)^\perp=S$ supplies an $a_0\in S^\perp$
with $a_0\cdot v=1$.  The terms indexed by $a$ and $a+a_0$
then have opposite signs and cancel, so the sum is zero.

Let $U$ be uniform on the finite set $\Omega$.  Every $\omega\in A$
satisfies $\Phi(\omega)\in S$, even if $\Phi$ is noninjective.
Consequently,
\begin{equation}\label{pre:eq:span-density}
 \Pr[\Phi(U)\in S]\ge\Pr[U\in A]
 =\frac{|A|}{|\Omega|}\ge\alpha.
\end{equation}
Substitute $v=\Phi(U)$ into
\eqref{pre:eq:span-orthogonality} and average over $U$.  Since
the zero character has expectation one, we obtain the \emph{exact}
identity
\begin{align}
 \Pr[\Phi(U)\in S]
 &=2^{-c}\sum_{a\in S^\perp}
       \mathbb E_U(-1)^{a\cdot\Phi(U)}\notag\\
 &=2^{-c}+2^{-c}\sum_{a\in S^\perp\setminus\{0\}}
       \mathbb E_U(-1)^{a\cdot\Phi(U)}.
       \label{pre:eq:span-character-sum}
\end{align}
The hypothesis bounds the absolute value of each nonzero-character
expectation by $\beta$.  There are exactly $2^c-1$ such terms,
so \eqref{pre:eq:span-density} and
\eqref{pre:eq:span-character-sum} imply
\[
 \alpha\le 2^{-c}+(1-2^{-c})\beta
       =\beta+(1-\beta)2^{-c}.
\]
Because $\beta\le\alpha/2$ and $\alpha\le1$, we have $\beta<1$.
Rearranging gives the slightly sharper bound
\[
 2^{-c}\ge\frac{\alpha-\beta}{1-\beta}
           \ge\frac\alpha2.
\]
Taking base-two logarithms now yields
\[
 c\le\log_2(2/\alpha)=\log_2(1/\alpha)+1,
\]
as claimed.  The proof only uses the density of $A$ in $\Omega$;
it does not assume that $A$ is affine or that $\Phi$ is injective.
\end{proof}

\begin{lemma}[Whole-field odd-power features]\label{pre:oddfeatures}
For $K=\F_{2^N}$ and a fixed integer $J\ge1$, define
\[
 \Psi_J(z)=(z,z^3,\ldots,z^{2J-1})\in K^J.
\]
Every nonzero binary character of $\Psi_J(U_K)$ has bias at most
$\beta_J=(2J-2)2^{-N/2}$. For every fixed $0<\tau<1/2$ and sufficiently
large $N$ (depending on $J,\tau$), each $A\subseteq K$ with
$|A|\ge2^{(1-\tau)N}$ satisfies
\[
 \codim_{\F_2}\Span\Psi_J(A)\le\tau N+1
 \quad\text{inside the $JN$-dimensional binary space }K^J.
\]
\end{lemma}
\begin{proof}
We view $K^J$ as a binary vector space of dimension $JN$.  By
nondegeneracy of the absolute-trace pairing on $K$, every
$\F_2$-linear functional $\ell:K\to\F_2$ is uniquely of the form
$\ell(x)=\Tr_{K/\F_2}(ax)$ for some $a\in K$.
Indeed, for $a\ne0$ the functional $x\mapsto\Tr(ax)$ is
nonzero, because multiplication by $a$ is a permutation of $K$
and the absolute trace is a nonzero binary-linear functional.
The resulting injective map from $K$ to its binary dual is an
isomorphism by equality of dimensions.  Applying this fact
coordinatewise, every binary character of $K^J$ has the unique
form
\begin{equation}\label{pre:eq:odd-character}
 \chi_{\mathbf a}(v_1,\ldots,v_J)
 =(-1)^{\Tr_{K/\F_2}(\sum_{j=1}^J a_jv_j)},
 \qquad\mathbf a=(a_1,\ldots,a_J)\in K^J.
\end{equation}
It is nonzero as a binary-linear character precisely when
$\mathbf a\ne0$.  Its evaluation on $\Psi_J(z)$ is
\[
 \chi_{\mathbf a}(\Psi_J(z))
 =(-1)^{\Tr_{K/\F_2}P_{\mathbf a}(z)},
 \qquad
 P_{\mathbf a}(z)=\sum_{j=1}^J a_jz^{2j-1}.
\]
Choose the largest index $j_0$ with $a_{j_0}\ne0$.
The exponents $1,3,\ldots,2J-1$ are distinct, so
$P_{\mathbf a}$ has positive odd degree
$e=2j_0-1\le2J-1$ and nonzero leading coefficient.

If $e=1$, then $P_{\mathbf a}(z)=a_1z$ with $a_1\ne0$.
Its absolute trace is a nonzero binary-linear functional,
which takes each of the values $0$ and $1$ on exactly half of $K$;
therefore its character expectation is exactly zero.
If $e\ge3$, Lemma~\ref{pre:oddweil} gives
\[
 \left|\mathbb E_{z\sim U_K}
 (-1)^{\Tr_{K/\F_2}P_{\mathbf a}(z)}\right|
 \le(e-1)2^{-N/2}
 \le(2J-2)2^{-N/2}.
\]
The odd leading degree is significant: a nonconstant polynomial
of the form $R(z)^2+R(z)+c$ has even degree, whereas such
Artin--Schreier phases are the exceptional case for additive
character sums in characteristic two.  Combining the linear
and higher-degree cases proves the first assertion, uniformly
for every nonzero binary character of the entire $JN$-bit feature
vector.

Next fix $0<\tau<1/2$.  When $J=1$, the preceding character
bias is zero, and the following argument works for every $N$.
For $J\ge2$, since $J$ and $\tau$ are fixed, choose $N$ large
enough that
\begin{equation}\label{pre:eq:odd-N-choice}
 (1/2-\tau)N\ge1+\log_2(2J-2).
\end{equation}
This is equivalent to
\[
 \beta_J=(2J-2)2^{-N/2}\le2^{-\tau N-1}.
\]
Let $A\subseteq K$ satisfy $|A|\ge2^{(1-\tau)N}$.
Its density in $K$ is at least
\[
 \frac{|A|}{|K|}\ge2^{-\tau N}=:\alpha_0.
\]
Take $\Omega=K$ and $\Phi=\Psi_J$, with a fixed binary-coordinate
identification $K^J\cong\F_2^{JN}$.  The character bound just proved
says that every nonzero binary character of $\Phi(U_K)$ has bias
at most $\beta_J\le\alpha_0/2$.  Lemma~\ref{pre:smallbias-span}
therefore applies and gives
\[
 \codim_{\F_2}\Span_{\F_2}\Psi_J(A)
 \le\log_2(1/\alpha_0)+1=\tau N+1.
\]
All dimensions here are \emph{binary} dimensions inside $K^J$;
we do not take the $K$-linear span of $\Psi_J(A)$.
The strict inequality $\tau<1/2$ makes the condition
\eqref{pre:eq:odd-N-choice} achievable for every fixed $J$.
\end{proof}

\subsection{Approximate duality and its rank-sensitive form}\label{sec:ad}
\begin{lemma}[Set form of approximate duality]\label{pre:setAD}
\textnormal{\emph{Reference.} \cite[Lemma~7.9]{BDL}, using the polynomial Freiman--Ruzsa theorem~\cite{GGMT}.}
There is an absolute $C$ such that the following holds. For finite
$A,B\subseteq\F_2^R$ with
$|\E_{A\times B}(-1)^{\langle a,b\rangle}|\ge\varepsilon$, and every
fixed positive integer $t$, there is a constant subrectangle with relative
density at least $2^{-q}$ on both sides, where
\begin{equation}\label{pre:eq:setAD}
 q\le C\left[\frac Rt+2^t\log\frac2\varepsilon
                  +(t+1)\log(R+2)\right].
\end{equation}
\end{lemma}
This is the characteristic-two consequence of
Bhowmick--Dvir--Lovett~\cite[Lemma 7.9 and its proof]{BDL}, with doubling
parameter $K=2^{R/t}$, using the polynomial Freiman--Ruzsa theorem proved
by Gowers--Green--Manners--Tao~\cite{GGMT}. The proof of the cited estimate
bounds the two relative densities separately and applies to unequal set
sizes. Its polynomial factors become logarithmic terms as displayed above.
Bounded $R$ and integer rounding are absorbed into $C$. We need its
rank-sensitive indexed version, so repetitions are removed explicitly rather
than charged by appending all raw input labels.
\begin{lemma}[Biased indexed matrix]\label{pre:AD}
Let $M$ be a binary matrix indexed by nonempty sets $A,B$ of sizes at most $2^n$, of rank $r\ge1$. Suppose
\[
 \left|\E_{A\times B}(-1)^{M(x,y)}\right|\ge\varepsilon.
\]
For every fixed positive integer $L$, there is a monochromatic rectangle $A'\times B'$ whose relative densities on both sides are at least $2^{-q}$, with
\begin{equation}\label{pre:eq:AD}
 q\le C\frac rL+
 C_L\left(\log\frac2\varepsilon+\log(n+2)+\log(r+2)\right).
\end{equation}
Here $C$ is absolute and $C_L$ depends only on $L$.
\end{lemma}
\begin{proof}
Factor $M(x,y)=\langle a_x,b_y\rangle$ in $\F_2^r$. Split the distribution of $a_X$, for uniform $X\in A$, into dyadic classes of its nonzero atom probabilities. Since those probabilities are at least $2^{-n}$, at most $n+1$ classes are needed. Do the same for $b_Y$.

The signed average is a sum of at most $(n+1)^2$ class-pair contributions. Some contribution has absolute value at least $\varepsilon/(n+1)^2$. Both class probabilities and the absolute conditional bias on that class pair are at least $\varepsilon/(n+1)^2$.

Within a class with $t$ distinct labels the conditional probabilities differ by at most a factor two. They are at most $2/t$, so the distribution is a convex combination of uniforms on $\max\{1,\lfloor t/2\rfloor\}$-element sets. These sizes are at least $t/3$ (singletons are harmless). Independent decompositions on the two sides produce one pair of flat label sets with absolute bias at least $\varepsilon/(n+1)^2$.

Apply~\eqref{pre:eq:setAD} to that pair. If its monochromatic label subsets have relative densities $\rho$, each selected subset has conditional mass at least $\rho/6$ in its entire dyadic class: every label there has probability at least $1/(2t)$. Lifting to the original indices consequently gives relative masses at least
\[
 \frac{\varepsilon\rho}{6(n+1)^2}
\]
on both sides. Insert $\varepsilon/(n+1)^2$ into~\eqref{pre:eq:setAD} and include the logarithm of this last lifting factor. For fixed $L$, the result is~\eqref{pre:eq:AD} after changing constants.
\end{proof}

\begin{lemma}[Asymmetric rank-to-bias upgrade]\label{pre:asymAD}
Let two input lengths $n_L,n_R$ lie between fixed positive constant
multiples of $n$. Let a family of Boolean matrices have rank at most $An$.
Suppose every member is nonconstant on all permitted subrectangles with
side sizes at least $2^{a_0n_L}$ and $2^{b_0n_R}$, and permitted supports
are closed under separate restrictions. For fixed
$0<a_0<a_1\le1$ and $0<b_0<b_1\le1$, there is $\eta>0$ such that every
member has absolute bias at most $2^{-\eta n}$ on independent sources
of entropies at least $a_1n_L,b_1n_R$ in that class.
\end{lemma}
\begin{proof}
Take independent flat components using Lemma~\ref{pre:flat}. If the bias
were at least $2^{-\eta n}$, Lemma~\ref{pre:AD} would give a constant
subrectangle with logarithmic density loss at most
\[
 CA n/t+C_t\bigl(\eta n+O(\log(n+2))\bigr).
\]
First choose fixed $t$ so the first term is smaller than one quarter of
both available entropy gaps. Then choose $\eta$ so the coefficient of
$n$ in the second term is also smaller than one quarter of both gaps.
For large $n$ the logarithmic terms and one-bit floors fit the remaining
slack. The resulting rectangle is above the two forbidden thresholds,
a contradiction. Uniformity of the constants and convexity of absolute
bias allow averaging the independent flat decompositions.
\end{proof}

\begin{corollary}[Linear-rank dispersion upgrade]\label{pre:upgrade}
For fixed $A,\rho,\xi>0$ with $\rho+\xi\le1$, there is $\eta>0$ such
that every Boolean $\rho n$-two-source disperser of rank at most $An$ is
an ordinary extractor at side entropy $(\rho+\xi)n$ with error
$2^{-\eta n}$. The exponent is uniform over all functions satisfying
these two hypotheses.
\end{corollary}
\begin{proof}
Apply Lemma~\ref{pre:asymAD} with equal input lengths and equal entropy
gaps, then use that one-bit statistical distance is half its character
bias. The low-rank disperser-to-extractor principle originates with
Ben-Sasson--Ron-Zewi~\cite[Lemma~2.17]{BenSZ11}. Its exponentially
small-error consequence originally used the polynomial
Freiman--Ruzsa conjecture, now proved by
Gowers--Green--Manners--Tao~\cite{GGMT}. See also
Alrabiah--Goodman--Mosheiff--Ribeiro~\cite[Theorem~9]{AGMR}
for this unconditional formulation.
\end{proof}

\begin{lemma}[Uniform tolerance of a low-rank transcript]
\label{pre:transcript}
Fix constants $0<\rho<\rho+\xi\le1$, $A,\kappa\ge0$ and $\omega>0$.
Let $U(x,y)\in\F_2^u$, $V(x,y)\in\F_2^v$ be functions of two
$N$-bit inputs. Suppose, uniformly for $\alpha\ne0$ and $\beta$,
the phase $\alpha\cdot U+\beta\cdot V$ has global binary rank at most
$AN$ and restricted rank greater than $(\kappa+\omega)N$ on every
rectangle of side entropy at least $\rho N$.
Let $T(x,y)\in\F_2^z$ have every parity of global rank at most
$\kappa N$. There is $\eta>0$, depending only on the fixed constants,
such that, for sufficiently large $N$, independent sources of
entropy at least $(\rho+\xi)N$, and $u+v+z\le\eta N$,
\[
 (U,V,T)\approx_{2^{-\eta N/2}}(U_u,V,T).
\]
The ideal distribution retains the actual joint marginal of $(V,T)$.
\end{lemma}
\begin{proof}
For $\alpha\ne0$, adding a transcript parity leaves restricted rank
at least $\omega N>1$ and global rank at most $(A+\kappa)N$.
Corollary~\ref{pre:upgrade}, after decreasing its exponent to absorb
the bias/distance factor two, gives uniform bias $2^{-\eta N}$.
Only characters with $\alpha\ne0$ contribute to comparison with
$(U_u,V,T)$. Lemma~\ref{pre:fourier} gives distance at most
$2^{(u+v+z)/2-\eta N}/2\le2^{-\eta N/2}$.
No value of the jointly computed transcript is conditioned on.
\end{proof}

\subsection{Published seeded primitives and entropy-efficient extraction}\label{sec:seeded}
We state seeded results in the convention that the seed is \emph{not}
counted in the output length. Where a source counts it, we delete its
fixed seed coordinates explicitly.

\begin{theorem}[Two strong seeded parameter regimes]\label{pre:guv-basic}
\textnormal{\emph{Reference.} \cite[Theorems~5.12 and 5.14 and Remark~5.15]{GUV}; see also~\cite[Lemma~4.10]{ChattopadhyayGL21}.}
There are absolute constants with the following properties.
For an $n$-bit source of entropy $k$ and error $\zeta$, put
$s=\log(1/\zeta)$.
\begin{enumerate}[label=(\roman*),leftmargin=*]
\item A strong seeded extractor outputs at least $k-2s-O(1)$ bits with
seed length
$\log(n+2)+O(\log(k+2)\log((k+2)/\zeta))$.
\item Whenever $k\ge B(\log(n+2)+s+1)$ for a sufficiently large absolute
constant $B$, a strong seeded extractor outputs $\lfloor k/2\rfloor$
bits with seed length $O(\log(n+2)+s+1)$.
\end{enumerate}
Both constructions are polynomial-time in $n+s$. An endpoint $k=n$
can be treated by appending one fixed input bit when invoking a statement
formulated for $k<n$.
\end{theorem}
Part (i) is Theorem~5.14 and Remark~5.15 of GUV~\cite{GUV}, after deleting
the seed coordinates. Part (ii) follows from their fixed-fraction theorem
and strongness remark; the displayed entropy lower bound absorbs the seed
coordinates and all additive constants. It is also recorded in a
convenient half-output form in \cite[Lemma 4.10]{ChattopadhyayGL21}.

\begin{lemma}[Strong seeded extraction with near-total output]\label{pre:seed-efficient}
There are absolute $A,B>0$ such that, for every $n$, integer $k\le n$,
and $0<\zeta\le1/4$, there is a strong seeded extractor with error at most
$\zeta$, output length at least $\max\{0,k-A(s+1)\}$ up to an integer
floor, and seed length at most
\begin{equation}\label{pre:eq:seed-efficient}
 B\seedcost{n}{s},\qquad s=\log(1/\zeta).
\end{equation}
It is deterministic polynomial-time in $n+s$. Ignored seed padding and
output truncation preserve the conclusion.
\end{lemma}
\begin{proof}
If $s\le\log(n+2)$, use part (i) of Theorem~\ref{pre:guv-basic}.
Its seed length is $O(\log^2(n+2))$ and its entropy loss is $O(s+1)$.
If the claimed output is nonpositive, use the empty output.

Suppose now $s>\log(n+2)$. Let
$T=\lceil\log_2(n+2)\rceil+2$, set the error of each extraction stage to
$\zeta_0=\zeta/(8T)$, and put
\[
 t=\lceil s+\log(8T)\rceil+1.
\]
Start with cumulative output length $M=0$. At a stage define
$h=k-M-t$. If $h<H_*$, stop, where
$H_*=B_*(s+\log(n+2)+\log(8T)+1)$ and $B_*$ is a sufficiently large
absolute constant. Otherwise apply part (ii) with threshold $h$ and error
$\zeta_0$, using a fresh independent seed, output $\lfloor h/2\rfloor$
bits, and add that length to $M$.

All thresholds and lengths are deterministic functions of $n,k,\zeta$.
Condition on any fixing of all previous seeds. The original source still
has min-entropy $k$ before its previous output is disclosed. Since that
output has exactly $M$ bits, Lemma~\ref{pre:disclosure} shows that, outside
conditional probability at most $2^{-t}$, the source conditioned on its
previous output has entropy at least $h$. Averaging over the previous
seeds gives the same exception bound. The strong guarantee at the current
stage therefore makes its output uniform independently of its seed and
the actual preceding output/seed transcript, to error at most
$\zeta_0+2^{-t}$.

Iterating this comparison proves that the complete output, jointly with
all seeds, is within the sum of the stage errors of uniform. A formal
induction compares the actual prefix/current-output distribution first with the
actual prefix times a fresh uniform current output, then applies the
previous induction hypothesis. Future seeds are independent and can be
attached throughout. No replacement of the source distribution after an
earlier hybrid is needed.

The quantity $h$ is replaced by $\lceil h/2\rceil$ at each stage, so there
are at most $T$ stages and in fact
$O(\log(2+n/(s+1)))$ stages. Their total error is at most $\zeta$.
At termination $k-M\le t+H_*+O(1)=O(s+1)$, since $s>\log(n+2)$.
Each seed has length $O(s+1)$ in this regime, so the total is
$O((s+1)\log(2+n/(s+1)))$. This proves the unified bound. At most
$O(\log n)$ polynomial-time seeded calls are made.
\end{proof}

\begin{remark}[Why this seed length is sufficient]\label{rem:seedcost}
The additive-loss output proofs require that the generated
seed fit within a sufficiently small linear number of bits when
$s=\log(1/\eps)\le b_\delta n$. The function in
\eqref{pre:eq:seed-efficient} has this property because
$b\log(2+1/b)\to0$ as $b\to0$. This explicit derivation from GUV avoids
needing a stronger entropy-efficient seeded parameter statement. Its
somewhat larger internal seed is harmless because that seed is retained
in the final output.
\end{remark}

\subsection{Constant-seed strong condenser}\label{pre:sec:strongcondenser}
\begin{theorem}[Constant-seed strong condenser]\label{pre:strongcondenser}
\textnormal{\emph{Reference.} \cite[Definition~2.10 and Theorem~4.3]{Zuckerman}.}
For fixed $0<\rho\le1$ and $0<\beta,\eps<1$, there are a constant seed length $d$ and
an efficient map
\[
 \Cond:\bits^n\times\bits^d\longrightarrow\bits^{m_0},
 \qquad m_0=\Omega(n),
\]
such that, for every $X$ of min-entropy at least $\rho n$,
\[
 2^{-d}\sum_{i\in\bits^d}
 \min_{Z_i:\ H_\infty(Z_i)\ge(1-\beta)m_0}
 \TV(\Cond(X,i),Z_i)\le\eps.
\]
\end{theorem}
We use the strong-condenser convention of Zuckerman~\cite[Definition 2.10 and Theorem 4.3, attributed to Raz]{Zuckerman}. The maps need not be linear. For the small target deficiencies used below their output lengths may be taken at most $2n$ for sufficiently large $n$, by the support-size bound on a uniform input.

\subsection{Uniform quantitative results for a growing number of rows}\label{lr:sec:inputs}
The uniformity in the next three statements is essential. Their constants
are absolute; their algorithms do not require the entropy rate or the
number of rows to be fixed.

\begin{lemma}[Quantitative approximate duality]\label{lr:thm:ad}
\textnormal{\emph{Reference.} \cite[Lemma~1.11]{BenSassonLR14}, with the PFR theorem~\cite{GGMT}.}
There is an absolute constant $C_{\rm AD}$ such that the following holds.
If $A,B\subseteq\F_2^a$ are nonempty and
\[
 \left|\E_{x\sim U_A,y\sim U_B}(-1)^{x\cdot y}\right|\ge2^{-v},
\]
then, for every integer $u\ge1$, there are nonempty $A'\subseteq A$ and
$B'\subseteq B$ on which $x\cdot y$ is constant and both relative sizes
are at least
\begin{equation}\label{lr:eq:ad}
 2^{-C_{\rm AD}\{a/u+2^u(v+1)+(u+1)\log(a+2)\}}.
\end{equation}
\end{lemma}
This is Lemma 1.11 of Ben-Sasson--Lovett--Ron-Zewi~\cite{BenSassonLR14}, with its
PFR hypothesis supplied by~\cite{GGMT}. In that lemma take its parameter
$K=2^{a/u}$, so its parameter $t=a/\log K$ is exactly $u$. The negative
logarithms of its two density bounds are at most an absolute multiple of
$a/u+2^u(v+1)+u\log(4a)+\log a$. This gives~\eqref{lr:eq:ad}, with the
small dimensions covered by increasing $C_{\rm AD}$. We use the full
tradeoff, not only the special case $v\le\sqrt a$. In particular, $u$ may
grow with $n$.

\begin{theorem}[Uniform linear somewhere condenser]\label{lr:thm:cond}
\textnormal{\emph{Reference.} \cite[Theorem~31 and Algorithm~4]{LZ}, using~\cite{GGMT}.}
There are absolute constants $A\ge1$, $C_c\ge1$, $c_c>0$, and
$\beta_0>0$ such that, in the parameter ranges used below, for
$0<\delta\le1/2$ there are uniformly explicit binary-linear maps
\[
 C_1,\ldots,C_J:\F_2^n\to\F_2^m
\]
with
\begin{equation}\label{lr:eq:cond}
 J\le C_c\delta^{-A},\qquad
 C_c^{-1}n\delta^A\le m\le n,
\end{equation}
whose tuple on every source of min-entropy at least $\delta n$ is
$2^{-c_cm}$-close to a mixture having a designated row of min-entropy
at least $(1/2+\beta_0)m$. The polynomial exponents and the constants
are independent of $\delta$.
\end{theorem}
This is the general-weak-source version of Li--Zhong's linear condenser,
Theorem 31 and Algorithm 4 of~\cite{LZ}. Its $O(\log(1/\delta))$ stages
have a constant branching factor and a constant length reduction per
stage, giving the absolute polynomial bounds in~\eqref{lr:eq:cond}.
The general-weak-source analysis uses the proved PFR theorem. Enlarging
$A,C_c$ accommodates integer rounding and the finitely many initial
lengths. We use this statement only when $m$ grows as $n$ divided by a
fixed power of $\log n$, so all stage and error estimates are in their
large-length range. This is stronger than simply writing $O_\delta(1)$
rows and $\Theta_\delta(n)$ bits without tracking the rate dependence.

\paragraph{Table conventions.}
An elementary \emph{somewhere-uniform} distribution on $T$ rows of $s$ bits
has at least one exactly uniform row marginal; its other rows can be
arbitrarily correlated with that row. A somewhere-uniform distribution is a
positive mixture of elementary distributions, with the good index allowed to
vary across components. An \emph{affine somewhere-random} distribution is
uniform on an affine subspace whose projection onto some row is all
of $\F_2^s$. Neither convention assumes independence among rows.

\begin{theorem}[Affine somewhere-random extraction with growing row count]
\label{lr:thm:sr}
\textnormal{\emph{Reference.} \cite[Lemma~3.6(iii)]{LZdraft}; compare~\cite[Theorem~39]{LZ}.}
There is an absolute integer $A_0\ge1$ such that, with
\[
 H(T)=(2T)^{A_0\log(2T)},\qquad a_s=s/H(T),
\]
the following holds whenever $a_s\ge(2T)^4$. There is a uniformly explicit
map on $T$ rows of $s$ bits which extracts at least $\lfloor a_s\rfloor$
bits, with error at most $2^{-a_s}$, from every uniform affine source
whose projection onto at least one row is all of $\F_2^s$.
Evaluation is polynomial-time in this parameter range.
\end{theorem}
We use Li's affine somewhere-random extractor in the quantitative form of
Li--Zhong~\cite[Lemma 3.6(iii)]{LZdraft}; see also~\cite[Theorem 39]{LZ}.
The original construction has $O(\log T)$ recursive stages and a loss
$T^{O(\log T)}$, with absolute exponents. No internal independence of the
rows is required. For $T=(\log n)^{O(1)}$, one has
\begin{equation}\label{lr:eq:srsubpoly}
 H(T)=2^{O((\log\log n)^2)}=n^{o(1)}.\end{equation}
A theorem only for fixed $T$, with unspecified $T$-dependent constants,
would not be enough. No degree bound on this final affine extractor is
needed in our application.

\subsection{Weighted approximate duality for arbitrary common features}
\label{lr:sec:weighted}
For the decreasing-rate two-source application the features are the
raw input bits (and their source-local tampered copies). We state
the analysis for arbitrary separate feature maps because the
regularization theorem applies equally to nonlinear precompositions.

\begin{lemma}[Weighted rank-projection lemma]\label{lr:lem:weighted}
Let $X,Y$ be independent flat sources on subsets of $\F_2^n$.
Suppose a Boolean phase is affine-bilinear in separate feature variables
of at most $P$ bits per side, and its bilinear part on the two current
feature affine hulls has rank $r$. If its bias exceeds $2^{-H}$, then
for every integer $u\ge1$ there are nonempty raw subsets $A',B'$ on
which the phase is constant, with both relative probabilities at least
$2^{-g(r,H,u)}$, where
\begin{equation}\label{lr:eq:g}
 g(r,H,u)=C_w\left\{\frac{r+2}{u}
       +2^u(H+1)+(u+1)\log(n+P+2)\right\}.
\end{equation}
The constant $C_w$ is absolute. The feature maps need not be injective,
linear in the raw inputs, or uniformly distributed.
\end{lemma}
\begin{proof}
In affine coordinates on the two feature hulls write the phase as
\[
 f(x,y)=L(x)\cdot R(y)+a(x)+b(y)+c,
\]
where the rank factors have $r$ coordinates. Here all functions are
composed with their respective raw feature maps. The independent
projections
\[
 Z=(L(X),a(X),1),\qquad W=(R(Y),1,b(Y)+c)
\]
lie in $\F_2^{r+2}$ and satisfy $f(X,Y)=Z\cdot W$. Their positive atom
probabilities are integer multiples of $1/|\supp X|$ and
$1/|\supp Y|$, hence at least $2^{-n}$.

Set $\eps=2^{-H}$ and choose a sign so the signed correlation is
larger than $\eps$. Divide each projected support into at most
$n+1$ dyadic probability buckets. If $p_i,q_j$ are the bucket masses
and $c_{ij}$ their conditional signed correlations, then
$\sum p_iq_jc_{ij}>\eps$. The bucket pairs with
$c_{ij}>\eps/2$ have total weight greater than $\eps/2$:
otherwise the sum is at most
$\eps/2+(1-\eps/2)\eps/2<\eps$.
Some such pair has
\begin{equation}\label{lr:eq:bucket}
 p_iq_j>\frac{\eps}{2(n+1)^2}.
\end{equation}
Each individual bucket mass is also at least this quantity.

Inside a bucket of size $N_1$, every conditional atom lies between
$1/(2N_1)$ and $2/N_1$. Put $m_1=\max\{1,\lfloor N_1/2\rfloor\}$.
These atoms are at most $1/m_1$, so Lemma~\ref{pre:flat} decomposes the
conditional distribution into flat distributions on $m_1$ points, where $m_1\ge N_1/3$.
Do this independently on the right. Some pair of flat components has
signed correlation above $\eps/2$. Apply Lemma~\ref{lr:thm:ad}
to that pair in dimension $r+2$ with exponent $H+1$.
If the monochromatic sets have relative flat density at least $2^{-b}$,
then their conditional probability under their original buckets is at
least $2^{-b}/6$ on either side. Taking their entire raw preimages and
using~\eqref{lr:eq:bucket} gives relative raw probability at least
\[
 \frac{\eps}{12(n+1)^2}
 2^{-C_{\rm AD}\{(r+2)/u+2^u(H+2)+(u+1)\log(r+4)\}}.
\]
Taking negative logarithms and using $r\le P$ proves~\eqref{lr:eq:g}
after increasing one absolute constant. Pulling back the monochromatic
sets preserves the phase identity. It does not require the projected
sources to have been flat before bucket selection.
\end{proof}

\subsection{A uniform shared-feature regularization theorem}
\label{lr:sec:regularization}
Let
\[
 \mathcal F:\F_2^n\times\F_2^n\to\F_2^D.
\]
Assume fixed maps $u_L,u_R:\F_2^n\to\F_2^P$ represent every character
by a bilinear form:
\begin{equation}\label{lr:eq:features}
 \lambda\cdot\mathcal F(x,y)
       =u_L(x)^{\mathsf T}M_\lambda u_R(y)
       \qquad(\lambda\in\F_2^D).
\end{equation}
Padding features by zero coordinates permits one common length $P$.
The same maps must work for \emph{all} $\lambda$.

\begin{theorem}[Uniform positive affine regularization]\label{lr:thm:reg}
There is an absolute $C_0$ such that the following holds. For independent
flat inputs, integers $1\le r_*\le P$, $u\ge1$, and real $H\ge0,t\ge1$,
there is a finite partition into raw product rectangles. On every leaf,
the conditional output is within $2^{D/2-H-1}$ of uniform on its exact
affine hull. Outside total leaf mass at most $2^{1-t}$, each source
loses at most
\begin{align}\label{lr:eq:loss}
 \mathcal L=C_0\biggl[&\frac P u+
 \frac P{r_*}\bigl(2^u(H+1)+(u+1)\log(n+P+2)\bigr)\notag\\
 &+D(r_*+2)+t\biggr]
\end{align}
bits of min-entropy. The comparison is a positive mixture of uniform
affine distributions. The partition is proof-only and need not be computed.
\end{theorem}

\subsubsection{The progress rule}
On a current rectangle $A\times B$, let $U,V$ be the affine hulls of
$u_L(A),u_R(B)$ and let $W$ be the affine hull of
$\mathcal F(A,B)$. If every character nonconstant on $W$ has bias at
most $2^{-H}$, terminate. Otherwise select an offending character and
let $r$ be the rank of its bilinear part on the directions of $U,V$.

When $r>r_*$, use Lemma~\ref{lr:lem:weighted} to obtain a monochromatic
subrectangle $A'\times B'$. If $U',V'$ are its feature affine hulls,
then
\begin{equation}\label{lr:eq:drop}
 \operatorname{codim}_{U}U'+\operatorname{codim}_{V}V'\ge r.
\end{equation}
Indeed, the four-point difference of a constant affine-bilinear phase
vanishes. Thus its bilinear part vanishes on the two direction spaces
of $U',V'$. Restriction to codimensions $c_1,c_2$ deletes at most
$c_1+c_2$ from matrix rank, proving~\eqref{lr:eq:drop}.
Query membership, on one raw input, in the preimage of a feature affine
hull that has codimension at least $r/2$. Its yes probability is at least
$2^{-g(r,H,u)}$. It is less than one since the queried feature hull is
proper, and both outcomes remain raw product rectangles.

When $r\le r_*$, write the character in current affine coordinates as
$L(x)\cdot R(y)+a(x)+b(y)+c$. Reveal $(L(x),a(x))$, at most $r+1$ bits,
then the right bit $L(x)\cdot R(y)+b(y)+c$. This makes the character
constant, using at most $r_*+2$ declared message bits. Because the
character was nonconstant on $W$, every resulting nonempty output hull
has dimension at least one smaller.

All choices can be fixed for each rectangle, making a deterministic
finite tree. Every completed action strictly reduces the raw rectangle
cardinality on each branch. At high rank both membership outcomes are
proper; at low rank a phase-value cell cannot be the whole support of
a nonconstant phase. The tree is finite even though no efficient bound
on its depth is asserted.

\subsubsection{One shared potential controls every high-rank character}
The sum of the two current feature-affine-hull dimensions starts at most
$2P$ and never increases. On high-rank yes answers it decreases by at
least $r/2$. Therefore, pathwise,
\begin{equation}\label{lr:eq:budgets}
 \sum_{\rm high\ yes}r_j\le4P,
 \qquad N_+\le4P/r_*.
\end{equation}
The output hull loses dimension on every low-rank action, so there are
at most $D$ such actions. Their total declared message length is at most
\begin{equation}\label{lr:eq:messages}
 B_{\rm ex}\le D(r_*+2).
\end{equation}
Since high ranks are at least two, $r+2\le2r$ there. Equations
\eqref{lr:eq:g} and~\eqref{lr:eq:budgets} give a deterministic bound on
affirmative information:
\begin{equation}\label{lr:eq:yescost}
 I_+\le C_1\left[\frac P u+
 \frac P{r_*}\bigl(2^u(H+1)+(u+1)\log(n+P+2)\bigr)\right].
\end{equation}
Different character factorizations alone would not prove this bound;
it is the fixed feature pair~\eqref{lr:eq:features} that makes the
potential common to the whole path.

\subsubsection{Two surprisal supermartingales}
Let $I_-$ be the total information of high-rank no answers. At a query
whose yes probability is $p$, the conditional multiplier of
$2^{I_-/2-N_+}$ has mean
\[
 (1-p)(1-p)^{-1/2}+p/2=\sqrt{1-p}+p/2\le1.
\]
Other actions leave it unchanged. It is a nonnegative supermartingale
starting at one, and the tree is finite. Markov's inequality implies
\begin{equation}\label{lr:eq:notail}
 \Pr[I_->2N_++2t]\le2^{-t}.
\end{equation}
This controls arbitrarily many rejected queries without counting them.

Let $I_{\rm ex}$ be the information of all low-rank exact messages,
and let $B$ be their accumulated declared bit lengths. At an $a$-bit
message with positive outcome probabilities $p_z$, the conditional
multiplier of $2^{I_{\rm ex}-B}$ has expectation
\[
 \sum_{z:p_z>0}p_z\,(2^ap_z)^{-1}
   =\frac{|\{z:p_z>0\}|}{2^a}\le1.
\]
This process is unchanged at high-rank queries. Equations
\eqref{lr:eq:messages} and Markov's inequality give
\begin{equation}\label{lr:eq:extail}
 \Pr[I_{\rm ex}>D(r_*+2)+t]\le2^{-t}.
\end{equation}
Neither estimate assumes independent or unbiased transcript bits.

\subsubsection{Entropy and positive affine completion}
For a leaf $A'\times B'$ within the initial flat rectangle $A\times B$,
its probability is $(|A'|/|A|)(|B'|/|B|)$. Answer probabilities telescope,
so its negative logarithm is exactly $I_++I_-+I_{\rm ex}$.
Each individual source's entropy loss is nonnegative and at most this
sum. Outside the union of~\eqref{lr:eq:notail} and~\eqref{lr:eq:extail},
combine~\eqref{lr:eq:yescost}, $2N_+\le8P/r_*$, and the message bound
to obtain~\eqref{lr:eq:loss}, with an absolute $C_0$.

At every terminal leaf all nonconstant characters of its affine hull
$W$ have bias at most $2^{-H}$. In affine coordinates on $W$, Parseval
and Cauchy--Schwarz imply
\[
 \TV(\mathcal F(X,Y)\mid\text{leaf},U_W)
 \le\tfrac12\sqrt{2^{\dim W}-1}\,2^{-H}
 \le2^{D/2-H-1}.
\]
Averaging these uniform distributions with the original leaf probabilities is a
positive mixture, and proves Theorem~\ref{lr:thm:reg}.
For bilinear raw maps, $u_L(x)=x,u_R(y)=y$ gives $P=n$. This is the
special economy used by the inverse-logarithmic two-source construction.


\section{Constant-rate extraction with small error}\label{un:linear}
All three constructions in this section have linear output and error exponentially small in the original input length. The fixed-$C$ theorem includes every fixed number of separate tamperings. For $\delta n\le k\le n$, an output length $c_\delta n$ is in particular at least $c_\delta k$.
Throughout the manuscript, for any field $F$ and $1\le D\le s$, define
\[
 \sigma_D(z_1,\ldots,z_s)
 :=\sum_{\substack{I\subseteq\{1,\ldots,s\}\\ |I|=D}}
       \prod_{i\in I} z_i,\qquad z_1,\ldots,z_s\in F.
\]
Thus $\sigma_D$ is the elementary symmetric polynomial of degree $D$;
equivalently, it is the coefficient of $T^D$ in
$\prod_{i=1}^{s}(1+z_iT)$.

\paragraph{Reading the construction boxes.}
Each numbered construction box starts with the previously established
primitives it uses, then separates \emph{public setup} from the steps
performed on an input. Public constants, field representations, row orders,
projections, and subroutine parameters are fixed from the stated lengths,
entropy thresholds, error target, and tampering count, never from a source
support or a tampering map. A proof may later certify that a sufficiently
small or large public constant works; evaluation does not search for that
constant on the input. Except for a seeded subroutine's displayed seed
argument, no external randomness is supplied. Good rows, subsources,
conditional representative tables, and approximate-duality partitions are
analysis objects, not hidden algorithmic steps. The boxes use a separate
counter, so all existing theorem and equation numbers are retained.

\paragraph{Evaluating the common field merger.}
Every call to $\sigma_D(w_1,\ldots,w_s)$ can be evaluated by the same
fixed recurrence: initialize $e_0=1$ and $e_1=\cdots=e_D=0$; for
$i=1,\ldots,s$, update $e_j\leftarrow e_j+w_i e_{j-1}$ in decreasing
order $j=\min\{i,D\},\ldots,1$; return $e_D$.
The descending update order uses the previous polynomial coefficients,
so after row $i$, $e_j$ is the coefficient of $T^j$ in
$\prod_{b=1}^i(1+w_bT)$, for $j\le D$. This is an evaluation rule,
not the proof's selection or forcing of good rows.

\subsection{Ordinary two-source extraction}\label{un:ordinarylinear}
\subsubsection{Ingredients and proofs: odd-power restricted rank and dense good rows}
Fix
\begin{equation}\label{cr2:eq:tau}
 \tau=2^{-10}.
\end{equation}
For $K=\F_{2^M}$, the kernel $H_J(u,v)=\sum_{j=1}^J u^{2j-1}v^{2j-1}$ is the field
inner product $H_J(u,v)=\langle\Psi_J(u),\Psi_J(v)\rangle_K$.

\begin{lemma}[Odd-power restricted rank]\label{cr2:lem:oddrank}
For every fixed $J\ge1$, sufficiently large $M$, and
$\lambda\in K^\times$, put
$h_\lambda(u,v)=\Tr_{K/\F_2}(\lambda H_J(u,v))$.
Its global binary matrix rank is at most $JM$. On arbitrary
$A,B\subseteq K$ with $|A|,|B|\ge2^{(1-\tau)M}$,
\begin{equation}\label{cr2:eq:oddrestricted}
 \rk_2[h_\lambda]_{A\times B}\ge(J-2\tau)M-2.
\end{equation}
The deficiency $\tau$ is independent of $J$.
\end{lemma}
\begin{proof}
The trace pairing $\Tr_{K/\F_2}(\lambda\sum_j p_jq_j)$ is nondegenerate
on $K^J$, of binary dimension $JM$. By Lemma~\ref{pre:oddfeatures},
the spans of $\Psi_J(A)$ and $\Psi_J(B)$ each have codimension at most
$\tau M+1$. Lemma~\ref{cr2:lem:rankrestriction} loses at most their
sum and proves the lower bound. Expanding the $J$ separated field
products and using trace-dual bases proves the global upper bound.
\end{proof}

\begin{lemma}[Designated-pair freshness]\label{cr2:lem:freshness}
Fix $J\ge1$ and $0\le\kappa<J-2\tau$. There is $\eta>0$ such that,
for sufficiently large $M$, any transcript $T:K^2\to\F_2^v$ whose
parities have rank at most $\kappa M$, any surjective binary-linear
map $\pi:K\to\F_2^u$, and independent sources $U,V$ of entropy at least
$(1-\tau/2)M$ satisfy
\begin{equation}\label{cr2:eq:freshness}
 (\pi H_J(U,V),T(U,V))\approx_{2^{-\eta M/2}}(U_u,T(U,V)),
 \qquad u+v\le\eta M.
\end{equation}
The exponent is uniform in $T$, $\pi$, and the sources.
\end{lemma}
\begin{proof}
Each nonzero projected character is an absolute-trace character with
nonzero coefficient. By Lemma~\ref{cr2:lem:oddrank}, it has global
rank at most $JM$ and restricted rank at least $(J-2\tau)M-2$.
Choose $\omega=(J-2\tau-\kappa)/2>0$ and increase the length threshold.
Apply Lemma~\ref{pre:transcript} with no current comparison coordinates,
$\rho=1-\tau$, and $\xi=\tau/2$. This proves the assertion.
\end{proof}

\subsubsection{Selecting subsources with many good rows}

For a distribution $P$ on $\bits^{m_0}$ and $0\le h\le m_0$, define
\[
 e_h(P)=\sum_z(P(z)-2^{-h})_+.
\]
This equals its distance to the closest $h$-source. Indeed, every such
source must remove this excess mass; conversely, clipping to the cap
$2^{-h}$ and distributing the removed mass into the unused capacity
achieves this distance.

\begin{lemma}[Many good rows on an actual subsource]\label{cr2:lem:subsource}
Suppose deterministic rows $C_i(X)\in\bits^{m_0}$, $1\le i\le r$, satisfy
$\sum_i e_h(C_i(X))\le\eps_c r$, where $\eps_c<1/8$.
There is a distribution $X^*$ obtained by conditioning $X$ together
with auxiliary randomness on an event of probability at least
$(1-8\eps_c)2^{-r}$, and an index set $I$ with $|I|\ge7r/8$, such that
\[
 H_\infty(C_i(X^*))\ge h-r-\log\frac1{1-8\eps_c}
 \qquad(i\in I).
\]
In particular $\supp X^*\subseteq\supp X$.
\end{lemma}
\begin{proof}
Let $p_i(z)=\Prb[C_i(X)=z]$. Conditional on $X=x$, sample acceptance
indicators $A_i$ with
\[
 \Prb[A_i=1\mid X=x]
 =\min\{1,2^{-h}/p_i(C_i(x))\}.
\]
Only positive-probability values are sampled, so this ratio is defined.
The indicators may be sampled independently conditional on $X$.
Then $\Prb[A_i=0]=e_h(C_i(X))$. For
$N_{\rm rej}=\sum_i(1-A_i)$, one has
$\E N_{\rm rej}\le\eps_c r$; hence Markov's inequality gives
\[
 \Prb[N_{\rm rej}>r/8]\le8\eps_c.
\]
On the complementary event there are at least $7r/8$ acceptances.
This is a statement about an integer count and needs no divisibility
assumption on $r$. Among the at most $2^r$ exact acceptance patterns,
some pattern with at least $7r/8$ acceptances has probability
$p\ge(1-8\eps_c)2^{-r}$. Condition on this pattern, and let $I$
be its accepted indices. The resulting marginal of $X$ is $X^*$.

For $i\in I$, the conditioning event implies $A_i=1$, and hence
\[
 \Prb[C_i(X^*)=z]
 \le\frac{\Prb[C_i(X)=z,A_i=1]}p
 =\frac{\min\{p_i(z),2^{-h}\}}p
 \le\frac{2^{-h}}p.
\]
Taking logarithms proves the entropy bound. Conditioning cannot add
points to the support of $X$.
\end{proof}

The auxiliary randomness creates no randomness in the extractor.
This conditioning is used only to prove that both original input
supports contain a product subsource on which the needed events occur.
Applying the lemma separately to independent $X$ and $Y$ preserves
independence of $X^*$ and $Y^*$.

\subsubsection{Main theorem and construction interface}

\begin{theorem}[Linear output and exponential error]\label{cr2:thm:main}
Fix $0<\delta\le1$. There are positive constants $c_\delta,a_\delta,C_\delta$
and a deterministic polynomial-time family
\[
 \Ext_n:\bits^n\times\bits^n\longrightarrow\F_2^{m(n)},
 \qquad m(n)\ge c_\delta n,
\]
such that independent sources $X,Y$ with
$H_\infty(X),H_\infty(Y)\ge\delta n$ satisfy
\[
 \TV(\Ext_n(X,Y),U_{m(n)})\le2^{-a_\delta n}.
\]
For every $\beta\in\F_2^{m(n)}\setminus\{0\}$,
\[
 \rk_2\bigl((x,y)\mapsto\beta\cdot\Ext_n(x,y)\bigr)\le C_\delta n.
\]
In addition, there are a field $F_n=\F_{2^{\ell(n)}}$ with
$\ell(n)=\Theta_\delta(n)$, an integer $R=O_\delta(1)$, explicit maps
$G_L,G_R:\bits^n\to F_n^R$, and a surjective binary-linear map
$\pi_n:F_n\to\F_2^{m(n)}$ such that
\begin{equation}\label{cr2:eq:mainform}
 \Ext_n(x,y)=\pi_n\!\left(\ip{G_L(x)}{G_R(y)}_{F_n}\right).
\end{equation}
Thus each encoding has $R\ell(n)=O_\delta(n)$ bits. A separately source-strong linear-output companion is proved in Corollary~\ref{cr2:cor:source-strong}, with the same rank form.
\end{theorem}

This is an independent pair-input construction: it is not obtained by
applying the sumset extractor to $x+y$. Enumerate the same $r$ strong-condenser
seeds on both inputs, with $r\ge4$ a power of two. Use only the corresponding
left/right rows and delete the fixed last index, leaving
$s=r-1=2D-1$ columns with $D=r/2$. At column $a$, interpret the
$a$th row of each input as an element of one common
field $K$ and use
\begin{equation}\label{cr2:eq:oddcore}
 H_{J_a}(u,v)=\sum_{j=1}^{J_a}u^{2j-1}v^{2j-1},\qquad
 W_a=\Tr_{K/F}H_{J_a}(u,v),\qquad J_a=2^{a-1}.
\end{equation}
Return a fixed binary-linear projection of
$Z=\sigma_D(W_1,\ldots,W_s)$. The common output field $F$ has
$\Theta_\delta(n)$ bits. Source-dependent good rows, subsources, and
conditional representative tables occur only in the analysis.

\subsubsection{Complete deterministic evaluator: parameter order and fields}

\begin{constructionbox}{Ordinary constant-rate two-source extractor}{cb:ordinary}
\cbingredients The strong condenser of Theorem~\ref{pre:strongcondenser};
the odd-power kernel and freshness bound of
Lemmas~\ref{cr2:lem:oddrank}--\ref{cr2:lem:freshness}; finite-field
representations and relative traces from Section~\ref{sec:prelim}.
The polynomial $\sigma_D$ is defined at the start of this section.

\cbsetup Fix $0<\delta\le1$ and $\tau=2^{-10}$. Instantiate the
condenser at rate $\delta/2$, target deficiency $\tau/4$, and error
$1/100$. Pad its seed by ignored bits so that its seed space has
$r=2^d\ge4$ elements, in a fixed order. Let $m_0=m_0(n)$ be its output
length. Choose the remaining constants in the following order:
\begin{equation}\label{cr2:eq:rowcount}
 s=r-1=2D-1,\qquad D=r/2.
\end{equation}
For $a=1,\ldots,s$, set
\begin{equation}\label{cr2:eq:hierarchy}
 J_a=2^{a-1},\qquad \kappa_a=\sum_{b<a}J_b=J_a-1.
\end{equation}
Let $\eta_a>0$ be the exponent supplied by Lemma~\ref{cr2:lem:freshness}
for $(J_a,\kappa_a)$, put $\eta=\min_a\eta_a$, and fix a power of two
\begin{equation}\label{cr2:eq:Lchoice}
 L\ge\max\{2,4s/\eta\}.
\end{equation}
For each sufficiently large input length $n$, set
\begin{equation}\label{cr2:eq:lengths}
 \ell=\lfloor m_0/L\rfloor,\qquad M=L\ell,\qquad
 K=\F_{2^M},\quad F=\F_{2^\ell}\subseteq K.
\end{equation}
Fix binary bases and a surjective binary-linear map
$\pi_n:F\to\F_2^{m(n)}$, where
$m(n)=\min\{\ell,\lfloor\nu_\delta n\rfloor\}$ and $\nu_\delta>0$
is the fixed sufficiently small constant certified in
Section~\ref{cr2:sec:final}. The choice is source independent.

\cbinput An ordered pair $(x,y)\in\bits^n\times\bits^n$.
\cbsteps
\begin{constructionsteps}
\item \textbf{Condense each input.} For every seed index $i\in[r]$, evaluate
the condenser on $x$ and on $y$. Retain the first $M$ bits of each result
and identify them with $U_i(x),V_i(y)\in K$.
\item \textbf{Use the fixed matched columns.} Keep the pairs
$(U_a(x),V_a(y))$ for $a=1,\ldots,s$. Omit index $r$ on both sides;
do not select columns based on their entropy.
\item \textbf{Evaluate every prescribed row kernel.} Compute
\begin{equation}\label{cr2:eq:Wa}
 W_a(x,y)=\Tr_{K/F}\sum_{j=1}^{J_a}
                      U_a(x)^{2j-1}V_a(y)^{2j-1}\quad (a\in[s]).
\end{equation}
\item \textbf{Merge in the field and then project.} Compute
\begin{equation}\label{cr2:eq:Zn}
 Z_n(x,y)=\sigma_D(W_1(x,y),\ldots,W_s(x,y)),
\end{equation}
and return $\pi_n(Z_n(x,y))$.
\end{constructionsteps}
\cboutput $m(n)=\Omega_\delta(n)$ bits, with the guarantee of
Theorem~\ref{cr2:thm:main}. For the finite initial range, use the one-bit completion
$x_1+y_1$ in Section~\ref{cr2:sec:final}.
\cbproofonly Acceptance patterns, high-entropy subsources, representative
tables, and monochromatic rectangles.
\end{constructionbox}

\subsubsection{Complete-prefix freshness on actual subsources}
Start with independent sources of entropy at least $(\delta/2)n$.
Apply Lemma~\ref{cr2:lem:subsource} separately on the two sides, with
$h=(1-\tau/4)m_0$. On the resulting product distribution $(X^*,Y^*)$, there
are accepted sets $I_X,I_Y\subseteq[r]$ of sizes at least $7r/8$.
Since $\eps_c=1/100$ gives
$\log(1/(1-8\eps_c))<1$, every $i\in I_X$ has, after truncation,
\begin{equation}\label{cr2:eq:rowentropy}
 H_\infty(U_i(X^*))\ge h-r-1-(m_0-M)\ge(1-\tau/2)M,
\end{equation}
and the analogous bound holds for $V_i(Y^*)$ when $i\in I_Y$,
for sufficiently large $n$. The last inequality uses $m_0-M<L$,
where $L,r$ are constants, while $M=\Theta_\delta(n)$.
A retained column $a$ is good when $a\in I_X\cap I_Y$. Therefore
\begin{equation}\label{cr2:eq:goodcount}
 \begin{split}
 |(I_X\cap I_Y)\cap[r-1]|
 &\ge |I_X|+|I_Y|-r-1\\
 &\ge 3r/4-1\ge r/2=D,
 \end{split}
\end{equation}
where the last inequality uses $r\ge4$. Thus the fixed deletion
preserves the required majority without choosing indices from the sources.

\begin{lemma}[Every good block is fresh relative to its complete prefix]
\label{cr2:lem:prefix}
On this same selected product distribution, every good column $a$ satisfies
\begin{equation}\label{cr2:eq:prefixfresh}
 (W_a,W_{<a})\approx_{2^{-\eta M/2}}(U_F,W_{<a}).
\end{equation}
\end{lemma}
\begin{proof}
Apply Lemma~\ref{pre:tables} to the original left input conditional on
$U_a$, and independently on the right conditional on $V_a$.
Fixing the two representative tables preserves these good marginals
exactly. Every earlier character is now a separate precomposition of
its own row kernel and has rank at most $J_bM$; every prefix parity
has rank at most $\kappa_aM$. On the designated row spaces, the
current character plus a prefix parity consequently has restricted rank
at least
\[
 (J_a-2\tau)M-2-(J_a-1)M=(1-2\tau)M-2>1.
\]
The relative trace is surjective. Its output and complete prefix have
combined length $a\ell\le s\ell\le\eta M/4\le\eta_aM$.
Lemma~\ref{cr2:lem:freshness} applies, and averaging over the independent
tables proves~\eqref{cr2:eq:prefixfresh}. The prefix itself is never fixed.
Different good columns may use different table representations; all
conclusions refer to the same original selected distribution. Neither the
representative-table argument nor rank subtraction requires an off-diagonal
pair of condenser indices. In particular, using the same deterministic
seed index on both inputs does not compromise their independence.
\end{proof}

\begin{corollary}[Joint uniformity of selected good blocks]\label{cr2:cor:gooduniform}
For any set $I$ of good columns,
$\TV(W_I,U_F^{\otimes|I|})\le |I|2^{-\eta M/2}$.
\end{corollary}
\begin{proof}
Order $I$ increasingly and project each complete-prefix estimate onto
the preceding selected coordinates. Induction and contraction of
statistical distance give the sum of the errors.
\end{proof}

\subsubsection{Majority forcing on original product supports}
\begin{lemma}[Forcing every field value]\label{cr2:lem:majority}
Let $D$ be a power of two, let $s=2D-1$, and let $F$ have
characteristic two. If any specified $D$ of $z_1,\ldots,z_s$ equal
$\alpha$, then
\begin{equation}\label{cr2:eq:forcing}
 \sigma_D(z_1,\ldots,z_s)=\alpha^D
\end{equation}
for arbitrary values of the other coordinates. On Boolean inputs,
$\sigma_D$ is strict majority.
\end{lemma}
\begin{proof}
The polynomial $\sigma_D$ is the coefficient of $t^D$ in
$\prod_i(1+z_it)$. Let \(H\subseteq[2D-1]\) be the specified set of \(D\) indices whose entries equal \(\alpha\), fixing the specified coordinates gives
\[
 \prod_i(1+z_it)=(1+\alpha t)^D
    \prod_{i\notin H}(1+z_it)
 =(1+\alpha^Dt^D)\prod_{i\notin H}(1+z_it).
\]
The remaining product has degree at most $D-1$, proving
\eqref{cr2:eq:forcing}. On a Boolean input of weight $w<2D$, the value is
$\binom wD$ modulo two. The coefficient of $t^D$ in $(1+t)^w$ is
the binary digit of $w$ at the power $D$, so it equals one precisely
when $w\ge D$.
\end{proof}

\begin{proposition}[Full image on every large original rectangle]
\label{cr2:prop:fullimage}
For sufficiently large $n$, $Z_n(A\times B)=F$ whenever
$A,B\subseteq\bits^n$ have sizes at least $2^{(\delta/2)n}$.
\end{proposition}
\begin{proof}
Start from $U_A,U_B$ and choose $D$ good columns on the actual product
subsource above. For every $\alpha\in F$ their simultaneous value
$(\alpha,\ldots,\alpha)$ has probability at least
$2^{-D\ell}-D2^{-\eta M/2}>0$, by~\eqref{cr2:eq:Lchoice}.
Lemma~\ref{cr2:lem:majority} forces $Z_n=\alpha^D$ there, regardless
of the other columns. Since $D$ is a power of two, $\alpha\mapsto
\alpha^D$ permutes $F$. Every witness is an actual point of $A\times B$.
\end{proof}

\subsubsection{Original-input rank and the final extraction step}\label{cr2:sec:final}
A trace-dual $F$-basis of $K$ expands each separated $K$-product into
at most $L=[K:F]$ separated $F$-products. Thus, including all condenser
preprocessing, column $a$ has separation length
\begin{equation}\label{cr2:eq:Ra}
 R_a\le L J_a.
\end{equation}
Expanding the fixed polynomial $\sigma_D$ gives, pointwise on the
original $n$-bit input domains,
\begin{equation}\label{cr2:eq:Zseparated}
 Z_n(x,y)=\sum_{t=1}^{R}A_t(x)B_t(y),\qquad
 R\le\sum_{|I|=D}\prod_{a\in I}LJ_a
 \le\prod_{a=1}^s(1+LJ_a)=O_\delta(1).
\end{equation}
Every absolute-trace character therefore has binary rank at most
$R\ell=O_\delta(n)$. Its nonconstancy on original rectangles follows
from Proposition~\ref{cr2:prop:fullimage}, not from a claim that a
restricted-rank lower bound survives arbitrary condenser preprocessing.

Corollary~\ref{pre:upgrade}, with original entropy gap $\delta n/2$,
gives a common $\theta_\delta>0$ such that every nonzero character has
bias at most $2^{-\theta_\delta n}$ on sources of entropy $\delta n$,
after decreasing the exponent and increasing the length threshold.
Choose a fixed positive rational $\nu_\delta\le\theta_\delta$ and set
\begin{align}
 m(n)&=\min\{\ell,\lfloor\nu_\delta n\rfloor\},\label{cr2:eq:outputlength}\\
 \Ext_n(x,y)&=\pi_n(Z_n(x,y)).\label{cr2:eq:finalextractor}
\end{align}
where $\pi_n$ is any fixed surjective binary-linear map of this rank.
By Lemma~\ref{pre:fourier},
\[
 \TV(\Ext_n(X,Y),U_{m(n)})
 \le\tfrac12 2^{m(n)/2-\theta_\delta n}
 \le2^{-\theta_\delta n/2}.
\]
The same character identity preserves the original-input rank bound.
Taking $G_L=(A_t)_t$, $G_R=(B_t)_t$ proves~\eqref{cr2:eq:mainform}.
The algorithm uses a constant number of fixed powers and field operations
and an explicit condenser; it never finds the proof-side subsources.

For the finitely many lengths below a fixed cutoff $N_\delta$, use
one bit $x_1+y_1$. Its error, including separately conditioned error,
is at most $1/2$, and its rank is at most two. It has the field-pairing
representation $(x_1,1)\cdot(1,y_1)$ over $\F_{2^n}$, followed by a
coordinate projection. Pad all encodings to one constant length and
decrease the output and error constants to at most $1/N_\delta$.
This completes Theorem~\ref{cr2:thm:main} at every positive length.

\subsubsection{Source-strong linear-output companion}
\begin{corollary}[Separate source strength at constant entropy rate]
\label{cr2:cor:source-strong}
For every fixed $\delta>0$ there are $c'_{\delta},a'_{\delta}>0$
and an explicit deterministic polynomial-time family
$\Ext^{\rm ss}_n:\bits^n\times\bits^n\to\bits^{r_n}$,
$r_n\ge c'_{\delta}n$, such that for independent sources of
min-entropy at least $\delta n$,
\[
 \TV((\Ext^{\rm ss}_n(X,Y),X),(U_{r_n},X)),\quad
 \TV((\Ext^{\rm ss}_n(X,Y),Y),(U_{r_n},Y))
 \le2^{-a'_{\delta}n}.
\]
Its output is a fixed linear projection of the field-valued
inner-product evaluator constructed above, and every nonzero output
parity still has binary truth-table rank $O_\delta(n)$.
\end{corollary}
\begin{constructionbox}{Source-strong ordinary linear-output companion}{cb:ordinary-strong}
\cbingredients The ordinary extractor of Theorem~\ref{cr2:thm:main}
(Construction~\ref{cb:ordinary}) and entropy-slack strongification,
Lemma~\ref{pre:lem:strongify-general} with zero comparisons.
\cbsetup Fix $\delta>0$ and put $h:=\delta n/4$.
Instantiate the ordinary family at rate $\delta/4$, equivalently at
component entropy threshold $h$; denote its output and error constants
by $c_0,a_0>0$.
Fix $0<b_\delta<\min\{c_0/2,a_0/4,\delta/16\}$ and set
$r_n=\lfloor b_\delta n\rfloor$ in the sufficiently large range.
\cbinput $(x,y)\in\bits^n\times\bits^n$.
\cbsteps
\begin{constructionsteps}
\item Evaluate the rate-$\delta/4$ ordinary extractor on $(x,y)$.
\item Return its first $r_n$ output bits. In the finite initial range,
return the base family's one-bit completion $x_1+y_1$.
\end{constructionsteps}
\cboutput The companion $\Ext_n^{\rm ss}(x,y)$ of
Corollary~\ref{cr2:cor:source-strong}; the \emph{same} function is strong
in either source separately. No additional seeded-extractor call or
conditioning step is performed.
\end{constructionbox}

\paragraph{Proof.}
For all sufficiently large $n$, let the base construction at entropy
$h=\delta n/4$ have at least $c_0n$ output bits and error at most
$2^{-a_0n}$.
Choose $0<b_\delta<\min\{c_0/2,a_0/4,\delta/16\}$.
For input entropy $k\ge\delta n$, the slack $k-h$ is at least
$3\delta n/4$. Lemma~\ref{pre:lem:strongify-general} with $t=0$
bounds both source-strong errors by
\[
2^{r_n}\left(2^{-a_0n}+2^{-3\delta n/4}\right)
 \le2^{-\Omega_\delta(n)}.
\]
Thus, for some fixed integer $N_\delta\ge1$ and $a_*>0$, the
error is at most $2^{-a_*n}$ and $r_n\ge b_\delta n/2$ for
$n\ge N_\delta$. Below $N_\delta$, the one-bit completion has
both strong errors at most $1/2$. Taking
\[
 c'_\delta=\min\{b_\delta/2,1/N_\delta\},\qquad
 a'_\delta=\min\{a_*,1/N_\delta\}
\]
proves the displayed bounds for every $n\ge1$. A fixed output
projection preserves the rank and separated-field representation,
proving the final assertions.
\hfill$\square$

\subsection{Sumset extraction}\label{un:sumsetlinear}

\subsubsection{Ingredients and their proofs: linear strong condenser and moment rank}
We recall the two sumset-specific cited theorems and the character-sum lemma; the rank-sensitive
approximate-duality upgrade is already proved in the preliminaries.
The reduction after these inputs is given below.

\begin{theorem}[Linear somewhere condenser]\label{ss:ext:lz}
\textnormal{\emph{Reference.} \cite[Theorem~31 and Lemma~27]{LZ}, with PFR supplied by~\cite{GGMT}.}
For each fixed $\rho>0$, there are fixed $t\ge1$, $\beta\in(0,1/4]$,
$c,c'>0$, and explicit binary-linear maps
\[
 P_1,\ldots,P_t:\F_2^n\to\F_2^{m},\qquad m=\Theta_\rho(n),
\]
such that, for every $\Hin(X)\ge\rho n$, the tuple $(P_iX)_{i\le t}$ is
within $2^{-c'n}$ of a positive mixture of distributions having some coordinate of
min-entropy at least $(1/2+\beta)m$.
\end{theorem}
This is the fixed-rate consequence of the general-weak-source construction
in Theorem 31 of Li--Zhong \cite{LZ}, obtained from the basic rate-boosting
Lemma 27 in Section 4 and constant-depth iteration. The polynomial Freiman--Ruzsa input
is supplied by \cite{GGMT}. Padding or truncation by a fixed number of bits
is handled below. The coordinates are jointly binary-linear functions of
the original input, not nonlinear source condensers.

\begin{theorem}[The constant-seed curve merger]\label{ss:ext:dkss}
\textnormal{\emph{Reference.} \cite{DW} and~\cite[Theorem~18]{DvirKSS09}.}
Let $\mathbb E=\F_q$, choose distinct $\gamma_1,\ldots,\gamma_t\in\mathbb E$,
and let $c_i(U)$ be their Lagrange basis polynomials of degree at most $t-1$.
Define, for any number $v$ of $\mathbb E$ coordinates,
\begin{equation}\label{ss:eq:curve}
 \operatorname{CM}((z_i)_{i\le t},u)=\sum_{i=1}^t c_i(u)z_i.
\end{equation}
If $(Z_i)_{i\le t}$ is a mixture with one uniform coordinate in
$\mathbb E^v$, and $U$ is independent uniform on $\mathbb E$, then for
$0<\vartheta,\eps<1$ and
\[
 q\ge\max\{t,(2t/\eps)^{1/\vartheta}\},
\]
$\operatorname{CM}(Z,U)$ is $\eps$-close to a distribution of min-entropy at
least $(1-\vartheta)v\log q$.
\end{theorem}
This is the Dvir--Wigderson curve construction \cite{DW} with
Dvir--Kopparty--Saraf--Sudan's Theorem 18 \cite{DvirKSS09}. Crucially the lower
bound on $q$ is independent of $v$. Each fixed-seed map is
$\mathbb E$-linear in the entire table, and commutes with every
$\mathbb E$-linear map on its coordinate vectors. We prove the high-entropy
rather than uniform-row consequence instead of assuming it.

The odd-degree additive-character estimate is Lemma~\ref{pre:oddweil}.

We use the uniform linear-rank disperser upgrade from
Corollary~\ref{pre:upgrade}. For one-bit characters, decrease its error
exponent if necessary to absorb the factor two between bias and
statistical distance.

Compatible finite fields and their operations can be constructed in
deterministic polynomial time in fixed characteristic \cite{Shoup}.
All extension ratios used by the final evaluator are fixed constants.

For a field $F$, a function $B(x,y)$ has separation length at most $r$ if
$B(x,y)=\sum_{j=1}^r a_j(x)b_j(y)$. Separate precomposition cannot increase
this length, sums add it, and products multiply it. If
$F=\F_{2^\ell}$, each absolute trace character of such a function has binary
rank at most $r\ell$, by a trace-dual basis for $F/\F_2$. Tracing a separated
product from an extension $K/F$ of degree $a$ uses at most $a$ separated
$F$ products.

\subsubsection{From a linear somewhere condenser to many good linear views}
For a finite alphabet $\Omega$ and $h\le\log|\Omega|$, let $\Kcal_h$ be
the distributions with every atom at most $2^{-h}$, and define
\[
 e_h(P)=\sum_{z\in\Omega}(P(z)-2^{-h})_+.
\]

\begin{lemma}[Distance to an entropy cap]\label{ss:lem:cap}
$e_h(P)=\min_{Q\in\Kcal_h}\TV(P,Q)$. Any subdistribution capped by
$2^{-h}$ can be extended to a probability distribution with that cap.
\end{lemma}
\begin{proof}
Every comparison distribution with atom cap $2^{-h}$ must remove the displayed
excess mass. Capping $P$ at that value achieves the removal, and its
missing mass fits in the remaining capacity since
$|\Omega|2^{-h}\ge1$. Redistribute the mass there.
\end{proof}

\begin{lemma}[Lifting a capped comparison through a function]\label{ss:lem:liftcap}
Let $Z$ take values in $\Omega$, and let $A:\Omega\to\Omega'$ be onto.
If $AZ$ is $\eps$-close to a distribution in $\Kcal_h$, then $Z$ is also
$\eps$-close to a distribution in $\Kcal_h$.
\end{lemma}
\begin{proof}
Lift a capped comparison distribution $Q$ of $AZ$ by the conditional distribution of
$Z$ on each nonempty actual fiber of $A$ (arbitrarily on empty fibers).
Each lifted atom is at most its image mass $Q(w)\le2^{-h}$, and
summing total variation over fibers preserves the distance.
\end{proof}

\begin{lemma}[A proof-only linear projection]\label{ss:lem:hash}
For $Z\in\F_q^v$ of min-entropy at least $h$ and an integer $r\le v$,
there is an $\F_q$-linear map $A:\F_q^v\to\F_q^r$ with
\[
 \TV(AZ,U_{\F_q^r})\le\tfrac12\sqrt{q^r2^{-h}}.
\]
If this bound is less than $1/2$, the selected map is onto.
\end{lemma}
\begin{proof}
For a uniformly random $r\times v$ linear matrix $A$,
\[
 \E_A\cp(AZ)=q^{-r}+(1-q^{-r})\cp(Z).
\]
Cauchy--Schwarz and the collision-to-distance bound show that the
average distance to $U_{\F_q^r}$ is at most
$\frac12\sqrt{q^r\cp(Z)}\le\frac12\sqrt{q^r2^{-h}}$.
Choose a matrix attaining this average bound. A nonsurjective matrix
has output supported on at most $q^{r-1}$ values, at distance
$\ge1-1/q\ge1/2$ from uniform.
\end{proof}

\begin{lemma}[High-entropy inputs to the curve merger]\label{ss:lem:highmerger}
Suppose $(Z_1,\ldots,Z_t)\in(\F_q^v)^t$ is a positive mixture having some
row of min-entropy at least $h$. Fix $0<\tau<1/2$ and choose
\[
 r=\left\lfloor\frac{h-2\log(1/\tau)}{\log q}\right\rfloor\ge1.
\]
Under the field-size condition in Theorem~\ref{ss:ext:dkss},
$\operatorname{CM}(Z,U_q)$ is $(\tau+\eps)$-close to
$\Kcal_{(1-\vartheta)r\log q}$.
\end{lemma}
\begin{proof}
Consider one mixture component with a high-entropy row $Z_g$.
Lemma~\ref{ss:lem:hash} supplies an onto $A$ such that $AZ_g$ is
within $\tau/2$ of uniform. Replace this marginal by uniform while
retaining the conditional distributions of all other projected rows. The
projected table changes by at most $\tau$, becomes somewhere random,
and the curve merger brings its output within $\tau+\eps$ of the
stated entropy cap. Since
\[
 A\operatorname{CM}(Z,u)=\operatorname{CM}((AZ_i)_i,u),
\]
Lemma~\ref{ss:lem:liftcap} transfers this capped comparison back to
the unprojected output. The projection may depend on the component:
it is proof-only. Averaging over components works because the
entropy-cap class is convex and the target entropy is common.
\end{proof}

\begin{lemma}[Disclosing a constant seed]\label{ss:lem:strongify}
Let $S$ be uniform on a set of size $Q$, with arbitrary joint distribution
with $Z$, and suppose $e_h(Z)\le\eps$. Then
\[
 \frac1Q\sum_s e_{h-\log Q}(Z\mid S=s)\le\eps.
\]
\end{lemma}
\begin{proof}
Put $p_{s,z}=\Pr[S=s,Z=z]$ and $c=2^{-h}$. By uniformity of $S$,
the left side is $\sum_{s,z}(p_{s,z}-c)_+$. For nonnegative $a_s$,
$\sum_s(a_s-c)_+\le(\sum_s a_s-c)_+$, so
\[
 \sum_{s,z}(p_{s,z}-c)_+
 \le\sum_z\left(\sum_s p_{s,z}-c\right)_+=e_h(Z).
\]
Use Lemma~\ref{ss:lem:cap} to identify excess mass with distance.
\end{proof}

\begin{theorem}[Constant-seed linear strong condenser above half]
\label{ss:thm:linearstrong}
Fix $\rho>0$ and a constant $\eps_*>0$. There are constants
$\beta>0$, a power of two $Q\ge16$, and explicit binary-linear maps
\[
 L_u:\F_2^n\to\F_2^M,\quad u\in\F_Q,\qquad M=\Theta_\rho(n),
\]
such that every $\Hin(X)\ge\rho n$ satisfies
\begin{equation}\label{ss:eq:strong}
 \frac1Q\sum_{u\in\F_Q}
 \min_{Z_u:\Hin(Z_u)\ge(1/2+\beta/2)M}
       \TV(L_uX,Z_u)\le\eps_*.
\end{equation}
All constants may also depend on $\eps_*$. The maps are linear for
each fixed seed, not just low-degree polynomials.
\end{theorem}

\begin{constructionbox}{Constant-seed linear strong condenser}{cb:linear-cond}
\cbingredients The linear somewhere condenser of Theorem~\ref{ss:ext:lz}
and the curve merger of Theorem~\ref{ss:ext:dkss}. Lemmas
\ref{ss:lem:highmerger} and~\ref{ss:lem:strongify} provide the
high-entropy and retained-seed analysis, not extra evaluation steps.
\cbsetup Given a fixed rate $\rho>0$ and constant error $\eps_*>0$,
instantiate Theorem~\ref{ss:ext:lz} to obtain $P_1,\ldots,P_t$, row length
$m$, and $0<\beta\le1/4$. Put
$\vartheta=\beta/4$ and $\eps=\min\{\eps_*/4,1/8\}$.
Fix a power of two $Q=2^s\ge\max\{16,t,(2t/\eps)^{1/\vartheta}\}$,
distinct $\gamma_1,\ldots,\gamma_t\in\F_Q$, and their Lagrange basis
polynomials $c_i(U)$. Put $M=s\lceil m/s\rceil$.
\cbinput $z\in\F_2^n$ and a numerical seed $u\in\F_Q$.
\cbsteps
\begin{constructionsteps}
\item Evaluate $P_i(z)$ for all $i\in[t]$ and append zeros to give each
row exactly $M$ bits. Parse each as a vector of $M/s$ elements of $\F_Q$.
\item Compute the coordinatewise field sum
$L_u(z)=\sum_{i=1}^t c_i(u)P_i(z)$ and return its $M$ binary coordinates.
\end{constructionsteps}
\cboutput One $M$-bit linear view. To obtain the full deterministic table,
run the same procedure at every $u\in\F_Q$ in the fixed field order;
no seed is sampled in that use. The guarantee is Theorem~\ref{ss:thm:linearstrong}.
\cbproofonly The source-dependent linear projection in
Lemma~\ref{ss:lem:hash}; it is not part of $L_u$.
\end{constructionbox}

\begin{proof}
Use Theorem~\ref{ss:ext:lz}, with its $t,m,\beta$ and decrease
$\beta$ if necessary to ensure $\beta\le1/4$. Choose
$\vartheta=\beta/4$ and
$\tau=\eps=\min\{\eps_*/4,1/8\}$.
Choose a fixed power of two $Q=2^s\ge16$ satisfying
\[
 Q\ge\max\{t,(2t/\eps)^{1/\vartheta}\}.
\]
Pad each $P_i$ by fewer than $s$ zero bits to length
$M=s\lceil m/s\rceil$, and regard it as a vector in $\F_Q^{M/s}$.
Define the actual maps by
\begin{equation}\label{ss:eq:linearviews}
 L_u(z)=\sum_{i=1}^t c_i(u)P_i(z),\qquad u\in\F_Q,
\end{equation}
with the padding understood. Multiplication by a fixed element of
$\F_Q$ is binary-linear, so all these maps are binary-linear.

The input comparison table has row entropy $h=(1/2+\beta)m$.
Lemma~\ref{ss:lem:highmerger} makes the output with an independent uniform
seed close to entropy
\[
 H=(1-\vartheta)s
       \left\lfloor\frac{h-2\log(1/\tau)}s\right\rfloor.
\]
The total error is at most $2^{-c'n}+\tau+\eps\le\eps_*$
for sufficiently large $n$. Apply Lemma~\ref{ss:lem:strongify}: the average
fixed-seed distance to entropy $H-s$ has the same bound.
Finally,
\[
 H-s\ge(1-\beta/4)(1/2+\beta)m-O_{\rho,\eps_*}(1)
       \ge(1/2+\beta/2)M
\]
for sufficiently large $n$. The last inequality has a positive linear
margin, since
$(1-\beta/4)(1/2+\beta)-(1/2+\beta/2)=3\beta/8-\beta^2/4>0$.
This proves~\eqref{ss:eq:strong}.
\end{proof}

\begin{remark}[Which error is required]
The seed space $\F_Q$ is fixed for constant $\rho,\eps_*$.
This theorem does not claim exponentially small strong-condenser error
with constant seed length. Only the small constant error in
\eqref{ss:eq:strong} will be used in a dispersion proof.
\end{remark}

\subsubsection{One actual product subsource with a common majority}
\begin{lemma}[Capped-pattern selection]\label{ss:lem:pattern}
Suppose $Q$ marginal views $L_iX$ satisfy
$Q^{-1}\sum_i e_h(L_iX)\le\eps_*$, with
$\eps_*\le1/32$. There is an actual source-local conditional distribution
$X^*$, supported in $\supp X$, and a set $G_X$ of at least $15Q/16$
indices such that
\[
 \Hin(L_iX^*)\ge h-Q-1\qquad(i\in G_X).
\]
The selected conditioning event, allowing private auxiliary coins,
has probability at least $2^{-Q-1}$.
\end{lemma}
\begin{proof}
Given a row value $v$, independently accept index $i$ with probability
$\min\{1,2^{-h}/p_i(v)\}$, where $p_i$ is the distribution of $L_iX$.
The expected number of rejections is at most $Q\eps_*$.
Markov's inequality gives probability at least
$1-16\eps_*\ge1/2$ of accepting $15Q/16$ indices.
Among at most $2^Q$ acceptance patterns, one such pattern has
probability $p\ge2^{-Q-1}$. Condition on it. At any accepted index,
\[
 \Pr[L_iX=v\mid\text{pattern}]
 \le\frac{\Pr[L_iX=v,\ i\text{ accepted}]}p
 \le 2^{-h}/p\le2^{-h+Q+1}.
\]
The pattern is source-local, and conditioning adds no support points.
\end{proof}

Apply this lemma separately to independent $X,Y$. The private coins and
selection events are on their own sides, so $X^*,Y^*$ remain independent.
At least $7Q/8$ indices belong to $G_X\cap G_Y$. Delete one fixed index
from the algorithm's list and write
\begin{equation}\label{ss:eq:columncount}
 T=Q-1=2D-1,\qquad D=Q/2.
\end{equation}
Since $Q\ge16$, at least $7Q/8-1\ge D$ retained columns are good for
both sources. The same map is used on each summand at a given column;
there is no unknown pairing of different left and right row indices.

Later we pad the maps from length $M$ to
$N=a\lceil M/a\rceil$, where $a$ is a fixed constant. With
$h=(1/2+\beta/2)M$, the good viewed sources then have, for sufficiently
large $n$, min-entropy at least
\begin{equation}\label{ss:eq:goodrate}
 \alpha_1N,\qquad \alpha_1=1/2+\beta/4.
\end{equation}
Indeed their entropy is at least $h-Q-1$, and padding adds fewer than
$a$ bits. All these losses are constants after the rate has been fixed.
We need only one positive-weight product distribution in the support of each
original product source; no statistical replacement by this distribution is made.

\subsubsection{The moment--trace-code restricted-rank bound}
Write $K=\F_{2^N}$, let $h\ge3$ and $d=2^h-1$ be fixed,
and assume $N>2h$ is sufficiently large. All traces in this
subsubsection are absolute.

\begin{proposition}[Moment--trace-code restricted rank]\label{ss:prop:rank}
Let $p\ge1$, $2p\le2^{h-2}$, $\lambda\in K^\times$,
$A,B\subseteq K$, $|A|\ge p$, and
$|B|\ge2(d-1)2^{N/2}$. Then
\begin{equation}\label{ss:eq:ranklower}
 \rk_2[\Tr(\lambda(x+y)^d)]_{A\times B}
 \ge p\log|A|-p\log p+\log|B|-N-1.
\end{equation}
The full binary matrix has rank at most $(d+1)N=2^hN$.
\end{proposition}
\begin{proof}
\emph{The trace-code restriction loss.}
Let $\mathcal C_d$ be the binary space of functions
\[
 f(z)=\varepsilon+
 \Tr\!\left(\sum_{\substack{1\le j\le d\\j\text{ odd}}}a_jz^j\right),
 \qquad \varepsilon\in\F_2,\ a_j\in K.
\]
Every polynomial trace of degree at most $d$ belongs to this space:
write each positive exponent as $2^vj$ with $j$ odd and use
$\Tr(az^{2^vj})=\Tr(a^{2^{N-v}}z^j)$.
For a nonzero coefficient tuple, the largest nonvanishing odd term
has positive odd degree at most $d$. The odd-degree character bound
(Lemma~\ref{ss:ext:weil}) makes its absolute bias strictly
less than one for sufficiently large $N$. Hence the coefficients
(including $\varepsilon$) are unique, and every nonconstant
member of $\mathcal C_d$ has bias at most $(d-1)2^{-N/2}$.

Let $r$ be the dimension of the kernel of restriction
$\mathcal C_d\to\F_2^B$. This kernel does not contain the constant
function one, and all its members vanish on $B$. Orthogonality gives
\[
 \frac{|B|}{2^N}\le 2^{-r}
    \sum_{f\text{ in the kernel}}\E_z(-1)^{f(z)}
 \le 2^{-r}+(d-1)2^{-N/2}.
\]
The lower bound on $|B|$ therefore implies
\begin{equation}\label{ss:eq:restriction}
 r\le N-\log|B|+1.
\end{equation}

\emph{Moments in the unrestricted row span.}
By Lucas's theorem, all binomial coefficients in
$(x+y)^d$ are one modulo two. For each odd $j>d/2$, the coefficient
of $y^j$ in the unique odd-exponent trace representation of
$\Tr(\lambda(x+y)^d)$ is exactly $\lambda x^{d-j}$:
no higher even exponent can reduce to $j$, since $2j>d$.
Taking $j=d-2v$ for $0\le v<2^{h-2}$, then dividing by $\lambda$
and applying inverse Frobenius, recovers the moment vector
\[
 (1,x,x^2,\ldots,x^{2^{h-2}-1})
\]
by binary-linear operations on the unrestricted row-function space.
For two distinct $p$-element subsets of $A$, equality of the sums
of these vectors would give a nontrivial binary relation among at most
$2p$ distinct Vandermonde columns. The first $2p$ coordinates rule
out such a relation. Thus the full-domain row span has dimension
at least
\[
 \log\binom{|A|}{p}\ge p\log|A|-p\log p.
\]

\emph{Restriction and global rank.}
All row functions belong to $\mathcal C_d$. Restricting the right
argument to $B$ loses at most the kernel dimension in
\eqref{ss:eq:restriction}, yielding~\eqref{ss:eq:ranklower}.
The global upper bound follows either by expanding $(x+y)^d$ into
$d+1$ separated $K$-products or by the single-product realization
\eqref{ss:eq:largeip} and a trace-dual basis.
\end{proof}

\begin{corollary}[Arbitrary mask margin at a fixed rate]\label{ss:cor:mask}
Fix $1/2<\alpha_0<\alpha_1<1$ and $\kappa\ge0$. Choose
\[
 p=\left\lceil\frac{\kappa+4}{\alpha_0}\right\rceil,
 \qquad h\ge3,\quad 2^{h-2}\ge2p.
\]
There is $\theta>0$, uniform over $N$, $\lambda\in K^\times$ and every
Boolean kernel $T:K^2\to\F_2$ of global rank at most $\kappa N$, such that
for independent $U,V$ of entropy at least $\alpha_1N$,
\begin{equation}\label{ss:eq:maskedbias}
 \left|\E(-1)^{\Tr(\lambda(U+V)^{2^h-1})+T(U,V)}\right|
 \le2^{-\theta N}
\end{equation}
for sufficiently large $N$.
\end{corollary}
\begin{proof}
On rectangles of side entropy at least $\alpha_0N$, the size condition
on $B$ holds for large $N$, because $\alpha_0>1/2$. Proposition~\ref{ss:prop:rank}
gives rank at least
\[
 (p\alpha_0+\alpha_0-1)N-p\log p-1>(\kappa+2)N.
\]
Adding $T$ leaves rank greater than one, so the phase is a disperser at
that threshold. Its global rank is at most $(2^h+\kappa)N$.
Corollary~\ref{pre:upgrade}, with entropy increment
$\alpha_1-\alpha_0$, gives the bias estimate after decreasing its exponent.
The uniformity of the cited lemma gives uniformity over all masks and
nonzero coefficients. The permitted entropy rates are fixed before the
requested margin $\kappa$; only $p,h,\theta$ depend on it.
\end{proof}

\subsubsection{Theorem}
All logarithms are base two. Entropy is binary min-entropy. Statistical
distance is $\TV(P,Q)=\frac12\sum_z|P(z)-Q(z)|$. The original input length is
$n$; a sumset extractor sees only $z=x+y\in\F_2^n$.

\begin{theorem}[Constant-rate sumset extraction]\label{ss:thm:main}
For every fixed $0<\delta\le1$, there are positive constants
$c_\delta,e_\delta,C_\delta$ such that, for all sufficiently large $n$, a
uniformly polynomial-time computable map
\[
 \Ext_n:\F_2^n\longrightarrow\F_2^{m_n},\qquad m_n\ge c_\delta n,
\]
satisfies
\[
 \TV(\Ext_n(X+Y),U_{m_n})\le2^{-e_\delta n}
\]
for every independent pair with $\Hin(X),\Hin(Y)\ge\delta n$.
For every nonzero output functional $\lambda$, the matrix
\[
 [\lambda\cdot\Ext_n(x+y)]_{x,y\in\F_2^n}
\]
has rank over $\F_2$ at most $C_\delta n$. More precisely, the output is a
binary-linear projection of a field-valued expression having a constant
number of separated products over a field of binary dimension
$\Theta_\delta(n)$. Thus explicit separated encodings of total binary
length $O_\delta(n)$ are available for its lifted characters.
\end{theorem}

Here ``uniformly polynomial-time'' is for each fixed rate: the constants in
the algorithm are fixed independently of the source distributions and of
$n$. The construction is not a search for successful coefficients.

\subsubsection{Complete sum-only construction}
The evaluator has the form
\begin{equation}\label{ss:eq:algorithm}
 \Ext_n(z)=\pi\,\sigma_D\left(
   \Tr_{K/F}((L_1z)^{2^{h_1}-1}),\ldots,
   \Tr_{K/F}((L_{2D-1}z)^{2^{h_{2D-1}}-1})\right).
\end{equation}
Here $L_i:\F_2^n\to K$ are fixed binary-linear maps,
$D$ is a fixed power of two, $K=\F_{2^N}$,
$F=\F_{2^\ell}$ with $N=a\ell=\Theta_\delta(n)$ and fixed $a$, and
$\sigma_D$ is the elementary symmetric polynomial of degree $D$.
The moment exponents $h_i$ increase according to a fixed rank hierarchy.
The map $\pi$ is a fixed binary-linear output projection.

\paragraph{One large-field product per block.}
For the rank analysis only, write $z=x+y$ and put $d_i=2^{h_i}-1$.
Take an extension $E_i/K$ of degree $d_i+1$, with trace-dual
$K$-bases $(e_{i,j})_{0\le j\le d_i}$ and
$(e^*_{i,j})_{0\le j\le d_i}$. Define the source-local encodings
\[
 A_i(x)=\sum_{j=0}^{d_i}(L_ix)^j e_{i,j},\qquad
 B_i(y)=\sum_{j=0}^{d_i}(L_iy)^{d_i-j}e^*_{i,j}.
\]
Since every binomial coefficient of $(x+y)^{2^{h_i}-1}$ is odd,
trace duality gives the exact identity
\begin{equation}\label{ss:eq:largeip}
 \Tr_{K/F}\bigl((L_i(x+y))^{d_i}\bigr)
 =\Tr_{E_i/F}\bigl(A_i(x)B_i(y)\bigr).
\end{equation}
The field degree $[E_i:F]=a2^{h_i}$ is constant once the entropy
rate and column have been fixed. This is an algebraic representation,
not a change to the sum-only evaluator. In particular, its binary
characters have global rank at most $2^{h_i}N$.

The new input-side step is an explicit linear strong condenser with a
constant seed space and output rate above one half. A small constant error
is sufficient: it is used to select one actual product subsource for a
support argument, and is not carried into the final statistical error.
All seeds of this condenser are enumerated by the evaluator.

\subsubsection{Parameter schedule, native fields and actual row algorithm}

\begin{constructionbox}{Constant-rate sumset extractor}{cb:sumset}
\cbingredients The linear strong condenser of
Theorem~\ref{ss:thm:linearstrong} (Construction~\ref{cb:linear-cond});
the moment-mask margin of Corollary~\ref{ss:cor:mask}; the polynomial
$\sigma_D$ and its forcing identity, Lemma~\ref{cr2:lem:majority};
relative traces from Section~\ref{sec:prelim}.
\cbsetup Fix $\delta>0$. Instantiate the condenser at
$\rho=\delta/2$ and $\eps_*=2^{-12}$, obtaining $Q$ linear views of
length $M=M(n)$ and the constant $\beta>0$. Enumerate the seeds once,
omit the fixed last seed, and call the remaining maps
$L_1,\ldots,L_T$, where $T=Q-1=2D-1$ and $D=Q/2$.
Put $\alpha_0=1/2+\beta/8$ and $\alpha_1=1/2+\beta/4$.
In increasing column order, define
\begin{equation}\label{ss:eq:hierarchy}
 \kappa_i=\sum_{j<i}2^{h_j},\qquad
 p_i=\left\lceil\frac{\kappa_i+4}{\alpha_0}\right\rceil,
\end{equation}
and take the smallest $h_i\ge3$ with $2^{h_i-2}\ge2p_i$.
Let $\theta_i>0$ be the resulting exponent in Corollary~\ref{ss:cor:mask},
and put $\theta=\min_i\theta_i$. Fix an integer
\begin{equation}\label{ss:eq:a}
 a\ge\max\{2,4T/\theta\}.
\end{equation}
Set $\ell=\lceil M/a\rceil$, $N=a\ell$, and fix compatible fields
$F=\F_{2^\ell}\subseteq K=\F_{2^N}$ and binary bases. Fix a surjective
binary-linear output projection $\pi:F\to\F_2^{m_n}$ with
$m_n=\min\{\ell,\lfloor\zeta_\delta n\rfloor\}$, for the fixed bias
constant $\zeta_\delta$ certified in the proof of Theorem~\ref{ss:thm:main}.
\cbinput Only $z\in\F_2^n$, interpreted in the theorem as $X+Y$.
\cbsteps
\begin{constructionsteps}
\item Evaluate $L_i(z)$ for every $i\in[T]$, append $N-M$ zero bits,
and identify the result with an element of $K$.
\item For every retained column, compute
\begin{equation}\label{ss:eq:rows}
 W_i(z)=\Tr_{K/F}\bigl((L_i z)^{2^{h_i}-1}\bigr)\in F.
\end{equation}
\item Compute $Z_n(z)=\sigma_D(W_1(z),\ldots,W_T(z))$ in $F$,
and return $\pi(Z_n(z))$.
\end{constructionsteps}
\cboutput $m_n=\Omega_\delta(n)$ bits with the guarantee of
Theorem~\ref{ss:thm:main}. All preprocessing maps are binary-linear;
no decomposition of $z$ as $x+y$ is supplied to the evaluator.
\cbproofonly Good-column selection, capped patterns, representative
tables, and the extension fields $E_i$ in the separated-product proof.
\end{constructionbox}

The coefficient field $\F_Q$ used for the linear maps need not be a
subfield of $K$: it specifies binary-linear preprocessing. The trace
fields $F\subseteq K$ have the displayed compatible representation.

For every output character of $W_j(x+y)$, trace transitivity and
Proposition~\ref{ss:prop:rank} give the global binary rank bound
$2^{h_j}N$. Hence every parity of the full earlier prefix
\[
 \Pi_i=(W_1(X+Y),\ldots,W_{i-1}(X+Y))
\]
has rank at most $\kappa_iN$. This bound is independent of the later
choice of $a$: the row output is shortened, but the native field core still acts on $N$ bits. All asymptotic estimates in the local theorem
hold uniformly for these field degrees.

\subsubsection{Freshness on the actual product source}
We first give the elementary representative-table device used to transfer
a local rank estimate without fixing a joint prefix.

\begin{lemma}[Separate representative tables]\label{ss:lem:tables}
Let $X,Y$ be independent and let $U=L(X)$, $V=R(Y)$. There are independent
random deterministic maps $\mathsf T,\mathsf S$, independent of $U,V$,
such that
\[
 (\mathsf T(U),\mathsf S(V))\stackrel{\rm distribution}=(X,Y),
 \qquad L(\mathsf T(u))=u,\quad R(\mathsf S(v))=v
\]
on the supported values. Separate precomposition by each fixed map pair
cannot increase the binary rank of any raw kernel.
\end{lemma}
\begin{proof}
Sample each table entry $\mathsf T(u)$ from the conditional distribution
$X\mid U=u$, before independently drawing $U$; then
$\mathsf T(U)$ reproduces $X$, with $L(\mathsf T(u))=u$ on its support.
Repeat independently on the right. For fixed tables, separate
precomposition selects or repeats matrix rows and columns, so it
cannot increase binary rank.
\end{proof}

\begin{lemma}[Complete-prefix freshness]\label{ss:lem:prefix}
Let $X^*,Y^*$ be an actual independent product distribution for which the
$i$th viewed marginals have entropy at least $\alpha_1N$. Then the
rows~\eqref{ss:eq:rows} satisfy
\begin{equation}\label{ss:eq:fresh}
 \TV((W_i(X^*+Y^*),\Pi_i),(U_F,\Pi_i))
 \le\frac12\,2^{i\ell/2-\theta_iN}
 \le2^{-\theta N/2}
\end{equation}
for sufficiently large $n$. No hypothesis is imposed on the entropy
of the earlier rows.
\end{lemma}
\begin{proof}
Fix a nonzero character of the current field output and any binary
character of the prefix. In the native field, their combined phase is
\[
 \Tr_{K/\F_2}(\lambda(L_i x+L_i y)^{2^{h_i}-1})
       +T_{\rm pre}(x,y),\qquad \lambda\in F^\times\subset K^\times.
\]
The raw mask has rank at most $\kappa_iN$. Use
Lemma~\ref{ss:lem:tables} for the current two views. For every fixed
representative-table pair it is still a mask of rank at most
$\kappa_iN$ on $K^2$, while the two viewed distributions remain
independent and retain their entropies. Corollary~\ref{ss:cor:mask}
bounds the phase bias by $2^{-\theta_iN}$. Average over the table
pairs; this is the same original joint output/prefix distribution.

Compare that distribution with a uniform current field value independent of
the actual prefix. Fourier coefficients with zero current character
cancel exactly. At most $(2^\ell-1)2^{(i-1)\ell}$ coefficients remain.
Parseval and Cauchy--Schwarz give the first bound in~\eqref{ss:eq:fresh}.
Since $a\ge4T/\theta$, one has $i\ell/2\le\theta N/8$, which implies
the second, weaker bound. We have not conditioned the two sources on
$\Pi_i$; prefix freshness was proved by mixed characters on the
independent source pair.
\end{proof}

\subsubsection{Majority forcing and the final error bound}
The forcing identity from Lemma~\ref{cr2:lem:majority} applies to
the same $\sigma_D$ over the sumset output field $F$.

\begin{proposition}[Full output support on every lower-rate rectangle]
\label{ss:prop:fullsupport}
For the fixed construction above, put
\[
 Z_n(z)=\sigma_D(W_1(z),\ldots,W_{2D-1}(z))\in F.
\]
For any $A,B\subseteq\F_2^n$ with
$|A|,|B|\ge2^{(\delta/2)n}$, one has $Z_n(A+B)=F$ for sufficiently
large $n$.
\end{proposition}
\begin{proof}
Apply the strong linear condenser to $U_A$ and $U_B$. Use
Lemma~\ref{ss:lem:pattern} separately and obtain one actual product distribution
$X^*,Y^*$ supported in $A\times B$, with at least $D$ common good
retained columns. Fix a set $H$ of $D$ such indices in their algorithmic
order. Their viewed entropies are at least $\alpha_1N$ by
\eqref{ss:eq:goodrate}, so Lemma~\ref{ss:lem:prefix} applies to every index.
Project each earlier prefix to the earlier selected indices and use
successive replacements. This gives
\[
 \TV((W_i(X^*+Y^*))_{i\in H},U_F^{\otimes D})
       \le D2^{-\theta N/2}.
\]
The elementary induction here first replaces the current coordinate
against the actual preceding selected marginal, then replaces that
preceding marginal by uniform using the previous induction bound.
It does not require independence of bad or future columns.

By~\eqref{ss:eq:a}, $D\ell\le\theta N/4$. Hence, for large $n$,
\[
 D2^{-\theta N/2}<2^{-D\ell}=|F|^{-D}.
\]
For every $u\in F$, the event that all selected columns equal $u$
therefore has positive probability under the actual product distribution.
Lemma~\ref{cr2:lem:majority} makes the merged value $u^D$ on that event.
The map $u\mapsto u^D$ is a Frobenius automorphism of $F$, so every
field value occurs. All these pairs belong to $A\times B$.
\end{proof}

\begin{lemma}[Rank of the actual sum-only composition]\label{ss:lem:finalrank}
Each $W_i(x+y)$ has separation length at most $a2^{h_i}$ over $F$.
Consequently $Z_n(x+y)$ has separation length at most
\begin{equation}\label{ss:eq:Rstar}
 R_*:=\sum_{\substack{I\subseteq[2D-1]\\|I|=D}}
                       \prod_{i\in I}a2^{h_i},
\end{equation}
a constant depending only on the fixed rate. Every absolute trace
character of $Z_n(x+y)$ has binary rank at most $R_*\ell=O_\delta(n)$.
\end{lemma}
\begin{proof}
Apply the exact field-product identity~\eqref{ss:eq:largeip} to
$W_i(x+y)$. Trace duality over $E_i/F$ yields at most
$[E_i:F]=a2^{h_i}$ separated $F$-products. A fixed product of rows
multiplies their separation lengths, and sums add them; expanding
$\sigma_D$ therefore gives~\eqref{ss:eq:Rstar}. Finally, each
separated $F$-product contributes at most $\ell$ binary rank after
absolute trace. All identities hold for the actual unconditioned
sum-only evaluator.
\end{proof}

\begin{proof}[Proof of Theorem~\ref{ss:thm:main}]
For every nonzero $\lambda\in F$, the Boolean function
\[
 f_\lambda(x,y)=\Tr_{F/\F_2}(\lambda Z_n(x+y))
\]
is nonconstant on every rectangle of side entropy $(\delta/2)n$, by
Proposition~\ref{ss:prop:fullsupport}. Lemma~\ref{ss:lem:finalrank} gives a
uniform $O_\delta(n)$ rank bound. Corollary~\ref{pre:upgrade}, now on
the original $n$-bit summands with entropy increment $\delta/2$, yields
some $\zeta_\delta>0$ such that every such character has bias at most
$2^{-\zeta_\delta n}$ on all independent entropy-$\delta n$ inputs.
Take any fixed surjective binary-linear projection to
\[
 m_n=\min\{\ell,\lfloor\zeta_\delta n\rfloor\}=\Omega_\delta(n)
\]
bits. Nonzero output parities are nonzero absolute-trace characters of
$F$. Thus
\[
 \TV(\pi Z_n(X+Y),U_{m_n})
 \le\tfrac12\sqrt{2^{m_n}-1}\,2^{-\zeta_\delta n}
 \le\tfrac12\,2^{-\zeta_\delta n/2}.
\]
This gives the error and output claims and retains the rank bound.

The linear maps, powers, fields, and symmetric polynomial are fixed
independently of sources. Finite-field construction and arithmetic are
polynomial-time for fixed characteristic; the number of views, moment
orders, field-extension ratio, and merger degree are fixed constants.
The symmetric polynomial can also be evaluated by its elementary
coefficient recurrence. No good-column selection, hash projection used
in Lemma~\ref{ss:lem:highmerger}, acceptance pattern, representative table,
or approximate-duality rectangle is computed by the evaluator.
All length roundings cost only constants at the relevant stage and are
absorbed for sufficiently large $n$. This proves uniform explicitness.
\end{proof}


\subsubsection{Ingredients and proofs for the one-sided non-malleable sumset seed}
\begin{lemma}[Rank and Fourier conventions]\label{ns:lem:basic}
Separate precomposition cannot increase the binary rank of a function matrix. Ranks are subadditive. If $F=\F_{2^\ell}$ and $H(x,y)=\sum_{j=1}^r a_j(x)b_j(y)$ over $F$, then every absolute-trace character of $H$ has binary rank at most $r\ell$. A fixed-degree relative trace costs at most its degree in separated products.

If $(Z,T)\in\F_2^q\times\F_2^v$ and every mixed character with nonzero coefficient on $Z$ has bias at most $\omega$, then
\[
 \TV((Z,T),(U_q,T))\le\tfrac12\sqrt{(2^q-1)2^v}\,\omega.
\]
For a pair of $q$-bit outputs this applies with $v=q$.
\end{lemma}
\begin{proof}
The rank assertions are the common separation and trace-duality facts
of Section~\ref{sec:prelim}; the distance bound is
Lemma~\ref{pre:fourier} with only honest-bearing characters retained.
\end{proof}

\begin{lemma}[Actual conditional representatives]\label{ns:lem:represent}
For a deterministic map $U=T(X)$, one can write the joint distribution of $(X,U)$ as a mixture of $(R(U),U)$ with $R$ a random deterministic table independent of $U$, and $T(R(u))=u$ for all positive-probability $u$. Separate such representations for independent sources preserve their independence and do not increase any separately lifted matrix rank after fixing the tables.
\end{lemma}
\begin{proof}
Apply Lemma~\ref{pre:tables}. Its representatives lie in their actual
fibers, and separate precomposition cannot increase matrix rank.
\end{proof}

\subsubsection{Moment secants against one-sided tampering}\label{ns:sec:secants}
Fix $h\ge3$, $d=2^h-1$, $K=\F_{2^N}$, and $N>2h$, taking $N$ sufficiently large below. Put $L=2^{h-2}$.

\begin{lemma}[The trace-code restriction kernel]\label{ns:lem:code}
Let
\[
 \mathcal C_d=\left\{\eps+\Tr_{K/\F_2}\sum_{\substack{1\le j\le d\\ j\text{ odd}}}c_j y^j\right\}.
\]
Its displayed coefficients are unique, and every nonconstant member has bias at most $(d-1)2^{-N/2}$. If $B\subseteq K$ has size at least $2(d-1)2^{N/2}$, then
\[
 \dim\ker(\mathcal C_d\to\F_2^B)\le N-\log|B|+1.
\]
\end{lemma}
\begin{proof}
These are precisely the unique-coefficient and restriction-kernel
claims established in the first part of the proof of
Proposition~\ref{ss:prop:rank}; equation~\eqref{ss:eq:restriction}
gives the displayed dimension bound. That argument uses only the
right set $B$ and applies unchanged to secant row functions.
\end{proof}

\begin{lemma}[A forest of moment secants]\label{ns:lem:forest}
Let $A$ be a finite subset of a field $K$ of characteristic two and $f:A\to K$ satisfy $f(x)\ne x$. For nonzero $a,b\in K$, put
\[
 v_L(x)=(1,x,\ldots,x^{L-1}),\qquad w(x)=a v_L(x)+b v_L(f(x)).
\]
If $4p\le L$ and $|A|\ge\max\{12,4p\}$, then
\[
 \dim_{\F_2}\Span\{w(x):x\in A\}
 \ge\log\binom{\lfloor|A|/3\rfloor}{p}
 \ge p\log|A|-p\log(4p).
\]
\end{lemma}
\begin{proof}
Draw the undirected edge $\{x,f(x)\}$ for every $x\in A$. There are no loops; an undirected edge is represented at most twice, so there are at least $|A|/2$ distinct edges. The directed graph has out-degree at most one (vertices outside $A$ have out-degree zero), hence each undirected component has at most one cycle. After duplicate reciprocal edges are identified, every cycle has length at least three. Delete one edge per cycle. The remaining forest has at least two thirds of the distinct edges, hence at least $|A|/3$ edges. Choose one domain representative for each of $\lfloor|A|/3\rfloor$ retained edges.

Sums of $w(x)$ over different $p$-element subsets of these representatives are distinct. Otherwise their symmetric difference is a nonempty forest of at most $2p$ edges and at most $4p$ vertices. At a leaf vertex its coefficient in the proposed relation is either $a$ or $b$, and is nonzero. The relation among the corresponding at most $4p\le L$ Vandermonde columns is therefore nontrivial, a contradiction. There are $\binom{\lfloor|A|/3\rfloor}{p}$ distinct sums in the binary span. Finally $\lfloor|A|/3\rfloor\ge|A|/4$ for $|A|\ge12$, and $\binom up\ge(u/p)^p$.
\end{proof}

\begin{proposition}[Mixed restricted-rank bound]\label{ns:prop:mixedrank}
For $\lambda\in K^\times$, $\mu\in K$, and fixed-point-free $f:A\to K$, let
\[
 M(x,y)=\Tr_{K/\F_2}\bigl(\lambda(x+y)^d+\mu(f(x)+y)^d\bigr).
\]
If $4p\le L$, $|A|\ge\max\{12,4p\}$, and $|B|\ge2(d-1)2^{N/2}$, then
\begin{equation}\label{ns:eq:mixedrank}
 \rk_2 M[A,B]\ge p\log|A|-p\log(4p)+\log|B|-N-1.
\end{equation}
The full-domain matrix, for any extension of $f$, has rank at most $2^{h+1}N$.
\end{proposition}
\begin{proof}
All row functions lie in $\mathcal C_d$. For $0\le t<L$, the exponent $d-2t$ is odd and exceeds $d/2$. In the binomial expansion, no other exponent can reduce to it under trace, so its coefficient is exactly
\[
 \lambda x^{2t}+\mu f(x)^{2t}.
\]
Coefficient extraction followed by inverse squaring is binary-linear and gives $\sqrt\lambda\,v_L(x)+\sqrt\mu\,v_L(f(x))$. If $\mu\ne0$, Lemma~\ref{ns:lem:forest} lower-bounds the full row span. If $\mu=0$, sums over distinct $p$-subsets of $A$ are distinct by ordinary Vandermonde independence; the same weaker bound follows. Lemma~\ref{ns:lem:code} bounds the dimension lost on restricting to $B$.

Each of the two binomial expansions has $d+1=2^h$ separated $K$-products; Lemma~\ref{ns:lem:basic} gives the global bound. The one-output character bound is $2^hN$.
\end{proof}

\begin{corollary}[Fixed-rate masks of any prescribed linear rank]\label{ns:cor:masked}
Fix $1/2<\alpha_0<\alpha_1<1$ and a constant $\kappa>0$. One can choose a fixed $h$ such that, for all sufficiently large $N$, every fixed-point-free $f:K\to K$, every $\lambda\ne0$, every $\mu$, and every Boolean mask $T$ of global rank at most $\kappa N$ satisfy
\[
 \left|\E(-1)^{\Tr(\lambda(U+V)^{2^h-1}+\mu(f(U)+V)^{2^h-1})+T(U,V)}\right|
 \le2^{-\theta N}
\]
for independent $U,V$ of min-entropy at least $\alpha_1N$, where $\theta>0$ is uniform over $f,\lambda,\mu,T$. The same conclusion for $\mu=0$ requires no condition on $f$.
\end{corollary}
\begin{proof}
Choose $p$ with $(p+1)\alpha_0-1>\kappa+2$, then $h$ with $4p\le2^{h-2}$. On every side-$\alpha_0N$ rectangle, Proposition~\ref{ns:prop:mixedrank} leaves rank greater than one after the mask is added. The full rank is at most $(2^{h+1}+\kappa)N$. Corollary~\ref{pre:upgrade}, with entropy increment $\alpha_1-\alpha_0$, proves the result. A map known to be fixed-point-free only on the actual left support may be extended outside it by a nonzero translation; all distributions in the application are unchanged.
\end{proof}

\subsubsection{One-sided non-malleable sumset seed: main theorem}
The same source-compatible architecture has a strengthening used in the \emph{sumset} entropy-recycling theorem of Section~\ref{un:sumsetoptimal}. Only the first summand is tampered. This does not assert two-sided non-malleability of a function of the sum. The same seed also supplies the one-sided interface used by the ordinary entropy-recycling wrapper in Section~\ref{un:standard}. The independent ordinary constant-rate construction in Section~\ref{un:ordinarylinear} does not use this seed.

\begin{theorem}[One-sided non-malleable sumset seed]\label{ns:thm:seed}
For every fixed $0<\sigma\le1$, there are $c_\sigma,e_\sigma>0$ and a deterministic polynomial-time map $G_n:\F_2^n\to\F_2^{q_n}$, $q_n\ge c_\sigma n$, such that independent sources of min-entropy at least $\sigma n$ and every $f$ with no fixed points on $\supp X$ satisfy
\begin{equation}\label{ns:eq:nmseed}
 (G_n(X+Y),G_n(f(X)+Y))
 \approx_{2^{-e_\sigma n}}(U_{q_n},G_n(f(X)+Y)).
\end{equation}
Every output character has global lifted binary rank $O_\sigma(n)$, as does every one-comparison mixed character. Ordinary extraction follows by choosing any nonzero translation $f$ and projecting the honest output. The family is defined for every positive integer $n$, using the finite-length completion specified below.
\end{theorem}

\paragraph{Construction.}
Augment the linear strong-condenser views by polynomial-evaluation fingerprints of the original input. Enumerate a constant-size evaluation set, so every nonzero raw difference survives in almost every fingerprint. On each augmented view apply a moment power of order chosen in a fixed rank hierarchy, trace to a common field, and merge with the same degree-$D$ elementary symmetric polynomial. The larger moment orders protect the honest-bearing secant coefficients against every earlier-prefix character. A final binary-linear projection supplies the seed. Only the sum input is read by this evaluator; the tampering map occurs solely in the proof.
\paragraph{Proof.}
We first establish the secant-rank estimate and then give the full parameter and support-forcing argument.

\subsubsection{Complete one-sided construction and its proof}

\begin{constructionbox}{One-sided non-malleable sumset seed}{cb:sumset-seed}
\cbingredients The linear strong condenser of
Theorem~\ref{ss:thm:linearstrong}; the secant-mask bound of
Corollary~\ref{ns:cor:masked}; the majority polynomial of
Lemma~\ref{cr2:lem:majority}; fixed field representations and relative traces.
\cbsetup Fix the target component rate $\sigma>0$.
Instantiate the condenser at rate $\nu=\sigma/64$ and error $2^{-16}$,
obtaining maps $L_u$, $u\in\F_Q$, row length $M$, and constant $\beta$.
Put $r=\lfloor\beta M/16\rfloor$, $E=\F_{2^r}$, and
$b_n=\lceil n/r\rceil$. Fix a constant $b_*$ bounding $b_n$ in the
large-length range, a power of two $R\ge64b_*$, and distinct
$v_1,\ldots,v_R\in E$. Order $\F_Q\times[R]$ lexicographically and
delete its last pair. The retained number is $T=QR-1=2D-1$, $D=QR/2$.

Set $\alpha_0=1/2+\beta/8$ and $\alpha_1=1/2+\beta/4$.
For each retained index $i$, in order, put
$\kappa_i=1+2\sum_{j<i}2^{h_j}$ and choose $h_i$ by
Corollary~\ref{ns:cor:masked}. Concretely, its proof permits an integer
$p_i$ with $(p_i+1)\alpha_0-1>\kappa_i+2$, followed by
$h_i\ge3$ with $2^{h_i-2}\ge4p_i$.
Let $\theta=\min_i\theta_i>0$ be the supplied bias exponent and fix
$a\ge\max\{2,16T/\theta\}$. Put
$\ell=\lceil(M+r)/a\rceil$, $N=a\ell$, and
$F=\F_{2^\ell}\subseteq K=\F_{2^N}$.
Fix a surjective projection to
$q_n=\min\{\ell,\lfloor cn/2\rfloor\}$ bits, using the constant $c>0$
certified by the final global lift below.
\cbinput Only $z\in\F_2^n$.
\cbsteps
\begin{constructionsteps}
\item Pad $z$ to $b_nr$ bits, parse it as $z_0,\ldots,z_{b_n-1}\in E$,
and form $P_z(V)=\sum_{j<b_n}z_jV^j$.
\item For every retained pair $(u,v_j)$, concatenate
$T_{u,v_j}(z)=(L_u z,P_z(v_j))$. Append zeros to reach $N$ bits;
identify the result with $T_i(z)\in K$ in the fixed column order.
\item Compute $W_i(z)=\Tr_{K/F}(T_i(z)^{2^{h_i}-1})$ for all $i\in[T]$.
\item Compute $Z_n(z)=\sigma_D(W_1(z),\ldots,W_T(z))$ and return its
fixed $q_n$-bit projection $G_n(z)$.
\end{constructionsteps}
\cboutput The seed in Theorem~\ref{ns:thm:seed}. Its non-malleability
allows tampering with the first summand only, with the second unchanged. For the finite initial range, return one zero bit as specified after the proof.
\cbproofonly The map $f$, its functional graph and forest, acceptance
patterns, and the decision whether a comparison image has a heavy atom.
They do not affect the evaluated list of views or exponents.
\end{constructionbox}

\label{ns:sec:seed}

\paragraph{Linear fingerprints prevent local fixed points}
Use Theorem~\ref{ss:thm:linearstrong} at raw threshold $\nu n$ with $\nu=\sigma/64$ and error $\varepsilon_c=2^{-16}$. Let $M,Q,\beta$ be its parameters. Put
\[
 r=\lfloor\beta M/16\rfloor,\qquad E=\F_{2^r},\qquad b_n=\lceil n/r\rceil.
\]
For large $n$, $b_n\le b_*$ for a fixed $b_*$. Pad $z$ to $b_n$ field coefficients and write $P_z(V)=\sum_{j<b_n}z_jV^j$. Choose a power of two $R\ge64b_*$ and fixed distinct $v_1,\ldots,v_R\in E$. Define all augmented linear views
\begin{equation}\label{ns:eq:augmented}
 T_{u,v}(z)=(L_u z,P_z(v))\in\F_2^{M+r}.
\end{equation}
Every distinct raw pair has different fingerprints at all but $b_n-1$ of these indices, by the polynomial root bound. Thus at least $63R/64$ fingerprints detect $x\ne f(x)$, pointwise in $x$.

\paragraph{Actual common good views}
Suppose $X,f(X),Y$ all have entropy at least $\nu n$. Let $K_t=(1/2+\beta/2)M$. For each $L_uX$ and $L_uf(X)$ introduce a source-local acceptance bit, accepting value $z$ with probability $\min\{1,2^{-K_t}/P(z)\}$. The rejection probability equals the excess above the entropy cap and is at most the distance in Theorem~\ref{ss:thm:linearstrong}. The expected fraction of indices with either left rejection is at most $2\varepsilon_c$. Hence with probability at least $1-32\varepsilon_c>1/2$, at most $Q/16$ indices are left-bad.

Also record the $R$ deterministic fingerprint-difference bits. There are at most $2^{2Q+R}$ full left patterns. A pattern satisfying the rejection bound has probability at least $2^{-2Q-R-1}$. Fix it. Each accepted row marginal of both $X$ and $f(X)$ is capped by $2^{-K_t+2Q+R+1}$. The set of detecting fingerprint indices is fixed and has size at least $63R/64$. Independently on the right, a positive pattern makes at least $15Q/16$ rows of $Y$ capped at entropy $K_t-O(Q)$. Denote the resulting independent actual source distributions by $X^*,Y^*$.

At least $7Q/8$ indices $u$ are good for all three marginals. Pair them with the detecting fingerprint indices. Among the $QR$ augmented views, more than $3QR/4$ have both honest and comparison left views, and the right view, of entropy at least $K_t-O(Q+R)$, and satisfy
\begin{equation}\label{ns:eq:localnofix}
 T_{u,v}(x)\ne T_{u,v}(f(x))\quad\text{on the selected left support}.
\end{equation}
Delete one fixed grid index and write
\[
 T=QR-1=2D-1,\qquad D=QR/2.
\]
Then $D$ is a power of two and at least $D$ retained columns are common good. For a single execution, use the same argument without the comparison cap or difference bits: more than half the columns are good whenever $X,Y$ have entropy at least $\nu n$.

Let
\[
 \alpha_0=1/2+\beta/8,\qquad \alpha_1=1/2+\beta/4.
\]
After padding by any fixed number of bits to a native length $N=M+r+O(1)$, every common good view has entropy at least $\alpha_1N$ for sufficiently large $n$. Indeed $(1/2+\beta/2)>(1/2+\beta/4)(1+\beta/16)$, with a fixed positive margin; all pattern and padding losses are constants.

\paragraph{Hierarchy, actual prefixes, and field lengths}
Order the retained columns once. Recursively set
\[
 \kappa_i=1+2\sum_{j<i}2^{h_j}
\]
and choose $h_i$ by Corollary~\ref{ns:cor:masked} with that mask budget and the fixed rates $\alpha_0,\alpha_1$. Let $\theta_i>0$ be its exponent and $\theta=\min_i\theta_i>0$. Choose a fixed integer $a\ge\max\{2,16T/\theta\}$, then put
\[
 \ell=\lceil(M+r)/a\rceil,\qquad N=a\ell,\qquad
 F=\F_{2^\ell}\subseteq K=\F_{2^N}.
\]
Pad every view by fewer than $a$ zero bits and identify it with $K$. The actual field rows are
\begin{equation}\label{ns:eq:seedrows}
 W_i(z)=\Tr_{K/F}\bigl(T_i(z)^{2^{h_i}-1}\bigr).
\end{equation}
Each character has rank at most $2^{h_i}N$. The complete pair prefix
\[
 \Pi_i=(W_j(X^*+Y^*),W_j(f(X^*)+Y^*):j<i)
\]
has every parity of rank below $\kappa_iN$ on the original raw domain.

\begin{lemma}[Pair freshness for a common good column]\label{ns:lem:pairfresh}
On the one actual product distribution selected above, every common good column obeys
\[
 \TV((W_i,W'_i,\Pi_i),(U_F,U'_F,\Pi_i))\le2^{-\theta N/2}
\]
for sufficiently large $n$, where $W'_i=W_i(f(X^*)+Y^*)$. For a single execution the analogous prefix distribution holds with no larger error.
\end{lemma}
\begin{proof}
Take $U=T_iX^*$ and $V=T_iY^*$. By Lemma~\ref{ns:lem:represent}, fix independent conditional representative tables for the two raw inputs. On attained $u$, the comparison view is a deterministic map $f_i(u)$ distinct from $u$, by~\eqref{ns:eq:localnofix}. Extend $f_i$ outside the attained support without fixed points. Every prefix parity, after this separate substitution, still has rank below $\kappa_iN$. The native honest core is $(U+V)^{2^{h_i}-1}$; trace transitivity sends every nonzero $F$ coefficient to a nonzero native coefficient.

Apply Corollary~\ref{ns:cor:masked} and average the representative tables. Lemma~\ref{ns:lem:basic} gives
\[
 \TV((W_i,W'_i,\Pi_i),(U_F,W'_i,\Pi_i))
 \le\tfrac12 2^{i\ell-\theta_iN}.
\]
Now represent the raw left input conditional on its comparison view instead. That view has the same high entropy, the honest view is pointwise different, and the right source is unchanged. The same bound makes $W'_i$ fresh relative to $(W_i,\Pi_i)$. Both claims concern the original joint distribution, not a hybrid source. The operations replacing each current coordinate by independent uniform randomness commute and contract statistical distance. Their sum costs at most $2^{i\ell-\theta_iN}\le2^{-\theta N/2}$. No conditioning on the joint prefix is performed. The single-execution proof uses $\mu=0$.
\end{proof}

\paragraph{Forcing, heavy comparisons, and the global lift}
Define
\begin{equation}\label{ns:eq:fieldmerge}
 Z_n(z)=\sigma_D(W_1(z),\ldots,W_{2D-1}(z)).
\end{equation}
If $D$ entries equal $u$, then the degree-$D$ elementary symmetric polynomial equals $u^D$, regardless of the others: the generating polynomial has factor $(1+ut)^D=1+u^Dt^D$, while the remaining factor has degree at most $D-1$.

For a set of $D$ common good indices, Lemma~\ref{ns:lem:pairfresh} and successive replacement give distance at most $D2^{-\theta N/2}$ from $U_F^{\otimes2D}$. For sufficiently large $n$, this is less than $|F|^{-2D}$, since $2D\ell\le\theta N/8$. Therefore every event prescribing all honest selected entries to be $u$ and all comparison selected entries to be $v$ has positive probability. It forces $(Z_n,Z'_n)=(u^D,v^D)$. As $D$ is a power of two, this pair has full support $F^2$. The single-execution argument gives full support $F$ on every raw product support of side entropy at least $\nu n$.

It suffices first to treat globally fixed-point-free $f$. A map with no fixed points only on the original left support can be extended outside that support by a nonzero translation, without changing either evaluated distribution. To prove mixed dispersion at side entropy $(\sigma/2)n$, take uniform $X,Y$ on any corresponding supports, and a proposed constant mixed character with nonzero honest coefficient. If $f(X)$ has an atom of mass greater than $2^{-\sigma n/8}$, fix it on the left. This leaves honest left entropy at least $3\sigma n/8$. The comparison character is now a Boolean function only of $Y$; fix its value at cost at most one bit. The independent honest sources still exceed $\nu n$, so single-execution full support contradicts constancy. If no such atom exists, $X,f(X),Y$ all meet the condenser threshold. The common-good construction and pair forcing give the contradiction. The case with zero comparison coefficient is already single-execution dispersion.

Each row in~\eqref{ns:eq:seedrows} has field separation at most $a2^{h_i}$, because $T_i$ is linear. The separation of~\eqref{ns:eq:fieldmerge} is at most
\[
 R_*:=\sum_{|I|=D}\prod_{i\in I}a2^{h_i}=O_\sigma(1).
\]
A mixed character of $Z_n(x+y),Z_n(f(x)+y)$ has rank at most $2R_*\ell=O_\sigma(n)$. Corollary~\ref{pre:upgrade}, now on the original $n$-bit sources, raises the threshold from $\sigma n/2$ to $\sigma n$ and gives all honest-bearing biases at most $2^{-c n}$, uniformly over $f$. Project to $q_n=\min\{\ell,\lfloor cn/2\rfloor\}$ bits. Lemma~\ref{ns:lem:basic} gives one-sided non-malleability error at most $2^{q_n-1-cn}=2^{-\Omega_\sigma(n)}$. This proves Theorem~\ref{ns:thm:seed} for all sufficiently large $n$. All parameter choices are fixed before the source is known.

For the remaining finite set of positive lengths, return one zero bit
and decrease $c_\sigma,e_\sigma$ as in the
\hyperref[pre:finite-lengths]{finite-length convention in Section~\ref{un:tools}}.
This preserves one-sided non-malleability, ordinary extraction, and
the lifted rank bound, and completes the theorem at every positive length.

\begin{remark}[Only one summand is tampered]
Theorem~\ref{ns:thm:seed} does not allow simultaneous translations of both summands. Such translations preserve $X+Y$ and would make honest and comparison outputs equal, which is incompatible with ordinary non-malleability. There is no such contradiction when the second summand stays unchanged and $f$ has no fixed points.
\end{remark}

\subsection{Fixed-\texorpdfstring{$C$}{C} non-malleable two-source extraction}\label{un:nonmalleable}
\subsubsection{Tagged affine seeds and one-sided Gold recognition}
The starter below uses a short left prefix and a longer right row.
Its inverse argument recognizes only the left tampering. The outer
construction supplies distinct numerical tags by source-local
fingerprints of the original inputs.
\begin{theorem}[Near-full affine non-malleability]\label{crn:ext:affinenm}
\textnormal{\emph{Reference.} \cite[Theorem~1.12 and Section~3]{Li23}.}
There are absolute constants $\delta_A,c_A,e_A>0$ and an explicit map $V_N$ on $N$ bits, with at least $c_A N$ output bits, which is an affine non-malleable extractor at entropy $(1-\delta_A)N$ and error $2^{-e_A N}$. It also has the ordinary affine-extraction guarantee at that threshold.
\end{theorem}
This is Li's Theorem~1.12 and its construction in Section~3 of~\cite{Li23}. The no-fixed-point convention can be used on an active affine support by the next elementary extension lemma.

\begin{lemma}[Extending an affine comparison off its active support]\label{crn:lem:affineextension}
If $A\subseteq\F_2^N$ is a nonempty affine space and $f$ is affine with $f(x)\ne x$ for every $x\in A$, there is a globally fixed-point-free affine $\widetilde f$ agreeing with $f$ on $A$.
\end{lemma}
\begin{proof}
Choose an affine retraction $P:\F_2^N\to A$ and set $\widetilde f(x)=f(Px)+x+Px$. Then $\widetilde f(x)+x=f(Px)+Px\ne0$ for every $x$, and on $A$ one has $Px=x$.
\end{proof}

\begin{lemma}[Equality partition]\label{crn:lem:equalitypartition}
For uniform affine $Z\in\F_2^N$, an affine map $f$, and an integer $t\ge1$, there is an affine partition, apart from probability at most $2^{-t}$, on whose parts $f(Z)=Z$ identically or $f(Z)\ne Z$ everywhere. Every part lowers the entropy of $Z$ by at most $t$.
\end{lemma}
\begin{proof}
The image of $f(Z)+Z$ is affine. If it omits zero, no partition is needed. Otherwise it is a linear space of dimension $r$, on which the image is uniform. If $r\le t$, fix the complete image. If $r>t$, fix a surjective $t$-bit linear projection of it and discard the zero value. All nonzero values guarantee inequality. Every retained fixing uses at most $t$ independent affine equations; the discarded probability in the latter case is $2^{-t}$.
\end{proof}

\begin{lemma}[An affine-in-tag seed family]\label{crn:lem:taggedseed}
There is an absolute integer $A_*\ge1$ such that, for $N=A_*s$ and $F=\F_{2^s}$, one can explicitly compute two source-local functions $V_0,V_1:\F_2^N\to F$ for which
\begin{equation}\label{crn:eq:tagseedformula}
 S_a(z)=V_0(z)+aV_1(z)\qquad(a\in F)
\end{equation}
satisfies, for every affine $f$ and distinct $a,a'\in F$,
\begin{equation}\label{crn:eq:cubicseed}
 (S_a(U_N),S_{a'}(f(U_N)))
 \approx_{2^{-2s}}(U_F,S_{a'}(f(U_N))).
\end{equation}
Every field element is an allowed tag. The parameter $A_*$ is independent of all subsequent moment widths and rank margins.
\end{lemma}
\begin{proof}
Choose $A_*$ large enough that $4s\le\delta_A N/2$, $c_A N\ge2s$, and $e_A N\ge8s$. Truncate $V_N$ to $2s$ bits and identify them with $(V_0,V_1)\in F^2$. Apply Lemma~\ref{crn:lem:equalitypartition} with $t=4s$ to $U_N,f(U_N)$. The honest entropy on each retained part is at least $N-4s\ge(1-\delta_A)N$.

On a noncopy part, extend its comparison by Lemma~\ref{crn:lem:affineextension}. Theorem~\ref{crn:ext:affinenm} makes $(V_0,V_1)$ uniform independently of the comparison vector to error $2^{-8s}$. Its nonzero linear evaluation at $a$ is therefore uniform independently of the comparison evaluation. On a copy part, ordinary affine extraction makes the coefficient pair uniform to the same error. The two evaluation vectors $(1,a),(1,a')$ are independent, so their two values are exactly independent uniform under that comparison distribution. Restoring the actual comparison marginal costs at most another $2^{-8s}$. Averaging the parts and charging the discarded mass gives at most $2\cdot2^{-8s}+2^{-4s}\le2^{-2s}$ for sufficiently large $s$.
\end{proof}

We identify $\F_2^N$ with $L=\F_{2^N}$ and retain $F\subseteq L$. In this subsection and the next ones, a subscript on $S_a$ denotes a field tag, not a uniform seed supplied to the final evaluator. Formula~\eqref{crn:eq:tagseedformula} will account for the actual input-dependent advice without assuming any degree bound on $V_0,V_1$.

\paragraph{Gold defect and graph energy.}
For $z\in L$ put $\Gamma_c(z)=(z,z^{1+2},\ldots,z^{1+2^c})$.
For a finite subset $S$ of a binary vector space, write
$E(S)=|\{(z_1,z_2,z_3,z_4)\in S^4:z_1+z_2+z_3+z_4=0\}|$.
Let $A\subseteq V=\F_2^N$ and let $f:A\to W=\F_2^M$
be arbitrary. Define the graph of the restriction of $f$ to $A$ by
\[
 \operatorname{Graph}(f|_A)
 :=\{(x,f(x)):x\in A\}\subseteq V\times W.
\]
Here $f$ already has domain $A$; the restriction notation is retained
for compatibility with the subsequent graph-energy statements.
Define its normalized graph energy by
\[
 \eps_f=2^{-3N}E(\operatorname{Graph}(f|_A)).
\]
All probabilities below use uniform variables in the whole
ambient spaces; membership in $A$ is part of the event.

\begin{lemma}[Restricted homomorphism testing]\label{crn:ext:homomorphism}
\textnormal{\emph{Reference.} \cite[Lemma~3.3]{ACDG}.}
There are absolute $D_0,d_0\ge1$ such that, for all positive integers
$N,M$, every $A\subseteq\F_2^N$, every map $f:A\to\F_2^M$, and
$0<\rho\le1$,
\begin{equation}\label{crn:eq:noninjective_homomorphism}
 E(\operatorname{Graph}(f|_A))\ge\rho2^{3N}
 \quad\Longrightarrow\quad
 |\{x\in A:f(x)=\ell(x)\}|\ge D_0^{-1}\rho^{d_0}2^N
\end{equation}
for some affine map $\ell:\F_2^N\to\F_2^M$.
\end{lemma}
This is \cite[Lemma 3.3]{ACDG}. Only its existential assertion is used; its algorithmic complexity is irrelevant here.

Fix the binary identification $L\cong\F_2^N$. The binary coordinate
span of the Gold encoding is the following space of functions on $L$:
\[
 \mathcal C_c=
 \left\{z\longmapsto\Tr_{L/\F_2}
 \left(a_0z+\sum_{j=1}^c a_jz^{1+2^j}\right):
                  a_0,\ldots,a_c\in L\right\}.
\]
This agrees with the span of the binary coordinate functions of
$\Gamma_c$ by nondegeneracy of the trace pairing. For $A\subseteq L$
and $f:A\to L$, its pullback space is
\[
 (\mathcal C_c\circ f)|_A
 :=\{x\mapsto\varphi(f(x)):\varphi\in\mathcal C_c\}
 \le\F_2^A.
\]
Here $\circ$ denotes function composition. The space
$\mathrm{RM}(2,N)|_A$ consists of restrictions to $A$ of Boolean
polynomials of degree at most two in the fixed binary coordinates
of $L$; it includes the constant functions. Define the full Gold
defect by
\begin{equation}\label{crn:eq:golddefect}
 d_c(f;A)=\dim_{\F_2}\left(
 \frac{(\mathcal C_c\circ f)|_A+\mathrm{RM}(2,N)|_A}
      {\mathrm{RM}(2,N)|_A}\right).
\end{equation}
Put
\[
 \mathcal U_f:=(\mathcal C_c\circ f)|_A,
 \qquad \mathcal Q_A:=\mathrm{RM}(2,N)|_A.
\]
Both are binary vector subspaces of $\F_2^A$, but $\mathcal Q_A$
need not be contained in $\mathcal U_f$.  The symbol $+$ in
\eqref{crn:eq:golddefect} denotes their vector-space sum, so that the
denominator $\mathcal Q_A$ is a subspace of the numerator.  The
fraction denotes the \emph{quotient vector space}: two functions are
identified when their difference on $A$ belongs to $\mathcal Q_A$.
In particular, adding a quadratic function does not change the
quotient class.  The second isomorphism theorem gives
\begin{equation}\label{crn:eq:golddefect-dimension}
 \frac{\mathcal U_f+\mathcal Q_A}{\mathcal Q_A}
 \cong \frac{\mathcal U_f}{\mathcal U_f\cap\mathcal Q_A},
 \qquad
 d_c(f;A)=\dim_{\F_2}\mathcal U_f
       -\dim_{\F_2}(\mathcal U_f\cap\mathcal Q_A).
\end{equation}
Thus $d_c(f;A)$ counts the independent binary Gold-coordinate
combinations which remain after identifying combinations that agree
with quadratic functions on $A$.  No injectivity of $f$ is assumed.
In particular, $d_c(f;A)=0$ if $f$ is binary-affine, since the Gold
coordinates of an affine map have binary degree at most two.

\paragraph{Equivalent quadratic-plus-small-error representation.}
The quotient definition has the following concrete consequence, which
will be used in the proof of Theorem~\ref{crn:thm:noninjective_recognition}.
Put $B=L^{c+1}$, viewed as an $N(c+1)$-dimensional binary vector
space, and write $r=d_c(f;A)$.  There exist a vector-valued
polynomial map $Q_f:L\to B$ of binary degree at most two, a function
$e_f:A\to B$, and a \emph{fixed} binary linear subspace $W_f\le B$
of dimension $r$, such that
\begin{equation}\label{crn:eq:golddefect-decomposition}
 \Gamma_c(f(x))=Q_f(x)+e_f(x),\qquad
 e_f(x)\in W_f\quad(x\in A).
\end{equation}
Indeed, consider the binary-linear map
\[
 T:B^*\longrightarrow\F_2^A/\mathcal Q_A,
 \qquad
 T(\lambda)=\bigl[x\longmapsto\lambda(\Gamma_c(f(x)))\bigr].
\]
The image of $T$ is precisely $(\mathcal U_f+\mathcal Q_A)/\mathcal Q_A$,
so $\operatorname{rank}T=r$.  Choose a basis
$\lambda_1,\ldots,\lambda_{\dim B-r}$ of $\ker T$ and extend it to
a basis of $B^*$.  For each basis functional in $\ker T$, choose a
Boolean polynomial $q_i$ of degree at most two on $L$ agreeing on
$A$ with $x\mapsto\lambda_i(\Gamma_c(f(x)))$; for the remaining
$r$ basis functionals put $q_i=0$.  Because the $\lambda_i$ form a
basis of $B^*$, there is a unique map $Q_f:L\to B$ satisfying
$\lambda_i(Q_f(x))=q_i(x)$ for every $i$.  Its binary coordinates are
linear combinations of degree-at-most-two polynomials, so $Q_f$ is
quadratic.  Now set $e_f(x)=\Gamma_c(f(x))-Q_f(x)$ on $A$.
Every $\lambda\in\ker T$ annihilates $e_f(x)$, whence
\[
 e_f(A)\subseteq W_f:=(\ker T)^\perp
 :=\{z\in B:\lambda(z)=0\text{ for all }\lambda\in\ker T\},
 \qquad \dim_{\F_2}W_f=r.
\]
This proves \eqref{crn:eq:golddefect-decomposition}.  In particular,
$e_f$ takes at most $2^r$ distinct values; the quadratic map is
defined on all of $L$, while the decomposition is asserted only on $A$.

For a real function $g$ on a finite binary vector space $V$, write
\[
 \|g\|_{U^3}^{8}
 :=\mathbb E_{x,h_1,h_2,h_3\sim U_V}
       \prod_{\omega\in\{0,1\}^3}
        g(x+\omega_1h_1+\omega_2h_2+\omega_3h_3),
\]
where the four sampled vectors are independent. This is the third
Gowers uniformity norm (for real functions). Repeated
Cauchy--Schwarz gives $|\mathbb E g|\le\|g\|_{U^3}$.
For $g=\1_D$, the displayed average is exactly the probability that
all eight vertices of the random affine three-cube lie in $D$.
The directions need not be linearly independent, so degenerate
cubes and repeated vertices are included. In the next proof,
$V=L^p$, viewed as a binary vector space.

\begin{theorem}[Amortized recognition for arbitrary maps]
\label{crn:thm:noninjective_recognition}
For integer $p\ge1$, $c\ge8p-1$, $N>2c$, $|A|=2^{N-a}$ and $r=d_c(f;A)$,
some affine map agrees with $f$ on a subset of $A$ of size at least
$2^{N-O(a+r/p+p)}$. The constant is absolute.
\end{theorem}
\begin{proof}
We give the two algebraic and probabilistic facts used in the argument,
and then explain how taking $p$ copies amortizes the defect $r$.
Throughout, every cube is sampled by choosing its base point and its
three directions independently and uniformly; directions may be zero.

\smallskip\noindent
\emph{Step 1: a short Gold relation has small binary span.}
Suppose that $z_1,\ldots,z_m\in L$ satisfy
\begin{equation}\label{crn:eq:rec-goldzero}
 \sum_{\nu=1}^m\Gamma_c(z_\nu)=0,
 \qquad m\le 8p\le c+1.
\end{equation}
Put $d=\dim_{\F_2}\spn\{z_1,\ldots,z_m\}$ and form the
$m\times m$ Moore matrix over $L$,
\[
 \mathsf M=(z_\nu^{2^i})_{0\le i<m,\,1\le\nu\le m}.
\]
For $i<j$, the $(i,j)$ entry of $\mathsf M\mathsf M^{\mathsf T}$ is
\[
 \sum_\nu z_\nu^{2^i+2^j}
   =\left(\sum_\nu z_\nu^{1+2^{j-i}}\right)^{2^i}=0,
\]
because $1\le j-i\le m-1\le c$.  The entries with $i>j$
vanish by symmetry, and the diagonal entries vanish since
\[
 \sum_\nu z_\nu^{2^{i+1}}
 =\left(\sum_\nu z_\nu\right)^{2^{i+1}}=0.
\]
Thus $\mathsf M\mathsf M^{\mathsf T}=0$.  The Moore determinant
criterion, including its rectangular-rank form, says that
$\rk_L\mathsf M=d$: the rank is the $\F_2$-dimension of the span of
the labels, even if some labels repeat.  Since the image of
$\mathsf M^{\mathsf T}$ is contained in the kernel of $\mathsf M$,
\[
 d=\rk\mathsf M=\rk\mathsf M^{\mathsf T}
 \le\dim\ker\mathsf M=m-d.
\]
Consequently
\begin{equation}\label{crn:eq:rec-span}
 \dim_{\F_2}\spn\{z_1,\ldots,z_m\}\le m/2.
\end{equation}
No distinctness assumption on the $z_\nu$ is used here.

\smallskip\noindent
\emph{Step 2: a proper relation on one three-cube is a graph-energy
test.}
Let $I$ be any nonempty proper subset of $\F_2^3$, and fix $b\in L$.
We claim that a uniform affine three-cube $(x_\omega)_\omega$ in
$L$ satisfies
\begin{equation}\label{crn:eq:rec-corner}
 \Pr\bigl[x_\omega\in A\ (\forall\omega),\quad
             \sum_{\omega\in I}f(x_\omega)=b\bigr]
 \le \eps_f^{1/4}.
\end{equation}
Choose $\omega_*\in I$ and $\omega_\circ\notin I$.
Their two-point affine line in $\F_2^3$ is the simultaneous
$1$-fiber of two independent affine Boolean functions
$\beta_1,\beta_2$.  Complete $1,\beta_1,\beta_2$ by one more
affine Boolean function $\beta_3$ to a basis of all affine functions
on $\F_2^3$.  A uniform affine cube can consequently be written
\[
 x_\omega=z_0+\beta_1(\omega)U
                 +\beta_2(\omega)V+\beta_3(\omega)z_3,
\]
where $z_0,z_3,U,V$ are independent uniform elements of $L$.
Fix $z_0,z_3$ and let $F(U,V)$ be the indicator of the event in
\eqref{crn:eq:rec-corner}.  If $\rho=\E_{U,V}F(U,V)$,
the two-variable Cauchy--Schwarz (box) inequality gives
\[
 \rho^4\le
 \E_{U,U',V,V'} F(U,V)F(U',V)F(U,V')F(U',V'),
\]
for independent uniform $U,U',V,V'$.
If all four events on the right hold, add their four output equations.
In characteristic two the four copies of $b$ cancel.  For any vertex
$\omega\in I$ other than $\omega_*$, at least one of
$\beta_1(\omega),\beta_2(\omega)$ is zero, so its contribution occurs
in identical pairs and cancels as well.  The only remaining terms
are the four values of $f$ at $\omega_*$.  Their arguments are
\[
 z+U+V,\quad z+U'+V,\quad z+U+V',\quad z+U'+V',
 \qquad z=z_0+\beta_3(\omega_*)z_3.
\]
All four arguments belong to $A$, their sum is zero, and their
$f$-values sum to zero.  For every fixed $z$, three of these four
arguments are independent uniform elements of $L$; the fourth is
their sum.  The probability of this latter graph-parallelogram event
is therefore exactly
\[
 2^{-3N}E(\operatorname{Graph}(f|_A))=\eps_f.
\]
Hence $\rho^4\le\eps_f$ for each fixed $z_0,z_3$.
Averaging proves \eqref{crn:eq:rec-corner}.  The bound holds
uniformly in the chosen set $I$ and the constant $b$.

\smallskip\noindent
\emph{Step 3: represent the defect by an $r$-dimensional error space.}
Apply the quadratic-plus-small-error representation
\eqref{crn:eq:golddefect-decomposition} established immediately
following the definition of $d_c(f;A)$.  Write $Q=Q_f$, $e=e_f$ and
$W=W_f\le B=L^{c+1}$.  Thus
\begin{equation}\label{crn:eq:rec-error}
 \Gamma_c(f(x))=Q(x)+e(x),\qquad
 e(A)\subseteq W,\qquad \dim_{\F_2}W=r.
\end{equation}
The key point is that the error $e$ takes at most $2^r$ values,
whereas $Q$ has binary degree at most two on the full ambient input
space.  No assumption that $f$ is injective or globally polynomial
is involved.

Define $E_p:A^p\to W$ by
$E_p(x_1,\ldots,x_p)=\sum_{i=1}^p e(x_i)$.
Some fiber $\mathcal D=E_p^{-1}(w)$ has size at least $|A|^p/2^r$.
As a subset of the ambient space $L^p$, its density satisfies
\begin{equation}\label{crn:eq:rec-density}
 \alpha=\frac{|\mathcal D|}{2^{Np}}\ge 2^{-pa-r}.
\end{equation}
The $U^3$ inequality for a nonnegative indicator gives
\begin{equation}\label{crn:eq:rec-cubes}
 \Pr[\text{all eight vertices of a uniform three-cube in }L^p
                 \text{ belong to }\mathcal D]
  =\|\1_{\mathcal D}\|_{U^3}^{8}
  \ge\alpha^8\ge 2^{-8(pa+r)}.
\end{equation}
For every cube counted here, sum \eqref{crn:eq:rec-error} over
its eight vertices and over $i\in[p]$.  The eight identical values
$w$ cancel.  The sum of the values of $\sum_i Q(x_i)$ also vanishes,
because the third additive derivative of any binary quadratic
polynomial is zero.  Thus
\[
 \sum_{i=1}^p\sum_{\omega\in\F_2^3}
       \Gamma_c(f(x_{i,\omega}))=0.
\]
There are $8p$ output labels in this sum, and
\eqref{crn:eq:rec-span} implies that their binary span has dimension
at most $4p$.

\smallskip\noindent
\emph{Step 4: extract many proper relations and apply the four-corner
bound independently across blocks.}
Reveal the $p$ coordinate cubes one block at a time.  If $d_i$ is the
increment in the binary dimension of the output-label span when the
eight labels of block $i$ are added, then
\[
 0\le d_i\le8,\qquad \sum_{i=1}^p d_i\le4p.
\]
At most $4p/7$ blocks can have $d_i\ge7$.  Hence at least $3p/7$
blocks have $d_i\le6$.  For each such block, the eight labels,
considered modulo the span of all earlier labels, have a space of
binary relations of dimension at least two.  We may select a nonzero
relation supported on a nonempty \emph{proper} subset
$I_i\subsetneq\F_2^3$: a two-dimensional relation space cannot
consist only of the zero vector and the all-ones vector.
For some fixed binary coefficients $b_{i,h,\omega}$, this relation is
\begin{equation}\label{crn:eq:rec-block-relation}
 \sum_{\omega\in I_i} f(x_{i,\omega})
  =\sum_{h<i}\sum_{\omega\in\F_2^3}
                  b_{i,h,\omega}f(x_{h,\omega}).
\end{equation}
The set of selected blocks, the subsets $I_i$, and the coefficient
vectors in \eqref{crn:eq:rec-block-relation} have at most
$2^{O(p^2)}$ possible joint specifications.  This estimate is
independent of $N$ because all coefficients are binary and only
$8p$ labels occur.

Fix one specification with at least $3p/7$ selected blocks.
The coordinate cubes in $L^p$ are independent uniform three-cubes.
When we reveal them sequentially, at each selected block the right
side of \eqref{crn:eq:rec-block-relation} is already a fixed value of
$L$.  Therefore \eqref{crn:eq:rec-corner} bounds its conditional
success probability, including membership in $A$ of the eight
current vertices, by $\eps_f^{1/4}$.  Iterated conditioning bounds
the probability that all specified relations hold by
$\eps_f^{3p/28}$.  Every cube from \eqref{crn:eq:rec-cubes} admits
at least one such specification.  A union bound now gives
\begin{equation}\label{crn:eq:rec-final-energy}
 2^{-8(pa+r)}\le 2^{O(p^2)}\eps_f^{3p/28}.
\end{equation}
Taking logarithms (or treating $\eps_f=0$ separately, when the
inequality is impossible) yields
\[
 \eps_f\ge 2^{-O(a+r/p+p)}.
\]
Finally apply the restricted homomorphism-testing
Lemma~\ref{crn:ext:homomorphism}.  Its constants are absolute,
so it gives an affine map $\ell:L\to L$ agreeing with $f$ on at least
$2^{N-O(a+r/p+p)}$ points of $A$, as claimed.
\end{proof}

\subsubsection{An asymmetric tagged starter}
\begin{lemma}[Scaled polynomial evaluation]\label{nmnew:eval}
For $K=\F_{2^m}$ set $\Phi_J(u,v)=v(1,u,\ldots,u^{J-1})\in K^J$.
If $D\subseteq K^2$ has density $2^{-d}$ and
$(J-1)2^{-m}\le2^{-d-1}$, the binary affine span of $\Phi_J(D)$
has codimension at most $d+1$.
\end{lemma}
\begin{proof}
Put $q=|K|=2^m$ and let
$\chi(z)=(-1)^{\Tr_{K/\F_2}(z)}$.
Every binary linear functional on $K^J$ has the form
\[
 \Lambda(z_0,\ldots,z_{J-1})
   =\Tr_{K/\F_2}\left(\sum_{i=0}^{J-1}a_i z_i\right)
\]
for a unique vector $(a_0,\ldots,a_{J-1})\in K^J$, by
nondegeneracy of the trace pairing.  If $\Lambda\ne0$, the
polynomial $P(T)=\sum_{i=0}^{J-1}a_iT^i$ is nonzero and has degree
at most $J-1$.  Hence, for independent uniform $U,V\in K$,
\begin{align*}
 \E_{U,V}(-1)^{\Lambda(\Phi_J(U,V))}
 &=\E_U\E_V\chi(VP(U))\\
 &=\Pr_U[P(U)=0]\\
 &\le (J-1)q^{-1}=:\beta.
\end{align*}
The middle equality is exact: if $P(u)\ne0$, the map
$v\mapsto\Tr_{K/\F_2}(vP(u))$ is a nonzero binary functional and
its character average is zero; if $P(u)=0$, the average is one.

Let $H=\operatorname{aff}_{\F_2}(\Phi_J(D))$, and write its binary
codimension in $K^J$ as $b$.  If $H=z_*+W$ and
$W^\perp\le(K^J)^*$ is the annihilator of its direction space,
character orthogonality gives
\[
 \1_H(z)=2^{-b}\sum_{\Lambda\in W^\perp}
                (-1)^{\Lambda(z-z_*)}.
\]
Taking expectation at $z=\Phi_J(U,V)$, bounding the zero
character by one and the other $2^b-1$ characters by $\beta$, gives
\[
 \Pr[\Phi_J(U,V)\in H]
  \le 2^{-b}+(1-2^{-b})\beta
  \le 2^{-b}+\beta.
\]
But $\Phi_J(D)\subseteq H$, so the probability on the left is at
least $|D|/|K|^2=2^{-d}$.  Under the hypothesis
$\beta\le2^{-d-1}$, we have
\[
 2^{-b}\ge 2^{-d}-\beta\ge 2^{-d-1}.
\]
Thus $b\le d+1$.  Notice that the conclusion concerns the binary
\emph{affine} span; its difference space has the same dimension, and
we will use the bound for differences as well as for the full span.
\end{proof}

\paragraph{Bias and separate source strength.}
Call a character honest-bearing if its honest coefficient is nonzero.
If each such character has absolute bias at most $\epsilon$ for all
independent sources at threshold $k_0$, then at threshold $k_0+d$ its
average absolute conditional bias given either source is at most
$\epsilon+2^{1-d}$. Partition that source by the sign of its conditional
bias. Each class of mass at least $2^{-d}$ preserves the threshold and
independence; the other classes have total mass at most $2^{1-d}$.
For $q$ outputs of $m$ bits, conditional Fourier inversion and the
triangle inequality bound either strong error by
$2^{qm-1}(\epsilon+2^{1-d})$: the honest-free coefficients cancel
against uniform times the actual comparison marginal. We use the
linear-rank approximate-duality lift of Corollary~\ref{pre:upgrade}
with fixed linear entropy slack to obtain the initial bias from
nonconstancy on every lower-threshold rectangle.

\paragraph{Compatible $M$-bit rows.} In the following, we use this term to denote an arithmetic restriction on the \emph{length},
not a restriction on the distributions of the two rows or a promise
about the tampering maps.  For the starter extractor, choose the absolute
integer $A_*$ from Lemma~\ref{crn:lem:taggedseed}, and a fixed
output-shortening divisor $D_1=12A_*t_*$, where $t_*$ is a fixed
positive integer chosen in the proof.  A starter length $M$ is
\emph{compatible} if it can be written
\[
 M=6n_0,\qquad n_0=A_*s,
 \qquad 2t_*\mid s,
\]
with $s$ sufficiently large for the fixed constants $c,J$, and with
fixed representations of the tower
\[
 K=\F_{2^{s/2}}\subset F=\F_{2^s}\subset L=\F_{2^{n_0}}.
\]
Then the right row parses into two $L$-words and $J=8A_*$
$K$-words, since $2n_0+J(s/2)=6n_0=M$, and the final trace can
land in the subfield of degree $s/(2t_*)=M/D_1$.
Every input bit has a specified coordinate in these fixed bases.
There is no random choice or extra entropy condition hidden in the
word ``compatible.''  The set of admissible starter lengths contains
all sufficiently large multiples of the fixed integer
$12A_*t_*$.

\begin{lemma}[Tagged starter extractor]\label{mc:lem:starter}
For every sufficiently large compatible row length $M$ as defined above, there is a fixed-tag extractor against one
arbitrary separate tampering with a distinct tag, strong in the left
source, at entropy $(1-\tau_1)M$ for an absolute $\tau_1>0$. It has output length $\Omega(M)$ and 
error $2^{-\Omega(M)}$. Furthermore, it admits constant
field separation over compatible subfields for the actual advice below.
\end{lemma}

\begin{constructionbox}{Asymmetric tagged starter extractor}{cb:starter}
\cbingredients The affine-in-tag seed of Lemma~\ref{crn:lem:taggedseed}
and field polynomial evaluation.
\cbsetup Write $n_0=A_*s$, $K=\F_{2^{s/2}}\subset F=\F_{2^s}
\subset L=\F_{2^{n_0}}$, $J=8A_*$ and $M=6n_0$.
Use the fixed width $c$ chosen in the proof.
\cbinput Two arbitrary $M$-bit rows (with a compatible public length) and a fixed numerical tag $a\in F$.
\cbsteps
\begin{constructionsteps}
\item Denote the complete left and right input rows by
$z,w\in\F_2^M$. Take the first $n_0$ bits of $z$ as
$x=\operatorname{prefix}_{n_0}(z)\in L$, and parse the complete right
row as $w=(r,t,Y)\in L^2\times K^J$. Write
$S_a(x)=(u,v)\in K^2$.
\item For a coefficient vector $D$, put $P_D(T)=\sum_{i\ge0}D_iT^i$
over its indexed coordinates starting at zero. Compute 
\[
 E_a=\Tr_{L/K}\!\big(tP_{\Gamma_c(x)}(r)\big)+vP_Y(u).
\]
\end{constructionsteps}
\cboutput Return a relative
trace of $E_a$ to an appropriate output subfield specified in the proof below. The length of the output field word is a fixed positive fraction of $M$.
\end{constructionbox}

\begin{proof}
We prove the fixed-tag statement by ruling out a constant mixed
character on every rectangle of sufficiently small side deficiency.
The error and the separate left-source guarantee then follow from the
rank-to-bias theorem and entropy slack.  Throughout this proof the tags
$a\ne a'$ are fixed field elements; the later outer construction is
responsible for fixing its current input-dependent tags.

\smallskip\noindent
\emph{The input rectangle and the two terms.}
Fix supports of size at least $2^{M-\alpha M}$ on both sides, where
$\alpha>0$ will be chosen after the other constants.  Let
$z,w\in\F_2^M$ denote the \emph{complete} left and right rows, and
let $T_L,T_R:\F_2^M\to\F_2^M$ denote the arbitrary separate
comparison maps.  The starter uses only the first $n_0$ bits of
$z$; fix its unused suffix at a most frequent value $s_0$ and write
$z=(x,s_0)$.  Its resulting prefix support $A\subseteq L$ has
$|A|\ge2^{n_0-a_0}$ with $a_0\le\alpha M$.  The comparison's
\emph{short left prefix} is then the deterministic, possibly
noninjective function
\[
 f(x):=\operatorname{prefix}_{n_0}(T_L(x,s_0)),\qquad x\in A.
\]
The tampering $T_L$ is arbitrary and need not preserve the prefix.
Let $B$ denote the support of the \emph{complete} right row $w$,
$|B|\ge2^{6n_0-b_0}$ with $b_0\le\alpha M$.  Parse
$w=(r(w),t(w),Y(w))$ and its tampered value
$T_R(w)=(r'(w),t'(w),Y'(w))$; $T_R$ remains entirely arbitrary.
Thus, after fixing $s_0$, the restricted product rectangle is
parameterized by $(x,w)\in A\times B$.  From now on, $x$ always
means the short left prefix and $w$ the complete right row.

Put $m_K=s/2$, so that $K=\F_{2^{m_K}}$ and
$Jm_K=4n_0$.  Consider the absolute-trace character of the two
untruncated starter outputs, with nonzero honest coefficient
$\lambda\in K^\times$ and arbitrary comparison coefficient
$\mu\in K$:
\[
 \chi_{s_0}(x,w)=\Tr_{K/\F_2}\!\left(
  \lambda E_a(x;r(w),t(w),Y(w))
 +\mu E_{a'}(f(x);r'(w),t'(w),Y'(w))\right).
\]
Suppose it is constant on $A\times B$.  Here $E_a$ is the
untruncated prefix-level formula from Construction~\ref{cb:starter},
not a function whose left argument is the entire $M$-bit row.
For the honest terms, abbreviate
\[
 E_a(x;w)=G(x;w)+T_a(x;w),\qquad
 G(x;w)=\Tr_{L/K}(t(w)P_{\Gamma_c(x)}(r(w))),\qquad
 T_a(x;w)=vP_{Y(w)}(u),
\]
the tagged term is the pairing of the $J$ coordinates of
$\Phi_J(u,v)$ with the $J$ coefficients of $Y$.  After any separate
precomposition on the two sides, its binary character matrix has
rank at most $Jm_K=4n_0$.  Thus the sum of the two tagged terms,
including a possible constant matrix, has rank at most
\begin{equation}\label{mc:eq:starter-rank-budget}
 k_G:=2Jm_K+1=8n_0+1.
\end{equation}
Constancy of the full mixed character forces the corresponding
mixed Gold term, restricted to $A\times B$, to have matrix rank
at most $k_G$.  This assertion uses ordinary binary matrix rank:
for a function $h:A\times B\to\F_2$ its matrix is
$(h(x,y))_{x\in A,y\in B}$; a product of a left function and a
right function contributes rank one.

\smallskip\noindent
\emph{The dense spans of the two Gold feature families.}
Let $D_G=(c+1)n_0$, the binary dimension of $L^{c+1}$.
For a nonzero binary character $\Lambda$ of $\Gamma_c(Z)$,
where $Z$ is uniform in $L$, write its Boolean phase as
\[
 q(z)=\Lambda(\Gamma_c(z))
 =\Tr_{L/\F_2}\!\left(\gamma_0z+
          \sum_{i=1}^{c}\gamma_i z^{1+2^i}\right),
 \qquad \gamma_i\in L.
\]
If $\gamma_1=\cdots=\gamma_c=0$, then $q$ is a nonzero binary-linear
function and its character bias is zero.  Otherwise let $j\le c$
be the largest index for which $\gamma_j\ne0$.  We spell out why the
polar radical of $q$ is small.  The associated polar form is
\[
 \begin{aligned}
 B_q(z,h)&:=q(z+h)+q(z)+q(h)\\
 &=\Tr_{L/\F_2}\!\left(
    \sum_{i=1}^j \gamma_i\bigl(zh^{2^i}+hz^{2^i}\bigr)\right).
 \end{aligned}
\]
Indeed, $(z+h)^{2^i}=z^{2^i}+h^{2^i}$ in characteristic two,
so expanding $(z+h)^{1+2^i}$ leaves precisely the two displayed
cross terms; the linear term $\gamma_0z$ disappears.  The absolute
trace is Frobenius-invariant.  Thus, for any $w,z\in L$,
\[
 \Tr_{L/\F_2}(wz^{2^i})
  =\Tr_{L/\F_2}\bigl(w^{2^{-i}}z\bigr),
 \qquad w^{2^{-i}}:=w^{2^{n_0-i}},
\]
where the exponent $2^{-i}$ denotes the inverse $i$-fold
Frobenius automorphism, not a rational exponent.  Applying this
identity to $w=\gamma_i h$ rewrites the polar form as
\[
 B_q(z,h)=\Tr_{L/\F_2}\!\left(zR(h)\right),\qquad
 R(h):=\sum_{i=1}^{j}
 \left(\gamma_i h^{2^i}+(\gamma_i h)^{2^{-i}}\right).
\]
By nondegeneracy of the binary trace pairing on $L$, the polar
radical is exactly
\[
 \operatorname{Rad}(q)
 :=\{h\in L:B_q(z,h)=0\ \text{for every }z\in L\}
 =\{h\in L:R(h)=0\}.
\]
All the Frobenius maps appearing in $R$ are $\F_2$-linear, so
$\operatorname{Rad}(q)$ is a binary subspace.  To bound its size,
raise the equation $R(h)=0$ to the $2^j$th power.  As the
Frobenius automorphism is bijective, this is equivalent to
\[
 \begin{aligned}
 0=R(h)^{2^j}
 &=\sum_{i=1}^j
   \left(\gamma_i^{2^j}h^{2^{i+j}}
         +(\gamma_i h)^{2^{j-i}}\right).
 \end{aligned}
\]
Here $n_0>2c\ge2j$ ensures that all exponents $i+j$ are
strictly below $n_0$, so no Frobenius powers wrap around.
The term with $i=j$ in the first sum is
$\gamma_j^{2^j}h^{2^{2j}}$, with nonzero coefficient, and every
other monomial has strictly smaller degree.  Therefore this is
a nonzero ordinary polynomial in $h$ of degree exactly
$2^{2j}$, and it has at most $2^{2j}$ roots in $L$.  Hence
\[
 |\operatorname{Rad}(q)|\le2^{2j}.
\]
Finally, put $\beta=\E_{z\in L}(-1)^{q(z)}$.  Squaring the
bias and making the change of variables $(z,z')=(z,z+h)$,
then averaging first over $z$, gives
\[
 \begin{aligned}
 \beta^2
 &=\E_{z,h\in L}(-1)^{q(z)+q(z+h)}\\
 &=\E_{h\in L}(-1)^{q(h)}
       \E_{z\in L}(-1)^{B_q(z,h)}\\
 &=\E_{h\in L}\bigl[\1_{\operatorname{Rad}(q)}(h)
                (-1)^{q(h)}\bigr]
 \le\frac{|\operatorname{Rad}(q)|}{|L|}
 \le2^{2j-n_0}.
 \end{aligned}
\]
The penultimate equality again uses nondegeneracy of the trace
pairing: the average of $(-1)^{B_q(z,h)}$ over $z$ is one if
$R(h)=0$ and zero otherwise.  Taking square roots yields
\begin{equation}\label{mc:eq:starter-gold-bias}
 \left|\E_{Z\in L}(-1)^{\Lambda(\Gamma_c(Z))}\right|
  \le 2^{j-n_0/2}\le2^{c-n_0/2}
\end{equation}
for every nonzero coordinate character $\Lambda$.
For fixed $c$ and sufficiently large $n_0$, the last bound is at most
$2^{-a_0-1}$.  If the binary affine span of $\Gamma_c(A)$ has
codimension $u$, expanding the indicator of that span in its
$2^u$ annihilator characters, exactly as in
Lemma~\ref{nmnew:eval}, gives
\[
 2^{-a_0}\le 2^{-u}+2^{c-n_0/2}.
\]
Hence $u\le a_0+1$.  The coordinate functions of the honest
Gold features, restricted to $A$, consequently have binary span
of dimension at least $D_G-a_0-1$.

On the right, the projection of $B$ to $(r,t)\in L^2$ has at least
$2^{2n_0-b_0}$ elements: every pair $(r,t)$ has at most
$|K|^J=2^{4n_0}$ preimages.  The projection thus has density at
least $2^{-b_0}$ in $L^2$.  Apply Lemma~\ref{nmnew:eval} over $L$
with width $c+1$.  Its root-bound hypothesis holds for fixed $c$
and sufficiently large $n_0$.  We obtain affine-span codimension
at most $b_0+1$ for
\[
 \Phi_{c+1}(r,t)=t(1,r,\ldots,r^c)\in L^{c+1}.
\]

\smallskip\noindent
\emph{If the comparison Gold coefficient is nonzero, recognize the
left tampering.}
Assume $\mu\ne0$.  Let $\mathsf U$ be the binary feature matrix
whose row indexed by $x\in A$ concatenates the coordinate vectors of
$\Gamma_c(x)$ and $\Gamma_c(f(x))$.  Let $\mathsf V$ similarly
concatenate, for each right input in $B$, the feature vectors
$\Phi_{c+1}(r,t)$ and $\Phi_{c+1}(r',t')$, where $(r',t')$ are the
first two $L$-words of the tampered right input.
The mixed Gold character matrix is
\[
 \mathsf U\,\mathsf P\,\mathsf V^{\mathsf T},
\]
where $\mathsf P$ is a nondegenerate block-diagonal binary trace
pairing on a space of dimension $2D_G$; its two blocks are
nondegenerate because $\lambda,\mu\ne0$.  In particular
$\mathsf P$ is invertible.

The honest block of $\mathsf U$ has rank at least
$D_G-a_0-1$.  Every binary coordinate of $\Gamma_c(x)$ is a
quadratic Boolean function of $x$, whereas the span of the
coordinates of $\Gamma_c(f(x))$ modulo all quadratics on $A$ has
dimension $d_c(f;A)$.  Thus adjoining the tampered block increases
rank by at least this defect:
\[
 \rk\mathsf U\ge D_G-a_0-1+d_c(f;A).
\]
The honest block alone gives
$\rk\mathsf V\ge D_G-b_0-1$.
Sylvester's rank inequality, applied to the product above, yields
\begin{align*}
 k_G
 &\ge\rk(\mathsf U\mathsf P\mathsf V^{\mathsf T})\\
 &\ge\rk\mathsf U+\rk\mathsf V-2D_G\\
 &\ge d_c(f;A)-a_0-b_0-2.
\end{align*}
Therefore
\begin{equation}\label{mc:eq:starter-defect}
 d_c(f;A)\le k_G+a_0+b_0+2.
\end{equation}
By Theorem~\ref{crn:thm:noninjective_recognition}, there is a
subset $A_1\subseteq A$ on which $f$ agrees with an affine map
$\ell:L\to L$ and whose ambient deficiency satisfies
\begin{equation}\label{mc:eq:starter-affine-subset}
 a_1:=n_0-\log_2|A_1|
 \le C_{\rm rec}\left(a_0+
     \frac{k_G+a_0+b_0+2}{p}+p+1\right)
\end{equation}
for an absolute $C_{\rm rec}$.  Choose the constant $p$ large
compared with $A_*$ and $C_{\rm rec}$, then set $c=8p$; only after
these choices take $\alpha>0$ sufficiently small.  Since
$k_G=8n_0+1$ and $s=n_0/A_*$, the right side of
\eqref{mc:eq:starter-affine-subset} is at most $s/8$ for all
sufficiently large compatible lengths.  This choice also ensures
$n_0>2c$ and the span estimates above.

When $\mu=0$, the comparison Gold block is absent and no recognition
is needed.  We may take $A_1=A$ and use ordinary affine extraction of
$S_a$ in the argument below; this case retains at least as much
right entropy as the $\mu\ne0$ case.

\smallskip\noindent
\emph{Fix one comparison seed without destroying the honest seed
span.}
First take $\mu\ne0$.  On $A_1$ we have $f=\ell$, where $\ell$
is affine.  For the distinct tags $a,a'$, the affine-in-tag seed
lemma gives on the ambient uniform $Z\in L$
\[
 \bigl(S_a(Z),S_{a'}(\ell(Z))\bigr)
 \approx_{2^{-2s}}
 \bigl(U_{K^2},S_{a'}(\ell(Z))\bigr).
\]
Put $\beta=|A_1|/|L|=2^{-a_1}$.  For each possible comparison
seed value $z'$, let $T_{z'}$ be the set of honest seed values
$S_a(x)$ occurring for $x\in A_1$ with
$S_{a'}(\ell(x))=z'$.  Then
\[
 \beta\le
 \Pr\bigl[S_a(Z)\in T_{S_{a'}(\ell(Z))}\bigr]
 \le\max_{z'}\frac{|T_{z'}|}{2^s}+2^{-2s}.
\]
Since $a_1\le s/8$, some $z'$ has
\begin{equation}\label{mc:eq:starter-many-seeds}
 |T_{z'}|\ge2^{s-a_1-1}
\end{equation}
for large $s$.  Restrict the left source to the corresponding
nonempty fiber and, if necessary, choose one left input for each
honest seed in $T_{z'}$.  The comparison seed is now constant but
the honest seeds still realize a set of density at least
$2^{-a_1-1}$ inside $K^2$. We now use the estimates below to arrive at a contradiction.
For $\mu=0$ there is no comparison seed to fix.  Ordinary affine
extraction gives $S_a(U_L)\approx U_{K^2}$; applied to the set
$A_1=A$, the same counting argument supplies a set of at least
$2^{s-a_0-1}$ honest seed values.  The estimates below also apply to this case by using
$a_1=a_0$ with no tampered terms.

\smallskip\noindent
\emph{Fix the Gold coefficients on the right and obtain the
contradiction.}
On the right restrict to a largest fiber of the honest pair
$(r,t)\in L^2$ and of the tampered pair
$(r',t')\in L^2$ (omit the latter if $\mu=0$).
These four field words reveal at most $4n_0$ bits, so the remaining
coefficient-vector set $\mathcal Y\subseteq K^J$ has size at least
\begin{equation}\label{mc:eq:starter-y-size}
 |\mathcal Y|\ge 2^{6n_0-b_0-4n_0}
                  =2^{2n_0-b_0}.
\end{equation}
This is a bound on the number of distinct $Y$ coefficient vectors:
for each choice of $Y$ and fixed $(r,t)$ there is only one honest
right input, so the projection is injective on that fiber.
In the chosen rectangle, both Gold terms depend only on the left
input, because their coefficient pairs have been fixed.  The tampered
tagged term depends only on the right input, because the comparison
seed is fixed.  Only the honest tagged term remains genuinely mixed.

To make the cancellation precise, let $\mathcal X$ be the
retained set of short left prefixes in the selected comparison-seed
fiber, and let $\mathcal B$ be the retained set of \emph{complete}
right rows in the chosen coefficient fiber.  On the product
$\mathcal X\times\mathcal B$, the preceding fixings let us write
the supposedly constant mixed Boolean character in the form
\begin{equation}\label{mc:eq:starter-four-corner-decomposition}
 \chi_{s_0}(x,w)
 =L_{\mathrm{loc}}(x)+R_{\mathrm{loc}}(w)
  +\Tr_{K/\F_2}\!\left(
     \lambda\,\Phi_J(u(x),v(x))\cdot Y(w)\right)
 =c_\chi\in\F_2.
\end{equation}
Here $(u(x),v(x))=S_a(x)$, $Y(w)$ is the honest $K^J$ coefficient
vector of the complete right row $w$, and $\cdot$ is the ordinary
$K$-inner product on $K^J$.  The function $L_{\mathrm{loc}}$
contains the two Gold contributions (whose honest and tampered
right coefficient pairs have been fixed); $R_{\mathrm{loc}}$
contains the tampered tagged contribution (whose comparison seed
has been fixed).  If $\mu=0$, the corresponding comparison terms
are simply absent.  Thus no assumption on the right tampering map
is needed for this decomposition.

Choose arbitrary $x_1,x_2\in\mathcal X$ and
$w_1,w_2\in\mathcal B$, and abbreviate
\[
 \begin{gathered}
  (u_i,v_i)=S_a(x_i),\qquad
  \Phi_i=\Phi_J(u_i,v_i),\qquad Y_j=Y(w_j),\\
  L_i=L_{\mathrm{loc}}(x_i),\qquad
  R_j=R_{\mathrm{loc}}(w_j),\qquad
  H_{ij}=\Tr_{K/\F_2}(\lambda\Phi_i\cdot Y_j).
 \end{gathered}
\]
Constancy of $\chi_{s_0}$ on the full product rectangle yields
\begin{equation}\label{mc:eq:starter-four-corner-equations}
 \begin{aligned}
  L_1+R_1+H_{11}&=c_\chi,&
  L_1+R_2+H_{12}&=c_\chi,\\
  L_2+R_1+H_{21}&=c_\chi,&
  L_2+R_2+H_{22}&=c_\chi.
 \end{aligned}
\end{equation}
Add these four equalities in $\F_2$.  Each $L_i$ occurs
\emph{twice}, once with $w_1$ and once with $w_2$, so the
left-only contributions cancel.  Likewise each $R_j$ occurs
\emph{twice}, once with $x_1$ and once with $x_2$, and cancels.
The constant on the right contributes $4c_\chi=0$.
Consequently,
\[
 0=H_{11}+H_{12}+H_{21}+H_{22}
   =\Tr_{K/\F_2}\!\left(\lambda
     (\Phi_1\cdot Y_1+\Phi_1\cdot Y_2
      +\Phi_2\cdot Y_1+\Phi_2\cdot Y_2)\right).
\]
By bilinearity of the $K$-inner product, the four mixed terms
factor, giving
\begin{equation}\label{mc:eq:starter-orthogonality}
 \Tr_{K/\F_2}\left(
 \lambda\,\bigl[\Phi_J(u_1,v_1)+\Phi_J(u_2,v_2)\bigr]
            \cdot(Y_1+Y_2)\right)=0.
\end{equation}
Because the two left prefixes and the two right rows were arbitrary,
this is an orthogonality relation between \emph{every} difference
of honest scaled-evaluation features and \emph{every} difference
of retained right coefficient vectors.
The binary span of the differences of honest feature vectors has
dimension at least $Jm_K-a_1-2=4n_0-a_1-2$, by
Lemma~\ref{nmnew:eval} applied to the set of seed values in
\eqref{mc:eq:starter-many-seeds}; multiplication by
$\lambda\ne0$ preserves this dimension.
On the other hand the binary span of differences of vectors in
$\mathcal Y$ has dimension at least
$\log_2|\mathcal Y|\ge2n_0-b_0$.  Its annihilator under the
nondegenerate trace pairing on $K^J$ therefore has dimension at most
$4n_0-(2n_0-b_0)=2n_0+b_0$.
Equation~\eqref{mc:eq:starter-orthogonality} would force
\[
 4n_0-a_1-2\le2n_0+b_0,
\]
contradicting $a_1\le s/8$, $b_0\le\alpha M$ and a sufficiently
small fixed $\alpha$.  Thus no honest-bearing mixed character is
constant on the claimed rectangles.

\smallskip\noindent
\emph{Exponential error, left source strength, and prefix reuse.}
For fixed numerical tags, each Gold term is a sum of
$(c+1)n_0$ separated binary products, and each tagged term is a
sum of $Jm_K$ such products.  Thus every mixed character has global
binary rank $O(M)$ with a constant depending only on the fixed
starter parameters.  Apply the linear-rank approximate-duality lift
of Corollary~\ref{pre:upgrade} to the nonconstancy just proved,
using a fixed positive fraction of $\alpha M$ as entropy slack.
It gives character bias $2^{-\Omega(M)}$ at a slightly higher
min-entropy threshold.  Partitioning the honest left source by
the sign of its conditional character bias, as described before this
lemma, makes the bound hold on average \emph{with that entire source
retained}; the small sign classes cost at most an exponential tail.

Finally apply a relative trace $\Tr_{K/K'}$ to a subfield
$K'=\F_{2^{M/D_1}}\subseteq K$, choosing the fixed divisor
$D_1=12A_*t_*$ sufficiently large.  Every nonzero character of
$K'$ pulls back to a nonzero character of $K$, while the shorter
output length makes the Fourier inversion factor harmless.
The result has output length $M/D_1=\Omega(M)$ and error
$2^{-\Omega(M)}$, with left-source strength at entropy
$(1-\tau_1)M$ for a fixed $\tau_1>0$ (one may take
$\tau_1=\alpha/4$ after fixing the slack constants).

The same primitive may be applied to prefixes of longer source
words.  For clarity, let $Z$ be the honest left prefix and let
$X$ be the complete left word.  Independently of $Z$, sample a
\emph{table} that assigns to each prefix value $z$ a complete
word drawn from the conditional distribution $X\mid Z=z$.  Evaluating the
table at the sampled $Z$ reproduces the original distribution of $X$.
For each fixed table, a tampered prefix is a deterministic function
of the honest prefix; the two original sides still have independent
tables.  The left-strong guarantee for the prefix retains both $Z$
and its table and hence retains the complete left word it generates.
Averaging over the tables proves the needed statement for arbitrary
separate tamperings of the longer inputs.  This completion argument
will be used at every recursion level.
\end{proof}
\subsubsection{Doubling the number of comparisons}
\paragraph{The ordinary strong hash.}
For output length $v$, parse a source word into coefficients of $P$ over
$K=\F_{2^{4v}}$ (pad with zeros) and an $8v$-bit seed into $(u,t)\in K^2$.
Use $H_v=\Tr_{K/\F_{2^v}}(tP(u))$. For two distinct words of a fixed
length, collision probability on a uniform seed is at most
$2^{-v}+d_{\max}2^{-4v}$ by the root bound, where
$d_{\max}$ bounds the degrees used. For constant word/field length
ratios the degree bound is constant. Cauchy--Schwarz on the average
collision probability of the output conditional on the seed gives
strong error at most
$\tfrac12\sqrt{2^v(2^{-k}+O(2^{-4v}))}$ at min-entropy $k$.
In particular $k\ge3v$ suffices for error $O(2^{-v})$.

\paragraph{Compatible $N$-bit rows.} Similarly, in the following we use this term to denote an arithmetic restriction on the length. Specifically, once $C$ is fixed, the wrapper has only
$h=\max\{1,\lceil\log_2 C\rceil\}$ levels.  A length $N$ at a
wrapper level is \emph{compatible} when
\[
 p=\frac{N}{8q}\in\mathbb N,\qquad
 w=\operatorname{outlen}(\mathcal E_{\rm prev,p}),\qquad
 v=\frac{w}{256q}\in\mathbb N,
 \quad q=C+1,
\]
its $p$-bit prefixes are compatible at the preceding level, and
the fixed tower and output field at the current level have the
indicated integer degrees.  In particular $w=256qv$ makes the
splittings
\[
 \mathsf{R}\in\F_{2^{w/2}}^2,\qquad
 \mathsf{L}\in\F_{2^{4v}}^2,\qquad
 \F_{2^{4v}}\subseteq\F_{2^{w/2}}
\]
valid at the final level; the intermediate polynomial hash uses
$\F_{2^{4v}}$, whereas the hash producing $\mathsf{L}$ uses
$\F_{2^{32v}}$ and a $64v$-bit seed.  Since all ratios and the
number of levels are fixed once $C$ is fixed, we may select a
single fixed divisibility lattice for the input lengths and choose
the field embeddings in advance.  The integer $J$ below is chosen
\emph{after} this lattice and does not change it.

For the outer extractor, one does not assume an arbitrary condenser
row length already belongs to this lattice.  Instead, after fixing
its integer constants one takes a compatible prefix of length
$N=A_CHj\ell\le M$, where
$\ell=\lfloor M/(A_CHj)\rfloor$.  Thus
$0\le M-N<A_CHj=O_{\delta,C}(1)$, and discarding that bounded
number of bits costs only $O_{\delta,C}(1)$ in entropy.  The same
choice ensures the nested fields
$\F_{2^\ell}\subset\F_{2^{j\ell}}\subset
\F_{2^{Hj\ell}}$ used in the outer construction.  Any further
fixed tag-field embedding is arranged as part of this public
constant-divisibility setup.  ``Compatible'' never means that the
input data have a special algebraic form.

\begin{theorem}[Tagged row with adjustable mixed rank]\label{op:thm:main}
Fix $C\ge1$. For all sufficiently large compatible lengths $N$ as defined above, there are a deficiency
$\tau_C>0$ and native output degree $N/A_C$, both ratios fixed before
choosing a margin $K_*>0$. For each fixed margin the evaluator below
has the following property. On every rectangle with side sizes at least
$2^{(1-\tau_C)N}$ every honest-bearing mixed character of the row and
at most $C$ separate comparisons has binary rank $>K_*N$ whenever
the designated fixed tag differs from each comparison tag. Comparison
tags may coincide. Every character has global rank $O_{C,K_*}(N)$
and constant field separation over compatible fields, including after
the actual fingerprint substitution below.
\end{theorem}

For the doubling recursion, write $\mathsf{R}$ for the preceding
level's output and $\mathsf{L}$ for the new left-hash word.
These are intermediate outputs, not input subsets $A,B$.
Their names reflect the locality established \emph{after} the
transcript fixings in the proof below.

\begin{constructionbox}{Tagged row for a fixed number of comparisons}{cb:wrapper}
\cbingredients Construction~\ref{cb:starter}, $H_v$ and
Lemma~\ref{nmnew:eval}.
\cbsetup Put $q=C+1$, $h=\max\{1,\lceil\log_2 C\rceil\}$. Start with
the left-strong one-comparison starter. Apply the following recursion
for $h$ levels, using the last formula only at level $h$. All shrinking
ratios depend only on $C$. At a level of input length $N$, the preceding
level is called on length $p=N/(8q)$ and gives a word of length $w$.
Set $v=w/(256q)$. Require integrality by taking compatible lengths.
\cbinput Two rows $x,y$ and a numerical tag $a$ in the innermost starter field.
\cbsteps
\begin{constructionsteps}
\item Compute $\mathsf{R}$ by running the preceding evaluator with tag $a$ on
the two $p$-bit prefixes. Set $P=\operatorname{prefix}_{64v}(\mathsf{R})$
and $\mathsf{L}=H_{8v}(x;P)$.
\item At an intermediate level return $H_v(\mathsf{R};\mathsf{L})$.
At the final level split $\mathsf{L}=(u,b)\in K^2$, $\mathsf{R}=(r,t)\in L^2$,
where $K=\F_{2^{4v}}\subseteq L=\F_{2^{w/2}}$. Return
\[
 \Tr_{L/K}\big(bt(1+ur)^{J-1}\big).
\]
Choose $J$ a power of two with $4Jv-w/3-5>K_*N$.
\end{constructionsteps}
\cboutput A native field word of $4v=N/A_C$ bits. Only $J$ depends on
the requested margin.
\end{constructionbox}

\begin{proof}
We distinguish the \emph{statistical induction} used at all intermediate
levels from the \emph{restricted-rank argument} used only at the final
level.  The former is independent of $J$; in particular, no
margin-dependent entropy requirement is introduced during the
recursion.  We use the following notation: execution $0$ is the
honest pair, while executions $i\ge1$ are arbitrary separate
tamperings.  Their numerical tags are fixed and the honest tag
differs from each comparison tag.  Let $q=C+1$.

\smallskip\noindent
\emph{The induction hypothesis and entropy of the prefixes.}
Suppose the preceding evaluator has a $w$-bit output and is
left-source-strong against at least $t_0$ comparisons, at some fixed
near-full entropy rate, with error $2^{-\Omega(p)}$ on $p$-bit rows.
A current level has $N$ input bits on each side and uses
$p=N/(8q)$ prefix bits.  A source with entropy at least
$(1-\tau)N$ has a $p$-bit prefix with min-entropy at least
$p-\tau N$, because discarding $N-p$ bits reduces min-entropy by
at most $N-p$.  We fix $\tau>0$ small enough that
$\tau N\le\tfrac12\tau_{\rm prev}p$ and
$\tau\le1/16$, where $\tau_{\rm prev}>0$ is the preceding level's
allowed deficiency rate.  Thus the preceding evaluator applies
to the prefix distributions with a uniform linear entropy margin.
Only finitely many levels are used, so the minimum of these fixed
positive choices is again positive and does not depend on $J$.

Partition the at most $2t_0$ comparisons into $T$ and $S$, each of
size at most $t_0$.  At the current level compute
\[
 \mathsf{R}_i=\mathcal E_{\rm prev}(x_i^{\rm pre},y_i^{\rm pre}),
 \quad P_i=\operatorname{prefix}_{64v}(\mathsf{R}_i),
 \quad \mathsf{L}_i=H_{8v}(x_i;P_i),\qquad i=0,1,\ldots,t.
\]
The honest $\mathsf{R}_0$ is close to uniform given the $\mathsf{R}_i$ for either
of the two groups and given the entire honest left prefix.
To justify retaining the \emph{tampered} left prefixes as well,
use the completion-table argument at the end of
Lemma~\ref{mc:lem:starter}: conditioned on a sampled left
completion table, every tampered prefix is a deterministic function
of the honest prefix.  The preceding source-strength statement
can therefore retain their complete vector.  After averaging the
tables, for both $Q=T$ and $Q=S$ we have
\begin{equation}\label{op:eq:wrapper-A-fresh}
 (\mathsf{R}_0,\mathsf{R}_Q,Z)\approx_{2^{-\Omega(p)}}(U_w,\mathsf{R}_Q,Z),
\end{equation}
where $Z$ is the vector of all left prefixes, honest and tampered.
Its total length is at most $qp=N/8$.

\smallskip\noindent
\emph{The two-sided transcript and its locality.}
Once $Z$ is disclosed, the honest and tampered $\mathsf{R}_i$ depend only on
the original right source.  Disclose next
$\mathbf P=(P_0,\ldots,P_t)$, a tuple of at most
$64qv=w/4$ bits.  The tuple is right-local after $Z$ is fixed.
Now, conditional on $(Z,\mathbf P)$, the $\mathsf{L}_i$ depend only on
the original left source; the $\mathsf{R}_i$ remain right-local.
All these are deterministic functions of the corresponding
original source after the transcript values have been fixed.
Consequently every retained positive-probability transcript leaf
is a \emph{product} distribution on the two original sides.  This fact is
the reason for the order of disclosure.

We will repeatedly use a standard min-entropy tail rule.  If a
random variable $W$ has min-entropy at least $k$ and a message $M$
has at most $2^b$ values, then except on message mass at most
$2^{-s}$ the conditional min-entropy of $W$ is at least
$k-b-s$: indeed a message of probability at least $2^{-b-s}$
cannot raise any atom above $2^{-k+b+s}$.  The same rule applies
after any already-fixed product transcript; all error estimates
below are averaged over leaves rather than asserted for every
exceptionally unlikely leaf.

\smallskip\noindent
\emph{Freshness of the new left-local word $\mathsf{L}_0$.}
Fix $Z$ and $P_S$.  By \eqref{op:eq:wrapper-A-fresh} with $Q=S$,
$P_0$ is exponentially close to an independent uniform $64v$-bit
seed, even given $P_S,Z$.  Conditional on these values, the left
source and the remaining right-local seed tuple are independent.
Revealing $\mathsf{L}_S$ costs at most $8qv$ additional left-side bits,
and it does not restrict the right source because $\mathsf{L}_S$ is
left-local after $P_S,Z$ have been specified.

The honest left source initially has entropy $(1-\tau)N$.
On good transcript mass, the total cost of $Z$, $\mathsf{L}_S$, and two
$v$-bit tail allowances is at most $qp+8qv+2v$.
Thus its residual min-entropy is at least
\begin{equation}\label{op:eq:wrapper-left-budget}
 (1-\tau)N-qp-8qv-2v>24v
\end{equation}
for the stipulated fixed ratios and sufficiently large lengths.
The seed of $H_{8v}$ is $P_0$, with exactly $64v$ bits, and the
ordinary strong-hash calculation preceding this theorem gives
error $O(2^{-v})$ at the $24v$-bit entropy threshold.
Replace $P_0$ by an independent uniform seed, apply that strong
hash bound on each good left conditional distribution, and restore the true
seed distribution using \eqref{op:eq:wrapper-A-fresh}.

Importantly, we can also retain the \emph{whole} vector $\mathbf P$
in this comparison.  Given $(Z,P_S,\mathsf{L}_S)$, the conditional left-side
distribution is the same for every further right-local seed value; the
strong-hash error is averaged with respect to the honest seed
$P_0$ and is not increased by disclosing auxiliary information
about that seed.  One can formalize this by first adjoining the
entire right-local tuple to the ideal uniform-seed distribution, then
attaching the same left conditional distribution in each case.  We obtain
\begin{equation}\label{op:eq:wrapper-B-fresh}
 (\mathsf{L}_0,\mathsf{L}_S,Z,\mathbf P)
 \approx_{2^{-\Omega(v)}}(U_{8v},\mathsf{L}_S,Z,\mathbf P).
\end{equation}
This is the first half of the comparison doubling.

\smallskip\noindent
\emph{Enough entropy in $\mathsf{R}_0$ for the second ordinary hash.}
Use \eqref{op:eq:wrapper-A-fresh} with $Q=T$ to replace the
joint distribution $(\mathsf{R}_0,\mathsf{R}_T,Z)$ by the distribution with a uniform, independent
$\mathsf{R}_0$.  Attach the rest of the right input using the original
conditional distribution given $(\mathsf{R}_0,\mathsf{R}_T,Z)$, and then attach the
independent left side given $Z$.  This changes the full experiment
by at most its previous extraction error and preserves the
conditional product structure.  This smoothing viewpoint avoids
conditioning on the value of $\mathsf{R}_T$, which need not have high
probability in an individual leaf.

In the smoothed experiment $\mathsf{R}_0$ starts with $w$ bits of entropy
even given $\mathsf{R}_T,Z$.  Disclosing the entire $\mathbf P$ can cost
at most $w/4$ more bits.  After fixing $\mathsf{R}_T$ and $\mathsf{L}_S$ on their
respective sides, a comparison output in $T$ is left-local
(its $\mathsf{R}_i$ has been fixed), while an output in $S$ is right-local
(its $\mathsf{L}_i$ has been fixed).  The latter outputs expose at most
$qv$ further bits about $\mathsf{R}_0$.  Using a $2v$-bit tail slack, its
remaining min-entropy on the good product leaves satisfies
\begin{equation}\label{op:eq:wrapper-right-budget}
 H_\infty(\mathsf{R}_0\mid\text{current right-side messages})
 \ge w-w/4-qv-2v>3v.
\end{equation}
The inequality holds in the smoothed, good-leaf sense; the
exceptional mass and the smoothing cost are both exponentially
small in $v$.

By \eqref{op:eq:wrapper-B-fresh}, the honest seed $\mathsf{L}_0$ is fresh
against $\mathsf{L}_S,Z,\mathbf P$.  Conditional on the fixed product
transcript, it is left-local and the residual $\mathsf{R}_0$ is right-local.
Apply the strong hash $H_v(\mathsf{R}_0;\mathsf{L}_0)$ with the entropy bound
\eqref{op:eq:wrapper-right-budget}.  Keep the $T$-comparison
outputs, which are left-local on the fixed $\mathsf{R}_T$ leaf, and the
$S$-comparison outputs, which are right-local once $\mathsf{L}_S$ is fixed.
The seed-specific averaging argument allows us to retain the
entire original left input as well.  Replacing smoothed distributions by the
actual distributions and restoring the actual comparison marginal costs
only the preceding exponentially small errors.  Thus
\begin{equation}\label{op:eq:wrapper-intermediate}
 \bigl(H_v(\mathsf{R}_0;\mathsf{L}_0),\ (H_v(\mathsf{R}_i;\mathsf{L}_i))_{i=1}^t,\ X\bigr)
 \approx_{2^{-\Omega(v)}}
 \bigl(U_v,\ (H_v(\mathsf{R}_i;\mathsf{L}_i))_{i=1}^t,\ X\bigr).
\end{equation}
This completes the induction for every intermediate level,
and doubles the number of allowed comparisons.  A bounded number
of levels gives an exponentially accurate left-strong statistical
evaluator for all the groups needed at the final level.

\smallskip\noindent
\emph{The final level: supports rather than statistical extraction.}
At the last level, run the same transcript and freshness argument
on \emph{uniform distributions} on any rectangle of side sizes
$2^{(1-\tau_C)N}$.  We now extract support-size information.
From \eqref{op:eq:wrapper-A-fresh} with $Q=T$, the honest $\mathsf{R}_0$
is close to uniform conditioned on $(\mathsf{R}_T,Z)$.  The message
$\mathbf P$ takes at most $2^{w/4}$ values.  For fixed
$(\mathsf{R}_T,Z)$, the union over all values of $\mathbf P$ of those
conditional supports of $\mathsf{R}_0$ having fewer than $2^{2w/3}$
points has size at most
\[
 2^{w/4}2^{2w/3}=2^{11w/12}.
\]
A uniform $w$-bit word lies in this union with probability at
most $2^{-w/12}$; the statistical approximation adds its
exponentially small error.  Except on that small joint mass,
therefore,
\begin{equation}\label{op:eq:wrapper-A-support}
 |\supp(\mathsf{R}_0\mid \mathsf{R}_T,Z,\mathbf P)|\ge2^{2w/3}.
\end{equation}
Similarly \eqref{op:eq:wrapper-B-fresh} implies that, except on
exponentially small transcript mass, the support of the honest
$8v$-bit $\mathsf{L}_0$ given $(\mathsf{L}_S,Z,\mathbf P)$ contains at least
$2^{8v-1}$ points.  Indeed, on a leaf with a smaller support, its
conditional statistical distance from uniform is at least $1/2$;
the average distance in \eqref{op:eq:wrapper-B-fresh} is
exponentially small.  Consequently
\begin{equation}\label{op:eq:wrapper-B-support}
 |\supp(\mathsf{L}_0\mid \mathsf{L}_S,Z,\mathbf P)|\ge2^{8v-1}
\end{equation}
on all but exponentially small mass.

Choose a simultaneous good transcript, then fix the comparison
words $\mathsf{R}_T$ on the right and $\mathsf{L}_S$ on the left.  These are
opposite-side restrictions of a product leaf, so they do not
reduce the other side's conditional support.  We obtain a
nonempty product rectangle on which
\eqref{op:eq:wrapper-A-support} and
\eqref{op:eq:wrapper-B-support} both hold.  The final comparison
terms in $T$ depend only on the left variable, and those in $S$
only on the right variable.  In a binary character matrix their
total contribution therefore has rank at most two, regardless of
how many comparisons occur in either group.

\smallskip\noindent
\emph{Rank of the honest geometric-series hash.}
Write $\mathsf{L}_0=(u,b)\in K^2$ with $K=\F_{2^{4v}}$ and
$\mathsf{R}_0=(r,t)\in L^2$ with $L=\F_{2^{w/2}}$.
Because $J$ is a power of two, the polynomial identity
\[
 (1+z)^{J-1}=\sum_{i=0}^{J-1}z^i
\]
holds in characteristic two.  Hence, for any nonzero character
coefficient $\lambda\in K^\times$,
\begin{align*}
 \Tr_{K/\F_2}\left(\lambda\,
    \Tr_{L/K}(bt(1+ur)^{J-1})\right)
 &=\Tr_{L/\F_2}\left(
     \lambda\sum_{i=0}^{J-1}(bu^i)(tr^i)\right).
\end{align*}
This is the nondegenerate absolute-trace pairing between the
scaled evaluation vectors
\[
 \Phi_J(u,b)=b(1,u,\ldots,u^{J-1})\in K^J,
 \quad
 \Phi_J(r,t)=t(1,r,\ldots,r^{J-1})\in L^J.
\]
The pairing of $K^J$ with $L^J$ is nondegenerate on the $K^J$
side: for nonzero $z\in K^J$ the trace pairing against
$L^J$ is not identically zero.

The $\mathsf{L}_0$ support has density at least $1/2$ in $K^2$ by
\eqref{op:eq:wrapper-B-support}.  Lemma~\ref{nmnew:eval},
with deficiency $d\le1$, gives affine-span codimension at most two
in $K^J$ and hence linear-span dimension at least
\[
 4Jv-2.
\]
Likewise the $\mathsf{R}_0$ support has density at least $2^{-w/3}$
in $L^2$.  Applying the same lemma over $L$ gives affine-span
codimension at most $w/3+1$ in $L^J$.  Both applications satisfy
the lemma's root-bound hypothesis for sufficiently large compatible
lengths, since $J$ is a \emph{fixed} constant once the desired
margin has been selected.

Restricting a nondegenerate pairing to a short-side span of
dimension at least $4Jv-2$ and a long-side span of codimension
at most $w/3+1$ can decrease its rank by at most that latter
codimension.  Therefore the honest character matrix has rank at
least $4Jv-w/3-3$.  Subtracting the left-local and right-local
comparison contribution costs at most two additional ranks, giving
\begin{equation}\label{op:eq:wrapper-final-rank}
 \rk(\text{honest-bearing mixed character})
 \ge4Jv-\frac w3-5>K_*N,
\end{equation}
where $J$ was chosen in the public setup.  The rectangle was chosen
inside an arbitrary starting rectangle of the asserted size;
therefore that starting rectangle has rank greater than $K_*N$
as well.  This proves the adjustable restricted-rank assertion.

\smallskip\noindent
\emph{Global rank, actual tags, and field compatibility.}
For each fixed $C$ and $J$, the hashes use only fixed-degree
polynomial evaluations, powers, additions, multiplications, and
relative traces in fields whose extension ratios are fixed.
A fixed-tag binary character is consequently an explicit sum of a
constant number of products of one left field-valued function and
one right field-valued function.  In the outer construction the tag
is not fixed but has the form
\[
 a(x,y)=\iota(P_x(t),P_y(t))=a_L(x)+a_R(y),
\]
where the last equality uses the linearity of $\iota$.
Substituting it into $S_a(z)=V_0(z)+aV_1(z)$ gives a sum of
separated terms even though the functions $V_0,V_1$ themselves
need not be linear or low-degree in their source bits.  Subsequent
fixed powers and products increase the \emph{number} of separated
field terms by a constant factor depending on $C,J$, but not on the
input length.  Expanding each such field product using trace-dual
bases costs at most the binary degree of the common base field,
which is $O(N)$.  Thus every actual character has global binary
rank $O_{C,K_*}(N)$ and an explicit separated encoding.

The required extensions and trace maps are valid because of the
length convention preceding this theorem.  Taking all bit blocks in
fixed bases over the common subfield ensures that prefix operations
and restrictions to subfields preserve the separated representation.
The finite recursion permits shrinking $\tau_C$ once so that all
entropy allowances, smoothing errors, and field-size hypotheses
hold simultaneously.  Neither $\tau_C$ nor the native output length
$4v=N/A_C$ depends on the final width $J$.
\end{proof}
\subsubsection{Constant-rate non-malleability and separate source strength}
\begin{theorem}[Fixed-$C$ non-malleability with exponential error]
\label{nm:thm:mainconstant}\label{crn:thm:completion}
Fix $C\ge1$ and $\delta>0$ with $\delta\le1$. There is an explicit
polynomial-time map $G:\bits^n\times\bits^n\to\bits^m$ with
$m=\Omega_{\delta,C}(n)$ and $\epsilon=2^{-\Omega_{\delta,C}(n)}$
such that, for independent sources of min-entropy $k\ge\delta n$ and
any admissible $C$ separate tamperings, writing
$V=(G(f_i(X),g_i(Y)))_{i=1}^C$, one has simultaneously for the same map
\[
 \TV((G(X,Y),V,H),(U_m,V,H))\le\epsilon
 \qquad\text{for each }H\in\{X,Y\}.
\]
Here $U_m$ is independent of $(V,H)$. Dropping $H$ gives ordinary
non-malleability. Every output character has binary rank
$O_{\delta,C}(n)$ with explicit separated encodings of that length.
There is no entropy condition on comparison images.
\end{theorem}

\begin{constructionbox}{Constant-rate fixed-$C$ non-malleable extractor}{cb:outer-nm}
\cbingredients The strong condenser of Theorem~\ref{pre:strongcondenser},
the row evaluator of Theorem~\ref{op:thm:main}, and the forcing identity
of Lemma~\ref{cr2:lem:majority}.
\cbsetup Put $q=C+1$. Instantiate the condenser at rate $\delta/(16q)$,
deficiency $\tau_C/64$ and average row error at most $1/(1000q)$.
Write $M\ge c_Dn$ for its row length and $r$ for its power-of-two row
count (repeat rows by ignoring extra seed bits if necessary).
Choose a fixed integer $H\ge4$ with $4\mid H$ and
$q/(A_CH)\le\tau_C/64$, taking $H$ to be a sufficiently large
multiple of all fixed compatibility divisors in the row construction.
Specifically, for every sufficiently large integer $d$, require
$N=A_CHd$ to be a compatible row length, every working-field degree
and prefix or block length within the row evaluator to be an integer
multiple of $d$, and the innermost tag field to have binary degree at
least $2d$.  In particular, $v=Hd/4$ is an integer.  These conditions
involve only the fixed length ratios of the row construction, which
depend on $C$ and are independent of the adjustable column widths.
Thus $H$ can be fixed before those widths and before $j$.
For a fixed $j$ chosen below use
\[
 \ell=\lfloor M/(A_CHj)\rfloor,\quad d=j\ell,\quad N=A_CHd,
 \qquad F_0=\F_{2^\ell}\subset K_0=\F_{2^d}\subset F=\F_{2^{Hd}} .
\]
The row native field is $F$; the innermost tag field contains $K_0$
and has room for a fixed $K_0$-linear injection $\iota:K_0^2\to
\text{tag field}$. Use whole-subfield-block bases as in the row proof.
For large $n$, $M-N<A_CHj$ and $d\ge M/(2A_CH)$, so
$b=\lceil n/d\rceil\le b_*:=\lceil2A_CH/c_D\rceil$.
Choose a power of two $R\ge\max\{4,r,8\binom q2(b_*-1)\}$,
put $D=R/2$, and take distinct $t_a\in K_0$ for $1\le a\le R$.
Use $i(a)=1+((a-1)\bmod r)$. In increasing order set
$K_a=3+q\sum_{b'<a}C_{b'}$, where $C_{b'}N$ bounds the rank of
each actual advised character of column $b'$. Choose its width
by Theorem~\ref{op:thm:main}. These bounds can be taken before
choosing $j$: expand the actual expression in $K_0$-coordinates.
Each product of one left and one right field element costs at most $d$
binary rank after a character is applied. The ratios of field degrees
to $d$ are fixed, and projection to $F_0$ only pulls a character back
to $F$. Thus this finite symbolic expansion gives a valid $C_{b'}$.
Choose $j$ from the local bias exponent as in Step~7 of the proof.
After fixing $j$, bound the rank of the complete merged output and
choose the global bias exponent as in Step~8; only then choose the
final projection length as in Step~9.
\cbinput Two original words $x,y$ of length $n$.
\cbsteps
\begin{constructionsteps}
\item Compute the condenser rows on each source and keep their first $N$
bits. Pad each original input to $b$ coefficients in $K_0$ and denote
its coefficient polynomial by $P_x$ or $P_y$.
\item For each $1\le a<R$, form the numerical tag
$\alpha_a=\iota(P_x(t_a),P_y(t_a))$. Run the column-$a$ evaluator on
the two rows with index $i(a)$ and tag $\alpha_a$. Trace its output
from $F$ to $F_0$ to obtain $W_a$.
\item Return a fixed surjective binary projection of
$\sigma_D(W_1,\ldots,W_{2D-1})$ onto $m$ bits.
\end{constructionsteps}
\cboutput One extractor with both separate source-strength guarantees.
\end{constructionbox}

\begin{proof}
Fix independent sources $X,Y$ as in the statement and an admissible
tuple of separate tamperings $(f_i,g_i)_{i=1}^C$.  Write
$S_X=\supp(X)$ and $S_Y=\supp(Y)$.  Since every point of
$S_X\times S_Y$ has positive probability, admissibility means
\[
 (f_i(x),g_i(y))\ne(x,y)
 \qquad\text{for every }(x,y)\in S_X\times S_Y
 \text{ and every }i\in[C].
\]
This property is preserved when either support is restricted.

We first prove a \emph{dispersion statement}: every binary mixed
character with nonzero coefficient on the designated honest field
output is nonconstant on every product rectangle contained in
$S_X\times S_Y$ with side sizes at least $2^{\delta n/2}$.
We next bound the global binary rank of these actual
input-dependent-tag characters and apply the support-relative
approximate-duality rank-to-bias lemma.  Finally we use entropy slack
to retain either complete original source.  The proof-only restrictions
and representative tables below are not evaluations performed by $G$.

\smallskip\noindent
\emph{Step 1: remove comparison images that have heavy atoms.}
Fix a product rectangle $A_X\times A_Y\subseteq S_X\times S_Y$ with
$|A_X|,|A_Y|\ge2^{\delta n/2}$.  For Steps~1--7, start with the
independent uniform distributions on $A_X$ and $A_Y$ and apply the successive
source-local restrictions described below.  Suppose that a mixed character of
the honest and $C$ comparison \emph{field-valued} merger outputs,
with nonzero honest coefficient, is constant on this rectangle.
Write $q=C+1$ and set
\[
 h_{\rm raw}=\frac{\delta n}{4q}.
\]
If the left or right marginal of some active comparison has a
value with probability greater than $2^{-h_{\rm raw}}$, restrict
that original side to the preimage of this value.  The restriction
costs fewer than $h_{\rm raw}$ bits of min-entropy.  The chosen
comparison output, after this fixing, is a function of the other
original side only.  Fix the value of its one-bit character
contribution on that other side, at a further cost of at most one
bit, and remove this comparison from the character.  The remaining
character is still constant on a positive-weight product rectangle,
and its honest coefficient has not changed.

Recompute the surviving comparison marginals and repeat.  Each
iteration removes an active comparison, so at most $C$ iterations
occur.  The total entropy loss on either original side is at most
$Ch_{\rm raw}+C$.  Since $C/q<1$, the honest sources still have
min-entropy at least
\[
 \frac{\delta n}{2}-\frac{C\delta n}{4q}-C
 \ge\frac{\delta n}{4}-O_C(1).
\]
All surviving comparison marginals on both sides now have
min-entropy at least $h_{\rm raw}$.  This reduction is essential:
non-malleability never assumes that a tampering image retains
entropy.

\smallskip\noindent
\emph{Step 2: obtain fixed distinct execution pairs.}
On each original side separately choose a \emph{most probable}
pattern of the equality/inequality indicators for every pair among
the honest input and the remaining comparison inputs, and condition
on that pattern.  There are only $O(C^2)$ bits per side, so each
chosen pattern has conditional probability at least
$2^{-O(C^2)}$; this costs only $O(C^2)$ entropy bits and
preserves the product structure.  Any two executions now either
coincide on that original side everywhere on the restricted
support, or differ there everywhere.  If two full execution pairs
coincide, combine their coefficients in the mixed character.
Delete any representative whose combined coefficient becomes zero.
The honest representative is not merged with a tampered one,
because each admissible tampering has no fixed point as a pair on
the original product support.  Thus its coefficient remains
nonzero.  We obtain $r_0\le q$ surviving representatives with
pairwise distinct full input pairs, where every distinct pair
of representatives differs on at least one original side
throughout the current product distribution.

\smallskip\noindent
\emph{Step 3: choose common high-entropy condenser rows.}
The strong condenser was instantiated below the entropy threshold
of every remaining honest and comparison marginal, so for each
such marginal its seeded output table is, on average over the
$r$ condenser indices, close to a distribution capped at the
prescribed near-full row entropy $h$.
Here is the useful elementary form of the capping argument.
For a row distribution $P$ on $M$ bits, mark a value $z$ as
\emph{accepted} with conditional probability
\[
 a_P(z)=\min\{1,2^{-h}/P(z)\}
\]
(and accept with probability one at $P(z)=0$).
The accepted subdistribution assigns mass
$\min\{P(z),2^{-h}\}$ to each $z$.  Its rejection mass is
$\sum_z(P(z)-2^{-h})_+$, at most the distance of $P$ from the
class of distributions with min-entropy at least $h$.
Thus the condenser error bounds the average rejection
probability, including the random choice of row index.

Introduce these accept/reject indicators for each row and each
surviving execution, using coins local to the corresponding
original side.  The condenser parameters ensure that on either
side the expected \emph{fraction of indices with any rejection}
is at most $1/1000$ (after summing over the at most $q$
executions).  Markov's inequality shows that with probability at
least $1-16/1000$ at least $15r/16$ indices are accepted for
\emph{all} the executions on that side.
Fix one indicator pattern of probability at least
$(1-16/1000)2^{-qr}$ with this property.  Since $q,r$ are
constants, its probability is bounded below by a positive
constant.  Conditional on this pattern, every accepted row
marginal has maximum atom at most
\[
 \frac{2^{-h}}{\Pr[\text{chosen pattern}]},
\]
because its point mass before normalization is bounded by
$2^{-h}$.  It therefore loses only $O_{\delta,C}(1)$ bits of
min-entropy.  Do this independently on the two original sides;
all restrictions remain product restrictions.  The accepted rows
are \emph{common across all surviving executions} on each side.

\smallskip\noindent
\emph{Step 4: freeze fingerprint equality patterns, not all their
values.}
Every original word is represented by its fixed polynomial
$P_x(T)$ over $K_0$, of degree less than
$b\le b_*=O_{\delta,C}(1)$.  For each pair of surviving
executions, each original side, and each of the $R$ positions,
choose a most probable joint pattern of the Boolean bits telling
whether their two evaluations at $t_a$ coincide, and condition on
that pattern.  There are at most $R\binom q2$ such bits per side,
so the chosen pattern has probability at least
$2^{-R\binom q2}$ and costs only $O_{\delta,C}(1)$ entropy
while preserving independence.

For two different execution pairs, at least one of the original
sides has different full words everywhere on the restricted
support.  Let $Z_{ij}\subseteq[R]$ be the set of positions where
their fingerprints agree on that side.  Because the equality
indicators have been fixed, $Z_{ij}$ is the same for every input
in the conditioned support, although the difference polynomial
itself may vary with that input.  Choose any one input in this
support, and denote its two distinct full words by $u$ and $v$.
Their coefficient polynomials satisfy
\[
 Q_*(T):=P_u(T)-P_v(T)\not\equiv0,\qquad \deg Q_*\le b-1,
 \qquad Z_{ij}=\{a\in[R]:Q_*(t_a)=0\}.
\]
The fixed evaluation points $t_a$ are distinct, so the root bound
gives $|Z_{ij}|\le b-1$.  By injectivity of $\iota$, a collision
of the full tags requires agreement of the fingerprints on both
sides and hence can occur only in $Z_{ij}$.  Thus this pair has a
common set of at most $b-1$ possible tag-collision positions.
Taking the union over at most $\binom q2$ pairs, no more than
\[
 \binom q2(b_*-1)\le R/8
\]
column positions suffer any pairwise tag collision.

The balanced row schedule repeats each of the $r$ condenser
indices exactly $R/r$ times.  Since at least $15r/16$
indices are commonly accepted on each side, the bad data indices
exclude at most $R/16$ positions per side, or $R/8$ altogether.
After excluding those and the tag-collision positions, at least
$3R/4$ of the $R$ positions remain.  We do not use position $R$,
so among the actual $R-1=2D-1$ evaluated columns there are at least
\begin{equation}\label{nm:eq:outer-good-majority}
 \frac{3R}{4}-1\ge D=R/2
\end{equation}
positions which are simultaneously good for all surviving
executions and have pairwise distinct tags.  Fix any $D$ of them
in increasing order:
\[
 I=\{a_1<\cdots<a_D\}\subseteq[2D-1].
\]
This set is chosen \emph{once on one actual product subsource}; it
is not allowed to depend on the execution, the desired field
output, or a later Fourier character.

Truncating each accepted $M$-bit row to its public compatible
$N$-bit prefix loses at most $M-N=O_{\delta,C}(1)$ bits.
Together with the preceding constant losses, for sufficiently
large $n$ every selected row marginal has min-entropy at least
$(1-\tau_C/8)N$.

\smallskip\noindent
\emph{Step 5: fix only the numerical tags in the current column.}
Consider one selected index $a\in I$.  Disclose the $K_0$-valued
fingerprints $P_{x_i}(t_a)$ and $P_{y_i}(t_a)$ for all $r_0$
surviving executions.  The left disclosures depend only on the
left original input; the right disclosures only on the right.
Each side therefore reveals at most $qd$ bits.  An elementary
min-entropy tail bound says that, except on message mass at most
$2^{-t}$, conditioning on a message with at most $2^b$ values
lowers min-entropy by at most $b+t$.  Take
$t=\tau_CN/16$.  Since the public setup guarantees
\[
 \frac{qd}{N}=\frac{q}{A_CH}\le\frac{\tau_C}{64},
\]
we obtain, outside exponentially small total mass, the row entropy
bound
\begin{align*}
 H_\infty(\text{current designated data row}
               \mid\text{current fingerprint messages})
 &\ge (1-\tau_C/8)N-qd-\tau_CN/16\\
 &\ge(1-\tau_C/2)N.
\end{align*}
We fix these \emph{current} fingerprint values only.  All current
tags become fixed numerical tags with the honest-designated tag
distinct from every comparison tag.  We never condition on the
fingerprint values or evaluated outputs of preceding columns;
those remain nonconstant functions to be controlled by their
separated-rank bounds.

\smallskip\noindent
\emph{Step 6: last-active-column rank dominates the complete prefix.}
For each column $b$, the public symbolic expansion of the actual
advised evaluator supplies a global binary-character rank bound
$C_bN$, uniformly in the source and the tamperings.  The widths
were chosen successively so that, for column $a$,
\begin{equation}\label{nm:eq:outer-hierarchy}
 K_a=3+q\sum_{b<a}C_b
\end{equation}
and the row theorem gives restricted mixed rank greater than
$K_aN$ at that column.  These choices were made with fixed
parameters and do not depend on the subsources selected in this
proof.

Take a nonzero binary character of the complete vector of the
$r_0D$ selected field outputs.  Let $a\in I$ be its last column
with a nonzero coefficient and choose one execution whose
coefficient there is nonzero.  Regard that execution as the
designated row for the rank argument.  The row distributions of that
execution have entropy at least $(1-\tau_C/2)N$, and its
numerical tag differs from every other tag at position $a$.
Theorem~\ref{op:thm:main} therefore provides restricted rank
$>K_aN$ against the other at most $C$ executions, even if their
comparisons are arbitrarily dependent on the designated inputs.

We spell out the passage from the original source variables to the
designated condensed row.  For each row value $u$, sample an
original left word from the conditional distribution of the current
left subsource given its designated row $u$, and store these samples
in a table $T_L(u)$.  Sample an analogous independent right table
$T_R(v)$.  The tables are sampled independently of the designated
row values.  Drawing the row values first and then evaluating the
tables reproduces exactly the original product distribution.  For every
fixed pair of tables, each tampered row and each earlier-column
output is a separate deterministic function of the designated
left and right row values.  Separate precomposition cannot increase
matrix rank, so all columns preceding $a$ contribute together at
most
\[
 qN\sum_{b<a}C_b
\]
to the binary character matrix.  The current column has rank
$>K_aN$ on every row rectangle at the stipulated threshold.
By rank subadditivity, the complete character has restricted rank
strictly greater than
\[
 K_aN-qN\sum_{b<a}C_b=3N.
\]
The same reasoning applies after restricting to any subrectangle
of the designated row supports of side size at least
$2^{(1-\tau_C)N}$, because the available row entropy has the
fixed slack $\tau_CN/2$.

Apply the local linear-rank approximate-duality lift of
Corollary~\ref{pre:upgrade} on the designated product row distribution,
then average over representative tables and over the current
fingerprint leaves.  The exceptional rare-message mass is
$2^{-\Omega(N)}$.  There are only $D=O_{\delta,C}(1)$ selected
columns, each with $N=\Omega_{\delta,C}(n)$, so we obtain a
constant $\eta=\eta(\delta,C)>0$, chosen from the individual-column
character bounds independently of $j$, such that every nonzero
character $\chi$ of the joint selected-column output obeys
\begin{equation}\label{nm:eq:outer-fresh-bias}
 \left|\E(-1)^{\chi(\mathbf W_I)}\right|
 \le2^{-\eta n},
 \qquad
 \mathbf W_I\in F_0^{r_0D}.
\end{equation}
This is freshness relative to the \emph{complete prefix} of
selected columns, not merely to a separately chosen earlier
output or a fixed earlier transcript.

\smallskip\noindent
\emph{Step 7: Fourier inversion and majority forcing.}
Let $L_0=r_0D\ell$ be the binary length of the joint selected
vector.  For any prescribed $w\in F_0^{r_0D}$, Fourier inversion
and \eqref{nm:eq:outer-fresh-bias} give
\begin{equation}\label{nm:eq:outer-positive-atom}
 \Pr[\mathbf W_I=w]\ge
 2^{-L_0}\left(1-(2^{L_0}-1)2^{-\eta n}\right).
\end{equation}
Choose the constant divisor $j$ in the public field setup so that
$L_0\le qD\ell\le\eta n/2$ for sufficiently large $n$.
The right side of \eqref{nm:eq:outer-positive-atom} is then
strictly positive for \emph{every} $w$.  Thus one can prescribe
independently the values of all $D$ selected columns in all
$r_0$ executions while staying on the current product subsource.

Choose arbitrary $\beta_0,\ldots,\beta_{r_0-1}\in F_0$ and
prescribe $W_{i,a}=\beta_i$ for every $a\in I$ and every
execution $i$.  Since $D=R/2$ is a power of two, the coefficient
of $T^D$ in
\[
 (1+\beta_iT)^D\prod_{a\notin I}(1+W_{i,a}T)
\]
is exactly $\beta_i^D$: in characteristic two,
$(1+\beta_iT)^D=1+\beta_i^DT^D$, and the remaining
$D-1$ factors have degree less than $D$.  Consequently
\[
 \sigma_D(W_{i,1},\ldots,W_{i,2D-1})=\beta_i^D.
\]
The Frobenius map $\beta\mapsto\beta^D$ is a permutation of
$F_0$.  As the $\beta_i$ vary independently, the tuple of merged
field outputs visits \emph{every} point of $F_0^{r_0}$.  No
nontrivial honest-bearing mixed field character can therefore be
constant on the starting rectangle.  This is the required
dispersion conclusion.

\smallskip\noindent
\emph{Step 8: global rank and the exponential-error lift.}
It remains important that the evaluated tags depend on the
\emph{actual} inputs.  At each column,
\[
 \alpha_a(x,y)=\iota(P_x(t_a),P_y(t_a))
                  =\alpha_{a,L}(x)+\alpha_{a,R}(y).
\]
The affine-in-tag seed is
$V_0(x^{\rm pre})+\alpha_a(x,y)V_1(x^{\rm pre})$.
Even though $V_0,V_1$ can be nonlinear, this expression has a
bounded number of terms separated into left-local and right-local
field factors.  We now fix $j$ as in Step~7 and expand the completed
evaluator in coordinates over $F_0$.  In particular,
\[
 [K_0:F_0]=j,\qquad [F:F_0]=Hj.
\]
All working fields have extension degree over $F_0$ bounded by a
constant depending only on $\delta,C,j$.  The compatible field bases
make coordinate projections and relative traces $F_0$-linear, while
field multiplication has a bounded coordinate expansion.  The fixed
number of recursive field operations and the fixed-degree polynomial
$\sigma_D$ therefore give each merged field output a representation
\[
 \sum_{\nu=1}^{T_{\delta,C,j}} a_\nu(x)b_\nu(y),
 \qquad a_\nu(x),b_\nu(y)\in F_0,
\]
with $T_{\delta,C,j}$ independent of $n$.  The same separation bound
holds after composition with any separate left and right tampering.
For a binary character, each term becomes
$\operatorname{Tr}_{F_0/\F_2}(\lambda a_\nu(x)b_\nu(y))$ and has
binary matrix rank at most $\ell$ in trace-dual $\F_2$-bases of $F_0$.
Summing over the at most $q=C+1$ executions gives, for every actual
honest-bearing mixed character of the complete merged output tuple,
\[
 \operatorname{rank}_2[\chi(x,y)]_{x,y}
 \le qT_{\delta,C,j}\ell\le A_{\delta,C,j}n
\]
for a constant $A_{\delta,C,j}$ independent of $n$.

The individual-column constants $C_{b'}$ and the local exponent $\eta$
were chosen before $j$.  The constant $A_{\delta,C,j}$ for the
\emph{complete merger} is chosen only after $j$: multiplying the
shortened column outputs may introduce a separation width depending
on $[K_0:F_0]=j$.  Since Step~7 fixes $j$ as a function of $\delta,C$,
the resulting global rank bound is still $O_{\delta,C}(n)$.

For the approximate-duality application, take the permitted supports
to be all nonempty products $A\times B\subseteq S_X\times S_Y$.
This class is closed under separate restrictions, and every such
restriction preserves admissibility.  Steps~1--7 show that every
honest-bearing mixed character is nonconstant on each permitted
rectangle with $|A|,|B|\ge2^{\delta n/2}$.  Together with the global
rank bound $A_{\delta,C,j}n$, this verifies the hypotheses of
Lemma~\ref{pre:asymAD} with
\[
 n_L=n_R=n,\qquad a_0=b_0=\delta/2,
 \qquad a_1=b_1=3\delta/4.
\]
The entropy gap on each side is $\delta n/4>0$.  The lemma therefore
gives, uniformly for every such mixed character and every independent
pair of sources supported on $S_X$ and $S_Y$ with side min-entropy
at least $3\delta n/4$,
\begin{equation}\label{nm:eq:outer-global-bias}
 |\operatorname{bias}(\chi)|\le2^{-\zeta n}
\end{equation}
for a constant $\zeta=\zeta(\delta,C)>0$, uniform over the admissible
supports, tampering maps, and mixed characters.  This constant is fixed
only after $j$ and the complete-merger rank constant above have been chosen.
The final projection length is chosen afterward in Step~9.
A nonzero character after the surjective binary projection
$\pi:F_0\to\F_2^m$ pulls back to a nonzero field character, so
the same bound applies to the projected outputs.

\smallskip\noindent
\emph{Step 9: retain either original source and finish.}
Fix such a mixed character $\chi$, and first retain $X$.
Put
\[
 b(x)=\E_Y(-1)^{\chi(x,Y)}.
\]
Partition the $X$-space into the two sign classes of $b(x)$.
A class with probability at least $2^{-d_s}$, where
$d_s=\delta n/4$, has conditional min-entropy at least
$\delta n-d_s=3\delta n/4$; conditioning on that class affects
only the left source and preserves its independence from $Y$.
Its product support remains contained in $S_X\times S_Y$, so the
tamperings remain admissible and the restricted distribution belongs to the
permitted class from Step~8.  Apply
\eqref{nm:eq:outer-global-bias} to each such class.
The two classes of mass below $2^{-d_s}$ contribute at most
$2^{1-d_s}$.  This yields
\begin{equation}\label{nm:eq:outer-strong-bias}
 \E_X|b(X)|\le2^{-\zeta n}+2^{1-\delta n/4}.
\end{equation}
The argument is symmetric if we retain $Y$ instead.
Fourier inversion of the honest output and the $C$ actual
comparison outputs, conditional on the retained original source,
now gives
\[
 \TV\bigl((G(X,Y),V,H),(U_m,V,H)\bigr)
 \le2^{qm-1}
       \bigl(2^{-\zeta n}+2^{1-\delta n/4}\bigr),
 \qquad H\in\{X,Y\}.
\]
We used only characters having a nonzero coefficient on the honest
output; characters involving solely $V$ cancel against the actual
comparison marginal, which is retained unchanged.
Choose the public surjective projection length as in the setup,
\[
 m=\min\left\{\ell,
  \left\lfloor\frac{n\min\{\zeta,\delta/4\}}{2q}\right\rfloor
  \right\}.
\]
Since $\ell=\Theta_{\delta,C}(n)$ and $q$ is constant,
$m=\Omega_{\delta,C}(n)$, and the last statistical-distance bound
is $2^{-\Omega_{\delta,C}(n)}$.
The construction itself is polynomial-time for fixed $\delta,C$:
all publicly enumerated row schedules and field-operation counts
are constant, and the remaining operations are explicit
polynomial-time field arithmetic.
For the finitely many shorter lengths, the stipulated one-bit
constant completion has statistical error $1/2$ and character
rank zero; reducing the global positive constants covers this
finite range without changing the asymptotic statements.
\end{proof}

\begin{corollary}[Separate source strength with linear output]
\label{nm:cor:source-strong-constant}
Construction~\ref{cb:outer-nm} retains either original source
separately with the full comparison tuple, output
$\Omega_{\delta,C}(k)$ and error $2^{-\Omega_{\delta,C}(n)}$.
\end{corollary}
\begin{proof}
Use the two choices of $H$ in Theorem~\ref{nm:thm:mainconstant}.
\end{proof}

\section{Entropy-efficient output at constant rate}\label{un:optimal}
\par\smallskip
The construction boxes in this section distinguish the internally generated
seed from each seeded extractor's public seed interface. They record all
seed and payload lengths explicitly; the collision arguments that follow
are analysis, not evaluation steps.

This section first proves ordinary two-source output with additive entropy
loss, then source-strong $C$-tampering output, its $C=0$ ordinary
source-strong corollary, and finally sumset output with additive loss.
The ordinary concatenation and sumset wrappers share the one-payload
lemma and may share the seed of Theorem~\ref{ns:thm:seed}. The centered
non-malleable wrapper uses the fixed-$C$ theorem from
Section~\ref{un:nonmalleable}. Each nonempty-output
statement specifies its nontrivial exponentially small error range.
\subsection{Two-source output with additive entropy loss}\label{un:standard}
\subsubsection{Prerequisite: proof of the one-payload collision lemma}
The following weighted argument retains the actual other source. It
applies directly to arbitrary min-entropy distributions; no flat-source reduction
or assumption on an individual tampered image is required.

\begin{lemma}[One-sided source-strong collision recycling]
\label{os:lem:onepayloadstrong}
Let $X,Y$ be independent $n$-bit sources of min-entropy at
least $k$, put $K=2^k$, let $G$ be a $d$-bit seed, and put
$D=2^d$. Suppose that
\begin{enumerate}[label=(\roman*),leftmargin=*]
\item $G$ is one-sided non-malleable against every fixed-point-free
left map at component threshold $k-4$ with error $\gamma$,
and is an ordinary extractor there with error $\gamma$;
\item on the current product distribution,
$\TV((X,G(X,Y)),(X,U_d))\le\eta$;
\item $E:\bits^n\times\bits^d\to\bits^p$ is a strong seeded
extractor at entropy $k$, with error $\zeta$.
\end{enumerate}
Write $M=2^p$, and suppose $K\ge16M$.
For $0<\theta<1$ let
$\beta=3(1+\theta)\zeta/\theta+2\eta$.
Then
\begin{equation}\label{os:eq:strongcollisionbound}
 \TV((Y,G(X,Y),E(X,G(X,Y))),(Y,U_{d+p}))
 \le\beta+\frac12\sqrt{\theta+2\beta
           +2(1+\theta)D^2\gamma+DM/K}.
\end{equation}
The symmetric conclusion with $X$ retained follows by
exchanging the sources and using the right-tampering version
of (i).
\end{lemma}
\begin{proof}
Write $p_x=\Pr[X=x]$ and $q_y=\Pr[Y=y]$, so
$p_x,q_y\le1/K$. For a numerical seed $v\in\bits^d$,
partition the support of $X$ into hash fibers
$A_{v,a}=\{x:p_x>0,\ E(x,v)=a\}$, of masses
$w_{v,a}=\sum_{x\in A_{v,a}}p_x$. Retain the fibers whose
masses are at most $(1+\theta)/M$, and denote their union by
$A_v$. Write $r_v=\Pr[X\in A_v]$. If $r_v<1/2$, discard
the entire numerical seed $v$.

For an arbitrary probability distribution on $M$ atoms, the
mass of atoms larger than $(1+\theta)/M$ is at most
$(1+\theta)/\theta$ times its statistical distance from
uniform. The strong seeded-extractor guarantee, averaged over
numerical uniform seeds, therefore shows
\[
 D^{-1}\sum_v(1-r_v)\le(1+\theta)\zeta/\theta.
\]
Markov's inequality shows that the discarded numerical seeds
have uniform mass at most $2(1+\theta)\zeta/\theta$.
The two deletion events are measurable from $(X,v)$ and from
$v$, respectively. Hypothesis (ii), applied to the current
product distribution itself, transfers each one from a uniform $v$ to
$v=G(X,Y)$ at cost at most $\eta$.
Thus the distribution $P$ of $(Y,G(X,Y),E(X,G(X,Y)))$ has a surviving
subdistribution $\mu\le P$ of mass $t_0\ge1-\beta$.

We bound its \emph{conditional} collision energy, weighted by
the original $Y$-marginal:
\[
 \mathcal C_Y(\mu)
 =\sum_{y:q_y>0}\frac1{q_y}\sum_{v,a}\mu(y,v,a)^2.
\]
For a retained seed $v$ and a retained fiber $A_{v,a}$, put
\[
 e_{v,a}(y)=\sum_{x\in A_{v,a}}p_x
                         \mathbf1_{\{G(x,y)=v\}},
\]
and set $e_{v,a}(y)=0$ for discarded seeds or fibers.
Independence gives $\mu(y,v,a)=q_y e_{v,a}(y)$, and hence
\begin{equation}\label{os:eq:singleenergy}
 \mathcal C_Y(\mu)=\sum_yq_y\sum_{v,a}e_{v,a}(y)^2.
\end{equation}
The diagonal terms $x=x'$ contribute at most
$\sum_xp_x^2\le1/K$, since each raw pair $(x,y)$ has a
unique genuine seed.

For a retained numerical seed $v$, define the total weight of
ordered distinct pairs in a retained hash fiber by
\[
 N_v=\sum_{\substack{a:\ A_{v,a}\text{ retained}}}
          \sum_{\substack{x,x'\in A_{v,a}\\x\ne x'}}p_xp_{x'}.
\]
Because $r_v\ge1/2$ and there are at most $M$ hash values,
\[
 N_v\ge\frac{r_v^2}{M}-\sum_xp_x^2
       \ge\frac1{4M}-\frac1K\ge\frac1{8M},
 \qquad
 N_v\le\sum_{\substack{a:\ A_{v,a}\text{ retained}}}w_{v,a}^2
       \le\frac{1+\theta}{M}.
\]
Let $(X_v,X'_v)$ be such an ordered pair, sampled with
probability $p_xp_{x'}/N_v$. For $x$ in a retained fiber
$A_{v,a}$, its first marginal has point mass
\[
 \Pr[X_v=x]=\frac{p_x(w_{v,a}-p_x)}{N_v}
       \le\frac{8(1+\theta)}K\le\frac{16}K.
\]
By symmetry the same bound holds for $X'_v$; both marginals
have min-entropy at least $k-4$. The pair is sampled
independently of $Y$.

Represent the conditional distribution $X'_v\mid X_v$ by a random
fixed-point-free left map, sampled independently of $X_v$
and $Y$: at each point in the support of $X_v$, choose the
image according to this conditional distribution, and at other points
choose any different image. For each fixed map $f$, hypothesis
(i) implies
\[
 \Pr[G(X_v,Y)=v,G(f(X_v),Y)=v]
 \le D^{-1}\Pr[G(f(X_v),Y)=v]+\gamma.
\]
Average this inequality over the maps. Since the
\emph{averaged} marginal of $X'_v$ has entropy at least
$k-4$, ordinary extraction in hypothesis (i) then gives
\begin{equation}\label{os:eq:seedpair}
 \Pr[G(X_v,Y)=G(X'_v,Y)=v]
 \le D^{-2}+2\gamma.
\end{equation}
The second marginal need not have high entropy for each
\emph{fixed} map; ordinary extraction is applied only after
averaging. Independence from $Y$ holds throughout.

Expand the squares in~\eqref{os:eq:singleenergy}. For each
retained $v$, its off-diagonal contribution is $N_v$ times
the probability in~\eqref{os:eq:seedpair}. Set $N_v=0$ for
discarded seeds and sum over all $D$ numerical seeds:
\begin{align*}
 \mathcal C_Y(\mu)
 &\le\frac1K+\sum_vN_v(D^{-2}+2\gamma)\\
 &\le\frac1K+\frac{1+\theta}{DM}
                   +\frac{2(1+\theta)D\gamma}{M}.
\end{align*}
Weighted Cauchy--Schwarz against the probability measure
$P_Y\times U_{d+p}$, whose masses are $q_y/(DM)$, gives
\[
 \TV(\mu,t_0(P_Y\times U_{d+p}))
\le\frac12\sqrt{DM\mathcal C_Y(\mu)-t_0^2}.
\]
The distances from $P$ to $\mu$ and from the full ideal distribution
to $t_0$ times that ideal distribution cost a total of
$1-t_0\le\beta$. Finally,
\[
 t_0^2=1-2(1-t_0)+(1-t_0)^2\ge1-2\beta.
\]
Combining this inequality with the collision bound and
$DM/K=2^{d+p-k}$ proves~\eqref{os:eq:strongcollisionbound}.
The argument applies directly to the given source distributions and
retains the original $Y$-marginal exactly.
\end{proof}

\begin{lemma}[XOR of two separately source-strong payloads]
\label{os:lem:xorstrong}
Let $S=S(X,Y)$ be any $d$-bit function and let
$A=E_L(X,S)$, $B=E_R(Y,S)$ each have $p$ bits.
If $(Y,S,A)$ is $\xi_L$-close to $(Y,U_d,U_p)$ and
$(X,S,B)$ is $\xi_R$-close to $(X,U_d,U_p)$, then
$F=(S,A\oplus B)$ is separately source strong, with errors
$\xi_L$ upon revealing $Y$ and $\xi_R$ upon revealing $X$.
\end{lemma}
\begin{proof}
For fixed numerical $(y,s)$, XOR with $E_R(y,s)$ is a
permutation of $\bits^p$. Applying this deterministic
permutation to the honest and ideal distributions preserves their
distance, proving the first assertion. The second is the
same argument with $X$ retained and $E_L(X,S)$ fixed.
\end{proof}

\subsubsection{Theorem}

\begin{theorem}[Near-total output from two independent sources]\label{main:two}
There are absolute constants $C_0,C_1>0$ such that, for every fixed
$0<\delta\le1$, there are $b_\delta>0$ and $n_\delta$ with the following
property. For $n\ge n_\delta$, $\delta n\le k\le n$ integral, and
\[
 2^{-b_\delta n}\le\eps\le\tfrac14,
\]
there is a deterministic polynomial-time computable function
\[
 \Ext_{n,k,\delta,\eps}:\bits^n\times\bits^n\longrightarrow\bits^m
\]
such that, for independent $X,Y$ with $\Hin(X),\Hin(Y)\ge k$,
\[
 \TV(\Ext_{n,k,\delta,\eps}(X,Y),U_m)\le\eps,
 \qquad m\ge2k-C_0\log(1/\eps)-C_1.
\]
Its output has the form
\[
 \bigl(S,E_L(X,S),E_R(Y,S)\bigr),\qquad S=G(X,Y).
\]
The two seeded outputs have unequal lengths, and the seed $S$ is retained.
For each fixed $\delta$, construction and evaluation are polynomial in
$n+\log(1/\eps)$.
The $2k$-scale output is not source strong at that length, as shown by Lemma~\ref{pre:lem:strong-output-cap}; Corollary~\ref{os:thm:source-strong-optimal} separately gives $k-O(\log(1/\eps)+1)$ source-strong output.
\end{theorem}


\subsubsection{Detailed deterministic construction and seed parameters}

\begin{constructionbox}{Ordinary two-source output with additive entropy loss}{cb:total-two}
\cbingredients The one-sided sumset seed of Theorem~\ref{ns:thm:seed}
(Construction~\ref{cb:sumset-seed}); the strong seeded extractors of
Lemma~\ref{pre:seed-efficient}; entropy-slack strongification,
Lemma~\ref{pre:lem:strongify-general}. Lemma~\ref{os:lem:onepayloadstrong}
is the already proved collision interface used to analyze the first payload.
\cbsetup Fix $\delta>0$, integer $\delta n\le k\le n$, and the error
$\eps$ in Theorem~\ref{main:two}'s range. Put
$\sigma=\delta/4$, $s=\log(1/\eps)$, and $\zeta=\eps^4/2^{20}$.
Instantiate $G_{\Sigma,\sigma}$ from Theorem~\ref{ns:thm:seed} at
rate $\sigma$, with at least $c_\sigma n$ output bits.
With the absolute constants calibrated below, set
\[
 \begin{aligned}
 d&=\left\lceil C_1\seedcost{n}{s}\right\rceil,\\
 p&=\lfloor k-d-2s-6\rfloor,\qquad
 q=\lfloor k-A(s+1)\rfloor.
 \end{aligned}
\]
Instantiate $E_L,E_R$ from Lemma~\ref{pre:seed-efficient} at entropy
threshold $k$ and error $\zeta$, with a common $d$-bit seed after ignored
padding, and outputs truncated to $p$ and $q$ bits respectively.
The parameter verification below ensures $p,q\ge1$ and
$d\le c_\sigma n$. Here $q$ is a payload length, not a tampering count.
\cbinput The ordered pair $(x,y)\in\bits^n\times\bits^n$.
\cbsteps
\begin{constructionsteps}
\item Compute the internal seed
$S=\operatorname{prefix}_d G_{\Sigma,\sigma}(x+y)$.
\item Compute the $p$-bit left payload $A_L=E_L(x,S)$.
\item Using the \emph{same} seed, compute the $q$-bit right payload
$A_R=E_R(y,S)$.
\item Return the concatenation $(S,A_L,A_R)$, including all $d$ seed bits.
\end{constructionsteps}
\cboutput Exactly $m=d+p+q$ bits, with
$m\ge2k-(A+2)s-A-8$ and error at most $\eps$.
Only the seed is sum-compatible; the final evaluator reads $x,y$
separately and is not asserted source strong at this output length.
\cbproofonly Hash-fiber trimming, collision partners, and the
comparison experiment used to append the second payload. No additional
random seed is sampled during evaluation.
\end{constructionbox}

The independent ordinary constant-rate construction in
Section~\ref{un:ordinarylinear} is unchanged by this seed choice.
A shortened full two-sided one-tampering seed would satisfy the same
interface, but is not needed. The one-sided sumset seed supplies
non-malleability, its honest marginal supplies ordinary extraction,
and entropy slack supplies the required source strength.

\subsubsection{Appending the second payload}\label{os:sec:wrapper}
\begin{corollary}[Ordinary concatenation from one-payload recycling]
\label{os:thm:wrapper}
Assume Lemma~\ref{os:lem:onepayloadstrong} for the left payload of length
$p$, and let its distance bound be $\xi$. If $E_R$ is a strong seeded
extractor for $Y$ with $q$ output bits and error $\zeta_R$, then
\begin{equation}\label{os:eq:wrapper}
 \TV((S,E_L(X,S),E_R(Y,S)),U_{d+p+q})\le\xi+\zeta_R.
\end{equation}
\end{corollary}
\begin{proof}
Apply $(y,s,a)\mapsto(s,a,E_R(y,s))$ to the comparison
$(Y,S,E_L(X,S))\approx_\xi(Y,U_d,U_p)$. This gives
\[
 (S,E_L(X,S),E_R(Y,S))
 \approx_\xi(U_d,U_p,E_R(Y,U_d)).
\]
The independent seed on the right permits strong extraction from $Y$;
attach the independent $U_p$ and use the triangle inequality.
\end{proof}

\begin{corollary}[Common error calibration]\label{os:cor:errors}
For $0<\eps\le1/4$, the one-payload error is at most $\eps/4$
if
\[
 \theta=\eps^2/64,\quad \zeta\le\eps^4/2^{20},\quad
 \eta\le\eps^2/4096,\quad
 \gamma\le\eps^2/(512D^2),\quad DM/K\le\eps^2/64.
\]
If also $\zeta_R\le\eps^4/2^{20}$, the concatenated output has
error less than $\eps$. The bound on $DM/K$ implies $K\ge16M$.
\end{corollary}
\begin{proof}
Since $1+\theta<2$, the one-payload lemma has
$\beta\le\eps^2/1024$. Its radicand is at most
\[
 \eps^2(1/64+1/512+1/128+1/64)<\eps^2/16.
\]
Its distance is therefore at most $\eps^2/1024+\eps/8
<\eps/4$. Apply Corollary~\ref{os:thm:wrapper} for concatenation.
\end{proof}

\subsubsection{Final parameter substitution and entropy accounting}
\begin{corollary}[Short-output strongification]\label{os:lem:strongify}
If an ordinary $d$-bit two-source extractor works at component threshold
$h$ with error $\eta_0$, then at component entropy $k>h$ its error
strong in either original source is at most
$2^d(\eta_0+2^{h-k})$.
\end{corollary}
\begin{proof}
This is Lemma~\ref{pre:lem:strongify-general} with zero comparisons.
\end{proof}

Fix any $0<\delta\le1$ and an integer $\delta n\le k\le n$, and take
$G_{\Sigma,\sigma}$ from Theorem~\ref{ns:thm:seed} at the fixed rate
$\sigma=\delta/4$. Write $a=e_\sigma$, $c=c_\sigma$, and
$s=\log(1/\eps)$ for the exponential-error and output-rate constants
of this sum-compatible seed. Set
$G(x,y)=\operatorname{prefix}_dG_{\Sigma,\sigma}(x+y)$.
The seed theorem gives the required one-sided guarantee and
ordinary extraction with $\gamma\le2^{-an}$, with no additional loss.
For all sufficiently large $n$, $k-4\ge\sigma n$, as required for the
collision-partner sources. Corollary~\ref{os:lem:strongify} gives
\begin{equation}\label{os:eq:eta}
 \eta\le2^d(2^{-an}+2^{\sigma n-k})
 \le2^{d+1-a_*n},\qquad a_*:=\min\{a,\delta/2\}.
\end{equation}

Use Lemma~\ref{pre:seed-efficient} at error $\zeta=\eps^4/2^{20}$.
After changing absolute constants, its seed length is at most
$B\seedcost{n}{s}$ and its available output is at least $k-A(s+1)$.
Choose
\[
 d=\left\lceil C_1\seedcost{n}{s}\right\rceil
\]
with a sufficiently large absolute $C_1$. This dominates the actual seeded
extractor seed requirement, and we pad unused seed coordinates. Enlarge
$C_1$ also to ensure that the left output below is an admissible truncation.
Set
\begin{equation}\label{os:eq:outputs}
 p=\lfloor k-d-2s-6\rfloor,
 \qquad q=\lfloor k-A(s+1)\rfloor.
\end{equation}
Both lengths will be positive in the range now chosen. In particular,
$2^p\le2^k\eps^2/(64\cdot2^d)$, as required by
Corollary~\ref{os:cor:errors}.

It is enough to ensure
\begin{equation}\label{os:eq:budget-final}
 d\le cn,\qquad 2d+2s+14\le a_*n,\qquad p,q\ge1.
\end{equation}
Here $a_*,c$ are the already fixed positive constants of the
sum-compatible seed, not functions of $s$. The condition $d\le cn$ ensures that
the output of $G_{\Sigma,\sigma}$ is long enough for the specified prefix.
For $s\le b n$, monotonicity gives
\[
 \frac{\seedcost{n}{s}}n
 \le \frac{\log^2(n+2)}n+
       (b+1/n)\log\!\left(2+\frac1{b+1/n}\right).
\]
The right side tends to $b\log(2+1/b)$ as $n\to\infty$, and that expression
tends to zero as $b\downarrow0$. Choose a sufficiently small fixed
$b_\delta>0$ to satisfy every linear budget in \eqref{os:eq:budget-final},
including positivity of both output lengths. Then all budgets hold for
large $n$, uniformly for $2\le s\le b_\delta n$.

Equation~\eqref{os:eq:eta} and \eqref{os:eq:budget-final} imply the separate
strength requirement $\eta\le\eps^2/4096$ and the one-sided error
requirement $\gamma\le\eps^2/(512\cdot2^{2d})$. Apply the collision
wrapper and its calibration corollary. The output error is at most
$\eps$.

Finally, the retained seed cancels its own length charge:
\begin{equation}\label{os:eq:cancel}
 d+p+q\ge2k-(A+2)s-A-8.
\end{equation}
The constants $A+2$ and $A+8$ are absolute seeded-extractor constants.
This proves Theorem~\ref{main:two}. No $\log n$ term remains in the output loss.
The internal seed length is not claimed to be optimal. The one-sided sumset theorem supplies the seed. In
particular Theorem~\ref{main:two} still asserts ordinary extraction,
not non-malleability of its full long output.

\subsubsection{All-error interpretation and evaluation}\label{sec:allerrors}
For a fixed $\delta$, choose $C_\delta$ larger than the constants in
\eqref{os:eq:cancel} and larger than $2/b_\delta+2$. When
$s>b_\delta n$, the expression $2k-C_\delta(s+1)$ is nonpositive.
Returning the empty string is then exactly uniform and satisfies the
truncated all-error guarantee $m\ge\max\{0,\lfloor2k-C_\delta(s+1)\rfloor\}$.
Enlarge $C_\delta$ once more to cover the finite range $n<n_\delta$.
For $n\ge n_\delta$ and $s\le b_\delta n$, truncate the constructed
output, when necessary, to the claimed nonnegative length. Statistical
distance cannot increase under truncation.

For an arbitrary real accuracy request, it is enough to construct for
$2^{-\lceil\log(1/\eps)\rceil}$; the added entropy loss is an absolute
constant. Thus the algorithm can take the accuracy exponent as an integer.
Its seed evaluator consists of the fixed condenser, a constant number of
field powers and traces, and a fixed elementary symmetric polynomial.
Its second stages use Lemma~\ref{pre:seed-efficient}. All work is polynomial in
$n+\log(1/\eps)$ for fixed rate. Colorings, representative tables,
source pruning, and subsource selection occur only in the proof.

The final two-source wrapper is not asserted to preserve the seed's
$O(n)$ character rank or special inner-product form. That property is used
inside the seed construction, where its role has been proved explicitly.

\subsection{Source-strong non-malleable output with additive entropy loss}
\label{rec:sec:main}
Let $C\ge0$ be fixed. We use the $(2C+1)$-tampering extractor of
Theorem~\ref{nm:thm:mainconstant} at a smaller fixed entropy rate,
shorten it using Lemma~\ref{pre:lem:strongify-general}, and use the
strong seeded extractor of Lemma~\ref{pre:seed-efficient}, with its
absolute constants $A,B$. The case $C=0$ means an empty comparison
tuple; its short seed still uses one-comparison non-malleability.
The centered recycling lemma below is a standalone implication from
these seed and seeded-extractor interfaces.
\subsubsection{Prerequisite with proof: centered recycling under actual leakage}

We first prove a slightly more general statement allowing arbitrary
source-local payload leakage selected by the comparison seeds.
The large posterior distributions and collision graphs below
are proof objects; they are never computed by the evaluator.

\begin{lemma}[Centered recycling with actual leakage]\label{rec:lem:centered}
Let $X,Y$ be independent flat sources on sets
$\mathcal A,\mathcal B$ of size $K=2^k$. Put
\[
 D=2^d,\qquad M=2^p,\qquad L=DM=2^m,\qquad m=d+p.
\]
Suppose $k\ge h$ and $G:\bits^n\times\bits^n\to[D]$ is
non-malleable against up to $R=2C+1$ separate tamperings,
with error $\eta$, strong in either source separately, whenever
both source min-entropies are at least $h$.
Fix admissible $(f_i,g_i)_{i=1}^C$, and write
\[
 S_0=G(X,Y),\qquad
 \bar S=(G(f_i(X),g_i(Y)))_{i=1}^C.
\]
For each numerical $\bar s\in[D]^C$ let
$a_{\bar s}:\mathcal A\to[M]^C$ be arbitrary, and let
\[
 T=(\bar S,a_{\bar S}(X))
\]
be the actual leakage tuple. Let $0<\alpha,\beta<1$.
Suppose $E:\bits^n\times[D]\to[M]$ is a strong seeded
extractor of error $\zeta$ at entropy
\begin{equation}\label{rec:eq:kappa}
 \kappa=k-Cm-\log\frac1{\alpha\beta}.
\end{equation}
Then
\begin{equation}\label{rec:eq:centered-bound}
 \begin{split}
 &\TV\bigl((Y,T,S_0,E(X,S_0)),(Y,T,U_d,U_p)\bigr)\\
 &\quad\le
 2\beta+3\alpha+3\zeta+\frac{\eta}{2}
 +\frac12\sqrt{\Delta},
 \end{split}
\end{equation}
where
\begin{equation}\label{rec:eq:Delta}
 \boxed{\quad
 \Delta=(2C+1)\frac{L^{C+1}}K
       +\frac{2D^{C+2}L^C\,2^h}{\alpha^2K}
       +\frac{4\eta D^{2C+2}}{\alpha^2}.
 \quad}
\end{equation}
All variables $Y,T$ on the ideal side have their original
joint distribution.
\end{lemma}

\paragraph{Posteriors and an independently seeded reference}

Fix a numerical tail label $t=(\bar s,\bar a)$. Define
\[
 A_t=\{x\in\mathcal A:a_{\bar s}(x)=\bar a\},
 \qquad a_t=|A_t|.
\]
For each fixed $\bar s$ these sets partition $\mathcal A$, so
\begin{equation}\label{rec:eq:At-sum}
 \sum_t a_t=D^C K.
\end{equation}
For $x\in A_t$ put
\[
 w_t(x)=\Pr_Y[\bar S(x,Y)=\bar s],
 \qquad
 \lambda_t=\Pr[T=t],
 \qquad
 N_t(y)=|\{x\in A_t:\bar S(x,y)=\bar s\}|.
\]
Use $w_t(x)=0$ outside $A_t$. The basic identities are
\begin{equation}\label{rec:eq:posterior-identities}
 \sum_x w_t(x)=K\lambda_t,\qquad
 \Pr[X=x\mid T=t]=\frac{w_t(x)}{K\lambda_t},\qquad
 \Pr[Y=y,T=t]=\frac{N_t(y)}{K^2}.
\end{equation}
We omit zero-probability labels. Write $P_t$ for the posterior
distribution $X\mid T=t$.

Let $P$ be the actual distribution of $(Y,T,S_0,E(X,S_0))$.
Construct a reference distribution $\mathcal R$ by sampling the original
$(X,Y,T)$, then an independent uniform $S\in[D]$, and outputting
$(Y,T,S,E(X,S))$. Finally, let
\[
 Q(y,t,s,a)=\frac{\Pr[Y=y,T=t]}{DM}
           =\frac{N_t(y)}{K^2L}.
\]
Thus both $\mathcal R$ and $Q$ preserve the actual $(Y,T)$
marginal. The task is to compare $P$ first with $\mathcal R$.

Call $t$ good if
\begin{equation}\label{rec:eq:goodt}
 \lambda_t\ge\beta L^{-C}.
\end{equation}
Bad labels have total mass at most $\beta$. On a good label,\[
 \max_x P_t(x)\le(K\lambda_t)^{-1}
 \le 2^{-(k-Cm-\log(1/\beta))}.
\]
In particular, $P_t$ has min-entropy at least $\kappa$.
For a good $t$ define
\begin{equation}\label{rec:eq:psi}
 \psi_t(y)=\1[N_t(y)\ge\alpha K\lambda_t].
\end{equation}
The actual mass of pairs with $N_t(y)<\alpha K\lambda_t$
is at most $\alpha$:
\[
 \sum_t\sum_{y:N_t(y)<\alpha K\lambda_t}\frac{N_t(y)}{K^2}
 \le\sum_t\alpha\lambda_t=\alpha.
\]
For every surviving $(y,t)$, the conditional distribution
$X\mid(Y=y,T=t)$ is uniform on a set of size
\[
 N_t(y)\ge\alpha\beta K L^{-C}=2^\kappa.
\]
Strong seeded extraction therefore gives
\begin{equation}\label{rec:eq:R-Q}
 \TV(\mathcal R,Q)\le\zeta+\alpha+\beta.
\end{equation}
This step uses an independent seed and the actual tail fiber;
it does not apply $G$ to a conditioned two-source distribution.

\paragraph{Posterior trimming without a seeded-error multiplier}

Discard a raw pair $(X,T)$ if
\begin{equation}\label{rec:eq:w-cut}
 w_T(X)<\alpha D^{-C}.
\end{equation}
This costs at most $\alpha$. Indeed, for each fixed $x$ there
is only one possible $\bar a=a_{\bar s}(x)$ for each of the
$D^C$ numerical seed tuples $\bar s$. Hence
\[
 \frac1K\sum_x\sum_{t:w_t(x)<\alpha D^{-C}}w_t(x)\le\alpha.
\]
The estimate is for the actual $(X,T)$ distribution, which is shared by
$P$ and $\mathcal R$ before the honest seed is evaluated.

For good $t$ and numerical $s$ define the hash-fiber masses
\[
 r_{t,s}(a)=\Pr_{P_t}[E(X,s)=a].
\]
Discard points with $r_{t,s}(E(x,s))>2/M$. For any distribution
$r$ on $[M]$, the mass of its atoms exceeding $2/M$ is at most
$2\TV(r,U_M)$: on those atoms $r(a)-1/M>r(a)/2$.
Since $P_t$ has entropy at least $\kappa$, averaging over a
uniform $s$ shows that this deletion has probability at most
$2\zeta$ under $\mathcal R$, after summing over good $t$.

The same deletion under $P$ costs at most $2\zeta+\eta$.
To see this, use the source-strong seed comparison retaining $X$:
\[
 (X,\bar S,S_0)\approx_\eta(X,\bar S,U_d).
\]
Both $T=(\bar S,a_{\bar S}(X))$ and the heavy-fiber deletion
test are deterministic functions of the retained variables and
the honest seed. This is the only transfer needed for trimming.
In particular, no factor $D^C$ multiplies $\zeta$.

For good $t$ set
\begin{equation}\label{rec:eq:phi}
 \phi_{t,s}(x)=
 \1[x\in A_t]\,
 \1[w_t(x)\ge\alpha D^{-C}]\,
 \1[r_{t,s}(E(x,s))\le2/M].
\end{equation}
Let $\mu\le P$ and $\nu\le\mathcal R$ be the subdistributions
obtained by keeping good $t$, $\psi_t(y)=1$, and
$\phi_{t,s}(x)=1$, using the actual honest seed for $\mu$
and the independent seed for $\nu$. The mass losses satisfy
\begin{align}
 b_P:=1-\|\mu\|_1&\le\beta+2\alpha+2\zeta+\eta,\label{rec:eq:mu-mass}\\
 b_R:=1-\|\nu\|_1&\le\beta+2\alpha+2\zeta.\label{rec:eq:nu-mass}
\end{align}
The cutoffs defining $\phi_{t,s}$ are source local once $t,s$
are numerical. They use the posterior $P_t$, averaged over $Y$,
not the posterior conditioned on a particular $Y=y$.

\paragraph{The centered energy identity}

Put
\[
 h_s(x,y)=\1[G(x,y)=s]-D^{-1}.
\]
For good $t=(\bar s,\bar a)$ and $a\in[M]$ define
\[
B_{t,s,a}(y)=
 \sum_{\substack{x\in A_t\\E(x,s)=a}}
 \phi_{t,s}(x)\,
 \1[\bar S(x,y)=\bar s]\,h_s(x,y).
\]
Then, pointwise,
\[
 (\mu-\nu)(y,t,s,a)=K^{-2}\psi_t(y)B_{t,s,a}(y).
\]
Consequently,
\begin{equation}\label{rec:eq:centered-energy}
 \mathcal E:=
 \sum_{y,t,s,a}\frac{(\mu-\nu)(y,t,s,a)^2}{Q(y,t,s,a)}
 =\frac{L}{K^2}\sum_{t\ {\rm good}}\sum_{y:N_t(y)>0}
       \frac{\psi_t(y)}{N_t(y)}
       \sum_{s,a} B_{t,s,a}(y)^2.
\end{equation}
Zero-mass atoms are omitted. Weighted Cauchy--Schwarz gives
\begin{equation}\label{rec:eq:CS}
 \frac12\|\mu-\nu\|_1\le\frac12\sqrt{\mathcal E}.
\end{equation}
The reference is not uniform at this stage. Centering by
$D^{-1}$ is what makes the large-graph term vanish under
the non-malleable seed replacement.

\paragraph{Diagonal and cross-input exceptions}

Expand the squares in~\eqref{rec:eq:centered-energy} as ordered
pairs $(x,x')$. Call a pair exceptional if
\begin{equation}\label{rec:eq:exceptional}
 x=x',\qquad\text{or}\qquad
 x'=f_i(x),\qquad\text{or}\qquad
 x=f_i(x')\quad\text{for some }i.
\end{equation}
For every $x,x',y$,
\begin{equation}\label{rec:eq:centered-vector-bound}
 \sum_s|h_s(x,y)h_s(x',y)|
 \le
 \left(\sum_s h_s(x,y)^2\right)^{1/2}
 \left(\sum_s h_s(x',y)^2\right)^{1/2}
 =1-D^{-1}\le1.
\end{equation}
The hash equality and $\phi$ factors can only reduce this
absolute-value bound.

For each fixed $(y,t)$, the diagonal contributes at most
$N_t(y)$ ordered pairs. The relation $x'=f_i(x)$ also
contributes at most $N_t(y)$ pairs, by counting $x$;
the relation $x=f_i(x')$ contributes at most $N_t(y)$
by counting $x'$. This remains true for noninjective $f_i$.
There are at most $K L^C$ supported $(y,t)$.
Thus the exceptional contribution, bounded in absolute value,
is at most
\begin{equation}\label{rec:eq:exceptional-bound}
 \mathcal E_{\rm exc}
 \le(2C+1)\frac{L^{C+1}}K.
\end{equation}
The cross-input exceptions ensure that all subsequent
composed comparison maps are admissible.

\paragraph{Source-local collision graphs}

For each good $t$ and numerical $s$, form the symmetric directed
graph $\Gamma_{t,s}$ of ordered pairs $x,x'\in A_t$ satisfying
\[
 \phi_{t,s}(x)=\phi_{t,s}(x')=1,\qquad E(x,s)=E(x',s),
\]
with every exceptional pair in~\eqref{rec:eq:exceptional} removed.
Let $J_{t,s}=|\Gamma_{t,s}|$.

Every retained hash fiber has cardinality at most
\begin{equation}\label{rec:eq:degree-cap}
 B_t:=\frac{2D^C K\lambda_t}{\alpha M}.
\end{equation}
Indeed, each surviving $x$ has
$P_t(x)\ge\alpha/(D^C K\lambda_t)$, while the total $P_t$
mass of its retained fiber is at most $2/M$. Hence every
vertex of $\Gamma_{t,s}$ has degree at most $B_t$, and
\begin{equation}\label{rec:eq:edge-upper}
 J_{t,s}\le a_t B_t.
\end{equation}

Call a graph small if $J_{t,s}<2^h B_t$. On the retained
$(y,t)$,
$N_t(y)^{-1}\le(\alpha K\lambda_t)^{-1}$.
Using $|h_s h_s'|\le1$, the absolute contribution of one
small graph to~\eqref{rec:eq:centered-energy} is at most
\[
 \frac{LJ_{t,s}}{\alpha K^2\lambda_t}
 \le\frac{2D^{C+1}2^h}{\alpha^2 K}.
\]
There are at most $D L^C$ graph labels. Therefore
\begin{equation}\label{rec:eq:small-bound}
 \mathcal E_{\rm small}
 \le\frac{2D^{C+2}L^C\,2^h}{\alpha^2K}.
\end{equation}
No seed theorem is applied to these small graphs.

\paragraph{Large graphs and admissible extra comparisons}

For a large graph, sample $(V,V')$ uniformly from
$\Gamma_{t,s}$, independently of the original $Y$.
Its first marginal has maximum atom at most
$B_t/J_{t,s}\le2^{-h}$. The second marginal has the same
bound, although only the first is needed below.

Represent the conditional distribution $V'\mid V$ as a random
deterministic table $v\mapsto\rho(v)$, sampled independently
of $V,Y$. Each table chooses only neighbors of $v$ on the
support of $V$. For every fixed table the honest pair $(V,Y)$
has the following $2C+1$ comparisons:
\begin{equation}\label{rec:eq:extra-comparisons}
 (\rho(V),Y),\qquad
 (f_i(V),g_i(Y)),\qquad
 (f_i(\rho(V)),g_i(Y)),\quad 1\le i\le C.
\end{equation}
The first comparison is distinct because $V'\ne V$.
The middle comparisons satisfy the original admissibility
condition. The last comparisons have
$f_i(V')\ne V$ by~\eqref{rec:eq:exceptional}. All maps are
separate and are chosen independently of $Y$.

Apply strong non-malleability retaining $Y$, then average
over the tables. It replaces $G(V,Y)$ by an independent
uniform seed at distance at most $\eta$, while retaining
\[
 Y,\quad G(V',Y),\quad
 (G(f_i(V),g_i(Y)))_i,\quad
 (G(f_i(V'),g_i(Y)))_i.
\]
All the following factors except $h_s(V,Y)$ are functions of
these retained variables:
\[
 h_s(V',Y)\,
 \1[\bar S(V,Y)=\bar S(V',Y)=\bar s]\,
 \frac{\psi_t(Y)}{N_t(Y)}.
\]
Set the quotient to zero where $N_t(Y)=0$. Its absolute value
is at most $(\alpha K\lambda_t)^{-1}$. Under the independent
replacement, the expectation of $h_s(V,Y)$ is zero.
The standard bounded-test consequence of statistical distance
therefore gives
\begin{equation}\label{rec:eq:large-test}
 \left|\E\left[
 h_s(V,Y)h_s(V',Y)
 \1[\bar S(V,Y)=\bar S(V',Y)=\bar s]\,
 \frac{\psi_t(Y)}{N_t(Y)}
 \right]\right|
 \le\frac{2\eta}{\alpha K\lambda_t}.
\end{equation}
This is where the previous main-term difficulty disappears.
Neither the graph nor the random table is selected using $Y$.

Multiplying~\eqref{rec:eq:large-test} by $KJ_{t,s}$ restores
the sum over $y$ and ordered pairs. Hence
\[
 \mathcal E_{\rm large}
 \le\frac{2\eta L}{\alpha K^2}
       \sum_{t,s}\frac{J_{t,s}}{\lambda_t}.
\]
By~\eqref{rec:eq:At-sum}, \eqref{rec:eq:degree-cap},
and~\eqref{rec:eq:edge-upper},
\[
 \sum_{t,s}\frac{J_{t,s}}{\lambda_t}
 \le\sum_{t,s}\frac{a_t B_t}{\lambda_t}
 \le\frac{2D^{2C+1}K^2}{\alpha M}.
\]
It follows that
\begin{equation}\label{rec:eq:large-bound}
 \mathcal E_{\rm large}\le
       \frac{4\eta D^{2C+2}}{\alpha^2}.
\end{equation}
Only the exponentially accurate seed error incurs this
power of $D$.

\paragraph{Completion of the centered lemma}

Equations~\eqref{rec:eq:exceptional-bound},
\eqref{rec:eq:small-bound}, and~\eqref{rec:eq:large-bound} give
$\mathcal E\le\Delta$. Using half-$\ell_1$ distance also
for subdistributions, the triangle inequality yields
\begin{align*}
 \TV(P,Q)
 &\le \frac{b_P}{2}
       +\frac12\|\mu-\nu\|_1
       +\frac{b_R}{2}
       +\TV(\mathcal R,Q)\\
 &\le
 2\beta+3\alpha+3\zeta+\frac{\eta}{2}
       +\frac12\sqrt{\Delta}.
\end{align*}
This proves Lemma~\ref{rec:lem:centered}. The final ideal distribution
retains the original comparison marginal. Every deletion
used in the intermediate subdistributions is included in
the displayed error.

\subsubsection{Main theorem and complete retained-seed XOR evaluator}
\begin{theorem}[Source-strong endpoint]\label{rec:thm:main}
For every fixed $\delta>0$ and integer $C\ge0$
there is $b_{\delta,C}>0$ such that, for sufficiently large $n$,
integer $\delta n\le k\le n$, and
\[
 2^{-b_{\delta,C}n}\le\eps\le1/4,
\]
there is an explicit polynomial-time map $F:\bits^n\times
\bits^n\to\bits^m$ with
\[
 m\ge \frac{k}{C+1}-O(\log(1/\eps)+1)
\]
such that, for every independent min-entropy-$k$ pair and every
admissible $C$-tuple,
\begin{align}
 (F(X,Y),(F(f_i(X),g_i(Y)))_{i=1}^C,Y)
 &\approx_\eps (U_m,(F(f_i(X),g_i(Y)))_{i=1}^C,Y),\label{rec:eq:strongY}\\
 (F(X,Y),(F(f_i(X),g_i(Y)))_{i=1}^C,X)
 &\approx_\eps (U_m,(F(f_i(X),g_i(Y)))_{i=1}^C,X).\label{rec:eq:strongX}
\end{align}
The uniform variable is independent of everything else displayed.
The same function satisfies both comparisons. The additive
output-loss constants can be taken absolute; the error-range and
running-time constants may depend on $\delta,C$.
\end{theorem}

\begin{constructionbox}{Source-strong $C$-tampering retained-seed XOR}{cb:strong-total}
\cbingredients The fixed-comparison extractor of
Theorem~\ref{nm:thm:mainconstant} (Construction~\ref{cb:outer-nm});
Lemma~\ref{pre:lem:strongify-general}; the strong seeded extractor of
Lemma~\ref{pre:seed-efficient}; and the centered recycling interface
of Lemma~\ref{rec:lem:centered}.
\cbsetup Fix $\delta>0$, $C\ge0$, integer $\delta n\le k\le n$,
and target $\eps$ in Theorem~\ref{rec:thm:main}'s range.
First put $q=C+1$, $R=2C+1$, and $\sigma=\delta/(8q)$.
Instantiate a base extractor $G_0$ from Theorem~\ref{nm:thm:mainconstant}
with \emph{$R$ comparisons} at rate $\sigma$. Set
\[
 h_0=\left\lceil\frac{\delta n}{8q}\right\rceil,\quad
 h=\left\lceil\frac{\delta n}{4q}\right\rceil,\quad
 u=\lceil\log(1/\eps)\rceil+10,\quad r_C=\lceil\log(2C+1)\rceil.
\]
Using the constants $A,B$ of Lemma~\ref{pre:seed-efficient}, with $A$
a positive integer, choose
\[
 \begin{aligned}
 d&=\left\lceil B\seedcost{n}{u}\right\rceil,\\
 m&=\left\lfloor\frac{k-(A+4)(u+1)-r_C}{q}\right\rfloor,\\
 p&=m-d,\qquad \kappa=k-Cm-2u.
 \end{aligned}
\]
Define $G=\operatorname{prefix}_d G_0$. Instantiate $E$ at entropy
threshold $\kappa$ and error $2^{-u}$, pad its seed interface to $d$ bits,
and truncate its output to $p$ bits. The bounds
\eqref{rec:eq:param1}--\eqref{rec:eq:param3} certify that these calls fit.
Strongification from threshold $h_0$ to $h$ changes only the guarantee,
not the evaluation of $G$.
\cbinput $(x,y)\in\bits^n\times\bits^n$.
\cbsteps
\begin{constructionsteps}
\item Compute $S=G(x,y)$.
\item Compute $A_L=E(x,S)$ and $A_R=E(y,S)$, both of length $p$.
\item Return the retained seed concatenated with the bitwise XOR:
\begin{equation}\label{rec:eq:evaluator}
 S=G(x,y),\qquad F(x,y)=\bigl(S,E(x,S)\oplus E(y,S)\bigr).
\end{equation}
\end{constructionsteps}
\cboutput Exactly $m=d+p$ bits. Theorem~\ref{rec:thm:main} gives error
at most $\eps$, $m\ge k/(C+1)-O(\log(1/\eps)+1)$, and strength in
either source separately for this same function.
\cbproofonly The $2C+1$ auxiliary tamperings, actual-leakage posteriors,
collision graphs, and source-dependent bijections. The evaluator makes
one seed call and two seeded-extractor calls, not tampered-input calls.
\end{constructionbox}

For nonintegral $k\ge\delta n$, put $k'=\lfloor k\rfloor$ and
invoke the theorem with rate parameter $\delta/2$, not $\delta$:
for $n\ge2/\delta$, one has $k'\ge(\delta/2)n$. Use its error-range
constant $b_{\delta/2,C}$ and length cutoff. The output bound is
$k'/(C+1)-O(\log(1/\eps)+1)$, so the rounding costs at most
$1/(C+1)$ additional bit and is absorbed in the stated loss.

\subsubsection{Returning to the symmetric evaluator}

Let $F$ be~\eqref{rec:eq:evaluator}. Suppose $Y=y$ is retained.
For every output, honest or tampered, its seed is included
in the output. Therefore the transformations
\[
 (s,z)\longmapsto(s,z\oplus E(y,s))
\]
for the honest output and
\[
 (s_i,z_i)\longmapsto(s_i,z_i\oplus E(g_i(y),s_i))
\]
for comparison $i$ are bijections determined by the retained
$y$ and the displayed seeds. They transform the honest output
to $(S_0,E(X,S_0))$ and the comparison tuple to
\[
 T=\bigl(\bar S,(E(f_i(X),S_i))_{i=1}^C\bigr).
\]
This is precisely the leakage in Lemma~\ref{rec:lem:centered},
with $a_{\bar s}(x)=(E(f_i(x),s_i))_i$.
The honest uniform distribution is preserved by the same
bijection. Thus the lemma proves~\eqref{rec:eq:strongY}
with its displayed error.

To prove~\eqref{rec:eq:strongX}, exchange the two sources and
their tampering maps. Apply the same lemma to
$G^{\rm op}(x,y)=G(y,x)$, using the other source-strength
guarantee of $G$. The evaluator remains exactly
\eqref{rec:eq:evaluator}; only its analysis is exchanged.

For general independent sources of min-entropy at least
integer $k$, decompose each into a convex combination of
flat sources on $2^k$ points. Apply the bound to every product
component with the same evaluator and maps, then average.
The ideal distribution retains each component's actual source and
comparison tuple, so averaging gives their original joint
distribution. Admissibility persists on every component of positive
weight.

\subsubsection{Parameter calculation}

\paragraph{Obtaining the short source-strong seed}

Use the atomwise entropy-slack strongification in Lemma~\ref{pre:lem:strongify-general}.

Fix $q=C+1$, $R=2C+1$, and choose Theorem~\ref{nm:thm:mainconstant} at rate
$\sigma=\delta/(8q)$. Write its output and error constants
as $c_0,a_0>0$, and put
\[
 h_0=\left\lceil\frac{\delta n}{8q}\right\rceil,\qquad
 h=\left\lceil\frac{\delta n}{4q}\right\rceil,\qquad
 a=\min\left\{a_0,\frac{\delta}{8q}\right\}.
\]
Since $R+1=2q$, Lemma~\ref{pre:lem:strongify-general} gives, for
$d\le c_0n$,
\begin{equation}\label{rec:eq:eta-calibration}
 \eta\le D^{2q}(2^{-a_0n}+2^{h_0-h})
       \le 2^{\,2+2qd-an}.
\end{equation}
The same $G$ works at threshold $h$ for all comparisons
needed in the centered lemma.

\paragraph{Payload length and all three error terms}

Set
\[
 s=\lceil\log(1/\eps)\rceil,\qquad u=s+10,\qquad
 \alpha=\beta=\zeta=2^{-u}.
\]
Enlarge $A$ to a positive integer and put
\[
 r_C=\lceil\log(2C+1)\rceil,\qquad
 m=\left\lfloor\frac{k-(A+4)(u+1)-r_C}{q}\right\rfloor,
 \qquad p=m-d.
\]
Take $d$ to be the ceiling of the bound in~\eqref{pre:eq:seed-efficient}, with $s=u$.
Unused seeded-extractor seed bits can be ignored.
Use $E$ at threshold
\[
 \kappa=k-Cm-2u.
\]
Its output is available because
\begin{align*}
 \kappa-p
 &=k-qm+d-2u\\
 &\ge(A+4)(u+1)+r_C+d-2u
 \ge A(u+1).
\end{align*}
Choose the positive error-range constant $b_{\delta,C}$
sufficiently small, then $n$ sufficiently large, to ensure
\begin{align}
 d&\le c_0n,\qquad p\ge1,\label{rec:eq:param1}\\
 4qd+4u+4&\le an,\label{rec:eq:param2}\\
 k-Cm-(C+2)d-h&\ge4u+1.\label{rec:eq:param3}
\end{align}
These choices are simultaneous. Indeed, for
$u\le b_{\delta,C}n+O(1)$, the seed-length bound divided by
$n$ is at most
\[
 B\frac{\log^2(n+2)}n+
 B\left(b_{\delta,C}+O(1/n)\right)
 \log\left(2+\frac1{b_{\delta,C}+O(1/n)}\right)+O(1/n),
\]
which can be made arbitrarily small. Also,
$k-Cm\ge k/q\ge\delta n/q$ and
$h\le\delta n/(4q)+1$, leaving a fixed positive linear
margin in~\eqref{rec:eq:param3}. Finally $m/n$ stays bounded
away from zero when $b_{\delta,C}$ is sufficiently small.

The three terms in~\eqref{rec:eq:Delta} are each at most
$2^{-2u}$. For the first,
\[
 (2C+1)2^{qm-k}
 \le2^{-(A+4)(u+1)}\le2^{-2u}.\]
For the second,~\eqref{rec:eq:param3} gives
\[
 2^{1+2u+(C+2)d+Cm+h-k}\le2^{-2u}.
\]
For the third,~\eqref{rec:eq:eta-calibration} and
\eqref{rec:eq:param2} imply
\[
 \eta\le2^{-4u-2qd-2},
 \qquad
 4\eta\,2^{2u+2qd}\le2^{-2u}.
\]
Consequently Lemma~\ref{rec:lem:centered} has error at most
\[
 8\cdot2^{-u}+\frac{\eta}{2}
       +\frac{\sqrt3}{2}\,2^{-u}
 <10\cdot2^{-u}<\eps.
\]
The symmetric argument has the identical bound, proving
Theorem~\ref{rec:thm:main}.

Since $r_C\le C+1=q$, the output formula gives
\[
 m\ge\frac{k}{C+1}
       -\frac{(A+4)(u+1)}{C+1}-2
 =\frac{k}{C+1}-O(\log(1/\eps)+1),
\]
with absolute additive-loss constants. Evaluation calls
$G$ once and $E$ twice, hence is polynomial time.

\subsection{Corollary: ordinary source-strong output with additive entropy loss}
\begin{corollary}[The $C=0$ specialization]
\label{os:thm:source-strong-optimal}
For every fixed $\delta>0$, there is $b_{\delta,0}>0$ such that,
for sufficiently large $n$, integer $\delta n\le k\le n$, and
$2^{-b_{\delta,0}n}\le\eps\le1/4$, an explicit deterministic
polynomial-time two-source extractor has output
\[
 m\ge k-O(\log(1/\eps)+1)
\]
and is strong in either source separately:
\[
 (F(X,Y),X)\approx_\eps(U_m,X),\qquad
 (F(X,Y),Y)\approx_\eps(U_m,Y).
\]
The additive-loss constants are absolute, and the same function
satisfies both comparisons.
\end{corollary}

\begin{constructionbox}{Ordinary source-strong additive-loss specialization}{cb:strong-total-zero}
\cbingredients Construction~\ref{cb:strong-total} and
Theorem~\ref{rec:thm:main}, specialized to $C=0$.
\cbsetup Use the same rate and error inputs, but set
$C=0$, $q=1$, $R=1$, and $r_C=0$. Thus
$u=\lceil\log(1/\eps)\rceil+10$,
$d=\lceil B\seedcost{n}{u}\rceil$,
$m=\lfloor k-(A+4)(u+1)\rfloor$, $p=m-d$, and
$\kappa=k-2u$. The internal seed still uses \emph{one-comparison}
non-malleability.
\cbinput $(x,y)\in\bits^n\times\bits^n$.
\cbsteps
\begin{constructionsteps}
\item Compute the $d$-bit seed $S$ by Construction~\ref{cb:strong-total}
with these specialized public parameters.
\item Return $(S,E(x,S)\oplus E(y,S))$ using its $p$-bit payloads.
\end{constructionsteps}
\cboutput $m=d+p\ge k-O(\log(1/\eps)+1)$ bits, strong in either
source separately, as in Corollary~\ref{os:thm:source-strong-optimal}.
This is not a new amplification routine; the final comparison tuple
is empty, but the internal non-malleable seed is still needed.
\end{constructionbox}

\begin{proof}
Set $C=0$ in Theorem~\ref{rec:thm:main}. Then $q=1$, $R=1$,
and $r_C=\lceil\log(2C+1)\rceil=0$. The comparison tuple $T$
is empty; in the centered lemma it has one value, with
$\lambda_t=w_t(x)=1$ and $N_t(y)=K$. The exceptional pairs
are precisely the diagonal. All other steps and parameter inequalities
remain valid, and the energy bound specializes to
\[
 \Delta=\frac LK+\frac{2D^2\,2^h}{\alpha^2K}
                   +\frac{4\eta D^2}{\alpha^2}.
\]
The same evaluator retains $S$ and XORs its two seeded payloads,
with $m=\lfloor k-(A+4)(u+1)\rfloor$ and
$u=\lceil\log(1/\eps)\rceil+10$. This proves both strong
comparisons with the asserted loss.
\end{proof}

\subsection{Sumset output with additive entropy loss}\label{un:sumsetoptimal}
\subsubsection{Prerequisite: linear lossless condenser}

\begin{theorem}[Strong linear lossless condenser]\label{ns:in:lossless}
\textnormal{\emph{Reference.} \cite[Definition~4.6 and Lemma~4.7]{ChattopadhyayGL21}; see also~\cite{GUV,CI}.}
For every fixed $\alpha>0$, $k\le n$, and $0<\zeta<1/2$, there is an explicit strong linear lossless condenser $C:\F_2^n\times\F_2^{d_C}\to\F_2^v$ with error $\zeta$ and
\[
 d_C\le(1+1/\alpha)\log(nk/\zeta)+O_\alpha(1),\qquad
 v\le(1+\alpha)k+d_C+O_\alpha(1).
\]
Thus, conditional on an independent uniform seed, the output is on average $\zeta$-close to a source with min-entropy at least $k$. For fixed seed the map is binary-linear. Evaluation is polynomial-time in $n+\log(1/\zeta)$ for fixed $\alpha$.
\end{theorem}
This is the strong linear formulation in \cite[Definition 4.6 and Lemma 4.7]{ChattopadhyayGL21}, recording the GUV condenser and its linear implementation.

\subsubsection{Prerequisite: sum-compatible recycling and its proof}\label{ns:sec:recycling}
Write $K=2^k$, $D=2^d$, and $M=2^p$ for alphabet sizes.

\begin{lemma}[Entropy slack gives strength in a summand]\label{ns:lem:strongify}
If an ordinary $d$-bit sumset extractor $G$ has component threshold $h$
and error $\gamma$, then for independent $X,Y$ with
$\Hin(X)\ge k>h$, $\Hin(Y)\ge h$,
\[
 \TV((X,G(X+Y)),(X,U_d))\le2^d(\gamma+2^{h-k}).
\]
\end{lemma}
\begin{proof}
For each output value $u$, let
$B_u=\{x:\Pr_Y[G(x+Y)=u]>2^{-d}+\gamma\}$.
If $\Pr[X\in B_u]\ge2^{h-k}$, conditioning on $B_u$ preserves left
entropy $h$ and independence of $Y$, contradicting extraction at the
single atom $u$. Hence each $B_u$ has mass below $2^{h-k}$.
Outside $B_u$ the positive atom excess is at most $\gamma$, and
inside it at most one. Sum over the $2^d$ output atoms.
This is the one-sided entropy-slack argument underlying
Lemma~\ref{pre:lem:strongify-general}.
\end{proof}

\begin{corollary}[Retained-seed sumset recycling]\label{ns:thm:recycle}
Let $H_s:\F_2^n\to\F_2^p$ be linear for each $s$ and a strong
seeded $(k,\zeta)$ extractor. Let $G$ be strong in the first summand
at entropy $k$ with error $\eta$, and a left one-sided non-malleable
sumset seed at threshold $k-4$ with error $\gamma$, with the same
ordinary extraction guarantee. If $K\ge16M$, $0<\theta<1$, and
$\beta=3(1+\theta)\zeta/\theta+2\eta<1$, then for $S=G(X+Y)$,
\begin{equation}\label{ns:eq:recycle}
 \TV((Y,S,H_S(X+Y)),(Y,U_{d+p}))
 \le\beta+\tfrac12\sqrt{\theta+2\beta+
                     2(1+\theta)D^2\gamma+DM/K}.
\end{equation}
The ordinary sumset conclusion follows by discarding $Y$.
\end{corollary}
\begin{proof}
Apply Lemma~\ref{os:lem:onepayloadstrong} to
$\widetilde G(x,y)=G(x+y)$ and $E(x,s)=H_s(x)$.
It compares $(Y,S,H_S(X))$ with $(Y,U_d,U_p)$ at the displayed error.
Apply $(y,s,a)\mapsto(y,s,a+H_s(y))$. The real payload becomes
$H_s(X+Y)$ by linearity; a fixed translation preserves uniformity of
$U_p$ conditional on every $(y,s)$. No conditioning on $S$ is needed.
\end{proof}

\begin{corollary}[Sumset calibration]\label{ns:cor:calibration}
For $0<\eps\le1/4$, the bound in~\eqref{ns:eq:recycle} is at most
$\eps$ under the hypotheses of Corollary~\ref{os:cor:errors},
in particular when $DM/K\le\eps^2/256$ and its other parameters
are unchanged.
\end{corollary}
\begin{proof}
Apply the common calibration (the smaller diagonal term only helps).
\end{proof}

\subsubsection{Prerequisite: polynomial shear and seed-fit linear hashing}\label{ns:sec:hash}
A generic nonlinear strong seeded extractor cannot be substituted in Corollary~\ref{ns:thm:recycle}: its proof uses $H_u(x+y)=H_ux+H_uy$. A full-input universal hash, although linear, need not fit in the available sumset seed. The following elementary family gives the parameter regime we need after lossless condenser.

\begin{lemma}[Polynomial shear hashing]\label{ns:lem:shear}
Let $E=\F_{2^t}$ and $b\ge1$. For $u\in E$, define
\[
 \mathsf{Sh}_u(z_0,\ldots,z_b)
       =(z_1+u z_0,z_2+u^2z_0,\ldots,z_b+u^b z_0).
\]
For every distribution $Z\in E^{b+1}$ of min-entropy at least $k$, and an independent uniform $U\in E$,
\begin{equation}\label{ns:eq:shear}
 \TV((U,\mathsf{Sh}_U(Z)),U_{(b+1)t})
       \le\tfrac12\sqrt{b\,2^{bt-k}}.
\end{equation}
Every fixed-seed map is binary-linear and surjective. The seed length is $t$, not $(b+1)t$.
\end{lemma}
\begin{proof}
Use the nondegenerate trace pairing to index characters of $E^{b+1}$. A nonzero output character $\lambda=(\lambda_1,\ldots,\lambda_b)$ pulls back to
\[
 \bigl(\textstyle\sum_{i=1}^b\lambda_i u^i,\lambda_1,\ldots,\lambda_b\bigr).
\]
For fixed $\lambda\ne0$ and fixed first coefficient $\eta$, the equation $\sum_i\lambda_i u^i=\eta$ has at most $b$ roots, since its left side minus $\eta$ is a nonconstant polynomial of degree at most $b$. If $\widehat\mu(\xi)=\E\chi_\xi(Z)$, then
\[
 \frac1{|E|}\sum_u\sum_{\lambda\ne0}
 \left|\widehat\mu\left(\sum_i\lambda_i u^i,\lambda\right)\right|^2
 \le\frac b{|E|}\sum_\xi|\widehat\mu(\xi)|^2
 \le b\,2^{bt-k}.
\]
The last step is Parseval, $\sum_\xi|\widehat\mu(\xi)|^2=2^{(b+1)t}\cp(Z)$, and $\cp(Z)\le2^{-k}$. Fourier Cauchy--Schwarz bounds the distance for each fixed seed by half the square root of its nonzero-character sum. Average over seeds and use Jensen. Linearity is immediate; surjectivity follows by setting $z_0=0$.
\end{proof}

\begin{lemma}[Seed-fit linear extraction]\label{ns:lem:seedfit}
Fix $0<\delta\le1$ and $0<a<\delta/8$. There exists $b_{\delta,a}>0$ such that, for all sufficiently large $n$, all integer $\delta n\le k\le n$, and all integers $2\le s\le b_{\delta,a}n$, put
\[
 d=\lfloor an\rfloor,\qquad p=k-d-2s-8,\qquad \zeta=2^{-4s-20}.
\]
There is an explicit strong linear $(k,\zeta)$ seeded extractor with exactly $d$ seed bits and $p$ output bits. The constants and running time may depend on $\delta,a$, but $d+p=k-2s-8$ exactly.
\end{lemma}

\begin{constructionbox}{Seed-fit binary-linear payload extractor}{cb:linear-hash}
\cbingredients The strong linear lossless condenser of
Theorem~\ref{ns:in:lossless} and polynomial shear hashing,
Lemma~\ref{ns:lem:shear}.
\cbsetup Fix $\delta>0$, $0<a<\delta/8$, integer $\delta n\le k\le n$,
and integer $s$ in Lemma~\ref{ns:lem:seedfit}'s range. Set
$d=\lfloor an\rfloor$, $p=k-d-2s-8$, and $\zeta=2^{-4s-20}$.
Choose a fixed integer $b\ge\max\{2,16/a\}$ and $\alpha=1/(8b)$.
Instantiate the linear condenser $\mathcal C$ at entropy $k$,
parameter $\alpha$, and error $\zeta/2$; let its seed and output lengths
be $d_C,v$. Put
\[
 w=4s+21,\quad \Lambda=2w+\lceil\log b\rceil+2,\quad
 t=\left\lfloor\frac{k-\Lambda}{b}\right\rfloor,\quad E=\F_{2^t}.
\]
The proof below verifies $v\le(b+1)t$, $d_C+t\le d$, and $p\le bt$.
Fix the field's binary basis.
\cbinput $z\in\F_2^n$ and a $d$-bit numerical seed $S$.
\cbsteps
\begin{constructionsteps}
\item Parse the first $d_C$ seed bits as $S_C$, the next $t$ bits as
$U\in E$, and ignore the remaining $d-d_C-t$ bits.
\item Compute $\mathcal C(z,S_C)$, append zeros to reach $(b+1)t$ bits,
and parse the result as $(z_0,z_1,\ldots,z_b)\in E^{b+1}$.
\item Compute $(z_1+Uz_0,z_2+U^2z_0,\ldots,z_b+U^bz_0)$,
write the result as $bt$ bits, and return its first $p$ bits.
\end{constructionsteps}
\cboutput $H_S(z)\in\F_2^p$ with exactly $d$ supplied seed bits.
For every fixed $S$, the map in $z$ is binary-linear.
Lemma~\ref{ns:lem:seedfit} gives error $\zeta$ for an independent uniform
seed. In the sumset wrapper below $S$ is generated internally, and
Corollary~\ref{ns:thm:recycle}, not an independence assumption, justifies
that use.
\end{constructionbox}

\begin{proof}
Choose a fixed integer $b\ge\max\{2,16/a\}$ and put $\alpha=1/(8b)$. Apply Theorem~\ref{ns:in:lossless} with error $\zeta/2$. Write its seed length and output length as $d_C,v$. There is a constant $A_b$ such that
\[
 d_C\le A_b(\log(n+2)+s+1),\qquad
 v\le(1+\alpha)k+d_C+A_b.
\]
Define the integers
\[
 w=\log(2/\zeta)=4s+21,\qquad
 \Lambda=2w+\lceil\log b\rceil+2,
 \qquad t=\left\lfloor\frac{k-\Lambda}{b}\right\rfloor.
\]
Choose the positive constant $b_{\delta,a}$ sufficiently small. Since $\log n=o(n)$, uniformly over the specified $k,s$ and all sufficiently large $n$ we can ensure
\begin{align}
 d_C&\le an/4,\label{ns:eq:seedfit1}\\
 d_C+2\Lambda+A_b+b+1&\le k/(2b),\label{ns:eq:seedfit2}\\
 d+2s+8&\ge\Lambda+b,\qquad p\ge1.\label{ns:eq:seedfit3}
\end{align}
To make this explicit as a choice order, first fix $b,\alpha,A_b$, then choose $b_{\delta,a}$ so that the coefficients of $s/n$ in these three inequalities fit with a strict margin, and finally increase the length threshold to absorb the logarithmic and constant terms. All required right-side margins are positive multiples of $n$; no parameter depends on the input source.

The native hash input length fits because
\begin{align*}
 (b+1)t-v
 &\ge(1/b-\alpha)k-(1+1/b)\Lambda-(b+1)-d_C-A_b\\
 &\ge7k/(8b)-k/(2b)>0.
\end{align*}
Pad the condenser output to $(b+1)t$ bits with zeros and apply Lemma~\ref{ns:lem:shear} with an independent $t$-bit seed. Conditional on the condenser seed, its comparison source has min-entropy at least $k$. Since $bt\le k-\Lambda$, the shear error is at most
\[
 \tfrac12\sqrt b\,2^{-\Lambda/2}\le\zeta/8.
\]
The total error is less than $\zeta$. The combined map is linear for every fixed pair of seeds.

Its actual seed length obeys
\[
 d_C+t\le an/4+n/b\le5an/16\le\lfloor an\rfloor
\]
for large $n$. Pad with ignored seed bits to reach exactly $d$. Its available output satisfies $bt\ge k-\Lambda-b\ge p$ by~\eqref{ns:eq:seedfit3}; truncate to $p$ bits. All maps are explicit and their field operations and condenser evaluation are polynomial-time for fixed $b,\alpha$.
\end{proof}

\begin{remark}[Why a linear seed is an advantage here]
For a constant error, the construction may retain $d=\Theta_\delta(n)$ seed bits and produce only $k-d-O(1)$ hash bits. This is not a loss: the seed itself is included in the output. No claim of an optimal $O(\log n+\log(1/\eps))$ linear seed is used. Keeping $d$ fixed at a small linear fraction also gives enough room for the condenser and hash at every error exponent up to a sufficiently small linear fraction.
\end{remark}

\subsubsection{Theorem}

\begin{theorem}[Near-optimal sumset output]\label{ns:thm:main}
For every fixed $0<\delta\le1$ there are $b_\delta>0$ and $n_\delta$ such that, whenever $n\ge n_\delta$, $\delta n\le k\le n$ is an integer, and
\[
 2^{-b_\delta n}\le\eps\le\tfrac14,
\]
there is an explicit deterministic polynomial-time map
\[
 \SumExt_{n,k,\eps}:\F_2^n\longrightarrow\F_2^m,
 \qquad m=k-2\lceil\log(1/\eps)\rceil-8,
\]
whose output on $X+Y$ is $\eps$-close to uniform for all independent $X,Y$ with $\Hin(X),\Hin(Y)\ge k$. In particular,
\[
 m\ge k-2\log(1/\eps)-10.
\]
The constants in the running time and the nontrivial error range may depend on $\delta$, but the displayed output-loss constants do not.
\end{theorem}


\subsubsection{Complete sum-only retained-seed construction}

\begin{constructionbox}{Sumset output with additive entropy loss}{cb:sum-total}
\cbingredients The one-sided non-malleable sumset seed of
Theorem~\ref{ns:thm:seed} (Construction~\ref{cb:sumset-seed});
Lemma~\ref{ns:lem:seedfit} (Construction~\ref{cb:linear-hash});
and the retained-seed sumset recycling interface of
Corollary~\ref{ns:thm:recycle}.
\cbsetup Fix $\delta>0$, integer $\delta n\le k\le n$, and error
$\eps$ in Theorem~\ref{ns:thm:main}'s range.
Instantiate $G_0$ from Theorem~\ref{ns:thm:seed} at rate $\delta/8$,
with output and error constants $c,e>0$.
Fix $0<a<\min\{c/4,e/16,\delta/64,1/64\}$ and put
\[
 s=\lceil\log(1/\eps)\rceil,\qquad
 d=\lfloor an\rfloor,\qquad p=k-d-2s-8.
\]
Set $G=\operatorname{prefix}_dG_0$ and instantiate
Construction~\ref{cb:linear-hash} with these $n,k,\delta,a,s$.
Its $d$-bit seed interface supplies $p$ payload bits and error
$2^{-4s-20}$ under an independent uniform seed.
\cbinput Only $z\in\F_2^n$.
\cbsteps
\begin{constructionsteps}
\item Compute the $d$-bit seed $S=G(z)$.
\item Compute $H_S(z)$ by Construction~\ref{cb:linear-hash}, including
its seed split, condenser, polynomial shear, and final truncation.
\item Return
\begin{equation}\label{ns:eq:finalform}
 \SumExt(z)=(S,H_S(z)),\qquad S=G(z).
\end{equation}
\end{constructionsteps}
\cboutput Exactly $d+p=k-2\lceil\log(1/\eps)\rceil-8$ bits, with
error at most $\eps$ as in Theorem~\ref{ns:thm:main}.
The evaluator never receives either summand. Linearity of $H_S$ is
with respect to $z$ for each \emph{fixed} seed, not a claim that the
whole seed-dependent evaluator is linear.
\cbproofonly Summands, the shift by $H_S(y)$, collision partners,
and the independently seeded comparison experiment.
\end{constructionbox}

\subsubsection{Final parameters and proof of the main theorem}\label{ns:sec:final}
\begin{proof}[Proof of Theorem~\ref{ns:thm:main}]
Fix $\delta>0$ and use Theorem~\ref{ns:thm:seed} at rate $\sigma=\delta/8$. Let its output and error constants be $c,e>0$. Choose a positive rational
\[
 a<\min\{c/4,e/16,\delta/64,1/64\},\qquad d=\lfloor an\rfloor.
\]
Let $G$ be the first $d$ seed bits. Truncation preserves the one-sided guarantee and its ordinary marginal, with $\gamma\le2^{-en}$. For all sufficiently large $n$, $k-4\ge\sigma n$. Lemma~\ref{ns:lem:strongify} gives strength in the first summand with
\[
 \eta\le2^d\bigl(2^{-en}+2^{\sigma n-k}\bigr)
 \le2^{-(e-a)n}+2^{-(7\delta/8-a)n}.
\]

Put $s=\lceil\log(1/\eps)\rceil$ and $\varepsilon=2^{-s}\le\eps$. Choose $b_\delta>0$ sufficiently small that $s\le2b_\delta n$ lies in the range of Lemma~\ref{ns:lem:seedfit} for large $n$, and that, throughout this range,
\[
 2^{-en}\le2^{-2s-9-2d},\qquad
 2^{-(e-a)n}+2^{-(7\delta/8-a)n}\le2^{-2s-12}.
\]
These inequalities have fixed positive linear margins by the choice of $a$; decreasing $b_\delta$ secures them uniformly. The extra one from the ceiling of $s$ is absorbed by taking $n$ sufficiently large.

Use Lemma~\ref{ns:lem:seedfit} with error $\zeta=\varepsilon^4/2^{20}$ and payload length $p=k-d-2s-8$. With $K=2^k,D=2^d,M=2^p$ we have exactly
\[
 DM/K=2^{-2s-8}=\varepsilon^2/256.
\]
All hypotheses of Corollary~\ref{ns:cor:calibration} hold. Therefore the sum-only evaluator~\eqref{ns:eq:finalform} has error at most $\varepsilon\le\eps$. Its length is exactly
\[
 d+p=k-2s-8\ge k-2\log(1/\eps)-10.
\]
This proves the theorem, uniformly over all $k$ in the stated interval.
\end{proof}

\paragraph{Nontrivial error range and an all-error convention}
The meaningful construction above covers $\eps\ge2^{-b_\delta n}$. It is not an assertion that every smaller requested error still permits the same nonzero output bound with coefficient two. If an all-error expression with rate-dependent constants is desired, choose $C_\delta\ge\max\{12,1/b_\delta+2\}$. For $\eps<2^{-b_\delta n}$, the bound $k-C_\delta(\log(1/\eps)+1)$ is nonpositive because $k\le n$. Return the empty string, which is exactly uniform. Truncation of the main output in the nontrivial range and adjustment for finitely many short lengths give the conventional statement
\[
 m\ge\max\{0,\lfloor k-C_\delta(\log(1/\eps)+1)\rfloor\}
\]
for all errors, without claiming a nonempty output below the proved linear-exponent range.

\section{Decreasing entropy rates}\label{un:lowrate}
\par\smallskip
The evaluated pipeline is: the all-pairs table
(Construction~\ref{cb:low-table}), the head/advice/breaker short extractor
(Construction~\ref{cb:low-short}), and the retained-seed XOR output extension
(Construction~\ref{cb:low-final}). Regularization is used only to prove the
properties of this pipeline.

Fix $0<\vartheta<1/2$ and a nonnegative integer tampering count $t$ independently of $n$. This section proves separately source-strong multiple-tampering extraction at entropy rate $C/\log n$, with output $k/(t+1)-O(\log(1/\eps)+1)$. The ordinary near-$k$ statement is its sole corollary. We first prove the growing-row ingredients, then construct and analyze the short non-malleable seed, and finally give the centered-recycling calibration and complete long-output evaluator.
\subsection{Growing-row ingredients and their proofs}\label{app:lowtools}
The following lemmas and quoted quantitative theorems are the prerequisites for the inverse-logarithmic construction. Their full parameter bounds are given here before the evaluated
algorithm. Internal lemmas are proved below; imported primitives are
identified by version-specific theorem references.

\subsubsection{Cited primitives and their interfaces}\label{nm:sec:ingredients}
We use the quantitative approximate-duality, linear-condenser, and affine somewhere-random inputs from Section~\ref{un:tools}. For the local notation of this section, let $a_1$ be a sufficiently large absolute constant in the affine
somewhere-random input, and write $H(T)=(2T)^{a_1\log(2T)}$. The following correlation-breaker interface is used to generate the short seed. For final entropy-efficient output we use the strong seeded extractor of Lemma~\ref{pre:seed-efficient} and the centered recycling lemma from Section~\ref{un:optimal}.
\begin{theorem}[Quantitative affine correlation breaker]
\label{nm:thm:ACB}
\textnormal{\emph{Reference.} \cite[Lemma~3.16 and Section~3.6.2, Eq.~(3.61)]{LZdraft}; see also~\cite[Definition~3.4 and Theorem~3.5]{Li23}.}
Let $M=h+1\ge2$, $v=\lceil\log(2M)\rceil$, and let $u\ge m\ge1$
be integers. For data length $N$ and advice length $a$, put
$z=u+\lceil\log(N+2)\rceil$.
There is a uniformly explicit function
$\operatorname{ACB}:\{0,1\}^N\times\{0,1\}^d
\times\{0,1\}^a\to\{0,1\}^m$, with choices
\begin{align*}
 d&\le C\bigl(Ma+Mv^3z\bigr),\\
 K&\le C\bigl(Ma+(M^2+Mv^3)z\bigr),
\end{align*}
having the following property. Let $b\ge0$ be a seed-deficiency
allowance. Suppose $A_j,B_j,Q_j\in\F_2^N$ and $R_j\in\F_2^d$
are random variables, with $Q_j=A_j+B_j$ for $0\le j\le h$, and
\[
 (A_j)_{j=0}^h\ \perp\
       ((B_j)_{j=0}^h,(R_j)_{j=0}^h),\qquad
 \Hin(A_0)\ge K,\quad\Hin(R_0)\ge d-b .
\]
For fixed $\alpha_0,\ldots,\alpha_h\in\F_2^a$ with
$\alpha_0\ne\alpha_j$ for every $j>0$, put
$Z_j=\operatorname{ACB}(Q_j,R_j,\alpha_j)$. Then
\[
 (Z_0,(Z_j)_{j=1}^h)
 \approx_{256M^3 2^{b-u}}(U_m,(Z_j)_{j=1}^h).
\]
The comparison tuple has its actual joint marginal, and $U_m$ is
independent of that tuple. The conclusion allows arbitrary correlations within
each displayed tuple and no entropy in comparison inputs.
The parameter range here has $d,K\le N$.
\end{theorem}

This is the quantitative form of \cite[Lemma 3.16 and
Section 3.6.2, Eq.\ (3.61)]{LZdraft}. The loss $2^b$ for seed
deficiency follows there by retaining the seed in the conditioning
tuple and changing its marginal to uniform. The transformation
underlying the breaker is also stated in
\cite[Definition 3.4 and Theorem 3.5]{Li23}.
We use this theorem as a cited quantitative ingredient. In particular, its
independence requirement is on the residual $A$ part and the
seed-determined $B$ part, not on the full data and seed tuples.

For clarity, the deficiency transfer uses the cited theorem with
its conditioning tuple retained. Put
$\mathcal C=((B_j)_{j=0}^h,(R_j)_{j=0}^h)$. Replace the marginal of
$R_0$ by $U_d$, preserve the conditional distribution of the rest of
$\mathcal C$ given $R_0$, and keep the independent $A$ tuple
unchanged. Define arbitrary conditional distributions at previously unsupported
seed values. The original distribution is the new distribution reweighted by
$w(r)=2^d\Pr[R_0=r]\le2^b$. The retained tuple in the uniform-seed
theorem contains $R_0$, so reweighting both its real and ideal distributions
recovers the original retained marginal and increases their
half-$\ell_1$ distance by at most $2^b$. This proves the displayed
weak-seed error without assuming that $R_0$ is independent of the
other seeds or of the $B$ tuple.

\subsubsection{Elementary conditioning and Fourier facts}

\begin{lemma}[Flat-source decomposition]\label{nm:lem:flat}
For an integer $k$, every distribution with atoms at most $2^{-k}$
is a convex combination of uniform distributions on $2^k$-element sets.
\end{lemma}
\begin{proof}
Consider the polytope $0\le p_i\le2^{-k}$, $\sum_i p_i=1$.
A vertex cannot have two coordinates strictly between their bounds,
since a small opposite perturbation preserves the constraints.
It cannot have exactly one such coordinate, because the sum of all
other coordinates is an integer multiple of $2^{-k}$.
Every vertex is therefore uniform on exactly $2^k$ points.
\end{proof}

Decompose independent sources separately. The resulting component
pairs are still independent; statistical distance averages by convexity.
The tampering maps remain the original fixed maps on all components.

\begin{lemma}[Conditioning and truncation]\label{nm:lem:entropy}
An event of probability $p$ costs at most $\log(1/p)$ bits of
min-entropy. Deleting $b$ coordinates from a bit string costs at
most $b$ bits. Conditioning independent flat sources on a product
rectangle preserves their independence and flatness.
\end{lemma}
\begin{proof}
The first bound follows from
$\Pr[X=x\mid E]\le\Pr[X=x]/p$.
For the second, each retained string has at most $2^b$ preimages.
For the third, the conditional point probabilities factor into
the two uniform distributions.
\end{proof}

\begin{lemma}[Fourier approximation on the output hull]
\label{nm:lem:fourier}
Let $P$ be supported on an affine subspace $V\subseteq\F_2^D$.
If each ambient character nonconstant on $V$ has absolute expectation
at most $2^{-v}$, then
\[
 \TV(P,U_V)\le2^{\dim(V)/2-v-1}\le2^{D/2-v-1}.
\]
\end{lemma}\begin{proof}
Choose affine coordinates on $V$. Every nontrivial character in those
coordinates extends to an ambient character, up to a fixed sign.
For $d=\dim V$, Cauchy--Schwarz and Parseval give
\[
 \TV(P,U_V)
 \le\frac12\left(\sum_{\lambda\ne0}|\widehat P(\lambda)|^2\right)^{1/2}
 \le\frac12\sqrt{2^d-1}\,2^{-v}.
\]
The zero-dimensional case has distance zero.
\end{proof}

\subsubsection{Joint conversion through separate features}\label{nm:sec:reg}
The common regularization theorem applies to arbitrary separately computed features, including tuples of original and tampered inputs. We record its normalization here so that the subsequent parameter substitutions retain their original symbols.

Let $\phi:\F_2^n\to\F_2^{p_L}$ and $\psi:\F_2^n\to\F_2^{p_R}$ be fixed maps, put $P=p_L+p_R$, and let $F(x,y)=\mathcal B(\phi(x),\psi(y))\in\F_2^D$ for a fixed bilinear map $\mathcal B$. Constant feature coordinates may be included. The maps may be nonlinear, noninjective, and composed with arbitrary separate tamperings.
\begin{theorem}[Feature regularization]\label{nm:thm:regularize}
There is an absolute constant $K_0$ such that the following holds.
For independent flat sources $X,Y$, integers $R,u\ge1$, and
$v\ge0,z\ge1$, there is a finite deterministic product-rectangle
partition with these properties.
\begin{enumerate}[label=(\roman*)]
\item On every leaf $\tau$, the output distribution is within
$2^{D/2-v-1}$ of uniform on its output affine hull.
\item Except on leaves of total mass at most $2^{1-z}$, each input
loses at most
\begin{equation}\label{nm:eq:loss}
 \mathcal L=K_0\left[
 \frac P u+
 \frac P R\left(2^u(v+1)+(u+1)\log(n+P+2)\right)
 +D(R+2)+z\right]
\end{equation}
bits of min-entropy.
\end{enumerate}
The partition is used only in the proof and is not computed by the
extractor.
\end{theorem}

\begin{proof}
If $P=0$, the output is constant and the assertion is immediate.
Suppose first that $1\le R\le P$. Pad the two feature spaces to a
common dimension $P$ and apply Theorem~\ref{lr:thm:reg} with
$r_*=R$, $H=v$, and tail parameter $t=z$. The weighted duality
bound is Lemma~\ref{lr:lem:weighted}. Successful restrictions are
charged to the sum of the feature-hull dimensions, and exact
messages to the output-hull dimension. The two surprisal estimates
give the displayed loss after increasing the absolute constant $K_0$.
Every leaf is a product rectangle, and the Fourier error on its
exact output affine hull is at most $2^{D/2-v-1}$.

For completeness, if $R>P$, compute each of the $D$ output bits by
the low-rank protocol used in that theorem. Every bilinear part has
rank at most $P$, so at most $D(P+2)$ declared bits suffice in total.
The table is constant on each resulting rectangle and is therefore
exactly uniform on its singleton affine hull. A transcript with at
most $B_0=D(P+2)$ declared bits has at most $2^{B_0}$ leaves.
Leaves of probability below $2^{-B_0-z}$ have total mass at most
$2^{-z}$; on all others the joint surprisal, and hence each side's
entropy loss, is at most $B_0+z\le D(R+2)+z$. This also satisfies
\eqref{nm:eq:loss}. Thus all permitted cutoffs are covered, including
nonlinear and noninjective feature maps.
\end{proof}
\subsubsection{The condenser table and joint affine regularization}
\label{nm:sec:ordinaryproof}

\begin{constructionbox}{All-pairs field-inner-product condenser table}{cb:low-table}
\cbingredients The uniform linear somewhere condenser of
Theorem~\ref{lr:thm:cond} and the fixed finite-field conventions of
Section~\ref{sec:prelim}.
\cbsetup Given input length $n$ and an admissible working threshold $k_0$,
instantiate the condenser at that threshold, obtaining
$\mathcal C_1,\ldots,\mathcal C_L:\F_2^n\to\F_2^{B_0}$.
Choose integer $1\le s\le B_0$ and put
$B=s\lfloor B_0/s\rfloor$, $T=L^2$, and $F=\F_{2^s}$.
Fix a binary basis of $F$ and order pairs $(i,j)\in[L]^2$
lexicographically. In Proposition~\ref{nm:prop:regparameters}'s regime,
use $k_0=k/4$ and $s=\lfloor n^{1/2-\vartheta/2}\rfloor$.
\cbinput $(x,y)\in\bits^n\times\bits^n$.
\cbsteps
\begin{constructionsteps}
\item Evaluate every condenser row on each input and retain its first
$B$ bits. Parse each row into $B/s$ elements of $F$.
\item For \emph{every} pair $(i,j)\in[L]^2$, compute
\[
 W_{ij}(x,y)=\sum_{e=1}^{B/s}
             \mathcal C_i(x)_e\,\mathcal C_j(y)_e\in F.
\]
\item Return these $T$ field elements in the fixed pair order, each
represented by $s$ bits. Flattening the table gives $Ts$ bits.
\end{constructionsteps}
\cboutput The table $W(x,y)\in(\F_2^s)^T$; it is an intermediate
object, not itself the final extractor. Unlike the many-good-row
constant-rate constructions, this table retains all $L^2$ pairs.
\cbproofonly Joint affine regularization, affine hulls, a choice of good
row, and the tampered table tuple. They certify the table's distribution
but are not computed from $(x,y)$.
\end{constructionbox}

The table is binary bilinear in $x,y$, so its separate features have
total dimension $P=2n$, independently of $T$. For the proofs below,
write the condenser's good-row rate as $1/2+\beta$ and its error as
$\eps_c=2^{-cB_0}$, with the constants of Theorem~\ref{lr:thm:cond}.

\begin{lemma}[Field inner-product character bound]\label{nm:lem:IP}
If independent sources $U,V$ on $B$ bits have entropies $h_U,h_V$,
then every nontrivial binary character of their field inner product
has absolute bias at most
\[
 2^{(B-h_U-h_V)/2}.
\]
\end{lemma}
\begin{proof}
Each such character is a nondegenerate binary bilinear form on
$\F_2^B\times\F_2^B$. The corresponding sign matrix has orthogonal
rows of squared length $2^B$, and hence operator norm $2^{B/2}$.
The two source probability vectors have Euclidean norms at most
$2^{-h_U/2}$ and $2^{-h_V/2}$. Apply the operator-norm inequality.
\end{proof}

\begin{lemma}[Every sufficiently entropic rectangle has a full row]
\label{nm:lem:fullrow}
Assume $s=o(B_0)$ and $B_0\to\infty$.
For all sufficiently large $n$, if both sides of a flat rectangle
have entropy at least $k_0$, then some row projection of
$\aff(W(A,B))$ is all of $\F_2^s$.
\end{lemma}
\begin{proof}
Truncation costs fewer than $s$ entropy bits. Thus every designated
good row in the ideal condenser mixture has entropy at least
$(1/2+\beta/2)B$ for sufficiently large $n$.
The two ideal condenser mixtures can be sampled independently,
and their product distribution is within $2\eps_c$ of the actual
condenser product, where $\eps_c=2^{-cB_0}$.

Suppose every output row projection has a proper affine hull.
For each row pair $(i,j)$ choose a nontrivial binary character
constant, say equal to $b_{ij}$, on its actual output support.
Let $\mathcal E$ be the event that all these chosen equations hold.
Its actual probability is one.

In each product component of the ideal condenser mixtures, let
$(i,j)$ be the designated good row pair.
By Lemma~\ref{nm:lem:IP}, the probability of its selected equation
is at most $(1+2^{-\beta B/2})/2$, after weakening constants.
Thus the probability of $\mathcal E$ in every component is at most
that quantity, regardless of dependence among other rows.
Averaging and restoring the original distribution gives
\[
 1\le\frac{1+2^{-\beta B/2}}2+2\eps_c,
\]
a contradiction for large $n$.
\end{proof}

\paragraph{A reusable joint-conversion guarantee}

\begin{proposition}[Regularization parameters]\label{nm:prop:regparameters}
Fix $t$, and let $k$ be a working entropy threshold. Build the
condenser at threshold $k/4$, and let $T=L^2$ be the number of
field-product rows, each of length $s$.
For $k=C_{\vartheta,t}n/\log n$, with a sufficiently large constant,
take $s=\lfloor n^{1/2-\vartheta/2}\rfloor$ and
$T=(\log n)^{O(1)}$. Then, for every product of flat sources whose side entropies are at least
$k/2$, and every list of $t$ separate tamperings, the joint table distribution
is $2^{-s}$-close to a mixture of jointly affine tuples
$(Q^0,\ldots,Q^t)$, where $Q^j\in(\F_2^s)^T$ and some row
of $Q^0$ is exactly uniform. The good row may depend on the component.
\end{proposition}
\begin{proof}
For the $t+1$ executions, use separate features
$(x,f_1(x),\ldots,f_t(x))$ and
$(y,g_1(y),\ldots,g_t(y))$. Their total dimension is
$P=2(t+1)n$, and the output dimension is $D=(t+1)Ts$.
Apply Theorem~\ref{nm:thm:regularize} with
\[
 v=D+2s+2,\qquad z=s+2 .
\]
The total approximation and exceptional-leaf error is at most
$2^{-s}$. It suffices to make its entropy loss
$\mathcal L\le k/4$. Then every good rectangle has side entropies
at least $k/4$, and Lemma~\ref{nm:lem:fullrow} supplies the uniform row
in the honest affine projection. Replace exceptional leaves by any
valid affine tuple before averaging.

Take
$u=\lfloor(\vartheta/8)\log n\rfloor$ and
$R=\lceil n^{1/2+\vartheta/4}\rceil$. In \eqref{nm:eq:loss} the terms,
apart from constant factors depending on $t,\vartheta$, are bounded by
\begin{align*}
 P/u&=O_{\vartheta,t}(n/\log n),\\
 (P/R)2^u(v+1)&\le n^{1-5\vartheta/8+o(1)},\\
 (P/R)(u+1)\log(n+P+2)&\le n^{1/2-\vartheta/4+o(1)},\\
 D(R+2)&\le n^{1-\vartheta/4+o(1)},\\
 z&\le n^{1/2-\vartheta/2+o(1)} .
\end{align*}
All but the first are $o(n/\log n)$. Increasing $C_{\vartheta,t}$
therefore gives $\mathcal L\le k/4$.
Here $s=o(B_0)$, as required by the full-row lemma.
For the honest table alone, the same conclusion follows either
directly or by discarding the comparison coordinates.
\end{proof}

\subsubsection{Strong ordinary seeds and advice}
\label{nm:sec:advice}

We first make an ordinary seed strong with respect to either input
separately; no independence from both inputs jointly is claimed.

\begin{lemma}[Two separate strong guarantees]\label{nm:lem:bothstrong}
Suppose $E_0:\{0,1\}^n\times\{0,1\}^n\to\{0,1\}^q$
is an ordinary two-source extractor with error $\eps_0$
at entropy threshold $K$. If $X,Y$ are independent and each has
min-entropy at least $K+b$, then $S=E_0(X,Y)$ satisfies
\begin{align*}
 \TV((S,X),(U_q,X))&\le 2^q(\eps_0+2^{-b}),\\
 \TV((S,Y),(U_q,Y))&\le 2^q(\eps_0+2^{-b}).
\end{align*}
\end{lemma}
\begin{proof}
Fix an output value $s$ and put
$p_s(x)=\Pr[E_0(x,Y)=s]$.
Let $A_s=\{x:p_s(x)>2^{-q}+\eps_0\}$.
If $\Pr[X\in A_s]>2^{-b}$, the conditional source $X|A_s$
has entropy at least $K$ and is independent of $Y$.
The probability of output $s$ on this rectangle is strictly greater
than $2^{-q}+\eps_0$, contrary to extraction.
Thus $\Pr[X\in A_s]\le2^{-b}$.
Consequently
\[
 \E_X\bigl(p_s(X)-2^{-q}\bigr)_+
       \le\eps_0+2^{-b}.
\]
Sum over the $2^q$ output values to obtain the first inequality.
Interchange the roles of $X,Y$ for the second. Uniformity of either
source and any entropy condition on tampered sources are unnecessary.
\end{proof}

For an ordinary extractor with $q$ output bits and
$\eps_0\le2^{-4q}$, taking entropy slack $b=4q$ gives
\begin{equation}\label{nm:eq:strongseed-advice}
 \max\{\TV((S,X),(U_q,X)),\TV((S,Y),(U_q,Y))\}
       \le 2^{1-3q}.
\end{equation}
We will verify the required entropy slack for the seed used
in Section~\ref{nm:sec:construction}.

\paragraph{Polynomial fingerprints give separated advice}

Fix a representation of $F=\mathbb F_{2^q}$, and let
$h=\lceil n/q\rceil$. Pad an $n$-bit string $x$ with zeros to
$hq$ bits, interpret its consecutive blocks as
$x_0,\ldots,x_{h-1}\in F$, and define
\[
 P_x(T)=\sum_{j=0}^{h-1}x_j T^j\in F[T].
\]
The map $x\mapsto P_x$ is injective. Thus $P_x-P_{x'}$ is a
nonzero polynomial of degree at most $h-1$ whenever $x\ne x'$.
It has at most $h-1$ roots in $F$. For $h=1$ it has no roots.

\begin{theorem}[Advice generation from a two-sided strong seed]
\label{nm:thm:advice-fingerprint}
Let $S=S(X,Y)\in F$ be a deterministic function of two independent
sources and suppose
\[
 \TV((S,X),(U_q,X))\le\eps_s,\qquad
 \TV((S,Y),(U_q,Y))\le\eps_s.
\]
Define a single deterministic advice function by
\begin{equation}\label{nm:eq:fingerprint-advice}
 \operatorname{Adv}(x,y)=
       \bigl(S(x,y),P_x(S(x,y)),P_y(S(x,y))\bigr).
\end{equation}
It has exactly $a=3q$ bits. For any $t$ split-state tamperings
$(f_j,g_j)$ such that
$\Pr[(f_j(X),g_j(Y))=(X,Y)]=0$, its advice satisfies
\begin{equation}\label{nm:eq:fingerprint-collision}
 \Pr\bigl[\exists j:
 \operatorname{Adv}(X,Y)
       =\operatorname{Adv}(f_j(X),g_j(Y))\bigr]
 \le t\bigl(\eps_s+(h-1)2^{-q}\bigr).
\end{equation}
This applies to arbitrary tampering functions, with no entropy
requirement on their outputs. Evaluation is polynomial time if $S$
is polynomial time.
\end{theorem}
\begin{proof}
Fix a tampering $(f,g)$, and abbreviate
$S'=S(f(X),g(Y))$.
Independence of $X,Y$ gives
\[
 \Pr[f(X)=X,\ g(Y)=Y]
       =\Pr[f(X)=X]\Pr[g(Y)=Y].
\]
Since this product is zero, one of the factors is zero.
Suppose first that $\Pr[f(X)=X]=0$.
Equality of the advice strings implies $S=S'$ and then
\[
 P_X(S)=P_{f(X)}(S).
\]
This implication is used without conditioning on the event $S=S'$.
For every $x$ in the support of $X$, the set
\[
 B_x=\{s\in F:P_x(s)=P_{f(x)}(s)\}
\]
has size at most $h-1$. The strong-seed guarantee therefore gives
\begin{align*}
 \Pr[\operatorname{Adv}(X,Y)=\operatorname{Adv}(f(X),g(Y))]
 &\le\Pr[S\in B_X]\\
 &\le\Pr[U_q\in B_X]+\eps_s\\
 &\le(h-1)2^{-q}+\eps_s.
\end{align*}
The case $\Pr[g(Y)=Y]=0$ uses the third advice component and
the strong guarantee with respect to $Y$. Sum the resulting bounds
over $j\in[t]$. A field polynomial is evaluated with Horner's rule;
field arithmetic and a representation of $\mathbb F_{2^q}$ can be
implemented uniformly in polynomial time, as in the
field-product construction.
\end{proof}

This is the standard fingerprinting principle underlying sampled-code
advice. Its advantage here is a particularly transparent exact bound.
The proof does not condition on equal seeds, does not assume that
the tampered seed is uniform, and does not assume that it is
independent of either honest source.

\subsubsection{Joint affine conversion after fixing prefix-dependent advice}
\label{nm:sec:prefix}

This is the conditioning step that permits the affine correlation
breaker. It avoids both a feature expansion of the nonlinear advice
and an independence assumption on the resulting execution tuples.

\begin{lemma}[Branchwise conversion with separate side information]
\label{nm:lem:branchwise}
Let $W$ be the table in Proposition~\ref{nm:prop:regparameters}, and
write $Q=(Q^0,\ldots,Q^t)$ for its joint executions.
Let $\pi(Q)\in\{0,1\}^b$ be a fixed linear projection.
For each prescribed value $r\in\{0,1\}^b$, let
$L_r(x)$ and $R_r(y)$ be arbitrary separate functions, each
with at most $c$ output bits. They may depend on the fixed
tampering functions. Define the actual side information by
\[
 J=(L_{\pi(Q)}(X),R_{\pi(Q)}(Y)).
\]
Suppose $X,Y$ are independent flat sources of entropy at least $k$,
$c+2s\le k/2$, and $b\le s/4$. Then $(Q,J)$ is within
\begin{equation}\label{nm:eq:branch-error}
 \delta=2^{b+2-s}
\end{equation}
of a probability mixture with the following properties on each component:
$J$ and $\pi(Q)$ are fixed, $Q$ is jointly affine, and some honest
row has entropy at least $s-b$.
More specifically, that row is the projection of a uniform $s$-bit
row after at most $b$ affine constraints.
\end{lemma}
\begin{proof}
If $\delta=2^{b+2-s}\ge1$, take a uniform full table tuple, fix
$\pi(Q)=0$, and fix $J$ arbitrarily. The resulting affine distribution has an
honest row of entropy at least $s-b$, and its distance from the
actual distribution is at most $1\le\delta$. Hence assume $\delta<1$.

Fix $r$ temporarily, without conditioning on $\pi(Q)=r$.
Condition separately on the values of $L_r(X)$ and $R_r(Y)$.
These are product rectangles. On the left, labels of probability
less than $2^{-c-2s}$ have total mass at most $2^{-2s}$,
because there are at most $2^c$ labels. Outside them the conditional
entropy is at least $k-c-2s\ge k/2$. The same holds on the right.

On each good label rectangle apply
Proposition~\ref{nm:prop:regparameters} to the entire tuple $Q$.
On bad label rectangles use an arbitrary affine tuple having a
uniform honest row. Retain the fixed label values in either case.
We have constructed a probability distribution $\widehat P_r$ on $(Q,J_r)$,
where $J_r=(L_r(X),R_r(Y))$, such that
\[
 \TV\bigl(\mathcal L(Q,J_r),\widehat P_r\bigr)
       \le 2^{-s}+2^{1-2s}\le2^{1-s}.
\]
Every component of $\widehat P_r$ has fixed labels and a uniform
honest row.

Now restrict both distributions to the event $\pi(Q)=r$, without normalizing
these restricted measures, and sum over all $r$. The true measures
sum to the actual probability distribution of $(Q,J)$, since the prefix
events partition the sample space. Call the sum of the approximate
measures $\nu$ and its mass $w$.
The half-$\ell_1$ distance from the actual distribution to $\nu$ is at most
$2^{b+1-s}$. To bound its mass separately, write
$P_r=\mathcal L(Q,J_r)$ and $E_r=\{\pi(Q)=r\}$. Since $P_r$ and
$\widehat P_r$ are probability distributions before restriction,
\[
 |\widehat P_r(E_r)-P_r(E_r)|\le2^{1-s}.
\]
The true event probabilities sum to one, so
$|w-1|\le\sum_r|\widehat P_r(E_r)-P_r(E_r)|\le2^{b+1-s}$.
The mass estimate thus uses the event-probability bound for the
probability distributions before restriction; it is separate from the
half-$\ell_1$ estimate for the restricted measures.
Since $\delta<1$, we have $2^{b+1-s}=\delta/2<1/2$ and $w>0$.
Normalize $\nu$ by its mass. The additional half-$\ell_1$ cost is
$|w-1|/2$, and the total is at most
$\tfrac32 2^{b+1-s}\le2^{b+2-s}$.

Restricting a uniform affine distribution to $\pi(Q)=r$ gives a uniform
affine distribution on a nonempty fiber. At most $b$ independent affine
constraints were imposed, so the dimension of every projection
decreases by at most $b$. In particular, the formerly uniform
honest row retains entropy at least $s-b$.
The component weights are renormalized, but their affine supports
and their fixed labels are unchanged.
\end{proof}

The summation over $2^b$ choices is exclusively an analysis.
The evaluated extractor computes just its actual prefix and actual
advice. Thus no exponential running time is introduced.

\subsubsection{Affine correlation breaking and row merging}
\label{nm:sec:affine-merger}

\begin{lemma}[Decomposition relative to all seed rows]
\label{nm:lem:affine-decomp}
Suppose $(V_0,\ldots,V_h,R_0,\ldots,R_h)$ is jointly affine.
There is a representation $V_j=A_j+B_j$ such that
\[
 (A_0,\ldots,A_h)\ \perp\
       (B_0,\ldots,B_h,R_0,\ldots,R_h),
\]
and
\[
 \Hin(A_0)=H(V_0\mid R_0,\ldots,R_h)
       \ge H(V_0)-\sum_{j=0}^h |R_j|.
\]
Here $H$ denotes Shannon entropy in bits. For a uniform affine
distribution, its Shannon entropy and min-entropy both equal its affine
dimension. All nonempty fibers of the seed projection have the same
direction space, so the displayed conditional entropy is their common
affine dimension; $|R_j|$ denotes the bit length of the seed $R_j$.
\end{lemma}
\begin{proof}
Write the support as an affine translate of a vector space $V$.
The seed projection is a linear map on $V$.
Choose a complementary subspace to its kernel:
$V=V_{\rm ker}\oplus V_{\rm sec}$.
Independent uniform choices from these two subspaces generate
the uniform distribution on $V$. Project the kernel choice onto all data
coordinates to obtain $(A_j)_j$. Project the section choice,
including the affine offset, onto the data to obtain $(B_j)_j$;
the section choice also determines all seeds.
These two collections are independent. The image of the kernel
on the honest coordinate is exactly the common direction of
$V_0$ conditioned on the seed tuple. The entropy bound follows
from the dimension inequality for projection onto that tuple.
\end{proof}

\begin{lemma}[Advised affine somewhere-random merger]
\label{nm:lem:advised-merger}
Let $T\ge1$ and $t\ge0$ be integers with $(t+1)T\ge2$.
Let $Q^0,\ldots,Q^t\in(\F_2^s)^T$ be jointly affine, and suppose
some row of $Q^0$ has entropy at least $s-b$, where $b\ge0$.
Let $\alpha_0,\ldots,\alpha_t$ be fixed $a_0$-bit strings with
$\alpha_0\ne\alpha_j$ for every $j>0$.
Put
\[
 N=Ts,\quad M=(t+1)T,\quad
 v=\lceil\log(2M)\rceil,\quad a=a_0+\lceil\log T\rceil .\]
For desired output length $q\ge1$, instantiate
Theorem~\ref{nm:thm:ACB} with $M-1$ comparisons and primitive exponent
\[
 u=q+b+\lceil\log(1024M^3)\rceil .
\]
Let $d,K$ be its seed length and entropy threshold.
If $d\le s$ and
\begin{equation}\label{nm:eq:merger-entropy}
 s-b-Md\ge K,
\end{equation}
then the common function below has $t$-non-malleability error at
most $2^{-q}/4$ on this tuple. For $i\in[T]$, define
$\operatorname{bin}_T(i)$ to be the $\lceil\log_2T\rceil$-bit binary
encoding of $i-1$, padded with leading zeros (the empty string when
$T=1$). Using $\|$ for string concatenation, set
\begin{equation}\label{nm:eq:merger-map}
 \operatorname{Merge}(Q,\alpha)=
 \bigoplus_{i=1}^{T}
 \operatorname{ACB}\bigl(
 Q,\operatorname{prefix}_d(Q_i),
             \alpha\mathbin{\|}\operatorname{bin}_T(i)\bigr).
\end{equation}
Each table is flattened to $N$ bits. Every advice string in this
formula has the prescribed common length $a$, and distinct row
indices have distinct encodings.
\end{lemma}

\begin{constructionbox}{Advised affine somewhere-random merger}{cb:affine-merger}
\cbingredients The quantitative affine correlation breaker of
Theorem~\ref{nm:thm:ACB}. Lemma~\ref{nm:lem:affine-decomp} supplies
its independence decomposition only in the analysis.
\cbsetup Given $T,s,t,a_0,b$ and target output length $q$, put
\[
 N=Ts,\quad M=(t+1)T,\quad a=a_0+\lceil\log T\rceil,\quad
 u=q+b+\lceil\log(1024M^3)\rceil.
\]
Instantiate the breaker for data length $N$, advice length $a$,
$M-1$ comparisons, output length $q$, and exponent $u$.
Let $d,K$ be its seed length and entropy threshold.
Use only parameters satisfying $d\le s$ and $s-b-Md\ge K$.
\cbinput One table $Q=(Q_1,\ldots,Q_T)\in(\F_2^s)^T$ and one
$a_0$-bit advice string $\alpha$.
\cbsteps
\begin{constructionsteps}
\item Flatten the \emph{entire} table $Q$ to an $N$-bit data word.
\item For each $i\in[T]$, take $R_i=\operatorname{prefix}_d(Q_i)$
and append the $\lceil\log_2T\rceil$-bit encoding of $i-1$ to $\alpha$.
\item Compute
$Z_i=\operatorname{ACB}(Q,R_i,\alpha\mathbin{\|}\operatorname{bin}_T(i))$
for every $i\in[T]$, and return $\bigoplus_{i=1}^TZ_i$.
\end{constructionsteps}
\cboutput $q$ bits. Lemma~\ref{nm:lem:advised-merger} states the
non-malleability guarantee for jointly affine comparison tables and
fixed distinct honest/comparison advice. The algorithm evaluates
$T$ calls, although its analysis counts $M=(t+1)T$ calls across all
executions.
\cbproofonly The good row and the kernel--section decomposition.
No independence among rows is assumed by the evaluator.
\end{constructionbox}

\begin{proof}
Choose a good honest row $i_*$ in the analysis.
There are $M$ breaker calls, indexed by $(j,i)$ with
$0\le j\le t$, $1\le i\le T$.
The selected honest call is $(0,i_*)$; all other $M-1$ calls are
its comparisons, including other rows of the honest table.
Its advice differs from every comparison advice: unequal row
indices suffice, and for equal indices the given $\alpha$ strings
are distinct.

The data and seed tuple is jointly affine.
The honest $d$-bit prefix has entropy at least $d-b$, since deleting
$s-d$ coordinates loses at most that many entropy bits.
Apply Lemma~\ref{nm:lem:affine-decomp} to all $M$ calls.
The selected data coordinate $Q^0$ has entropy at least $s-b$,
so its independent residual part has entropy at least
$s-b-Md\ge K$.
The affine correlation breaker applies, without assuming that
the original data and seed tuples are independent.
Its error is at most
$256M^3 2^{b-u}\le2^{-q}/4$.

Thus the selected output is nearly uniform given every other
breaker output. XORing it with the other honest row outputs
preserves uniformity, and the comparison-table XORs are functions
of the already retained outputs. This proves the claim.
\end{proof}

If the fixed advice strings sometimes coincide in a mixture of such
components, discard those components. Their total probability
adds to the error. No affine noncollision condition on
$Q^0,Q^j$ is needed when the fixed advice strings already differ.

\subsection{Short-output non-malleability: construction and proof}\label{un:nmconstruction}
\label{nm:sec:construction}

The construction in this subsection is available for every fixed positive
comparison count $t$. In the final $t$-tampering theorem it will be
instantiated with comparison count $2t+1$ and a smaller source threshold.
The proofs here are independent of the later long-output theorem.

\begin{constructionbox}{Short non-malleable extractor at decreasing entropy rate}{cb:low-short}
\cbingredients The table in Construction~\ref{cb:low-table} and
Proposition~\ref{nm:prop:regparameters}; the affine somewhere-random
extractor of Theorem~\ref{lr:thm:sr}; the polynomial fingerprints of
Theorem~\ref{nm:thm:advice-fingerprint}; and the advised merger of
Lemma~\ref{nm:lem:advised-merger} (Construction~\ref{cb:affine-merger}).
\cbsetup Fix $t\ge1$, $0<\vartheta<1/2$, and a working entropy threshold
$k$ in the regime of Proposition~\ref{nm:prop:regparameters}.
Build the table with condenser threshold $k/4$ and
$s=\lfloor n^{1/2-\vartheta/2}\rfloor$. Let $T=L^2$ be its row count.
Put $N=Ts$, $M=(t+1)T$, and $v=\lceil\log(2M)\rceil$.
With the fixed sufficiently large absolute constant $C_*$ used below, set
\begin{equation}\label{nm:eq:mainlengths}
 \ell=\left\lfloor\frac{s}{C_*M^3v^3}\right\rfloor,\qquad
 q=\left\lfloor\frac{\ell}{8H(T)}\right\rfloor,\qquad b=M\ell,
\end{equation}
where $H(T)=(2T)^{a_1\log(2T)}$ comes from Theorem~\ref{lr:thm:sr}.
Instantiate that extractor on $T$ rows of $\ell$ bits and truncate its
output to $q$ bits. Instantiate Construction~\ref{cb:affine-merger}
with $a_0=3q$, deficiency allowance $b$, and output length $q$.
In particular its breaker uses $M-1$ comparisons and
$u=q+b+\lceil\log(1024M^3)\rceil$; denote its seed length by $d$.
The estimates below ensure $q\ge4\log(n+2)$,
$\ell/H(T)\ge(2T)^4$, $(t+1)q+2s\le k/2$, $b\le s/4$,
and $s-b-Md\ge K$.
\cbinput $(x,y)\in\bits^n\times\bits^n$.
\cbsteps
\begin{constructionsteps}
\item \textbf{Compute the table.} Form $W(x,y)$ by
Construction~\ref{cb:low-table}, retaining its fixed row order.
\item \textbf{Compute a short ordinary seed from the heads.}
For each row take its first $\ell$ bits to form $\operatorname{Head}(W)$.
Apply the chosen affine somewhere-random extractor and take its
$q$-bit prefix, denoted $S=S(W(x,y))$.
\item \textbf{Compute advice from the full original inputs.}
Over $\F_{2^q}$, pad $x,y$ to $\lceil n/q\rceil$ coefficients and
form their polynomials $P_x,P_y$ as in Theorem~\ref{nm:thm:advice-fingerprint}.
Interpret $S$ as a field element and set
$\alpha=(S,P_x(S),P_y(S))\in\bits^{3q}$.
\item \textbf{Run all breaker calls and XOR.} For every row $i$, use
as data the full flattened $Ts$-bit table, as seed its row prefix
$\operatorname{prefix}_d(W_i)$, and as advice
$\alpha\mathbin{\|}\operatorname{bin}_T(i)$.
Return
\begin{equation}\label{nm:eq:short-nm}
 E_{\rm short}(x,y)=\operatorname{Merge}(W(x,y),\alpha(x,y)).
\end{equation}
\end{constructionsteps}
\cboutput $q=n^{1/2-\vartheta/2-o(1)}$ bits in the inverse-logarithmic
regime, with error at most $2^{-q/2}$ by Theorem~\ref{nm:thm:short}.
The head seed $S$ and the per-row breaker seeds are \emph{different}
objects. Only the XOR of breaker outputs is returned here.
\cbproofonly Enumeration of hypothetical heads, advice conditioning,
regularization rectangles, and affine components. The evaluator computes
only its actual table, actual head seed, and actual advice.
\end{constructionbox}

\paragraph{Ordinary seed extraction and advice separation}

\begin{lemma}[The seed and advice guarantees]\label{nm:lem:seed-advice}
At source entropy $k/2$, $S(x,y)$ is an ordinary extractor with
error at most $2^{-4q}$.
At source entropy $k$, it is strong with respect to either original
source separately, with error at most $2^{1-3q}$.
For every admissible list of tamperings the advice collision
probability is at most
\begin{equation}\label{nm:eq:zeta-final}
 \zeta=t\left(2^{1-3q}
          +(\lceil n/q\rceil-1)2^{-q}\right).
\end{equation}
\end{lemma}
\begin{proof}
By Proposition~\ref{nm:prop:regparameters}, the honest table on
entropy-$k/2$ sources is $2^{-s}$-close to a mixture of affine
tables with a uniform row. Their heads have a uniform $\ell$-bit
row. The affine somewhere-random extractor therefore gives error
at most $2^{-\ell/H(T)}+2^{-s}\le2^{-4q}$.
This first argument extends from flat sources by convexity.
Since $k/2\ge4q$ for large $n$, Lemma~\ref{nm:lem:bothstrong}
with slack $4q$ gives the two strong guarantees.
Theorem~\ref{nm:thm:advice-fingerprint} now gives
\eqref{nm:eq:zeta-final}.
\end{proof}

\paragraph{Advice is fixed before the final joint conversion}

For fixed tamperings put $f_0=g_0=\mathrm{id}$ and
$Q^j=W(f_j(X),g_j(Y))$.
The concatenation of all $M$ row heads is a $b$-bit linear
projection of the full table tuple; denote it by $r=\pi(Q)$.
For a prescribed value $r$, all $q$-bit seeds
$s_j=S(Q^j)$ are determined. Define separate functions
\[
 L_r(x)=(P_{f_j(x)}(s_j))_{j=0}^{t},\qquad
 R_r(y)=(P_{g_j(y)}(s_j))_{j=0}^{t}.
\]
Each has $c=(t+1)q$ output bits.
The complete advice tuple is determined by
$r,L_r(x),R_r(y)$.

Apply Lemma~\ref{nm:lem:branchwise}. It gives an approximation, of
distance at most $\delta=2^{b+2-s}$, to the joint distribution of all tables
and all advice, in which every component has fixed advice,
jointly affine tables, and an honest row of entropy at least $s-b$.
Notice that the fingerprints did not have to be affine functions
of the tables: after a prescribed head is fixed, they are separate
functions of the original inputs, so their fixings precede the
rectangle regularization.

\paragraph{Checking the breaker's entropy inequalities}

Let $a=3q+\lceil\log T\rceil$ and
$u=q+b+\lceil\log(1024M^3)\rceil$.
The parameter conditions above imply
\[
 a=O(\ell),\qquad
 u+\log(N+2)=O(M\ell).
\]
Theorem~\ref{nm:thm:ACB} therefore supplies
\begin{align*}
 d&=O(M^2v^3\ell),\\
 K&=O\bigl((M^3+M^2v^3)\ell\bigr).
\end{align*}
Consequently
\[
 b+Md+K\le C_1M^3v^3\ell\le s
\]
for an absolute $C_1$, by choosing $C_*\ge2C_1$.
This also gives $d\le s$ and verifies
\eqref{nm:eq:merger-entropy}. Increasing $C_*$ further, if necessary,
absorbs all integer rounding.

\begin{theorem}[Short-output non-malleability]\label{nm:thm:short}
The function \eqref{nm:eq:short-nm} has $q$ output bits and
$t$-non-malleability error at most
\begin{equation}\label{nm:eq:short-error}
 \zeta+2^{-q}/4+3\cdot2^{b+2-s}.
\end{equation}
For sufficiently large $n$ this is at most $2^{-q/2}$.
\end{theorem}
\begin{proof}
First take flat sources and use the joint approximation just
constructed. Advice-collision probability in the ideal mixture
is at most $\zeta+\delta$, because this event is determined by
the retained advice tuple.
On every other component apply
Lemma~\ref{nm:lem:advised-merger}. The componentwise error is
at most $2^{-q}/4$.
Returning the honest/tampered output tuple to its actual distribution
costs $\delta$, and returning the ideal tail marginal to its
actual distribution costs another $\delta$. This proves
\eqref{nm:eq:short-error}.
Decompose arbitrary independent sources into flat sources
separately and average; all maps remain the original fixed maps.

Since $q\ge4\log(n+2)$, we have
$\zeta\le2t\,2^{-3q/4}$.
Also $b\le s/4$ and $q\le s/8$ imply
$2^{b+2-s}\le2^{2-6q}$.
For fixed $t$ and sufficiently large $q$, their sum in
\eqref{nm:eq:short-error} is at most $2^{-q/2}$.
\end{proof}

\paragraph{The inverse-logarithmic parameter calculation}

At rate $C_{\vartheta,t}/\log n$, $T=(\log n)^{O(1)}$ and
\[
 M^3v^3H(T)
    =2^{O_{\vartheta,t}((\log\log n)^2)}=n^{o(1)}.
\]
With $s=\lfloor n^{1/2-\vartheta/2}\rfloor$ this yields
\[
 q=n^{1/2-\vartheta/2-o(1)}
       \ge n^{1/2-\vartheta}
\]
for all sufficiently large $n$. In particular $q\gg\log n$,
$\ell/H(T)\ge(2T)^4$, and $(t+1)q+2s=o(k)$.
This verifies every size condition used above.

The evaluated algorithm computes one table, one ordinary seed,
two fingerprints, and $T$ correlation-breaker calls. All numbers
of rows, field sizes in bits, and subroutine input lengths are
polynomially bounded. The quantitative breaker is uniformly
polynomial time in its parameters. Neither regularization
partitions nor the $2^b$ prescribed heads are enumerated by
the algorithm.

\subsection{Source-strong output with additive entropy loss}
\label{nm:sec:lowrate-centered}
The short-output construction above is used only to generate a seed.
We now apply the centered recycling lemma from Section~\ref{un:optimal}.
The resulting evaluator retains its seed and XORs two source-local
payloads. In particular, the final source-strength claim retains all
actual long comparison outputs, not merely prefixes of them.

\subsubsection{Prerequisite: a finite-budget recycling statement}

\begin{lemma}[Centered recycling with explicit budgets]
\label{nm:lem:centered-budget}
Let $t\ge0$ be an integer, put $q=t+1$ and
$\rho_t=\lceil\log(2t+1)\rceil$, and let $A,B\ge1$ be absolute
constants in Lemma~\ref{pre:seed-efficient}, with $A$ an integer.
For integers $n,k\ge1$ and $u\ge2$ with $k\le n$, set
\begin{align}
 d&=\left\lceil B\seedcost{n}{u}\right\rceil+A(u+1),
       \label{nm:eq:budget-d}\\
 m&=\left\lfloor\frac{k-4u-\rho_t}{q}\right\rfloor,
 \qquad p=m-d,
 \qquad\kappa=k-tm-2u.\label{nm:eq:budget-lengths}
\end{align}
Suppose $1\le h\le k$, $p\ge1$, and a $d$-bit seed $G$ is
separately source-strong against up to $2t+1$ separate tamperings
at component entropy threshold $h$, with error $\eta$ satisfying
\begin{align}
 \eta&\le2^{-2qd-4u-4},\label{nm:eq:budget-seederror}\\
 k-tm-(t+2)d-h&\ge4u+2.\label{nm:eq:budget-smallgraph}
\end{align}
There is an explicit strong seeded extractor
$E:\bits^n\times\bits^d\to\bits^p$, and the map
\begin{equation}\label{nm:eq:centered-low-evaluator}
 F(x,y)=\bigl(G(x,y),E(x,G(x,y))\oplus E(y,G(x,y))\bigr)
\end{equation}
is separately source-strong $t$-non-malleable on all independent
min-entropy-$k$ inputs, with error less than $10\cdot2^{-u}$.
The same evaluator satisfies both source-strength comparisons and
retains the complete actual $t$-tuple of $m$-bit tampered outputs.
\end{lemma}
\begin{proof}
First check the seeded-extractor interface. The integer $\kappa$ obeys
$0<\kappa\le k\le n$, and
\begin{align*}
 \kappa-p
 &=k-qm+d-2u\ge d+2u+\rho_t\ge A(u+1).
\end{align*}
Thus Lemma~\ref{pre:seed-efficient} at threshold $\kappa$ and error
$\zeta=2^{-u}$ can supply $p$ output bits using at most
$\lceil B\seedcost{n}{u}\rceil\le d$ seed bits. Ignore unused seed
coordinates and truncate the output as necessary. The added
$A(u+1)$ seed coordinates in~\eqref{nm:eq:budget-d} are harmless
padding; they ensure the payload fits without putting the seeded
extractor's entropy-loss constant into the final output deficit.

Apply Lemma~\ref{rec:lem:centered} with its tampering count $C=t$,
$\alpha=\beta=\zeta=2^{-u}$, and its threshold equal to $h$.
Its three energy terms are, respectively,
\begin{align*}
 \Delta_1&=(2t+1)2^{qm-k}
       \le2^{-4u},\\
 \Delta_2&=2^{1+2u+(t+2)d+tm+h-k}
       \le2^{-2u-1},\\
 \Delta_3&=4\eta\,2^{2qd+2u}
       \le2^{-2u-2}.
\end{align*}
For the first inequality use $2t+1\le2^{\rho_t}$ and
$qm\le k-4u-\rho_t$; the other two are exactly
\eqref{nm:eq:budget-smallgraph} and
\eqref{nm:eq:budget-seederror}. In particular,
$\Delta_1+\Delta_2+\Delta_3\le3\cdot2^{-2u}$. The centered
lemma's distance is at most
\[
 8\cdot2^{-u}+\frac\eta2+\frac{\sqrt3}{2}2^{-u}
 <10\cdot2^{-u}.
\]

For completeness, when $Y=y$ is retained, subtract $E(y,s)$ from
the honest payload and $E(g_i(y),s_i)$ from comparison $i$.
These are bijections determined by $y$ and the displayed seed values.
They turn the honest output into $(S_0,E(X,S_0))$ and the
comparison tuple into
\[
 T=\bigl(\bar S,(E(f_i(X),S_i))_{i=1}^t\bigr),
\]
which is the actual source-local leakage allowed in
Lemma~\ref{rec:lem:centered}. Undoing the bijections preserves the
honest uniform distribution and restores the original comparison marginal.
Exchanging the sources and applying the same argument to
$G^{\rm op}(x,y)=G(y,x)$ proves the comparison retaining $X$;
the evaluated map~\eqref{nm:eq:centered-low-evaluator} does not change.
The proof for flat sources extends by independent flat-source
mixtures, preserving each component's actual comparison/source
marginal. When $t=0$ the comparison tuple is empty, but the seed
still has one-comparison non-malleability.
\end{proof}

\subsubsection{Prerequisite: calibrating the lower-entropy short seed}

\begin{lemma}[Inverse-logarithmic source-strong seed calibration]
\label{nm:lem:lowrate-seed-calibration}
Fix $t\ge0$ and $0<\vartheta<1/2$. There are constants
$c_0,C_{t,\vartheta}>0$ and an integer $n_0$ such that the following
holds for all $n\ge n_0$, $C_{t,\vartheta}n/\log n\le k\le n$,
and
\[
 2^{-n^{1/2-\vartheta}}\le\eps\le1/4.
\]
Put $q=t+1$, $R=2t+1$,
\begin{equation}\label{nm:eq:lowrate-working-thresholds}
 h_0=\left\lceil c_0n/\log n\right\rceil,
 \qquad h=2h_0,
 \qquad u=\lceil\log(1/\eps)\rceil+12,
\end{equation}
and choose $d,m,p,\kappa$ by
\eqref{nm:eq:budget-d}--\eqref{nm:eq:budget-lengths}.
Instantiate the short-output construction at threshold $h_0$
and comparison count $R$. If its output length is $q_0$, then
\[
 G(x,y)=\operatorname{prefix}_d E_{\rm short}(x,y)
\]
is well-defined and satisfies every seed and length hypothesis of
Lemma~\ref{nm:lem:centered-budget}, including both source-strength
comparisons at threshold $h$.
\end{lemma}
\begin{proof}
Use Theorem~\ref{nm:thm:short} and its inverse-logarithmic parameter
calculation with comparison count $R=2t+1$ and exponent parameter
$\vartheta$. Choose $c_0$ large enough for that construction at
threshold $h_0$. Its length and error satisfy
\begin{equation}\label{nm:eq:lowrate-base-seed}
 q_0=n^{1/2-\vartheta/2-o(1)},
 \qquad \gamma_0\le2^{-q_0/2}.
\end{equation}
All constants may depend on the fixed $t,\vartheta$ but not on
$k$ or $\eps$. This invocation is of the \emph{short-output}
theorem, not of the long-output theorem proved below.

Uniformly over the stated error range,
\begin{align}
 u&\le n^{1/2-\vartheta}+13,\notag\\
 d&=O\bigl(n^{1/2-\vartheta}\log n+\log^2(n+2)\bigr)
       =o(q_0).\label{nm:eq:lowrate-seed-fit}
\end{align}
Also $q_0=o(h_0)$, so $d=o(h_0)$ and in particular $d\le q_0$.
The strict exponent gap between
$1/2-\vartheta$ and $1/2-\vartheta/2$ absorbs the factor
$\log n$ in $d$ and all fixed comparison-count factors.
For all sufficiently large $n$ we therefore have
\begin{equation}\label{nm:eq:lowrate-two-seed-budgets}
 \min\{q_0/2,h_0\}\ge4qd+4u+5.
\end{equation}
Apply Lemma~\ref{pre:lem:strongify-general} to the first $d$ bits
of the $R$-tampering seed, raising its threshold from $h_0$ to
$h=2h_0$. Since $R+1=2q$, both strong errors are at most
\begin{align}
 \eta
 &\le2^{2qd}\bigl(2^{-q_0/2}+2^{-h_0}\bigr)\notag\\
 &\le2^{1+2qd-\min\{q_0/2,h_0\}}
 \le2^{-2qd-4u-4}.\label{nm:eq:lowrate-strong-seed-error}
\end{align}
The estimate is uniform over all admissible $R$ comparisons and
all independent sources with min-entropy at least $h$.
Thus it also applies to the auxiliary graph marginals and composed
comparisons in the centered recycling lemma; no entropy is required
in their tampered images.

Choose $C_{t,\vartheta}\ge16q c_0$ and enlarge $n_0$ to absorb the
ceilings. Then $h\le k/(4q)$. Moreover, $d+u=o(k)$ uniformly in
$k$ and $\eps$, and $m=k/q-O_t(u+1)$. Hence $p=m-d\ge1$.
Using $m\le k/q$ gives
\begin{align*}
 k-tm-(t+2)d-h
 &\ge\frac{k}{q}-(t+2)d-h\\
 &\ge\frac{3k}{4q}-(t+2)d
 \ge4u+2
\end{align*}
for all sufficiently large $n$. These are the remaining hypotheses
of Lemma~\ref{nm:lem:centered-budget}. All lower bounds on $n$ can
be chosen before the particular $k,\eps$, because the estimates
above use their worst allowed values.
\end{proof}

\subsubsection{Main theorem and complete retained-seed construction}

\begin{theorem}[Source-strong entropy-efficient output at rate $C/\log n$]
\label{nm:thm:mainlow}
Fix an integer $t\ge0$ and $0<\vartheta<1/2$. There are constants
$C_{t,\vartheta}>0$ and $n_0$ such that, for every $n\ge n_0$,
integer $C_{t,\vartheta}n/\log n\le k\le n$, and
\[
 2^{-n^{1/2-\vartheta}}\le\eps\le1/4,
\]
there is an explicit polynomial-time $t$-non-malleable two-source
extractor $F:\bits^n\times\bits^n\to\bits^m$ with
\[
 m\ge\frac{k}{t+1}-O\bigl(\log(1/\eps)+1\bigr).
\]
The implied additive-loss constant is absolute. For every independent
min-entropy-$k$ pair and every admissible $t$-tuple of separate
tamperings, put $\mathbf Z=(F(f_i(X),g_i(Y)))_{i=1}^t$. The same
function satisfies
\begin{align}
 (F(X,Y),\mathbf Z,Y)&\approx_\eps(U_m,\mathbf Z,Y),
       \label{nm:eq:lowrate-strongY}\\
 (F(X,Y),\mathbf Z,X)&\approx_\eps(U_m,\mathbf Z,X).
       \label{nm:eq:lowrate-strongX}
\end{align}
The uniform word is independent of the entire displayed marginal,
and $\mathbf Z$ consists of the actual $m$-bit comparison outputs.
For $t=0$ this is ordinary extraction strong in either original
source separately. No claim is made for a growing tampering count
or for the exact square-root error exponent with fixed constants.
\end{theorem}

\begin{constructionbox}{Source-strong low-rate output with additive entropy loss}{cb:low-final}
\cbingredients The short extractor of Theorem~\ref{nm:thm:short}
(Construction~\ref{cb:low-short}); the seed calibration of
Lemma~\ref{nm:lem:lowrate-seed-calibration}; the strong seeded
extractor of Lemma~\ref{pre:seed-efficient}; and the explicit-budget
recycling interface of Lemma~\ref{nm:lem:centered-budget}.
\cbsetup Fix $t\ge0$, $0<\vartheta<1/2$, integer
$C_{t,\vartheta}n/\log n\le k\le n$, and
$2^{-n^{1/2-\vartheta}}\le\eps\le1/4$.
Fix the absolute constants $A,B$ and calibration constant $c_0$.
First set
\[
 \begin{aligned}
 q&=t+1, & R&=2t+1, & \rho_t&=\lceil\log(2t+1)\rceil,\\
 h_0&=\lceil c_0n/\log n\rceil, & h&=2h_0,
 &u&=\lceil\log(1/\eps)\rceil+12.
 \end{aligned}
\]
Choose
\begin{align}
 d&=\left\lceil B\seedcost{n}{u}\right\rceil+A(u+1),\notag\\
 m&=\left\lfloor\frac{k-4u-\rho_t}{t+1}\right\rfloor,
 \qquad p=m-d,\qquad\kappa=k-tm-2u.
 \label{nm:eq:lowrate-final-lengths}
\end{align}
Instantiate Construction~\ref{cb:low-short} at threshold $h_0$ and
comparison count $R$, not $t$; denote its output length by $q_0$ to
distinguish it from $q=t+1$. Define $G=\operatorname{prefix}_d E_{\rm short}$.
Lemma~\ref{nm:lem:lowrate-seed-calibration} gives $d\le q_0$ and the
required source-strength guarantee at threshold $h$.
Instantiate $E$ at entropy $\kappa$, error $2^{-u}$, $d$ seed bits
(after ignored padding), and $p$ output bits.
\cbinput $(x,y)\in\bits^n\times\bits^n$.
\cbsteps
\begin{constructionsteps}
\item Evaluate that lower-threshold, $R$-comparison short extractor
once on $(x,y)$ and retain its first $d$ bits as $S$.
\item Compute the $p$-bit words $E(x,S)$ and $E(y,S)$.
\item Return
\begin{equation}\label{nm:eq:lowrate-final-evaluator}
 S=G(x,y),\qquad F(x,y)=\bigl(S,E(x,S)\oplus E(y,S)\bigr).
\end{equation}
\end{constructionsteps}
\cboutput Exactly $m=d+p$ bits with the two source-strength guarantees
of Theorem~\ref{nm:thm:mainlow}. The output includes the generated seed
and retains no intermediate table or advice. The error is at most
$10\cdot2^{-u}<\eps$ under the stated parameter range.
\cbproofonly Strongification, regularization, posteriors, collision
graphs, and the auxiliary comparison executions. These certify the
three evaluated calls and are not additional computations.
\end{constructionbox}

\begin{proof}
Lemma~\ref{nm:lem:lowrate-seed-calibration} verifies all the seed,
length, and entropy hypotheses of Lemma~\ref{nm:lem:centered-budget}.
The latter proves both comparisons with error
$10\cdot2^{-u}\le(10/4096)\eps<\eps$.
Since $\rho_t\le t+1$, the explicit length satisfies
\[
 m\ge\frac{k}{t+1}-\frac{4(\lceil\log(1/\eps)\rceil+12)}{t+1}-2
   =\frac{k}{t+1}-O(\log(1/\eps)+1),
\]
with an absolute constant. All selected lengths are positive for the
uniform sufficiently large $n$ in the calibration lemma. The short
construction and the two seeded calls are deterministic polynomial-time
for fixed $t,\vartheta$. No rank condition is needed for this final
nonlinear composition.
\end{proof}

At $\eps=2^{-n^{1/2-\vartheta}}$ the theorem gives
$m\ge k/(t+1)-O(n^{1/2-\vartheta})=(1-o(1))k/(t+1)$,
with both full-output source-strength guarantees. This replaces the
older asymmetric boost and its shorter-output source-strong companion.
The retained seed cancels its own length charge: only $p=m-d$ bits
are extracted in each payload, but all $d$ seed bits are returned.

\subsubsection{The ordinary source-strong corollary}

\begin{corollary}[Near-$k$ source-strong ordinary output at inverse-logarithmic rate]
\label{lr:thm:strongnear-low}
For each fixed $0<\vartheta<1/2$, there is $C_\vartheta>0$ such
that, for every sufficiently large $n$, every integer
$C_\vartheta n/\log n\le k\le n$, and every
\[
 2^{-n^{1/2-\vartheta}}\le\eps\le1/4,
\]
there is an explicit ordinary two-source extractor, strong in either
original source separately, with
\[
 m\ge k-4\lceil\log(1/\eps)\rceil-56,
 \qquad\text{error at most }\eps.
\]
The same evaluator satisfies both source-strength comparisons.
\end{corollary}

\begin{constructionbox}{Ordinary source-strong low-rate specialization}{cb:low-zero}
\cbingredients Construction~\ref{cb:low-final} and
Theorem~\ref{nm:thm:mainlow}, with $t=0$.
\cbsetup Set $q=1$, $R=1$, $\rho_0=0$, and
$u=\lceil\log(1/\eps)\rceil+12$. Keep the seed length $d$ and
lower seed threshold $h_0$ from Construction~\ref{cb:low-final}, and set
$m=k-4u$, $p=m-d$, $\kappa=k-2u$.
\cbinput $(x,y)\in\bits^n\times\bits^n$ in
Corollary~\ref{lr:thm:strongnear-low}'s parameter range.
\cbsteps
\begin{constructionsteps}
\item Compute $S$ from the one-comparison short seed at threshold $h_0$.
\item Return $(S,E(x,S)\oplus E(y,S))$ with these specialized lengths.
\end{constructionsteps}
\cboutput Exactly $k-4\lceil\log(1/\eps)\rceil-48$ bits before
any optional fixed truncation. This implies the corollary's stated
lower bound $k-4\lceil\log(1/\eps)\rceil-56$. The same evaluator
is strong in either source separately; the internal seed still handles
one comparison.
\end{constructionbox}

\begin{proof}
Set $t=0$ in Theorem~\ref{nm:thm:mainlow}. Its seed uses one
comparison, its leakage tuple is empty, and
$\rho_0=0$. The explicit output length in
\eqref{nm:eq:lowrate-final-lengths} is
$k-4\lceil\log(1/\eps)\rceil-48$, which is at least the stated
bound; an additional fixed truncation is harmless. Both strong
comparisons follow directly. There is no separate output-amplification
or collision argument in this corollary.
\end{proof}


\bibliographystyle{plain}
\bibliography{refs}
\end{document}